\documentclass[12pt]{article}

\pdfoutput=1

\usepackage{amsmath,amsfonts,amsthm,amssymb}
\usepackage{slashed}
\usepackage{xcolor}
\usepackage{graphicx}
\usepackage{epstopdf}
\usepackage{array,booktabs}
\usepackage{hyperref}
\usepackage{tabulary}
\usepackage{multirow}
\usepackage{array,arydshln}
\usepackage{bbm} 
\usepackage{cite}
\usepackage{mathtools} 
\usepackage{caption}
\usepackage{tikz}
\usetikzlibrary{automata}
\usetikzlibrary{arrows}
\usetikzlibrary{calc}
\usetikzlibrary{decorations.markings}
\usetikzlibrary{decorations.pathreplacing}
\usetikzlibrary{intersections}
\usetikzlibrary{positioning}
\usetikzlibrary{topaths}
\usetikzlibrary{shapes.geometric}
\usetikzlibrary{shapes.misc}
\tikzset{->-/.style = {
    decoration = {markings, mark = at position #1 with {\arrow{>}}},
    postaction = {decorate}}}
\tikzset{color-group/.style = {
    shape = circle,
    minimum size = 2.5ex,
    inner sep = .5ex,
    draw}}
\tikzset{flavor-group/.style = {
    shape = rectangle,
    minimum size = 2.5ex,
    inner sep = .5ex,
    draw}}
\tikzset{cf-group/.style = {
    shape = rounded rectangle,
    rounded rectangle right arc = none,
    draw}}
\tikzset{fc-group/.style = {
    shape = rounded rectangle,
    rounded rectangle left arc = none,
    draw}}
\tikzset{cross/.style={minimum width=1pt, path picture={
      \draw[black, very thick]
               (path picture bounding box.south east)
            -- (path picture bounding box.north west)
               (path picture bounding box.south west)
            -- (path picture bounding box.north east);
          }}}
\newcommand{\mathtikz}[2][]
  {\ensuremath{\vcenter{\hbox{%
          \begin{tikzpicture}[#1]#2\end{tikzpicture}}}}}

\hypersetup{linktoc = all}  
\hypersetup{pdfborderstyle={/S/U/W 0.5}}

\numberwithin{equation}{section}
\newcommand{\beq}{\begin{equation}}
\newcommand{\eeq}{\end{equation}}
\newcommand{\bea}{\begin{eqnarray}}
\newcommand{\eea}{\end{eqnarray}}

\newcommand{\address}[1]{\vbox{\center\em#1}}
\renewcommand{\title}[1]{\vbox{\center\LARGE{#1}}\vspace{5mm}}
\newcommand{\ch}{\textrm{ch}}
\newcommand{\sh}{\textrm{sh}}
\newcommand{\thh}{\textrm{th}}

\newcommand{\vev}[1]{{\left< {#1} \right>}}
\newcommand{\tr}{\textrm{Tr}}
\newcommand{\ti}{\widetilde}
\newcommand{\wat}{\widehat}
\newcommand{\lp}{\left(}
\newcommand{\rp}{\right)}

\newcommand{\N}{\mathcal{N}}

 \def\no{\nonumber}

\newcommand{\rf}[1]{\eqref{#1}}
\declareslashed{}{/}{-.08}{0}{V} 

\makeatletter
\newcommand*{\letterdef@}{}
\newcommand*{\letterdef}[3]{%
  \def\letterdef@##1{\expandafter\newcommand\csname #1\endcsname{#2{##1}}}%
  \@tfor\@tempa :=#3\do{\expandafter\letterdef@\expandafter{\@tempa}}}
\makeatother
\letterdef{bb#1}{\mathbb} {CHPRZ}    
\letterdef{c#1}{\mathcal}{ABCDEFGHIJKLMNOPQRSTUVWXYZ} 
\letterdef{rm#1}{\mathrm} {dDF} 

\newcommand{\bC}{\ensuremath{\mathbb{C}}}

\newcommand{\bN}{\ensuremath{\mathbb{N}}}

\newcommand{\bR}{\ensuremath{\mathbb{R}}}

\newcommand{\bZ}{\ensuremath{\mathbb{Z}}}

\newcommand{\scA}{\ensuremath{\mathcal{A}}}

\newcommand{\scF}{\ensuremath{\mathcal{F}}}

\newcommand{\scH}{\ensuremath{\mathcal{H}}}
\newcommand{\scI}{\ensuremath{\mathcal{I}}}

\newcommand{\scN}{\ensuremath{\mathcal{N}}}

\newcommand{\scR}{\ensuremath{\mathcal{R}}}
\newcommand{\scS}{\ensuremath{\mathcal{S}}}

\newcommand{\scW}{\ensuremath{\mathcal{W}}}

\newcommand{\scZ}{\ensuremath{\mathcal{Z}}}

\begin{document}
\bibliographystyle{JHEP}

\begin{titlepage}

\hfill{}
\\
\vspace{10mm}

\center{{\huge Mirror Symmetry And Loop Operators}}
\vspace{10mm}

\begin{center}
\renewcommand{\thefootnote}{$\alph{footnote}$}
Benjamin~Assel,\footnote{\href{mailto:benjamin.assel@gmail.com}{\tt benjamin.assel@gmail.com}}
Jaume Gomis\,\footnote{\href{mailto:jgomis@perimeterinstitute.ca}{\tt jgomis@perimeterinstitute.ca}}

\vspace{20mm}

\address{${}^{a}$Department of Mathematics, King's College London,\\
The Strand, London WC2R 2LS, United Kingdom}

\vspace{5 mm}
\address{${}^{b}$Perimeter Institute for Theoretical Physics,\\
Waterloo, Ontario, N2L 2Y5, Canada}

\renewcommand{\thefootnote}{\arabic{footnote}}
\setcounter{footnote}{0}

\end{center}

\vspace{10mm}

\abstract{
\medskip\medskip
\normalsize{

\noindent
Wilson loops in gauge theories pose a fundamental challenge for dualities. Wilson loops are labeled by a representation of the gauge group and should map under duality to loop operators labeled by the same data, yet generically, dual   theories  have completely different gauge groups. In this paper we resolve this conundrum for three dimensional mirror symmetry. We show that Wilson loops are exchanged under mirror symmetry with Vortex loop operators, whose microscopic definition in terms of  a supersymmetric quantum mechanics coupled to the theory  encode  in a non-trivial way a  representation of the original gauge group, despite that the gauge groups of mirror theories can be radically different. 
 Our predictions for the mirror map, which we derive  guided by   branes in string theory, are confirmed by the  computation of the exact expectation value of Wilson and Vortex loop operators  on the three-sphere.

}
}

\vspace{10mm}

\noindent
\today
\vfill

\end{titlepage}

\tableofcontents


\newpage

\section{Introduction}
\label{sec:intro}

Wilson loop operators \cite{Wilson:1974sk} play a central role in our understanding of gauge theories.  A Wilson loop   is specified by a curve $\gamma$ in spacetime and by the choice of a representation $\cR$ of the gauge group $G$
\beq
W_\cR(\gamma)=\hbox{Tr}_R P\exp{\oint_\gamma  i A_\mu dx^\mu\,. 
\nonumber}
\eeq
In a certain sense, they are the most fundamental observables in gauge theories. 
\smallskip

Wilson loops raise an immediate challenge to any conjectured duality, be it a    field theory duality  or a   gauge/gravity duality:  what is the dual description of Wilson loops? Even the more qualitative question of how the choice of a representation $\cR$ of the gauge group $G$ labeling a Wilson is encoded in the dual theory  is challenging, and in general the answer is unknown. Indeed,  the gauge groups of    theories participating in a field theory duality can be drastically different, and in gauge/gravity dualities there is not even  a Lie group $G$ in sight in the  dual bulk theory.

\smallskip

One may even argue that the existence of Wilson loops  actually introduces a puzzle for dualities. While global symmetries and $\text{'t Hooft}$ anomalies between dual theories {\it must} match, gauge symmetries between dual theories need not. In a sharp sense, gauge symmetries are not symmetries, but rather redundancies in our local description of particles of helicity one.
Nevertheless,  the gauge group $G$ is not void of important physical information about the theory:   Wilson loop operators  are labeled by a representation $\cR$ of $G$. And therefore, it is in this vain, that the gauge group is ``physical" and its elusive representations must be found in the dual.

\smallskip
 
\begin{figure}[h!]
\centering
\includegraphics[scale=0.7]{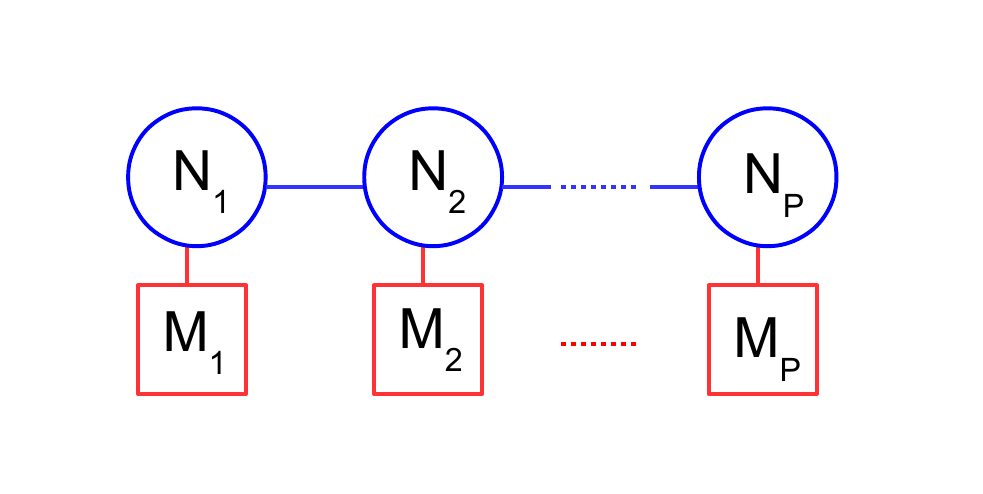} 
\includegraphics[scale=0.7]{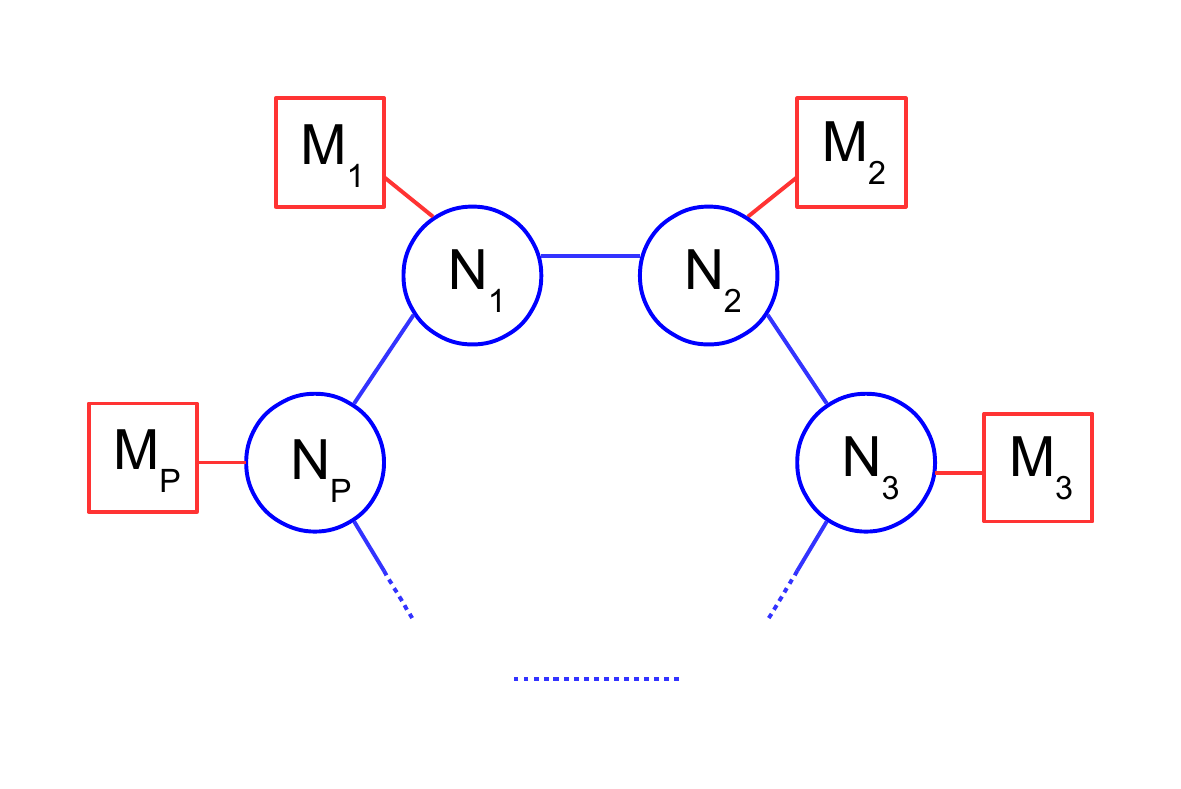}
\vskip -0.5cm
\caption{\footnotesize General linear and circular quiver diagrams.}
\label{circquiv}
\vskip -0.1cm
\end{figure}

  In this paper we identify  the  dual   description of half-supersymmetric Wilson loop operators in   gauge theories related by three dimensional mirror symmetry \cite{Intriligator:1996ex},  an infrared (IR) duality whereby
  two different ultraviolet (UV) 3d $\cN=4$ gauge theories flow in the IR to the same superconformal field theory (SCFT):
  \begin{center}
\begin{tikzpicture}
  \node     at (0,1.5) {UV};
  \node (A) at (2,1.5) {Theory A};
  \node (B) at (5,1.5) {Theory B};
  \node     at (0,0) {IR};
  \node (C) at (3.5,0) {SCFT};
  \draw[->] (A) -- (C);
  \draw[->] (B) -- (C);
\end{tikzpicture}
\end{center}
 We find that there is a rather intricate   ``mirror map" relating Wilson loop operators in one theory to Vortex loop operators in the mirror:
  \begin{center}
\begin{tikzpicture}
  \node (A) at (2,1.5) {Theory A};
  \node (C) at (2,0.5) {W};
   \node (D) at (2,-0.5) {V};
  \node (B) at (5,1.5) {Theory B};
   \node (E) at (5,0.5) {W};
     \node (F) at (5,-0.5) {V};
  \draw[<->] (C) -- (F);
  \draw[<->] (D) -- (E);
\end{tikzpicture}
\end{center}
The mirror map
in non-abelian gauge theories is rather subtle. In     abelian gauge theories    it follows from the 
mapping of  the abelian global symmetries of the mirror dual theories \cite{Kapustin:2012iw,Drukker:2012sr, ASG}.\footnote{Vortex loop operators were previously studied in \cite{Drukker:2008jm}. For Vortex loops in pure Chern-Simons theory see\cite{Moore:1989yh}.}   We determine the mapping of loop operators under mirror symmetry   for arbitrary  3d $\cN=4$ gauge theories encoded by a quiver diagram of linear or circular  topology (see figure \ref{circquiv}). 
 
\begin{figure}[h]
\centering
\includegraphics[scale=0.7]{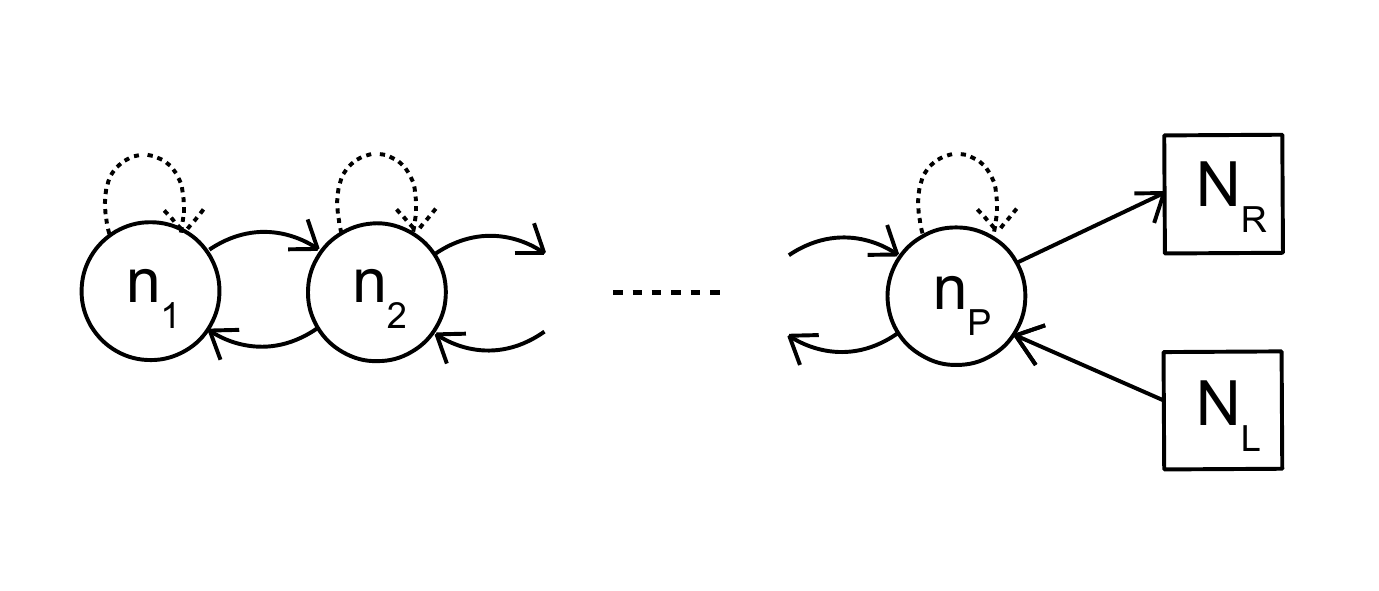}
\vspace{-0.8cm}
\caption{Dotted arrows reflect the possibility of adding an adjoint chiral multiplet in a gauge node.}
\label{VortexQM2}
\end{figure}

\smallskip
\noindent
We propose an explicit  UV  definition of the  Vortex loop operators exchanged with Wilson loops under mirror symmetry.  These Vortex loop operators are constructed by  coupling   1d $\cN=4$ supersymmetric  quantum mechanics (SQM) quiver gauge theories supported on the loop to the bulk 3d $\cN=4$ theory. The class of 1d $\cN=4$   gauge theories that enter in the description of the mirror map of Wilson loop operators is encoded by the quiver diagram of figure \ref{VortexQM2}.\footnote{Arrows encode bifundamental chiral multiplets of two gauge nodes (circles) or of a gauge and a flavor node (square). For the details of superpotential couplings see section \ref{sec:vortttlooops}.}
\vspace{-0.35cm}

\vspace{-0.2cm}
\noindent
Despite that the mirror dual 3d $\cN=4$  gauge theories -- denoted by Theory A and Theory B -- typically have completely different gauge groups $G_A$ and $G_B$, we are able to encode the choice of representation $\cR$ 
of a Wilson loop in one theory in the precise choice of the 1d quiver gauge theory in figure \ref{VortexQM2} describing the mirror dual Vortex loop operator. The 1d quivers in this class are characterized by the ranks of their gauge nodes, the number of fundamental and anti-fundamental chirals in the last node and the presence or not of an adjoint chiral in each node. In particular, to a quiver gauge theory with all Fayet-Iliopoulos (FI) parameters $\eta_i\neq 0$, for $i=1,\ldots,p$,   we associate a Wilson loop representation as follows:\footnote{See sections \ref{sec:Branes} and \ref{sssec:QMindex} for the general dictionary.}
\vspace{+0.15cm}
\begin{itemize}\setlength\itemsep{1.5ex}
\item $\cR=\cS_{n_1}\otimes \cS_{n_2-n_1}\otimes\ldots \otimes\cS_{n_p-n_{p-1}}$: for a  quiver   with one adjoint chiral in each gauge node\,,
\item $\cR=\cA_{n_1}\otimes \cA_{n_2-n_1}\otimes\ldots \otimes\cA_{n_p-n_{p-1}}$: for a quiver with  one adjoint chiral for $U(n_i)$ for $i=1,\ldots,p-1$, but no adjoint chiral in the $U(n_p)$ node,
 \end{itemize}
 \vspace{+0.15cm}
 where $\cA_k$ and $\cS_k$ denote the $k$-th antisymmetric and symmetric representations of $U(N)_L$, if $\eta_p<0$, or $U(N)_R$, if $\eta_p>0$. A generic product of symmetric and antisymmetric representations like
 $\cR= \cS_{n_1}\otimes \cA_{n_2-n_1}\otimes \cA_{n_3-n_2}\otimes\ldots \otimes\cS_{n_p-n_{p-1}}$
 is obtained by removing an adjoint chiral multiplet  in some nodes. We propose, but have  less quantitative evidence for, that an arbitrary  irreducible representation $\cR$ characterized by a Young diagram can be obtained from the family of quivers in figure 
   \ref{VortexQM2}  by setting all but one of the FI parameters  to zero -- that is $\eta_p\neq 0$ --  , i.e. $\eta_1=\eta_2=\ldots \eta_{p-1}=0$.\footnote{The same quiver but  for 2d $\cN=(2,2)$ gauge theories   
   was   shown to be labeled by an irreducible representation $\cR$ in \cite{Gomis:2014eya}, where they describe  M2-brane surface defects, which are indeed labeled by a representation $\cR$.}
      
\begin{figure}[h]
\centering
\includegraphics[scale=0.65]{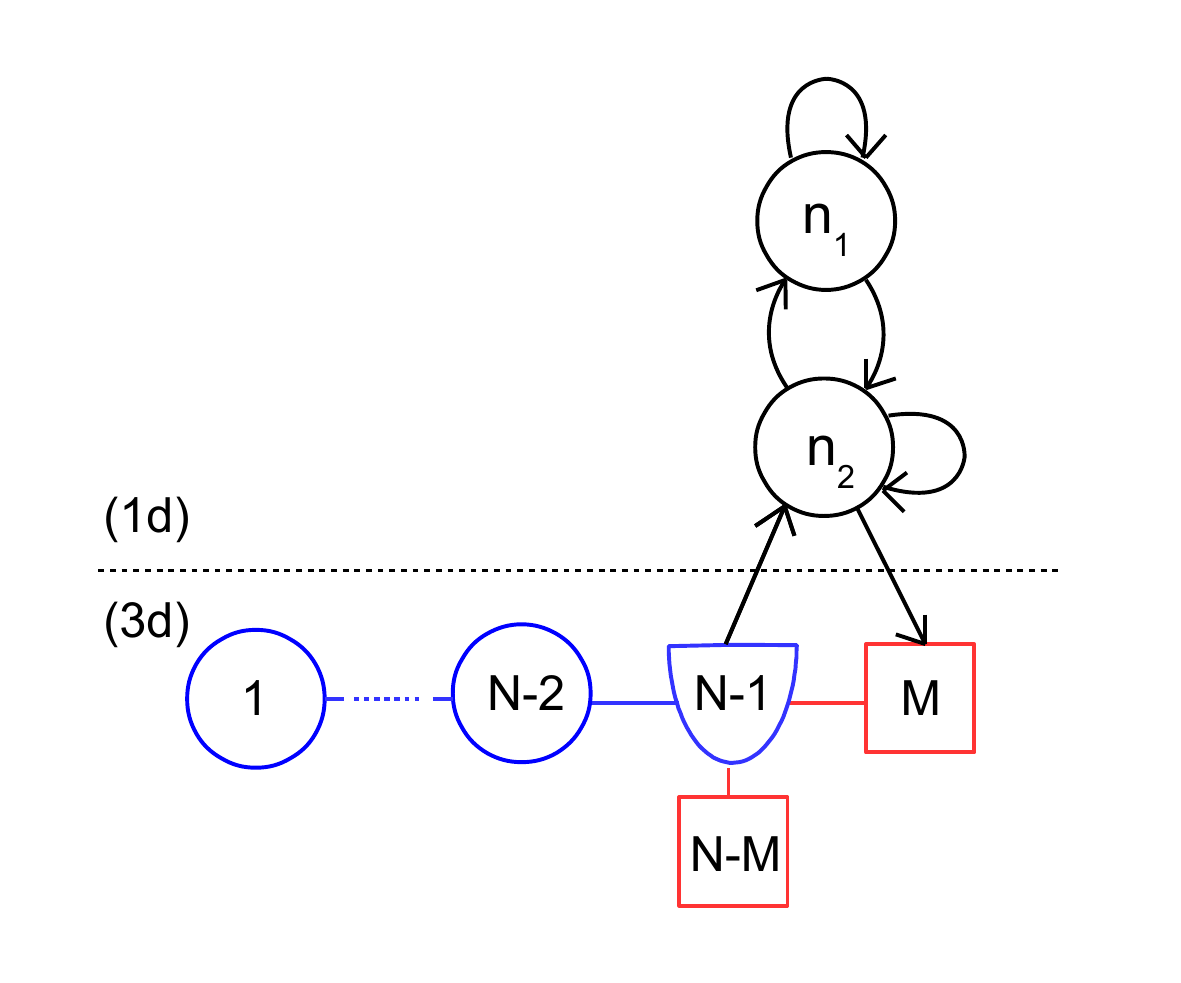}
\vspace{-0.5cm}
\caption{\footnotesize{3d/1d quiver realizing a Vortex loop in the $T[SU(N)]$ theory.}}
\label{TSUNVortex0}
\end{figure}

 \smallskip
    A Vortex loop associated to a given gauge group factor with $N$ fundamental hypermultiplets in the 3d $\cN=4$ gauge theory is labeled   by a representation $\cR$ of the gauge group as  well as by a splitting of $N$ into two factors: $N=M+(N-M)$.  A Vortex loop labeled by a representation $\cR$ and by the integer $M$  is constructed by gauging   flavour symmetries of the 1d $\cN=4$ quiver gauge theory with the bulk 3d $\cN=4$ theory. The explicit     3d/1d theory obtained in this way can be encoded by a mixed 3d/1d quiver diagram, see e.g figure \ref{TSUNVortex0}.

The mirror map between loop operators that we  uncover  is rather intricate and rich.  In particular, the   map depends strongly on the choice of integer $M$ labeling  a Vortex loop. As an illustrative example, the Vortex loop operator $V_{M, \scR}$ in $T[SU(N)]$ 
defined by the 3d/1d quiver in figure \ref{TSUNVortex0} maps under mirror symmetry to the following combination of Wilson loop operators in   $T[SU(N)]$ (this theory is self-mirror)\footnote{As we shall see in section \ref{ssec:OtherLoops} the Vortex loop operator $V_{N,\cR}$ maps to a Wilson loop for the $U(N-1)\subset U(N)$ flavour symmetry of $T[SU(N)]$.}
\begin{align}
\vev{V_{M, \scR}}  \ \xleftrightarrow{\text{\ mirror \ }}   \  \vev{\sum_{s \in \Delta_M}    W^{\rm fl}_{N,{\bf q}_s}  \  W^{U(M)}_{\wat \scR_s}}   \, \quad 0 \le M \le N-1  \, ,
\label{MirrorMapAllLoops0}
\end{align}
where $W^{U(M)}_{\wat \scR_s}$ is a Wilson loop in a representation ${\wat \scR_s}$ of $U(M)$,  $W^{\rm fl}_{N,{\bf q}_s}$ is an abelian flavour Wilson loop of charge $q_s$ and $\Delta_M$ is the set of representations that appear in the decomposition of $\cR=\oplus_{s \in \Delta_M} ({\bf q}_s,\wat\scR_s)$ under the embedding $U(1)\times U(M)\subset U(N-1)$.\footnote{$U(M)$ is embedded as $U(M)\times U(N-M-1)\subset U(N-1)$, and the $U(1)$ is the diagonal factor in  $U(N-M-1)$.} With the algorithm we put forward in this paper, the mirror map between loop operators in arbitrary linear and circular quivers can be constructed, and we provide explicit representative examples for both types of quivers.

\smallskip
The key insight that allows us to  construct the explicit mirror map between loop operators in linear and circular quivers is the identification of  the brane realization of Wilson and Vortex loop operators in the Type IIB Hanany-Witten construction of 3d $\cN=4$ gauge theories 
\cite{Hanany:1996ie}.\footnote{Our  brane realization of loop operators in  3d $\cN=4$ theories generalizes  the 
 brane realization of the defect field theory  description of  Wilson loop operators in 4d $\cN=4$ super-Yang-Mills of \cite{Gomis:2006sb,Gomis:2006im}. This brane construction was used in \cite{Gomis:2006sb,Gomis:2006im} to provide the bulk holographic description of Wilson loops in an arbitrary representation of $SU(N)$, following the dictionary put forward in \cite{Maldacena:1998im,Rey:1998ik} for the fundamental representation.} In the string theory construction, mirror symmetry is realized as S-duality in Type IIB string theory \cite{Hanany:1996ie}. 
 By understanding the detailed physics of branes in string theory, the action of S-duality on  our brane  realization of loop operators allows us to find an explicit map between brane configurations, which in turn  yields the mirror map between Wilson and Vortex loop operators. 

\smallskip

We provide quantitative evidence for our mirror maps by computing the {\it exact}   expectation value of circular Wilson and Vortex loop operators in  $\cN=4$ gauge theories on $S^3$.\footnote{The mapping of the $S^3$ partition functions themselves between mirror dual theories was initiated in \cite{Kapustin:2010xq, Benvenuti:2011ga} and proven in general for linear and circular unitary quiver theories in \cite{Assel:2014awa}.} The computation of the expectation value of Vortex loops combines in a rather interesting way the computations of the $S^3$ partition function in \cite{Kapustin:2009kz}  with the supersymmetric quantum mechanics index of  \cite{Hori:2014tda,Cordova:2014oxa}. The detailed matrix integral capturing the expectation value is obtained by understanding how to couple the 1d $\cN=4$ gauge theory defining the Vortex loop on an $S^1\subset S^3$ to the bulk gauge theory. All our computations confirm our brane-based predictions.

\smallskip

Our understanding of the action of duality on Wilson loops in the context of three dimensional mirror symmetry give us  ideas and renewed confidence that this problem can also be tackled in other interesting dualities, like four dimensional Seiberg duality \cite{Seiberg:1994pq}, 
where the gauge groups of the two dual theories are also different, and subject to the puzzles raised at the beginning.\footnote{Of course, in 4d $\cN=1$ only null Wilson lines are supersymmetric.}

\medskip

\begin{figure}[h!]
\centering
\includegraphics[scale=0.6]{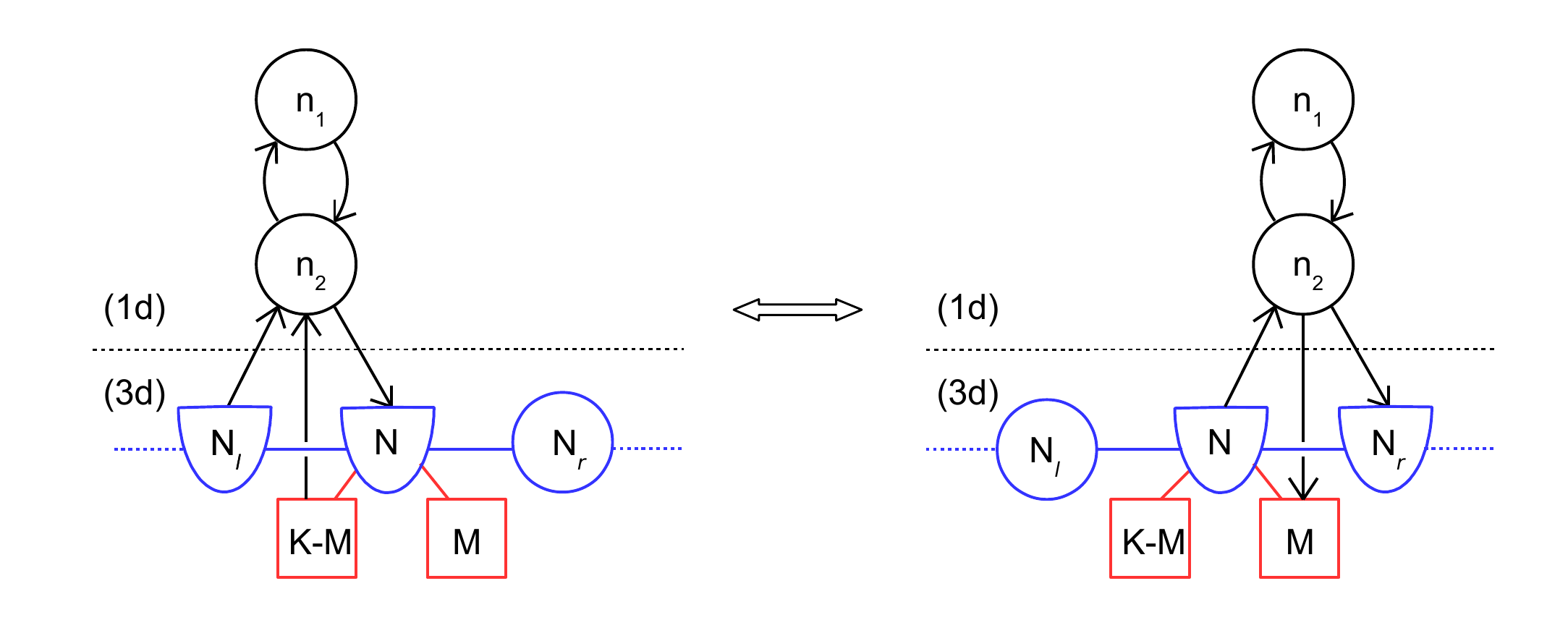}
\vspace{-0.5cm}
\caption{\footnotesize{3d/1d quiver theories realizing the same Vortex loop. }}
\label{Hopping0}
\end{figure}

The plan of the rest of the paper is as follows. In section \ref{sec:loops} we study the classes of loop operators that can be defined in a 3d $\cN=4$ SCFT as well as a UV gauge theory definition of the SCFT. This analysis leads to consider Wilson and Vortex loop operators. In section \ref{sec:loopsbranes}, after reviewing the brane construction of 3d $\cN=4$ gauge  theories we put forward a brane realization of Wilson and Vortex loop operators. We give at least two different UV descriptions for each Vortex loop operator, distinguished in particular    by the  gauging of the global symmetries of the 1d quiver gauge theory with bulk 3d fields. The explicit     3d/1d theories obtained this way can be encoded by mixed 3d/1d quiver diagrams as in figure \ref{Hopping0}.
We then develop a brane-algorithm that allows us to use the action of S-duality on brane configurations to construct the mirror map for loop operators between dual gauge theories. 
This algorithm can be applied to an arbitrary 3d $\cN=4$ gauge theory labeled by a linear or circular quiver. Section \ref{sec:examples} presents the detailed mirror map for representative classes of gauge theories, including $T[SU(N)]$, circular quivers with equal ranks on all nodes and supersymmetric QCD (SQCD). In section \ref{sec:LoopsOnS3} we consider  loop operators on $S^3$ and write down a matrix model representation of the exact expectation value of a Vortex loop operator in terms of the supersymmetric index of the SQM that defines it and the   matrix model for the 3d theory. We perform explicit computations of the expectation value of Wilson and Vortex loop operators and confirm our brane-based predictions for the mirror map.\footnote{In \cite{Kapustin:2012iw,Drukker:2012sr}  $\vev{V}$ was computed using a disorder definition of V for abelian theories, which we reproduce from our SQM perspective and which we extend to arbitrary non-abelian gauge theories.} We also show that the  distinct UV definitions of a given Vortex loop operator (see figure \ref{Hopping0}) give rise to the same operator in the IR, by showing that the expectation value of the two UV definitions coincide, and are related by hopping duality \cite{Gadde:2013dda}. Some technical details are relegated to the appendices.

\section{Loop Operators in 3d $\cN=4$ Theories}
\label{sec:loops}

An $\cN=4$ SCFT is invariant under the 3d $\cN=4$ superconformal symmetry $OSp(4|4)$. 
The bosonic generators comprise those in the $SO(3,2)\simeq Sp(4)$ conformal algebra and   those that generate the $SU(2)_C\times SU(2)_H$ \text{R-symmetry} of the SCFT. The supercharges in  $OSp(4|4)$ transform in the $(\bf{4},\bf{2},\bf{2})$  representation of   $SO(3,2)\times SU(2)_C\times SU(2)_H$. 
A UV Lagrangian definition of the IR SCFT preserves 
3d $\cN=4$ Poincar\'e supersymmetry, the subgroup of $OSp(4|4)$ that closes into the isometries of flat space.\footnote{The  3d $\cN=4$ Poincar\'e supersymmetry algebra is the fixed locus of an involution of $OSp(4|4)$.} Mirror symmetry is the statement that a pair of UV gauge theories flow in the IR to the same  SCFT with the roles of $SU(2)_C$ and $SU(2)_H$ exchanged:
\smallskip
\begin{center}
\begin{tikzpicture}
  \node     at (0,1.5) {UV};
  \node (A) at (2,1.5) {Theory A};
  \node (B) at (5,1.5) {Theory B};
  \node     at (0,0) {IR};
  \node (C) at (3.5,0) {SCFT};
  \draw[->] (A) -- (C);
  \draw[->] (B) -- (C);
\end{tikzpicture}
\end{center}

\smallskip
An $\cN=4$ SCFT admits half-supersymmetric line operators/defects supported on a straight line   in  flat space.\footnote{
Operators supported on   curves obtained by acting on a straight line by   broken conformal generators are also half-supersymmetric and  preserve an isomorphic symmetry algebra.
Under a broken conformal symmetry a  time-like line becomes a rectangular time-like hyperbola, a space-like line becomes a space-like hyperbola (which includes a circle)
and a null line remains a null line.} 
  There are two physically inequivalent classes of 
superconformal line defects in an $\cN=4$ SCFT, distinguished  by their preserved symmetries. Superconformal line defects supported on a time-like line are   invariant either  under an $U(1,1|2)_W$ or  $U(1,1|2)_V$ subalgebra  of 
$OSp(4|4)$.\footnote{$U(1,1|2)_W$ and $U(1,1|2)_V$ are   the fixed locus of an involution of $OSp(4|4)$}
 Let us exhibit this more explicitly.\footnote{A very similar analysis holds for a space-like line defect. For a null line defect, which we do not consider in this paper, the preserved symmetries are larger.} First, a straight time-like line in flat space preserves $SU(1,1)\times U(1)_\perp\subset SO(3,2)$. The $U(1,1|2)_W$ subalgebra is embedded as follows. Under the embedding of $SU(1,1)\times SU(2)_H\times U(1)_\perp\times U(1)_C$  in $SO(3,2)\times SU(2)_C\times SU(2)_H$, the supercharges generating $OSp(4|4)$
 decompose as $(\bf{2},\bf{2})_{++}\oplus (\bf{2},\bf{2})_{+-}\oplus (\bf{2},\bf{2})_{-+}\oplus (\bf{2},\bf{2})_{--}$.  The supercharges in $U(1,1|2)_W$   
 are $(\bf{2},\bf{2})_{++}\oplus (\bf{2},\bf{2})_{--}$.\footnote{The   other choice $(\bf{2},\bf{2})_{+-}\oplus (\bf{2},\bf{2})_{-+}$
 corresponds to the same line defect but with opposite orientation.}  For $U(1,1|2)_V$ the analysis is identical with the roles of $SU(2)_C$ and $SU(2)_H$   exchanged.
 A   defect preserving $U(1,1|2)_W$  is invariant under $U(1)_C\times SU(2)_H$ while    a defect    preserving
 $U(1,1|2)_V$ is invariant under $SU(2)_C\times U(1)_H$.
 The main goal of this paper is to give the UV   description of these two classes of line operators/defects and to identity how  the  UV descriptions are mapped under mirror symmetry.
 
\smallskip
A line defect in a   UV Lagrangian description of a SCFT   can be made  invariant under four of the supercharges in the 3d $\cN=4$   Poincar\'e supersymmetry algebra of the UV theory. The 3d $\cN=4$ Poincar\'e supercharges transform in the $(\bf{2},\bf{2})$ representation of the $SU(2)_C\times SU(2)_H$ R-symmetry of the UV theory and obey 
 \beq
 \{Q_{\alpha AA'},Q_{\beta BB'}\}=\left(\gamma_\mu C\right)_{\alpha\beta} P^\mu\epsilon_{AB}\epsilon_{A'B'}\,.
 \label{threePoincare}
 \eeq
We take the    $SO(2,1)$ $\gamma$-matrices $(\gamma_0,\gamma_1,\gamma_2)=(i\tau_3,\tau_1,\tau_2)$,  where $\tau_a$ are the Pauli matrices. The charge conjugation matrix is $C=\tau_2$. In Lorentzian signature  and in this basis the reality condition on the supercharges is   $Q^\dagger_{\alpha AA'}=\left(\tau_1\right)_\alpha^{\ \beta} \epsilon^{AB}\epsilon^{A'B'}Q_{\beta BB'}$.

\smallskip
 
There are two inequivalent   1d
 $\cN=4$ supersymmetric quantum mechanics (SQM) subalgebras of the 3d $\cN=4$   Poincar\'e superalgebra of a UV theory that can be preserved by a line defect. We   denote them by SQM$_W$ and SQM$_V$.  SQM$_W$  preserves $U(1)_C\times SU(2)_H$  and the supercharges $Q_{11A'}$ and $Q_{22B'}$, which anticommute to the generator of     translations $H$ along the defect 
 \beq
  \{Q_{11A'},Q_{22B'}\}=\epsilon_{A'B'} H\,.
 \eeq
  SQM$_V$   preserves $SU(2)_V\times U(1)_H$ and   the preserved supercharges $Q_{1A1}$ and $Q_{2B2}$ obey
\beq
  \{Q_{1A1},Q_{2B2}\}=\epsilon_{AB} H\,.
  \label{SQMV}
 \eeq
   A 1d $\cN=4$ SQM supersymmetry algebra  can be brought to the canonical form  
     \begin{align}
 \{Q_+\overline Q_+\}=H \qquad\qquad    [J_+, Q_+]=-Q_+\qquad\qquad   [J_+, \overline Q_+]=\overline Q_+\nonumber\\[+1ex]
  \{Q_-,\overline Q_-\}=H \qquad\qquad    [J_-, Q_-]=-Q_-\qquad\qquad   [J_-, \overline Q_-]=\overline Q_-
 \label{SQMcan}
  \end{align}
  with all other (anti)-commutators vanishing. For both SQM$_W$ and SQM$_V$ algebras   $Q_+=Q_{111}$ and $\overline Q_+=Q_{222} =Q_{111}^\dagger$. For SQM$_W$ $Q_-=Q_{112}$, $\overline Q_-= -Q_{221} =Q_{112}^\dagger$ while for SQM$_V$  $Q_-=Q_{212}$, $\overline Q_-=-Q_{121} =Q_{212}^\dagger$. The embedding of the SQM$_W$ and SQM$_V$ algebras in the 3d $\cN=4$  Poincar\'e superalgebra has a $U(1)$ commutant: $R_C-J_{12}$  for   SQM$_W$ and $R_H-J_{12}$  for SQM$_V$, where $J_{12}$ is  the $U(1)_\perp$ rotation generator  transverse to the defect and $R_C$ and $R_H$ are the Cartan generators of $SU(2)_C$ and $SU(2)_H$ respectively.\footnote{$R_C-J_{12}$ and $R_H-J_{12}$ is also the commutant of $U(1,1|2)_W$ and $U(1,1|2)_V$ in $OSp(4|4)$. They appear respectively in the anticommutator of the  $U(1,1|2)_W$ and $U(1,1|2)_V$ supercharges preserved by the corresponding defect.} 
  Therefore, up to   shifts by the ``flavour" symmetry $R_C-J_{12}$ for SQM$_W$ and by $R_H-J_{12}$ for SQM$_V$ the Cartan R-symmetry generators can be taken to be $J_+=-R_C -R_H$ and  $J_-=R_H -R_C$.  
 
  \smallskip
In summary, in a 3d $\cN=4$ gauge theory there are two   classes of line defects that can defined, one preserving SQM$_W$   and the other SQM$_V$.
These line defects   flow in the IR to superconformal line defects preserving $U(1,1|2)_W$ and $U(1,1|2)_V$ respectively. Our analysis can be succinctly summarized by the following diagram: 

\begin{center}
\begin{tikzpicture}
  \node     at (-0.5,1.5) {UV};
  \node (A) at (2,1.5) {3d $\cN=4$ Poincar\'e};
  \node (B) at (6,1.5) {SQM$_W$};
   \node (C) at (9,1.5) {SQM$_V$};
  \node     at (-0.5,-1) {IR};
  \node (D) at (2,-1) {$OSp(4|4)$};
  \node at (7.3,1.5){and};
    \node at (7.5,-1){and};
  \node (E) at (6,-1) {$U(1,1|2)_W$};
   \node (F) at (9,-1) {$U(1,1|2)_V$};
   \draw[->] (A) -- (D) node [midway,  right=+3pt] {\footnotesize RG};
  \draw[->] (A) -- (B)node [midway,above  ] {\footnotesize defect};
    \draw[->] (B) -- (E) node [midway,  right=+3pt] {\footnotesize RG};
      \draw[->] (C) -- (F) node [midway,  right=+3pt] {\footnotesize RG};
        \draw[->] (D) -- (E)node  [midway,above  ] {\footnotesize defect};
\end{tikzpicture}
\end{center}

Both classes of UV supersymmetric line defects in a UV 3d $\cN=4$ theory can be realized by coupling  different   1d  $\cN=4$ SQM theories with four supercharges to the bulk 3d $\cN=4$ theory. A canonical way of coupling a 1d $\cN=4$ SQM theory to a   3d $\cN=4$ theory is by gauging   flavour symmetries    of the SQM theory with 3d $\cN=4$ vector multiplets. 
 The gauging of the defect flavour symmetries with   bulk vector multiplets is made supersymmetric 
by embedding the defect vector multiplet of the supersymmetry algebra preserved by the defect 
  into  the bulk 3d $\cN=4$ vector multiplet  at the position of the line defect. The embedding is found by identifying which combination of fields in the higher dimensional vector multiplet transform as the fields of the defect vector multiplet under the supersymmetry preserved by the defect. Replacing the defect vector multiplet fields in the gauged   1d $\cN=4$ SQM theory with the proper combination of bulk vector multiplet fields ensures that the coupling of 1d fields to 3d fields is supersymmetric under the supersymmetry of the defect 1d $\cN=4$  SQM theory. Superpotential couplings between defect and bulk matter multiplets may also be added when defect matter multiplets can be embedded in bulk hypermultiplets. 
Such couplings gauge  defect flavour symmetries with bulk flavour  or gauge symmetries, depending on which symmetries of the bulk matter multiplet are global and which are  gauged. 
  
\smallskip
 We consider UV line defects invariant under SQM$_W$  and SQM$_V$   obtained by gauging  global symmetries of 1d $\cN=4$ SQM theories obtained 
by dimensional reduction of 2d $\cN=(0,4)$ and 2d $\cN=(2,2)$  theories 
   with 3d $\cN=4$ vector multiplets:
   \begin{itemize}
   \item SQM$_W$: 2d $\cN=(0,4)\rightarrow$ 1d $\cN=4$ SQM
    \item SQM$_V$: 2d $\cN=(2,2)\rightarrow$ 1d $\cN=4$ SQM
    \end{itemize}
   Superpotential couplings between defect and bulk fields also play an important role in the construction of defects.

\subsection{Wilson Loop Operators}

A  line defect in a  UV  3d $\cN=4$ gauge theory invariant under SQM$_W$ is 
the Wilson line operator, which is      labeled by a representation $\cR$ of the gauge group. It is given by
\beq
      W_\cR=\hbox{Tr}_\cR P\exp{\oint i \left( A_\mu \dot x^\mu+\sqrt{-\dot x ^2}\,\sigma\right)d\tau}\,,
      \label{susywilson}
      \eeq       
     where $\sigma\equiv \sigma_3$ is the scalar field in the $\cN=2$ vector multiplet inside the $\cN=4$ vector multiplet.  
This operator manifestly breaks the $SU(2)_C$ symmetry acting on the three scalars $\vec \sigma=(\sigma_1,\sigma_2,\sigma_3)$ in the ${\cal N}=4$   vector multiplet down to   $U(1)_C$,\footnote{The choice of scalar determines an embedding of $U(1)_C$ in $SU(2)_C$.} while preserving $SU(2)_H$.\  If the operator is supported on a straight line\footnote{When the scalar couples to the loop with constant charge a circular Wilson loop is not supersymmetric in the UV theory. See, however, discussion at the end of this subsection and   of circular Wilson loops on $S^3$ in section \ref{ssec:Wilsonsphere}.}, it preserves the 1d $\cN=4$ SQM subalgebra SQM$_W$ of the 3d $\cN=4$ theory and its $U(1)_C\times SU(2)_H$ R-symmetry. In the IR a  Wilson line operator flows to a conformal line operator in the SCFT preserving $U(1,1|2)_W$.
     
\smallskip

The  supersymmetric Wilson line operator  \rf{susywilson} can be realized by     coupling  a  1d $\cN=4$ SQM theory  living on the  line with the bulk  3d $\cN=4$ gauge theory. This coupling preserves the 
 SQM${_W}$ supersymmetry algebra. The defect 1d $\cN=4$ SQM that represents a  Wilson loop operator is the theory of a   1d $\cN=4$ fermi multiplet,  obtained by dimensional reduction of the  2d ${\cal N}=(0,4)$   fermi multiplet, which on-shell consists of a complex chiral fermion. 
The flavour symmetry of a 1d $\cN=(0,4)$ fermi multiplet can be gauged preserving supersymmetry with a 1d $\cN=(0,4)$ vector multiplet.
The     fields of the  1d $\cN=(0,4)$ vector multiplet  that couple to the fermi multiplet  can be embedded in the 3d $\cN=4$ vector multiplet as follows\footnote{The Fermi multiplet   couples only to a subset of fields in the $\cN=(0,4)$ vector multiplet, and those do admit an embedding into the 3d $\cN=4$ vector multiplet. This observation can be applied to  the   study of supersymmetric surface operators in 4d $\cN=2$, which can preserve either 2d $\cN=(2,2)$ or $\cN=(0,4)$ supersymmetry.   We can construct a surface operator by gauging a $\cN=(0,4)$ fermi multiplet with a bulk vector multiplet. These surface operators were studied in the context of $\cN=4$ SYM in \cite{Buchbinder:2007ar}. }
   \begin{align}
 a_0&=A_0\\
 \sigma_{1d}&=\sigma\,.
 \label{embedwilson}
  \end{align}
This embedding makes manifest that the coupling   of the fermi multiplet to the bulk ${\cN=4}$ vector multiplet
 preserves    $U(1)_C\times SU(2)_H$ R-symmetry and the  SQM${_W}$ algebra.\footnote{The purely 1d  $\cN=4$  
 theory of a gauged fermi multiplet is invariant under $SO(4)$ R-symmetry. The coupling of the fermi multiplet with the bulk through the embedding \rf{embedwilson} breaks the   R-symmetry down to $U(1)_C\times SU(2)_H$.} 
 
\smallskip
 The 1d $\cN=4$ SQM fermi multiplet on the defect can   be integrated out exactly and it results in the insertion of  a supersymmetric Wilson loop \rf{susywilson} in the bulk 3d $\cN=4$ theory. This representation of a supersymmetric Wilson loop  in 4d $\cN=4$ super-Yang-Mills (SYM) as  a coupling of a fermi multiplet with bulk fields appeared in  \cite{Gomis:2006sb,Gomis:2006im}, where the defect field theory was derived from brane intersections in string theory.  Inspired by \cite{Gomis:2006sb,Gomis:2006im}, a   brane realization of Wilson loop defects  in 3d $\cN=4$ gauge theories 
 will play a prominent role in section \ref{sec:loopsbranes}, where we will use S-duality of Type IIB   string  theory  to identify the mirror of Wilson loop operators. 
 
\smallskip
 
As an  aside, $1/4$-supersymmetric Wilson loops supported on an arbitrary curve $\gamma$ in $R^3$ can be defined mimicking the construction in \cite{Zarembo:2002an} of $1/16$-supersymmetric Wilson loops in 4d $\cN=4$ SYM. This requires tuning the coupling of  the loop to the three scalars in the vector multiplet. Explicitly, $1/4$-supersymmetric Wilson loops are given by
\beq
\hbox{Tr}_\cR P \exp{\oint_\gamma i \left(A_\mu+i \sigma_\mu\right){\dot x^\mu d\tau}}\,
\label{wilsontwist}
\eeq
and preserve two supercharges: $Q_{A'}^{H}\equiv Q_{\alpha AA'}\epsilon^{\alpha A}$. These Wilson loop operators  are in the cohomology of  the supercharges $Q_{A'}^{H}$ of the Rozansky-Witten twisted theory \cite{Rozansky:1996bq}\footnote{This is the 3d counterpart of the statement in 4d $\cN=4$ SYM that the 1/16-supersymmetric Wilson loop operators in \cite{Zarembo:2002an} are in the cohomology of a  supercharge of the Langlands twist \cite{Kapustin:2006pk}.} obtained by twisting spatial rotations with $SU(2)_C$. Half-supersymmetric Higgs branch operators are also in the cohomology of this twisted theory.

  \subsection{Vortex Loop Operators}
\label{sec:vortttlooops}

A supersymmetric line defect in a UV 3d $\cN=4$   theory preserving SQM$_V$ can   be constructed by coupling the   bulk theory to a 1d $\cN=4$  SQM theory with SQM$_V$ symmetry. For   line defects preserving  SQM$_V$, the  appropriate 1d   ${\cN=4}$ SQM theories  are obtained by dimensionally reducing 4d $\cN=1$ theories (or equivalently 2d $\cN=(2,2)$ theories).  $U(1)_H$ is the R-symmetry already present in 4d while $SU(2)_V$ emerges as an R-symmetry in the dimensional reduction down to 1d. Therefore the SQM$_V$ invariant 1d  $\cN=4$ SQM theories we consider are supersymmetric    gauge theories based on the familiar 4d $\cN=1$   vector multiplets and chiral multiplets.  
The same 4d $\cN=1$ theories dimensionally reduced to 2d define surface operators \cite{Gukov:2006jk} in 4d $\cN=2$ gauge theories. 

\smallskip
We can construct a  supersymmetric line defect in a 3d $\cN=4$ gauge theory by gauging   flavour symmetries of a 1d $\cN=4$ SQM$_V$ invariant  theory with bulk vector multiplets.  The embedding of the   bosonic fields in the 1d  vector multiplet $(a_3,\vec \sigma_{1d}, d)$, where $\vec \sigma_{1d}$ is a triplet of $SU(2)_V$,   in the 3d $\cN=4$ vector multiplet is (see appendix \ref{app:Embeddings})
\begin{align}
a_0&=A_0\\
\vec \sigma_{1d}&=\vec\sigma\\
d&=D+F_{12}\,.
\end{align} 
This embedding makes manifest that $SU(2)_V$ is preserved and that  $SU(2)_H$ is broken down to $U(1)_H$, as it  selects one of the auxiliary fields transforming as a triplet of $SU(2)_H$ in the 3d $\cN=4$ vector multiplet, which we have denoted by $D$.\footnote{The choice of auxiliary field  determines an embedding of $U(1)_H$ in $SU(2)_H$.} The coupled theory   preserves the SQM$_V$ algebra. 
  
\smallskip
  
  We can gauge   defect flavour symmetries  either with 3d $\cN=4$ fluctuating vector multiplets or background vector multiplets. Background vector multiplets for flavour symmetries are associated to canonical supersymmetric mass deformations in 3d $\cN=4$ and 1d $\cN=4$ theories.\footnote{Obtained by turning on  constant commuting values for the three scalars in the 1d and 3d $\cN=4$ vector multiplet.}  Gauging 1d flavour symmetries with background 3d vector multiplets means that 1d and 3d flavour symmetries are identified. 1d $\cN=4$   and 3d $\cN=4$ flavour symmetries are identified by   SQM$_V$-preserving defect cubic superpotential couplings  between       defect chiral multiplets and bulk hypermultiplets\footnote{The fields in the 1d $\cN=4$ chiral multiplet can be embedded in the 3d $\cN=4$ hypermultiplet. This embedding (see appendix \ref{app:Embeddings}), which we denote by $Q$, allows one to write supersymmetric couplings between defect chiral multiplets and bulk hypermultiplets.} 
  \beq
  W=\tilde q^i_I q^I_a Q_i^a\,,
  \label{supercoupl}
  \eeq
 where the index $I$ is a 1d gauge  index. The indices   $i$, $a$ are simultaneously  indices for 1d flavour symmetries and indices for either   3d flavour or gauge symmetries.
When $a$ (or $i$) is a 3d flavour index, the superpotential breaks the (otherwise independent) flavour symmetries acting on chiral multiplets $q_a$ (or $\tilde q^i$) and hypermultiplets  $Q^a_i$ to the diagonal flavour symmetry group.\footnote{When one of the indices is a 3d gauge index the superpotential    indeed enforces the gauging of 1d flavour symmetries with a 3d dynamical vector multiplet.}   The background   3d $\cN=4$ vector multiplet gives the same mass to the 1d chiral multiplets and 3d hypermultiplets that are acted on by the preserved diagonal flavour symmetry group.
   
\smallskip
   
   The 1d $\cN=4$ gauge theories that appear in the construction of the defects dual to Wilson loops can be encoded in a standard quiver diagram shown in figure \ref{VortexQM2}.\footnote{These quivers but in a 2d $\cN=(2,2)$ were used in \cite{Gomis:2014eya} to describe M2-brane surface operators.}

\noindent 
An adjoint chiral multiplet may be added to any $U(n_i)$ gauge group factor, an option which we denote by a dashed line. Each adjoint chiral multiplet is coupled to the neighbouring bifundamental chiral multiplets through a cubic superpotential, while nodes without an adjoint chiral multiplet have an associated  quartic superpotential coupling the corresponding bifundamental chiral multiplets.

\smallskip
   
\begin{figure}[th]
\centering
\includegraphics[scale=0.7]{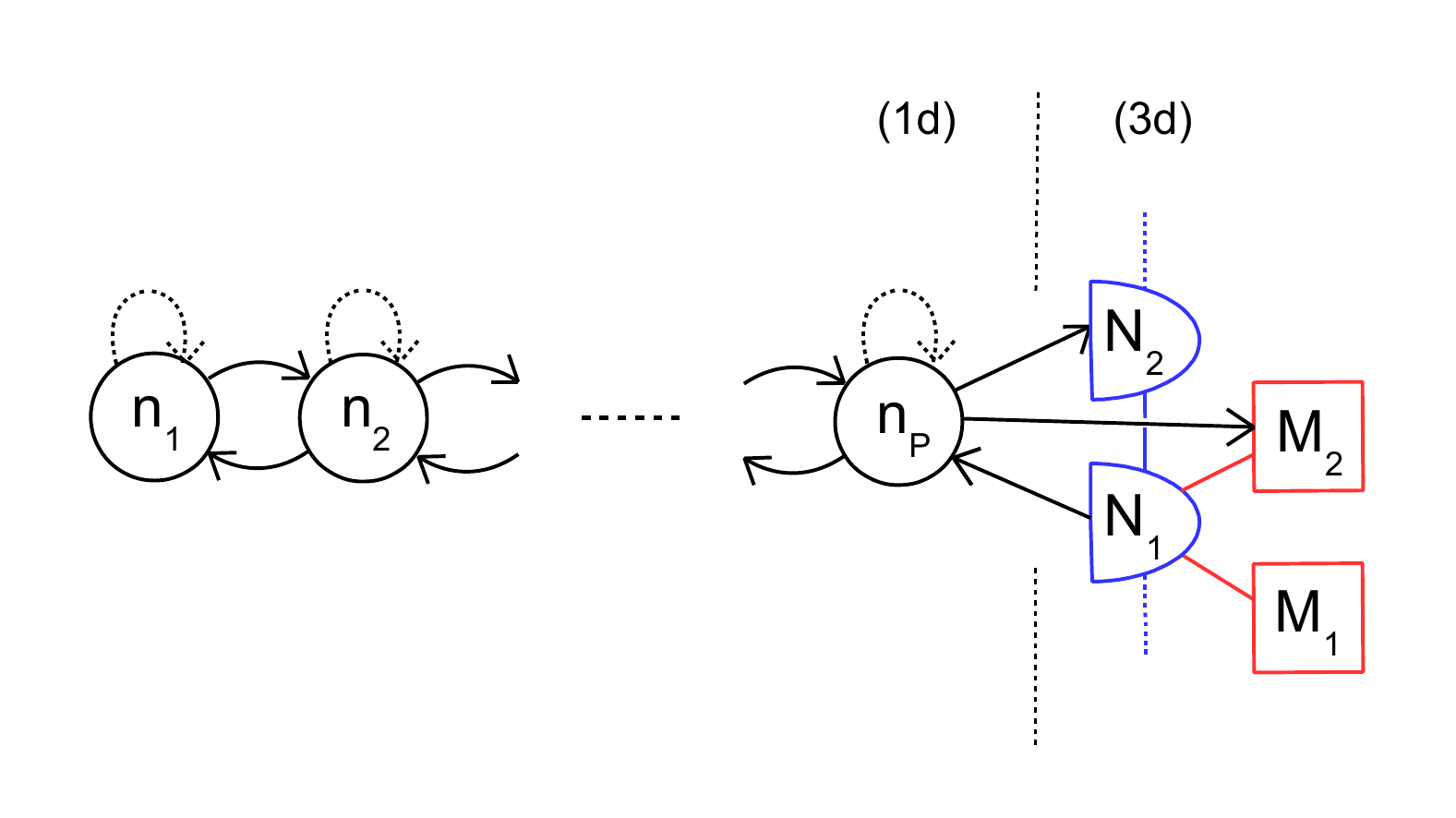}
\caption{\footnotesize{1d quiver theory coupled to  3d quiver theory by gauging 1d flavor symmetries with 3d vector mutiplets (dynamical or weakly gauged).}}
\label{VortexQM}
\end{figure}

 The specific couplings between 1d $\cN=4$   and 3d $\cN=4$ theories can be encoded in a combined 3d/1d quiver diagram (analogous 4d/2d quivers have appeared in   \cite{Gadde:2013dda} (see also \cite{Gomis:2014eya})).The quiver diagram makes explicit the 1d flavour symmetries which are gauged with bulk dynamical gauge fields and the flavour symmetries which are identified with 3d flavour symmetries, as shown in figure \ref{VortexQM}. We use the mixed circle and square notation of \cite{Gomis:2014eya} to denote the 1d flavour symmetries that are gauged with dynamical 3d vector multiplets. This 3d/1d quiver also assigns a    defect cubic superpotential coupling  between 1d chiral multiplets and 3d hypermultiplets for each triangle that can be formed with these fields. 

   \smallskip  
Demanding that the UV supersymmetric 3d/1d Lagrangian coupling 1d chiral multiplets with a 3d $\cN=4$ vector multiplet  is well-defined (finite) requires that \cite{Constable:2002xt,Gukov:2006jk,Gaiotto:2009fs}
\beq
F_{12}=g^2\mu\, \delta^2(x)\,,
\eeq
 where $\mu$ is the moment map for the flavour symmetry acting on the 1d chiral multiplets that is gauged with the bulk (dynamical) vector multiplet and $g$ is its 3d gauge coupling. 
 Therefore,  in the semiclassical UV  description, defect fields induce a singular {\it Vortex} field configuration on the 3d gauge fields. This justifies our use of  the subscript $V$  to describe this class of line defects, which we refer as Vortex line defects/operators. These UV Vortex line defects flow in the IR 
 to conformal line operators in the SCFT preserving $U(1,1|2)_V$.
\smallskip
 
 As another aside, we note that just as a $1/4$-supersymmetric Wilson loop supported on an arbitrary curve $\gamma$ in $R^3$ have been constructed in \rf{wilsontwist},  it should be possible to  construct a $1/4$-supersymmetric Vortex  loop on an arbitrary curve in $R^3$  by suitably adjusting the coupling of the 1d $\cN=4$ SQM to the bulk 3d $\cN=4$ theory. Such a Vortex loop would preserve  two supercharges: $Q_{A}^{C}\equiv Q_{\alpha AA'}\epsilon^{\alpha A'}$. These Vortex loop operators  are in the cohomology of  the supercharges $Q_{A}^{C}$ of the other version of the Rozansky-Witten twisted theory,  obtained by twisting spatial rotations with $SU(2)_H$. Half-supersymmetric Coulomb branch operators, that is monopole operators, are also in the cohomology of this twisted theory.

   \bigskip\medskip

Given two UV mirror theories that flow in the IR to the same SCFT, we   can construct both classes of line operators 
in each of the UV theories. How are line operators mapped   under mirror symmetry? Since mirror symmetry exchanges $SU(2)_C$ with $SU(2)_H$ in dual mirror theories,  Wilson line operators of one theory are mapped to Vortex line operators in the mirror and viceversa. This can be represented by the following diagram:
 \begin{center}
\begin{tikzpicture}
  \node (A) at (2,1.5) {Theory A};
  \node (C) at (2,0.5) {W};
   \node (D) at (2,-0.5) {V};
  \node (B) at (5,1.5) {Theory B};
   \node (E) at (5,0.5) {W};
     \node (F) at (5,-0.5) {V};
  \draw[<->] (C) -- (F);
  \draw[<->] (D) -- (E);
\end{tikzpicture}
\end{center}

Our immediate goal is to come up with an algorithm that yields the duality map between Wilson and Vortex loop operators in mirror dual theories.

    \section{Brane Realization Of Loop Operators and Mirror Map}
    \label{sec:loopsbranes}
    In this section we first briefly introduce the Type IIB string theory realization of 3d $\cN=4$ gauge theories of \cite{Hanany:1996ie} and recall how mirror symmetry gets realized as S-duality in string theory. Central to the main goal of this paper is the  
   brane realization of both types of line defects discussed in the previous section that we put forward in this section. We then devise an explicit algorithm using branes in string theory to identify the  map between loop operators in mirror dual theories.

\smallskip
    
3d $\cN=4$ supersymmetric   gauge theories admit an elegant realization as the low-energy limit of brane configurations in Type IIB string theory  \cite{Hanany:1996ie}.  This consists of an array of D3, D5 and NS5 branes oriented as shown in    table \ref{tab:probeconfigAAAA}.\footnote{For more details of these brane constructions see  \cite{Hanany:1996ie,Gaiotto:2008ak,Assel:2011xz,Assel:2012cj}.}
\begin{table}[h]
\begin{center}
\begin{tabular}{|c||c|c|c|c|c|c|c|c|c|c||}
  \hline
      & 0 & 1 & 2 & 3 & 4 & 5 & 6 & 7 & 8 & 9 \\ \hline
  D3  & X & X & X & X &   &   &   &   &   &   \\
  D5  & X & X & X &   & X & X & X &   &   &   \\
  NS5 & X & X & X &   &   &   &   & X & X & X \\ \hline
\end{tabular}
\caption{\footnotesize Brane array for three-dimensional quiver gauge theories.}
\label{tab:probeconfigAAAA}
\end{center}
\end{table}

\noindent
The gauge theory associated to a brane configuration is constructed by assigning: 

\begin{itemize}
\item A $U(N)$ vector multiplet to N D3-branes suspended between two NS5-branes
\item A hypermultiplet in the fundamental representation of $U(N)$ to a D5-brane intersecting N D3-branes stretched between two NS5-branes
\item A   hypermultiplet in the bifundamental representation of $U(N_1)\times U(N_2)$ to an NS5-brane  with $N_1$ D3-branes ending
on its left and $N_2$ branes ending on its right
\end{itemize}   
    
\begin{figure}[h!]
\centering
\includegraphics[scale=0.8]{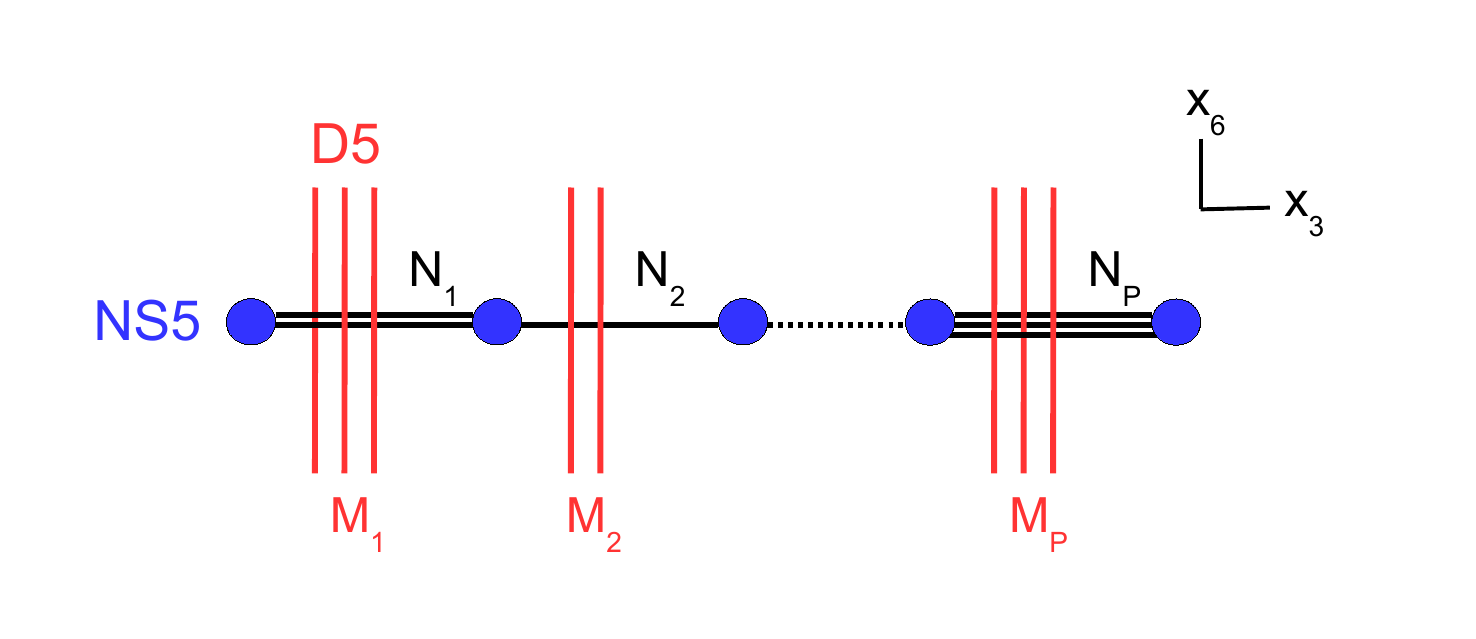} 
\vskip -1cm
\caption{\footnotesize Brane configuration realizing a linear quiver theory.}
\label{linquivbrane}
\end{figure}

    Depending on whether the $x^3$ coordinate takes values on the    line or is  circle valued, the 3d $\cN=4$ gauge theories engineered this way are described either  by quiver diagrams of linear topology or circular topology: linear and circular quiver diagrams respectively.
The quiver diagrams for linear and circular quiver theories are presented in figure \ref{circquiv}. The general brane configuration realizing a linear quiver theory is shown in figure \ref{linquivbrane}. For a circular quiver the $x^3$ direction is periodic and there are extra D3-branes stretched between the first and last NS5-branes.\footnote{The number inside a circle denotes the rank of a gauge group factor. The number inside a rectangle denotes the number of hypermultiplets in the fundamental representation of the gauge   group factor corresponding  to  the circle to which attaches. A line between two circles represents a bifundamental hypermultiplet of the two gauge group factors connected by the line.} 

\smallskip
We are interested in 3d $\cN=4$ gauge theories that flow in the IR to an irreducible, interacting SCFT. 
A 3d $\cN=4$ gauge theory flows to   such  a  SCFT   in the IR if  each gauge group factor $U(N_c)$ has a number of fundamental hypermultiplets $N_f$ obeying $N_f\geq 2N_c$ and  $\sum_{i} M_i\geq 2$. The second condition is automatically obeyed by linear quivers obeying the first condition, but it is an extra requirement for circular quivers. 
 When these conditions are satisfied  the gauge group of the quiver  can be completely Higgsed \cite{deBoer:1996mp,Gaiotto:2008ak} and there are no monopole operators hitting the unitarity bound \cite{Gaiotto:2008ak}.\footnote{When $N_f<2N_c$ or $\sum_{i=1}^{\hat P} M_i\leq 2$ for circular quivers  the IR theory is believed to contain a  decoupled sector.} Mirror symmetry is the statement that two different 3d $\cN=4$ gauge theories flow in the IR to the same irreducible SCFT.

\smallskip
 The  $SU(2)_C\times SU(2)_H$ R-symmetry of such an irreducible   SCFT  coincides with the $SU(2)_C\times SU(2)_H$ R-symmetry of the UV gauge theory.  In the brane construction the R-symmetry is realized geometrically as spacetime rotations: $SU(2)_C$ rotates $x^{789}$ and $SU(2)_H$ rotates $x^{456}$. 3d $\cN=4$ gauge theories admit moduli spaces of vacua: the Coulomb branch and the Higgs branch.\footnote{Mixed branches can emerge at submanifolds of the Higgs and Coulomb branch.} These are hyperk\"ahler manifolds   invariant under $SU(2)_H$
 and $SU(2)_C$ and acted on by a group of isometries $G_C$ and $G_H$ respectively.   $G_H$ is manifest in the UV definition of the SCFT  and is realized as  the  flavour symmetry acting on the hypermultiplets,  while only the Cartan subalgebra of $G_C$ is manifest in the UV. Each  $U(1)$ gauge group factor gives rise to a  manifest $U(1)$ global symmetry, known as a topological symmetry,   which acts  on the Coulomb branch. The  abelian symmetry acting on the Coulomb branch can be  enhanced  to a non-abelian $G_C$ symmetry when conserved currents associated to the roots of $G_C$ can be  constructed with monopoles operators.\footnote{$G_C$ maps to the flavour symmetry acting on the hypermultiplets of the mirror theory.}  
  The non-trivial, irreducible SCFT sits at the intersection of the Higgs and Coulomb branch where the R-symmetry is enhanced to $SU(2)_C\times SU(2)_H$. The IR SCFT   inherits a $G_C\times G_H$ global symmetry. In the brane realization, the Coulomb branch corresponds to the motion of D3-branes along $x^{789}$ while the Higgs branch to the motion of D3-branes along $x^{456}$.\footnote{The brane realization makes it clear why  $\sum_{i=1}^{\hat P} M_i\geq 2$ is required for complete Higgsing in circular quivers. Indeed, unless there are two D5-branes, D3-branes segments cannot be detached from the NS5-branes.}

\begin{figure}[h!]
\centering
\includegraphics[scale=0.7]{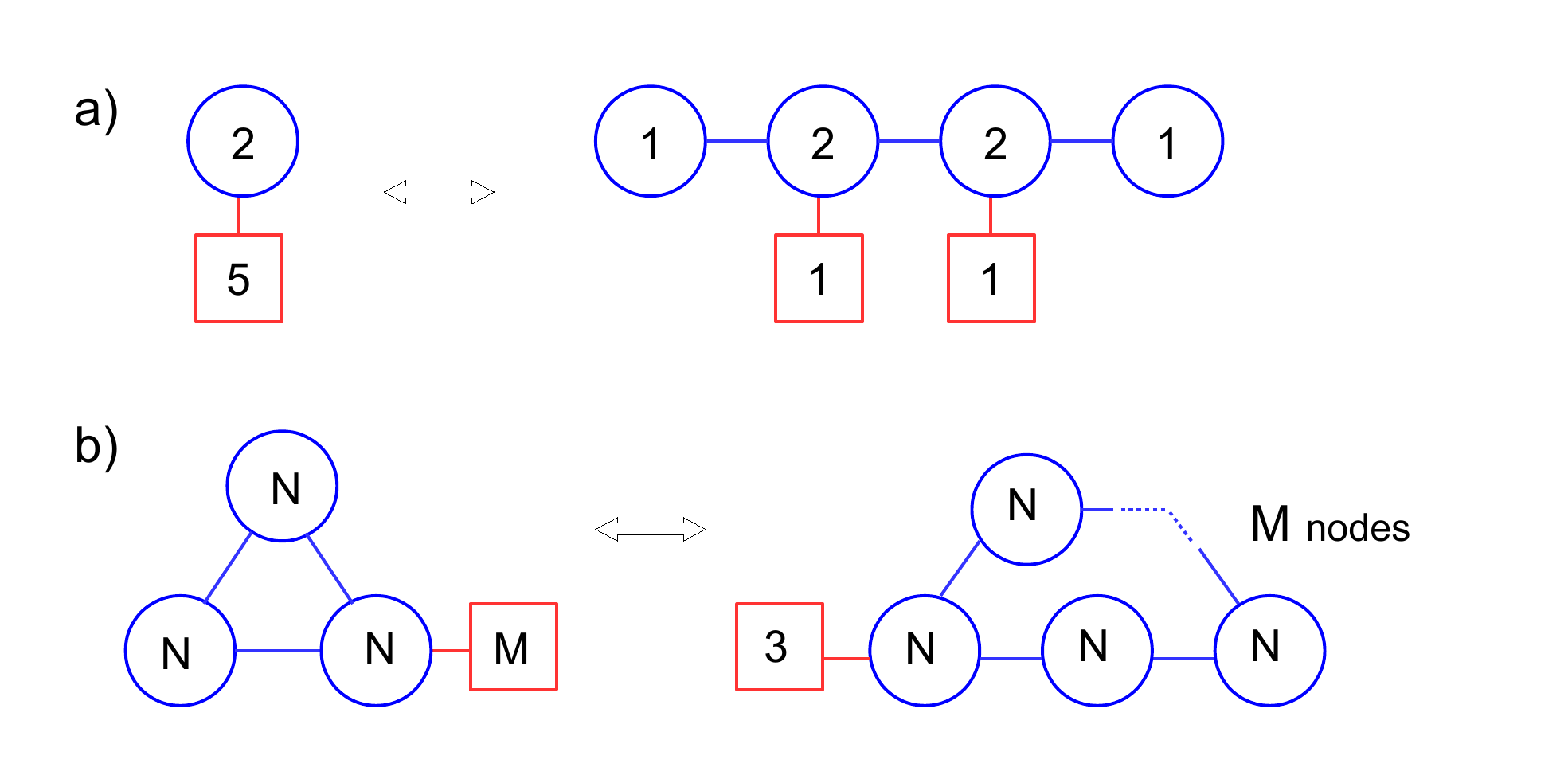} 
\vskip -0.5cm
\caption{\footnotesize a) A pair of mirror dual linear quivers. b) A pair of mirror dual circular quivers.}
\label{mirrorquiv_ex}
\end{figure}

\smallskip
 The brane description of 3d $\cN=4$ UV gauge theories  gives an elegant   realization of mirror symmetry \cite{Hanany:1996ie}, whereby two 
different UV gauge theories flow to the same nontrivial SCFT in the IR with the roles of $SU(2)_C$ and $SU(2)_H$ exchanged.
 Mirror symmetry  is realized as S-duality  in Type IIB string theory combined with   a spacetime rotation that sends $x^{456}$ to $x^{789}$ and $x^{789}$ to $-x^{456}$, which exchanges $SU(2)_C$ with $SU(2)_H$. The combined transformation, which we will refer as S-duality for brevity,  maps the class of brane configurations we have discussed to itself. Given the  brane configuration  corresponding to a UV 3d $\cN=4$  gauge theory, the mirror dual gauge theory is obtained by analyzing the low energy dynamics of the S-dual  brane  configuration. The mirror UV gauge theory can be read by rearranging the branes along the $x^3$ direction, possibly using Hanany-Witten moves \cite{Hanany:1996ie} involving the creation/annihilation of    a D3-brane  when an NS5-brane crosses a D5-brane,  to bring the S-dual brane configuration to a configuration where the low energy gauge theory can be read using the rules summarized above. 
 This transformation preserves the type of quiver, and thus the mirror of a linear quiver is a linear quiver and the mirror of a circular quiver is a circular quiver.\footnote{The irreducibility condition  of the IR SCFT is preserved under mirror symmetry, except for a circular quiver with a single node, whose mirror dual has a single fundamental hypermultiplet}.   Examples of mirror-dual pairs of quivers are given in figure \ref{mirrorquiv_ex}.
 
\smallskip
  An $\cN=4$  SCFT in flat space    admits  canonical relevant deformations   preserving 3d 
${\cal N}=4$ Poincar\'e supersymmetry. These deformations are associated to the flavour symmetries
$G_C\times G_H$  acting on the Coulomb  and  Higgs branches  of the SCFT. In a UV realization of the SCFT, these deformations   couple  to a triplet of mass  and FI parameters, which transform in the  $(\bf{3},\bf{1})$ and $(\bf{1},\bf{3})$ of $SU(2)_C\times SU(2)_H$. Mass and FI deformations are  obtained by deforming the UV theory with supersymmetric background vector multiplets in the Cartan of  $G_H$ and  supersymmetric  background twisted vector multiplets\footnote{In twisted multiplets   the roles of $SU(2)_C$ and $SU(2)_H$ are exchanged.} in the Cartan of  $G_C$ respectively. In the brane realization, these parameters are represented by the positions of five-branes. The position of the $i$-th D5-brane along $x^{789}$ corresponds to a mass deformation $\vec m_i$ while the position of the   $i$-th NS5-brane along $x^{456}$ corresponds to an FI parameter $\vec \xi_i$.  
Mass and FI parameters are exchanged between mirror dual theories. Indeed, in the brane realization of mirror symmetry through S-duality  the roles of the NS5 and D5-branes are exchanged.
 The positions of the 5-branes in the $x^3$ direction are irrelevant in the infrared 3d SCFT. For instance, the separation between two consecutive NS5-branes is inversely proportional to the coupling $g_{\rm YM}^2$ of the effective low-energy 3d SYM theory living on the D3-branes stretched between the two NS5-branes. In the deep IR, where the Yang-Mills coupling diverges,   the dependence on $g_{\rm YM}^2$ disappears. 
 
 \smallskip
 Linear quivers that flow to irreducible, interacting SCFT's can be labeled  by two partitions of $N$ -- $\rho$ and $\hat\rho$ -- and are denoted by $T^\rho_{\hat\rho}[SU(N)]$ \cite{Gaiotto:2008ak}. Circular quivers flowing to interacting SCFT's  are labeled  also by   two partitions of $N$ and a positive integer $L$, and can be denoted by $C^\rho_{\hat\rho} [SU(N),L]$ \cite{Assel:2012cj}.\footnote{In this paper we will not need the explicit mapping between the data of the quivers and $\rho$, $\hat\rho$ and $L$, but present it here for completeness. It is based on the linking numbers of the D5-branes  $l_i$ and NS5-branes $\hat l_j$, which obey $\sum_i^k l_i=\sum_j^{\hat k}\hat l_j=N$, with $k$ and $\hat k$ the total numbers of D5/NS5-branes. For linear quivers we have that $N=  l_1+\ldots+l_{k}=   \underbrace{1+\ldots+1}_{M_1}\, +\, \underbrace{2+\ldots+2}_{M_2}\, +\, \ldots$ and  $\hat l_j = N_{j-1} - N_{j} + \sum_{s=j}^{\hat k} M_s$, where $N_{0}=N_{\hat k}=0$ and $M_{\hat k}=0$. For circular quivers it is the same but with $N_{\hat k}=L$ and non-zero $M_{\hat k}$.}
 Under mirror symmetry
 \beq
 T^\rho_{\hat\rho}[SU(N)]\Longleftrightarrow T_\rho^{\hat\rho}[SU(N)]
 \eeq
 and
 \beq
 C^\rho_{\hat\rho}[SU(N),L]\Longleftrightarrow C_\rho^{\hat\rho} [SU(N),L]\,,
 \eeq
 and the role of the two partitions are exchanged.  The Coulomb branch of these theories, and by mirror symmetry the Higgs branch,  describe the moduli space of monopoles in the presence of Dirac monopole singularities for   linear quivers and the moduli space of instantons on a  vector bundle  over an   ALE space for circular quivers.

\smallskip
 
 In this paper we give a brane realization of both classes of loop operators discussed in section \ref{sec:loops} and put forward an algorithm that produces a map between  loop operators of  mirror dual theories.

 \subsection{Brane Realization Of Wilson Loop Operators}
\label{sec:Branes}

A key ingredient in our derivation of the mirror map of loop operators is identifying a brane realization of Wilson loop operators, which are  labeled by a representation $\scR$ of the gauge group. 
  Inserting a supersymmetric Wilson loop operator in a 3d $\cN=4$ linear or circular quiver gauge theory admits a simple brane interpretation, obtained by enriching the setup in  \cite{Hanany:1996ie}. The construction we   propose extends to 3d $\cN=4$ gauge theories the realization of Wilson loops by branes in 4d $\cN=4$ SYM in \cite{Gomis:2006sb}\cite{Gomis:2006im} (see also \cite{Drukker:2005kx}\cite{Yamaguchi:2006tq}).

\begin{figure}[th]
\centering
\includegraphics[scale=0.7]{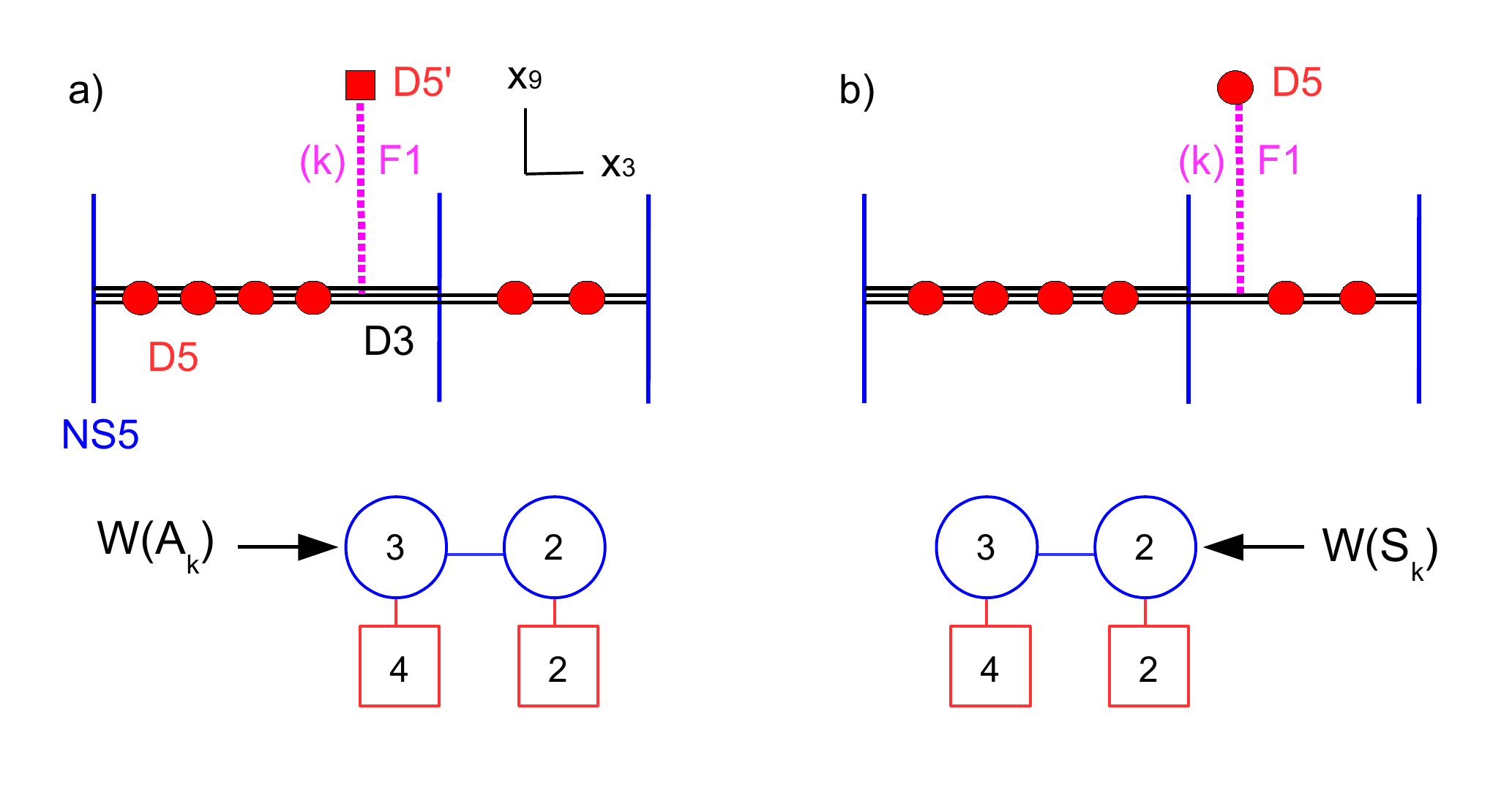}
\vspace{-0.5cm}
\caption{\footnotesize{a) Brane configuration with $k$ F1-strings ending on a D5'-brane, realizing the insertion of a Wilson loop in the $\scA_k$ representation of the $U(3)$ node of a linear quiver. Here $1 \le k \le 3$. b) Brane configuration with $k$ F1-strings ending on a D5-brane, realizing the insertion of a Wilson loop in the $\scS_k$ representation of the $U(2)$ node of the same linear quiver. }}
\label{WilsonLoopsEx}
\end{figure}

\smallskip
We start with the brane realization of a supersymmetric Wilson loop in the   $k$-th antisymmetric representation of a $U(N)$ gauge group factor in the quiver, which we denote  by ${\cal A}_k$. Such an operator insertion is realized by   adding $k$ F1 strings  stretched in the $x^9$ direction   ending at one end on the $N$ D3-branes where the $U(N)$ gauge group is supported  and at the other end on a D5' brane, defined as a D5-brane stretched in the  $x^{045678}$ directions.\footnote{Adding the D5'-brane does not break any further symmetries beyond those broken by the F1-strings.} The brane configuration realizing such a Wilson loop is given in table \ref{F1config}.
 The array of fundamental strings  ends between the two NS5-branes over which the N D3-branes are suspended. 
This brane setup is depicted in the example of figure \ref{WilsonLoopsEx}-a.
\smallskip

 \begin{table}[h!]
\begin{center}
\begin{tabular}{|c||c|c|c|c|c|c|c|c|c|c||}
  \hline
      & 0 & 1 & 2 & 3 & 4 & 5 & 6 & 7 & 8 & 9 \\ \hline
  D3  & X & X & X & X &   &   &   &   &   &   \\
  D5  & X & X & X &   & X & X & X &   &   &   \\
  NS5 & X & X & X &   &   &   &   & X & X & X \\
  F1  & X &   &   &   &   &   &   &   &   & X \\  
  D5'  & X &   &   &   &  X & X  & X  & X  &  X  &  \\  \hline
\end{tabular}
\caption{\footnotesize Brane Realization of Wilson Loops in 3d $\cN=4$ Gauge Theories.}
\label{F1config}
\end{center}
\end{table}

This enriched brane configuration is supersymmetric: it preserves the SQM$_W$ subalgebra of the 3d $\cN=4$ super-Poincar\'e algebra discussed in section \ref{sec:loops}. Quantization of the  open strings stretched between the D3 and the D5'-branes gives rise to 1d  complex fermions $\chi$ localized on the line defect    and   furnish the dimensional reduction of the 2d $\cN=(0,4)$ gauged Fermi multiplet
\beq
\int dt\, \chi^\dagger \left[i\partial_t+(A_0+\sigma - m)\right]\chi\,,
\eeq
where $x^9=m$ is the location of the D5'-brane.   The fermions are in the bifundamental representation of $U(N)\times U(1)$, where $U(N)$ is the gauge group on the D3-brane segment where the F1-strings end and the $U(1)$ is the flavour symmetry associated to the D5'-brane. The fermions can be integrated out exactly and yield\footnote{Here we put the system on a circle of length $\beta$.}
\beq
Z_0={1\over \sqrt{\hbox{det U}}}\sum_{l=0}^N e^{ - i \beta m l} \hbox{Tr}_{\scA_l} U\,,
\eeq
where $U=Pe^{i\int_0^\beta dt A_0+\sigma}$ is the supersymmetric holonomy operator. As it stands there is a global anomaly for the $U(1)\subset U(N)$, since under large gauge transformations   $Z_0\rightarrow -Z_0$. Our brane realization of the Wilson loop, however, engineers a bare supersymmetric Chern-Simons term at level $k=1/2$, that is $1/2 \int (A_0+\sigma)$, which precisely cancels the offending factor $\hbox{det(U)}^{-1/2}$, and the brane system is anomaly free.\footnote{This coupling is obtained by inserting the flux produced by the D5'-brane on  the non-abelian Chern-Simons term on the worldvolume of the D3-branes. This is T-dual to the haf-integral Chern-Simons term discussed in \cite{Danielsson:1997wq}.}
 Integrating out the fermions in the presence of the Chern-Simons term  therefore yields
\beq
Z=\sum_{l=0}^N e^{-i\beta ml} \hbox{Tr}_{\scA_l} U\,.
\eeq	

\smallskip
The presence of $k$ F1-strings stretched between the D3 and D5'-branes is represented in the gauge theory by the insertion of $k$ creation operators for these fermions in the past and   $k$ annihilation operators in the future. Physically, these operators insert a charged probe into the gauge theory. Integrating out these fermions inserts a supersymmetric Wilson loop operator in the $k$-th antisymmetric representation \cite{Gomis:2006sb}
\beq
{\vev{\left(\chi^\dagger(0)\right)^k \chi^k(\beta)}\over Z}=e^{-i\beta mk} \hbox{Tr}_{\scA_k} U\,.
\eeq

\begin{figure}[th]
\centering
\includegraphics[scale=0.6]{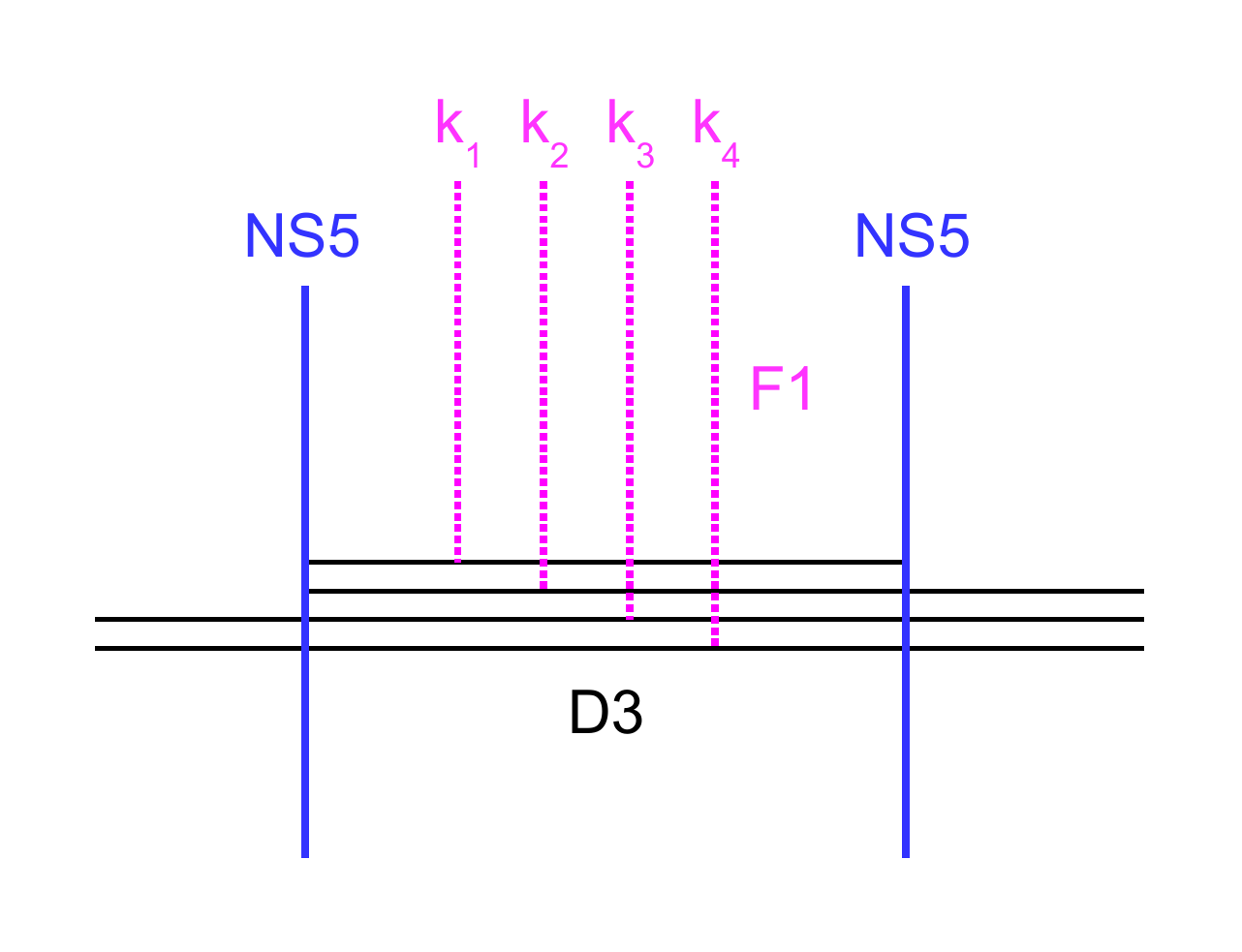}
\caption{\footnotesize{ Configuration with stacks of $k_j$ F1-strings, $1\le j \le 4$, ending on each of the four D3-branes, associated to a weight $(k_1,k_2,k_3,k_4)$ of a representation of $U(4)$. }}
\label{F1pattern}
\end{figure}

\smallskip
The weights of the  $k$-th  antisymmetric representation of $U(N)$ admit an elegant description in the brane construction. We must distribute $k$ F1-strings among $N$ D3-branes (all $k$ F1-strings terminate at the other end on a single D5'-brane). To a pattern of $k$ F1-strings where $k_j$ strings end on the $j$-th  D3-brane (see figure \ref{F1pattern}) we associate a set of $N$ non-negative integers $\{k_j\}$ obeying $k= k_1+k_2+ \cdots + k_N$, with $k_j \ge 0$ for all $j$. However, not all positive integers $k_j$ are allowed. There can be at most one F1-string stretched between a D3-brane and a D5'-brane. This is the so-called s-rule \cite{Hanany:1996ie}, and is a constraint that follows from Pauli's exclusion principle \cite{Bachas:1997kn}. Therefore, the allowed configurations are described by a collection of $N$ non-negative integers $\{k_j\}$ with the constraint that $k_j\leq 1$. This set of configurations is in one-to-one correspondence with the weights of the  $k$-th  antisymmetric representation of $U(N)$, i.e. of  ${\cal A}_k$.

We now turn to a Wilson loop in the $k$-th  symmetric representation of $U(N)$, which we denote by ${\cal S}_k$. Inserting a Wilson loop in the  $k$-th  symmetric representation  is realized by adding $k$ F1 strings  stretched in the $x^9$ direction   ending at one end on the $N$ D3-branes where the $U(N)$ gauge group is supported  and at the other end on a D5-brane stretched in the $x^{012456}$ directions and localized   in the $x^9$ direction. The array of fundamental strings   ends between the two NS5-branes over which the N D3-branes are suspended. This setup is illustrated in figure \ref{WilsonLoopsEx}-b. In this case the charged probe particle inserted by the Wilson loop can be though of as arising  from a very heavy hypermultiplet,  represented by adding a D5-brane to the theory and then 
 taking the D5-brane far away from the stack, thus giving it a large mass and making the hypermultiplet fields 
 nonrelativistic. 
Integrating out the heavy hypermultiplet in the presence of $k$ heavy  insertions yields a supersymmetric Wilson loop operator in the $k$-th symmetric representation  \cite{Gomis:2006sb,Gomis:2006im}.\footnote{In \cite{Gomis:2006im} the heavy charged particle was obtained by going to the Coulomb branch while
here by giving a large mass to a hypermultiplet.}
 
 \smallskip
The weights of the  $k$-th  symmetric  representation of $U(N)$ admit an elegant description in the brane construction. We again   distribute $k$ F1-strings among $N$ D3-branes (all $k$ F1-strings terminate at the other end on a single D5-brane). To a pattern of $k$ F1-strings where $k_j$ strings end on the $j$-th D3-brane (see figure \ref{F1pattern}) we associate a set of non-negative integers $\{k_j\}$ obeying $k= k_1+k_2+ \cdots + k_N$, with $k_j \ge 0$ for all $j$. In this case   an arbitrary number of F1-strings can be stretched between the  D5 and a D3-brane.
Therefore the set of configurations is in one-to-one correspondence with the weights of the  $k$-th  symmetric representation of $U(N)$, i.e. of  ${\cal S}_k$.

\smallskip

Our brane construction can be easily generalized to Wilson loops in the tensor product of an arbitrary number of symmetric and antisymmetric representations of $U(N)$: $\cR=\otimes_{a=1}^{d} \scS_{k^{(a)}}   \otimes_{b=1}^{d'} \scA_{l^{(b)}}$.
This requires considering F1-strings stretched between $d$ D5-branes and $d'$ D5'-branes  and   the $N$ D3-branes that support the gauge group.   For the above mentioned representation  $k^{(a)}$ F1-strings  must emanate  from the  $a$-th  D5-brane and $l^{(b)}$ F1-strings from the  $b$-th  D5'-brane. Integrating out the massive charged particles produced by this configuration yields a Wilson loop in the desired representation. Furthermore, the set of allowed F1-string configurations, labeled by $\{k_j^{(1)},\ldots,  k_j^{(d)},l_j^{(1)},\ldots,  l_j^{(d')} \}_{j=1..N}$ with $k^{(a)}_j \in \bN$, $l^{(b)}_j \in\{0,1\}$  and such that $\sum_j  k_j ^{(a)}=k^{(a)}$ and  $\sum_j  l_j ^{(b)}=l^{(b)}$ yields precisely all the weights in the representation $\cR=\otimes_{a=1}^{d} \scS_{k^{(a)}}   \otimes_{b=1}^{d'} \scA_{l^{(b)}}$ of $U(N)$. More precisely a configuration is associated to a weight $w = (w_1,w_2, \cdots , w_N)$ in the orthogonal basis with $w_j =  \sum_{a=1}^d  k_j ^{(a)} + \sum_{b=1}^{d'} l_j ^{(b)}$.  Our analysis can be summarized in table \ref{D5D5'arrays}. 
\begin{table}[h!]
\begin{center}
\begin{tabular}{|c|c|c|}
  \hline
     number of D5 & number of D5' & rep. $\cR$  \\ \hline
  1  & 0 & $\scS_{k}$     \\
  0  & 1 & $\scA_{k}$   \\
  0  & 2 & $\scA_{l^{(1)}} \otimes \scA_{l^{(2)}}$   \\
  1  & 1 &   $\scS_{k}\otimes  \scA_{l}$   \\  \hline
\end{tabular}
\caption{\footnotesize D5/D5'-brane arrays inserting a Wilson loop in a representation $\scR$ of $U(N)$. The integers $k^{(a)}$ and $l^{(b)}$ correspond to numbers of F1-strings stretched between a single 5-brane and the $N$ D3-branes as explained in the text.}
\label{D5D5'arrays}
\end{center}
\end{table}

\begin{figure}[th]
\centering
\includegraphics[scale=0.7]{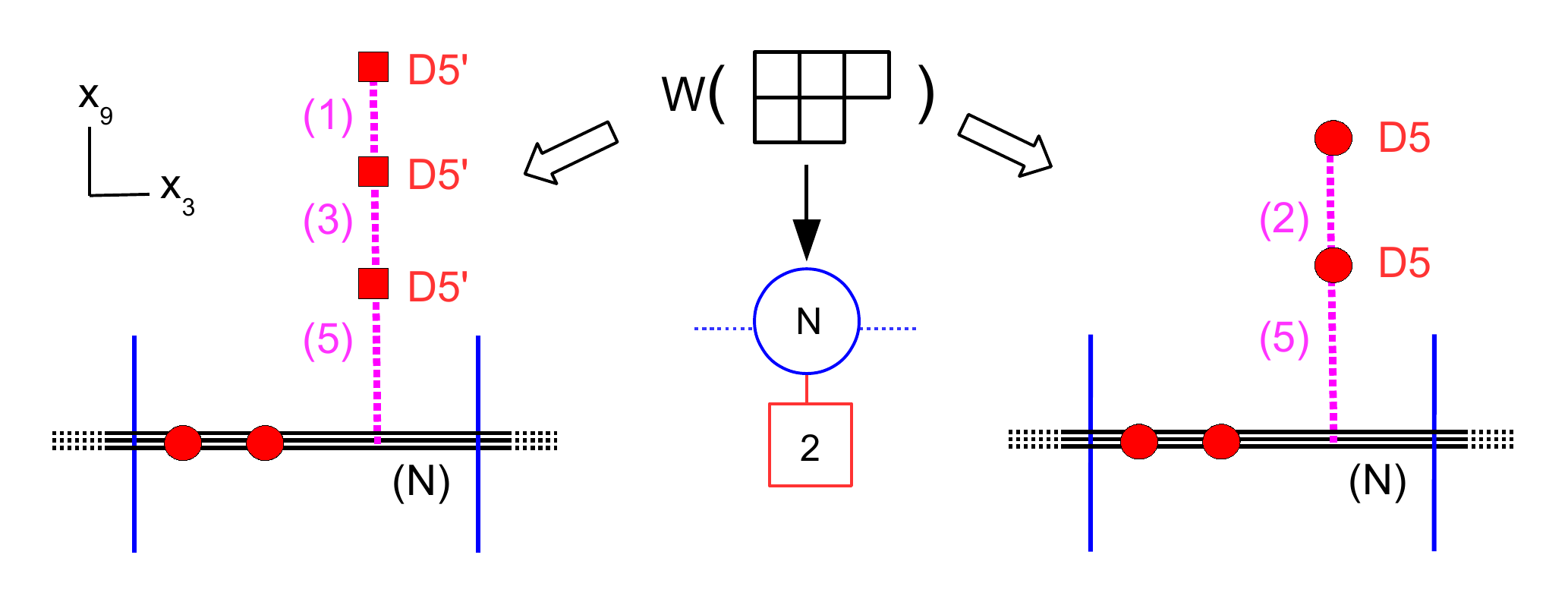}
\caption{\footnotesize{The Wilson loop in the representation of $U(N)$ labeled by the Young tableau of the figure can be realized  in two ways. On the left:  1, 2 and 2 strings emanate respectively from three coincident D5'-branes (along $x^3$). On the right: 2 and 3 strings emanate respectively from two coincident D5-branes. }}
\label{WGenRep}
\end{figure}

   In order to describe Wilson loops in the above mentioned representation we have to place the $d$ D5 and $d'$ D5'-branes at different positions in the $x^3$ and $x^9$ directions. The separation in the $x^9$ direction is not essential at this stage but plays a role when we identify  the S-dual brane configuration and the mirror dual Vortex loop. The separation in the $x^3$ direction is more crucial: if two D5-branes sit at the same $x^3$ position, the pattern of F1-strings is more complicated   since strings can now break and be stretched between the two D5-branes preserving the same amount of supersymmetry. In this case we expect that the brane configuration would insert a Wilson loop in an irreducible representation of $U(N)$, as in  \cite{Gomis:2006sb}\cite{Gomis:2006im}. An irreducible representation $\cR$ of $U(N)$ labeled by a Young diagram with $M$ rows and $L$ columns 
   \smallskip
   \begin{equation}
 \mathtikz[semithick,x=1em,y=1em]{
   \foreach \X/\Y in {2/0,2/-1,2/-2,2/-3,1/-4}  \draw ( 0,\Y) -- (\X,\Y);
   \foreach \X/\Y in {0/-4,1/-4,2/-3}       \draw (\X,0) -- (\X,\Y);
   \node          at (3,-1)          {$\cdots$};
   \foreach \X/\Y in {4/-2,5/-2,6/-1}       \draw (\X, 0) -- (\X,\Y);
   \foreach \X/\Y in {6/0,6/-1,5/-2}        \draw ( 4,\Y) -- (\X,\Y);
   \node [anchor=base] at (0.5,-4.9) {};
   \node [anchor=base west] at (1,-3.9) {};
   \node [anchor=base] at (4.5,-2.9) {};
   \node [anchor=base west] at (5,-1.9) {};
   \draw (0,2pt) -- (0,6pt) -- (6,6pt) node [midway,yshift=6pt] {$L$}
-- (6,2pt);
   \draw (-2pt,0) -- (-6pt,0) -- (-6pt,-4) node [midway,xshift=-6pt]
{$M$} -- (-2pt,-4);
 }
 \nonumber
\end{equation}  is realized either by having $|\cR|$ F1-strings end  on $M$ coincident D5-branes (in the $x^3$ direction)  or  $|\cR|$ F1-strings end   on $L$ coincident  D5'-branes
   away from the main stack, where $|\cR|$ is the number of boxes in the Young diagram corresponding to $\cR$.
Figure \ref{WGenRep} provides an example of the two possible brane configurations realizing a Wilson loop.

   \smallskip

In the brane realization   we must   specify between which two consecutive NS5-branes the F1-strings end. This determines  in which gauge group the Wilson loop is inserted. While the F1-strings cannot be moved across an NS5-brane without changing the Wilson loop operator,  they can be freely moved across a D5-brane  without changing the Wilson loop operator in the IR if the D5-brane has the same number of D3-branes ending on the left and the right, which is the case in the canonical brane configuration in figure \ref{linquivbrane}. The corresponding S-dual statement, that D1-branes can be moved across an NS5-brane  with the same number of D3-branes ending on the left and the right but that D1-branes cannot be moved across a D5-brane without changing the IR dynamics,  will play an important role in unraveling the mirror map of loop operators.

\subsection{ Mirror Of Wilson Loops From S-duality}
\label{ssec:LoopSduality}

After having found a brane realization of Wilson loops, we     now make use of the fact that mirror symmetry corresponds to  S-duality in the Type IIB brane realization to derive the mirror dual   of supersymmetric Wilson loop operators. Here we loosely call S-duality what is really S-duality combined with the rotation that sends $x^{456}$ to $x^{789}$ and $x^{789}$ to $-x^{456}$, so that D5-branes and NS5-branes get exchanged. We shall see that the information about the representation of the Wilson loop is encoded in the discrete data of a 1d $\cN=4$ SQM  quiver quantum mechanics gauge theory. 

\smallskip
We have found that   Wilson loop insertions can be realized by a brane configuration  with F1-strings stretched between the $N$ D3-branes and D5-branes and/or D5'-branes, oriented as in table \ref{F1config}. Under S-duality the D5 and D5'-branes become NS5 and NS5'-branes, while the F1-strings become D1-branes, oriented as in table \ref{D1config}.  This brane configuration   preserves the SQM$_V$ subalgebra of the 3d $\cN=4$ super-Poincar\'e algebra discussed in section \ref{sec:loops}.\footnote{S-duality in Type IIB string theory acts nontrivially on the supersymmetry charges.} A 1d $\cN=4$  SQM$_{V}$ invariant   theory is supported on the D1-branes \footnote{D1-branes span an interval in the $x^6$ direction. The low-energy effective theory is thus one-dimensional.}.
 The $U(1) \times SU(2)$ R-symmetry of SQM$_V$ is identified with   $SO(2)_{12} \times SO(3)_{789}$ rotations in spacetime.\footnote{For later convenience, we have combined the R-symmetry $R_H$ realized by $SO(2)_{45}$ with the with the $SO(2)_{12}-SO(2)_{45}$ commutant to define a new R-symmetry: $SO(2)_{12}$.} The remaining $SO(2)_{12}-SO(2)_{45}$ isometry (a diagonal $SO(2)$) is mapped to a flavor symmetry of the SQM theory, which is the commutant of SQM$_V$ in the 3d $\cN=4$ supersymmetry algebra.
 
   \begin{table}[h!]
\begin{center}
\begin{tabular}{|c||c|c|c|c|c|c|c|c|c|c||}
  \hline
      & 0 & 1 & 2 & 3 & 4 & 5 & 6 & 7 & 8 & 9 \\ \hline
  D3  & X & X & X & X &   &   &   &   &   &   \\
  D5  & X & X & X &   & X & X & X &   &   &   \\
  NS5 & X & X & X &   &   &   &   & X & X & X \\
  D1  & X &   &   &   &   &   &  X  &   &   &  \\  
  NS5'  & X &   &   &   &  X & X  &   & X  &  X  &  X \\  \hline
\end{tabular}
\caption{\footnotesize Brane array with D1-branes}
\label{D1config}
\end{center}
\end{table}

We will now exhibit    that Wilson loops in a UV 3d $\cN=4$ gauge theory are mirror to loop operators defined by     
1d $\cN=4$ SQM$_V$ invariant quiver gauge theories coupled   to the mirror 3d $\cN=4$ gauge theory. Likewise, the Wilson loops of the mirror theory are mapped to 1d $\cN=4$ quiver gauge theories coupled to the original 3d $\cN=4$ gauge theory. The 1d
 $\cN=4$ quiver gauge theories and the way they couple to the 3d $\cN=4$ gauge theory are found by identifying  the IR gauge theories living on the  D1-branes in the S-dual brane configuration.  

\smallskip

\begin{figure}[h!]
\centering
\includegraphics[scale=0.55]{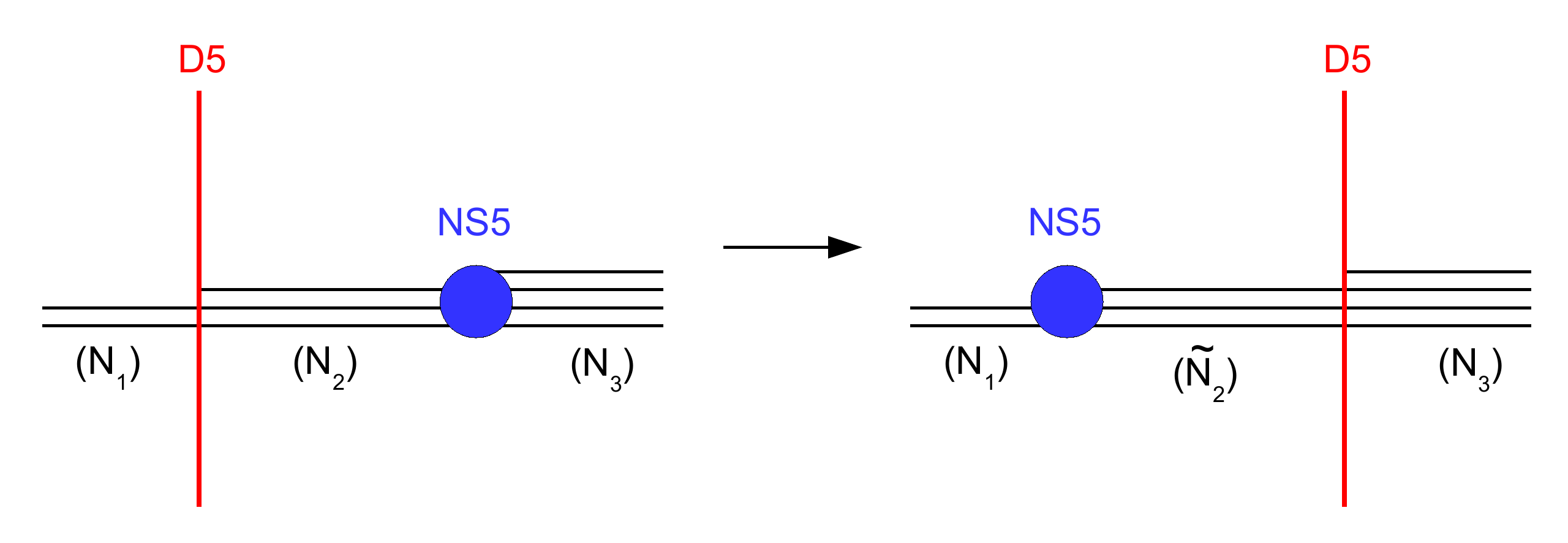}
\caption{\footnotesize{ Hanany-Witten five-brane move. Initially $N_1 < N_2$. After the D5 has crossed the NS5, the number of D3-branes between them is $\ti N_2 = N_1 + N_3 - N_2 + 1$. The D3-excess on the D5-brane sides goes down from $N_2 - N_1$ to $N_2 - N_1 - 1$.}}
\label{HWmove}
\end{figure}

How can the mirror operator to a given Wilson loop be constructed? First we perform S-duality on the brane configuration realizing a Wilson loop in a 3d $\cN=4$ gauge theory.  In the absence of the  F1-strings, the mirror 3d $\cN=4$ theory is found by rearranging the S-dual brane configuration so as to bring it to the canonical frame explained at the beginning of section \ref{sec:loopsbranes}, where the mirror gauge theory can be easily read. After S-duality, the number of D3-branes on the left and on the right of a D5-brane need not be the same  while that number is the same for every NS5-brane, as before S-duality the D5-branes had the same number of D3-branes on both sides. In order to bring the brane configuration to one where the gauge theory can be read we must move the D5-branes with an excess of D3-branes in the direction of the excess and make it pass through NS5-branes. Every such move, exchanging a D5-brane with an NS5-brane, results in the creation of a D3-brane \cite{Hanany:1996ie} on the side of the D5-brane that had a smaller number of D3-branes, thus diminishing the D3-brane excess. See figure \ref{HWmove}.  This process must be continued until the number of D3-branes on each of the D5-branes is the same on the left and on the right.  From this final configuration we can read the mirror dual gauge theory.

\smallskip

\begin{figure}[h!]
\centering
\includegraphics[scale=0.55]{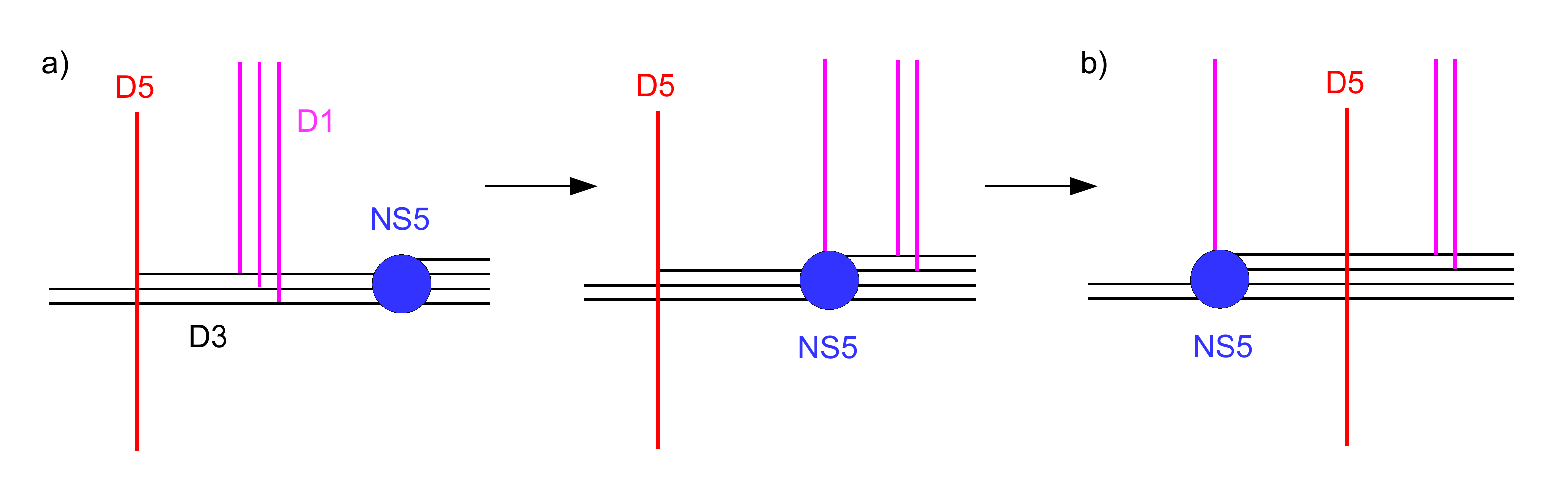}
\caption{\footnotesize{ a) The D5-brane has to be moved to the right of the NS5-brane to reach the canonical brane configuration. The D1-branes in between must first be moved across the NS5-brane themselves. b) After moving the D1 and D5-branes, stacks of D1-branes end up attached to the NS5-brane and moved away from the main brane configuration along transverse directions (transverse to the picture). }}
\label{D1figure}
\end{figure}

Insertion of  a Wilson loop operator in a gauge group factor requires adding   F1-strings ending between the two consecutive NS5-branes where the D3-branes supporting that gauge factor are suspended. After S-duality, we have the same S-dual brane configuration  as before but   now with extra D1-branes ending between a pair of consecutive D5-branes. It is from this S-dual brane configuration that we   read off the 1d $\cN=4$ quiver gauge theory living on the D1-branes and its coupling to the mirror dual gauge theory. In order to read   the mirror description of the original Wilson loop we once again move the D5-branes in the direction of D3-brane excess to bring the brane configuration   to the canonical one, where the 3d $\cN=4$ mirror gauge theory can be extracted.
During this process we do not allow a D5-brane to cross  a D1-brane. 
In the simplest situations,  the D5-branes can be moved to the canonical configuration of the mirror dual theory without having to move the D1-branes. Then we can directly read off the 1d $\cN=4$  quiver theory from the final brane configuration with D1-branes (see below).  In   more complicated examples, we have a situation when a D5-brane must be moved past an NS5-brane while   D1-branes stand between them. For definiteness let us imagine that the D5-brane must be moved from the left to the right of an NS5-brane, with the D1-branes in between, as in figure \ref{D1figure}-a.

In this situation the D1-branes  together with the NS5/NS5'-branes to which they are attached at the other end  must be moved first to the right of the NS5-brane. Here we must distinguish two situations: if the number  of D3-branes ending on both sides of the NS5-brane are equal, then the D1-branes can be moved across it without anything special happening (in the S-dual picture,   F1-strings can cross a D5-brane  with the same number  of D3-branes on both sides in the same trivial way). 
However the NS5-brane can also have an excess of D3-branes on the right.\footnote{The NS5-brane cannot have an excess of D3-branes on the left, where the D5-brane is coming from. This can be understood as follows. After S-duality of the original configuration, the number  of D3-branes on both sides of the NS5-brane are equal. During the brane rearrangement some D5-branes can move across it but only in one direction (because D5-branes do not cross between themselves during the rearrangement). In our example, D5-branes cross the NS5-brane from the left to the right. For each D5-brane crossing the NS5, the Hanany-Witten rule implies that the excess of D3-branes on the right of the NS5 increases by one, so that along the process  the NS5 develops an excess of D3-branes on the right. In the other case when D5-branes cross the NS5 from the right to the left, the NS5 develops an excess of D3-branes on the left. }   
Then the NS5-brane   develops a D3-brane spike \cite{Callan:1997kz} and some D1-branes may be moved along the spike  so that they end on the NS5-brane, and can then be moved far away from the main stack along transverse directions. If there were $k$ D1-branes in the original stack, they could split into $k_1$ D1-branes moved from the left to the right of the NS5-brane along the D3-branes and $k_2$ D1-branes moved along the NS5-brane spike and away from the main stack, with $k=k_1+k_2$. Then the D5-brane can be moved across the NS5-brane as usual, without encountering the D1-branes. This is illustrated in figure \ref{D1figure}-b. 

\smallskip

The choice of splitting $k=k_1+k_2$ may be constrained, but generically several splittings may be allowed.
The idea of the mirror map is that the initial Wilson loop is mapped to the sum of the Vortex loops realized by the possible final brane configurations, weighted with coefficients. The precise rules to derive which final brane configurations can be reached (and their weighting coefficients) are not obvious. Intuitively the difficulty is related to the fact that the D3-spike is ``on the wrong side" of the NS5-branes, so that we cannot directly move the D1-branes along the spikes. We will see shortly that the situation is  under much better control starting from the D1-branes configurations realizing Vortex loops and $S$-dualizing to configurations realizing the mirror Wilson loops.
 
\smallskip
  
What are the final brane configurations? In addition to the D3, D5, NS5 system realizing the mirror dual theory, there are D1-branes ending on D3-branes on one side and on NS5 and/or NS5'-branes situed far away from the main configuration. Moreover there are also D1-branes ending on NS5-branes in the main stack. The physical interpretation of having $q$  D1-branes ending on an NS5-brane is as a background Wilson loop of charge $q$ for a  $U(1)$ global symmetry associated to the NS5-brane, which is a combination of the so-called topological symmetries of the 3d theory. 
These flavor Wilson loops   combine  with the Vortex loop realized by the D1-branes ending on D3-branes, which we describe now.
 
\smallskip
 
A final brane configuration  with  D1-branes ending  on D3-branes in one end and on NS5 and/or NS5'-branes on the other  can be used to give at least two descriptions of the mirror of a Wilson loop. This brane configuration can be thought of as a deformation of two other brane configurations, both of which are   conducive to reading off the 1d $\cN=4$ quiver gauge theory living on the D1-branes and its couplings to the bulk 3d $\cN=4$ gauge theory. The D1-branes can be either moved to the nearest NS5-brane to the left or to   the nearest NS5-brane to the right. In order to reach this configuration, from which the gauge theory description of the mirror loop operators can be read off, we must again ensure that no D1-strings cross D5-branes. This means that we have to move the D5-branes between the D1-branes and the nearest NS5-brane to the other side of that NS5-brane. In summary, we can realize the mirror loop operator either as:
 \begin{enumerate}
 \item Deformation of the coupled 3d/1d theory realized by the D1-branes when they end on the NS5-brane on the left.  
  \item Deformation of the coupled 3d/1d theory realized by the D1-branes when they end on the NS5-brane on the right.   \end{enumerate}
 These yield two dual descriptions of the same operator.
  
  \smallskip
   Once we   move the stack of D1-branes so that   it ends on a neighbouring  NS5-brane, we can read off the 1d $\cN=4$ quiver gauge theory  and how it couples to the 3d $\cN=4$ gauge theory. The 1d $\cN=4$ gauge theory  associated to a brane configuration is constructed by assigning: 
\begin{itemize}
\item A $U(k)$ vector multiplet to $k$ D1-branes suspended between an  NS5-brane and an NS5'-brane.
\item A $U(k)$ vector multiplet and an adjoint chiral multiplet\footnote{The adjoint chiral multiplet describes the position of the D1-branes in the $x^{12}$ directions (for NS5-branes), or in the $x^{45}$ directions (for NS5'-branes).} to $k$ D1-branes suspended between two NS5-branes or  two NS5'-branes.
\item Two chiral multiplets in the bifundamental and anti-bifundamental representations of $U(k_1)\times U(k_2)$ \footnote{The bifundamental representation of $U(k_1)\times U(k_2)$ is $({\bf k_1}, {\bf \bar k_2})$ and the anti-bifundamental is its complex conjugate.}  to an NS5-brane or NS5'-brane with $k_1$ D1-branes ending
on its left and $k_2$ D1-branes ending on its right.
\item A chiral multiplet in the bifundamental of $U(N_R)\times U(k)$ and an chiral multiplet in the bifundamental of 
$U(k)\times U(N_L)$ to  $k$ D1-branes ending on an NS5-brane which has $N_L$ D3-branes  ending on its left and $N_R$ D3-branes ending on its right.
\end{itemize}   

\begin{figure}[th]
\centering
\includegraphics[scale=0.6]{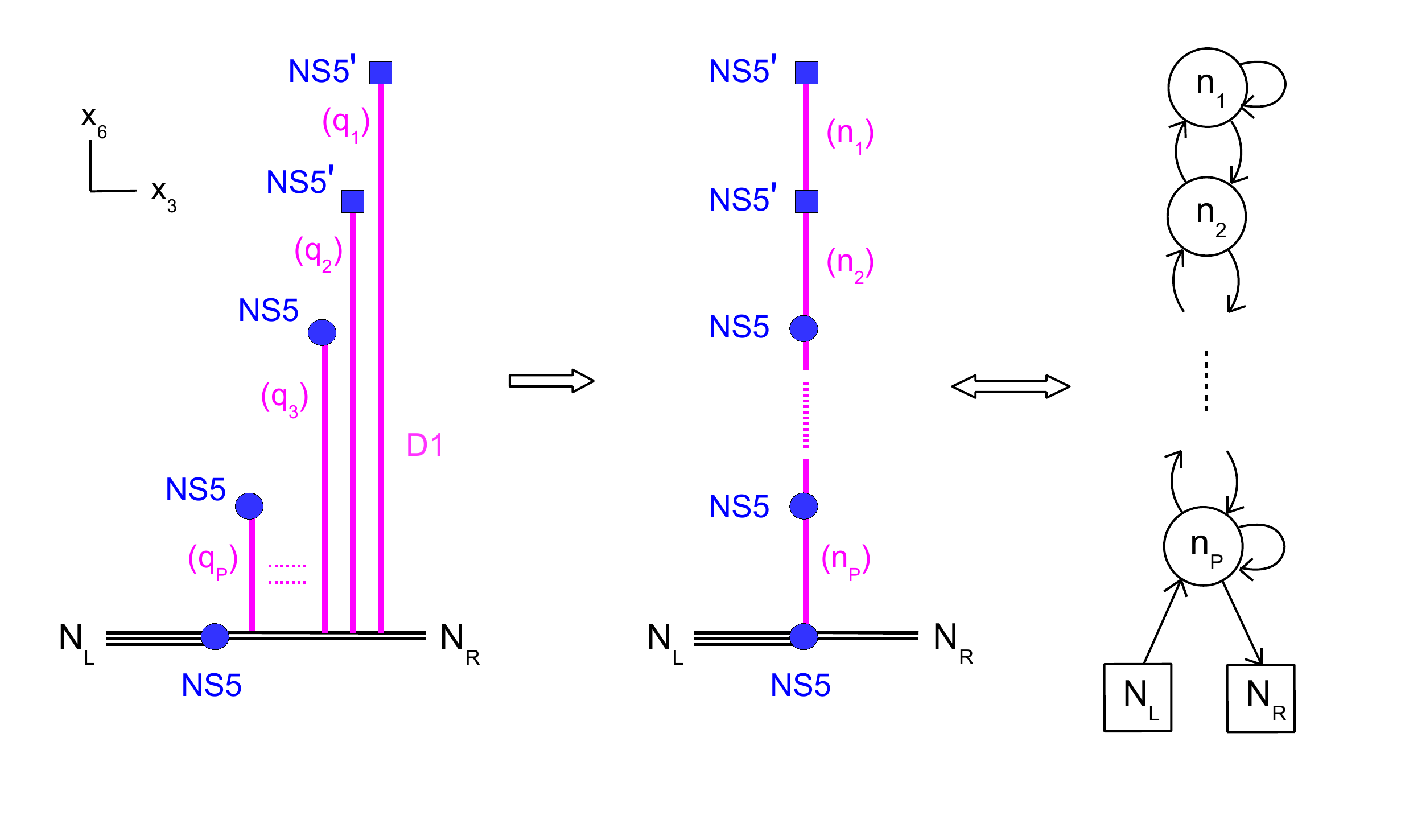}
\vspace{-1cm}
\caption{\footnotesize{To read the SQM quiver  we move the D1-branes to the closest NS5-brane on the left. The rank of nodes are given by $n_j = \sum_{i=1}^j q_i$, where $q_i$ is the number of D1-branes emanating from the $i$-th five-brane. The $q_i$ are trivially mapped to the $k^{(a)}$ and $l^{(b)}$ introduced in the text. NS5'-branes are denoted by blue squares, as a opposed to NS5-branes that are blue circles. }}
\label{QMrealization}
\end{figure}

\noindent
Figure \ref{QMrealization} summarizes the reading of the 1d $\cN=4$ quiver SQM gauge theory from the brane configuration, when moving the D1-strings to the nearest NS5-brane on the left.

\smallskip

The parameters of the 1d $\cN=4$ gauge theory living on the D1-branes admit a brane interpretation. The gauge coupling of  the vector multiplet  realized on D1-branes stretched between two five-branes (NS5 or NS5') is inversely proportional to the distance in the $x^6$ direction between the five-branes. The relative position in the $x^3$ direction between two consecutive five-branes (NS5 or NS5') determines the   1d $\cN=4$ FI parameter for the corresponding  gauge group factor. If we denote  the position of the NS5-brane in the main stack where the D1-strings originally end by $x^3_0$ and  $x^3_i$  the position of the $i$-th five-brane away from the main stack, the 1d FI parameters are given by $\eta_i= x_i^3-x_{i-1}^3$.
 In particular, if the D1-branes are moved to the right of  the  NS5-brane in the main stack along the D3-branes then the FI parameter for the first 1d gauge group factor is positive, while it is negative if we move the D1-branes to the left along the D3-branes.  Therefore, the 3d/1d gauge theories $1)$ and $2)$ described above   obtained after S-duality should be thought of as a deformation with $\eta_1>0$  of theory  $1)$ and a deformation with $\eta_1<0$ of theory $2)$.

\smallskip

  Consider for instance  the D1-brane configuration of figure \ref{QMrealization} near an NS5-brane in the main stack with $k^{(a)}$ D1-strings    emanating  from the $a$-th NS5-brane and $l^{(b)}$ D1-strings from the $b$-th NS5'-brane and ending on the $N_R$ D3-branes to the right of the NS5-brane in the main stack. There are also $N_L$ D3-branes to the left of this NS5-brane. 
To this brane configuration we can associate the  representation   $\cR=\otimes_{a=1}^{d} \scS_{k^{(a)}} \otimes_{b=1}^{d'} \scA_{l^{(b)}}$ of $U(N_R)$, where $d$ and $d'$ denote the numbers of NS5 and NS5'-branes from which the D1-branes emanate. This brane configuration can be thought of as a deformation of a 1d $\cN=4$ quiver gauge theory in figure \ref{QMrealization} with positive FI parameters. There are actually different dual descriptions of the 1d $\cN=4$ theories depending on the relative order of the $d$ NS5 and $d'$ NS5'-branes in the $x^6$ direction. Different relative positions give rise to different dual 1d $\cN=4$ descriptions of the same Vortex loop operator labeled by the representation  $\cR=\otimes_{a=1}^{d} \scS_{k^{(a)}}   \otimes_{b=1}^{d'} \scA_{l^{(b)}}$ of $U(N_R)$.  
Roughly speaking, these dual descriptions are related by a 1d $\cN=4$ version of Seiberg duality \cite{Seiberg:1994pq} (see also
\cite{Gadde:2013dda,Kim:2015fba}).
 
\smallskip
 
\begin{figure}[h]
\centering
\includegraphics[scale=0.7]{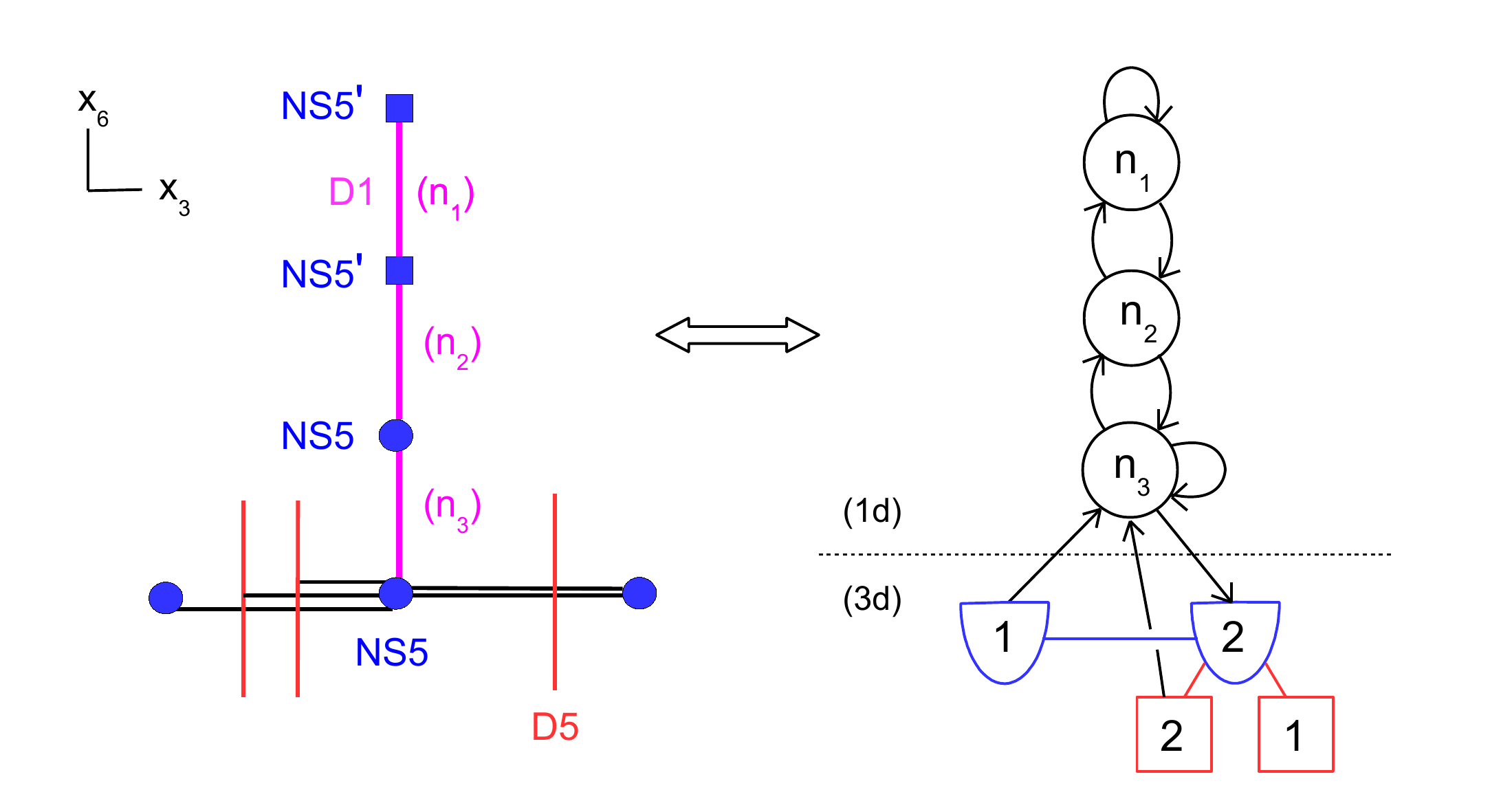}
\vspace{-0.5cm}
\caption{\footnotesize{Example of configuration with D1-branes and associated 1d $\scN=4$ quiver gauge theory, coupled to the 3d $\cN=4$ theory $T[SU(3)]$. Semi-squared nodes denote 3d gauge symmetries gauging 1d flavor symmetries. }}
\label{QMquiver}
\end{figure}

 We have described the 1d $\cN=4$ SQM$_V$ quiver theory living on the D1-branes. We must now explain how it is coupled to the 3d $\cN=4$ gauge theory. The idea is that the $U(N_L) \times U(N_R)$   flavor symmetry of the 1d $\cN=4$ theory is gauged with 3d bulk fields living on the D3-branes ending on the NS5-brane on the main stack. We must, however,     distinguish the D3-branes supporting dynamical gauge fields from the non-dynamical D3-branes stretched between the NS5 and a D5-brane. 
 To read the SQM theory we had to move D1-branes to the closest NS5-brane to left (or to the right). Suppose there are $n_{\rm D5}$ D5-branes standing between this NS5 and the D1-branes, then we have to move them first to the left of the NS5 and, by the Hanany-Witten effect, one D3-brane per D5 is created ending on the left  of the NS5-brane. In this case the $N_L$ D3-branes ending on the left of the NS5 decompose into $N_L = n_{\rm D5} + n_{\rm D3}$, where $n_{\rm D3}$ is the number of dynamical D3-branes, those supporting a $U(n_{\rm D3})$ 3d $\cN=4$ vector multiplet. On the other side, the $N_R$ D3-branes ending on the right of the NS5 support a $U(N_R)$ 3d $\cN=4$ vector multiplet. 
The  1d $\cN=4$ theory  is then coupled to the 3d $\cN=4$ theory by gauging the $U(N_R)$ and $U(n_{\rm D3}) \subset U(N_L)$ flavor symmetries on the defect with  dynamical  3d  $\cN=4$  vector multiplets.\footnote{More precisely the 3d $\N=4$ vector multiplet decomposes into multiplets of the subalgebra preserved by 
SQM$_V$, each multiplet containing fields at a given position in the plane orthogonal to the defect. A 1d $\cN=4$ SQM$_V$ vector multiplet embedded in the 3d $\cN=4$ vector multiplet lives at the position of the defect and gauges the corresponding flavor symmetry. }   
Furthermore,  there is a cubic superpotential as in \rf{supercoupl} which breaks the $U(n_{\rm D5})_{\rm 3d} \times U(n_{\rm D5})_{\rm 1d}$ flavor symmetry to the diagonal $U(n_{\rm D5})$, where $U(n_{\rm D5})_{\rm 1d} \subset U(N_L)$ and $U(n_{\rm D5})_{\rm 3d}$ is the 3d flavor symmetry acting on the $n_{\rm D5}$ hypermultiplets associated to the D5-branes. 
The 3d/1d coupling that can read from the brane picture is summarized in the example of figure \ref{QMquiver}.

\begin{figure}[h]
\centering
\includegraphics[scale=0.7]{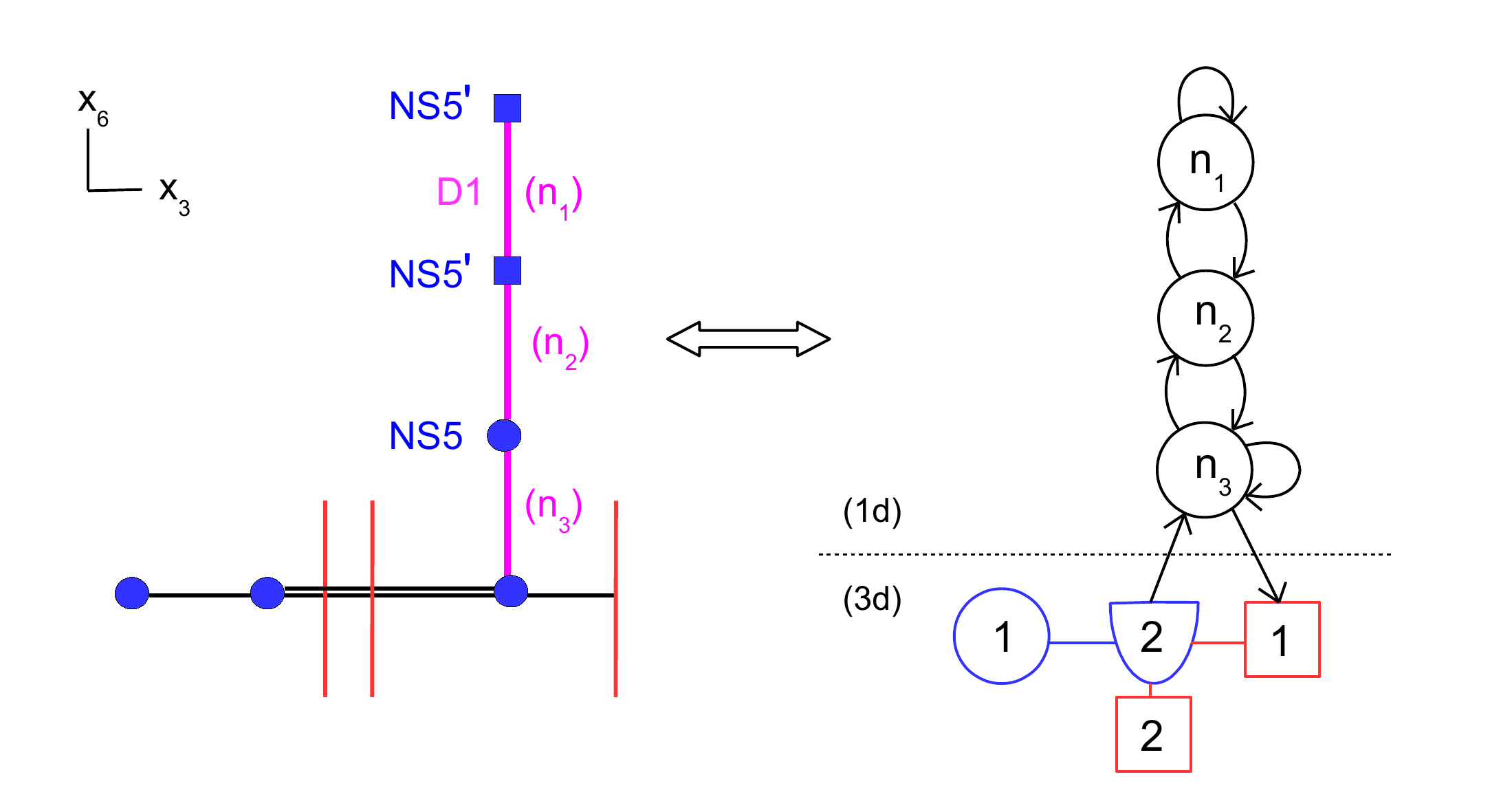}
\vspace{-0.5cm}
\caption{\footnotesize{Pushing the D1-branes to the NS5-brane on the right, we can read another 3d/1d theory, giving a second UV realization of the same Vortex loop.  }}
\label{QMquiver2}
\end{figure}

The 3d/1d coupling for the SQM$_V$ theories read from moving the D1s to the nearest NS5 on the right are found similarly, as shown in figure \ref{QMquiver2}.

\smallskip

\begin{figure}[h]
\centering
\includegraphics[scale=0.55]{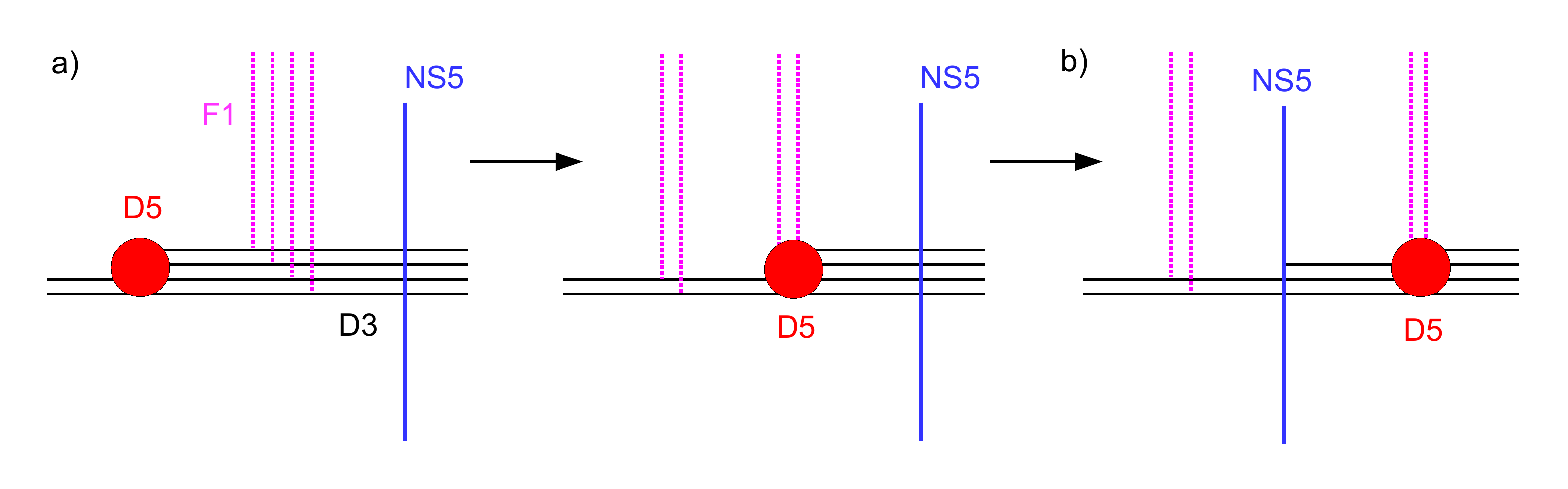}
\caption{\footnotesize{a) The D5-brane has to be moved to the right, across   F1-strings, to reach the canonical brane configuration. b) After moving the D5-brane, stacks of F1-strings end on the D5-brane and are moved along transverse directions (transverse to the picture) away from the main brane configuration. }}
\label{F1figure}
\end{figure}

As mentioned above, it turns out to be simpler to derive the mirror map if we consider the inverse problem of finding the combination of Wilson loops dual to a given Vortex loop.
Starting with a configuration of D1-branes realizing a Vortex loop, we can S-dualize  to obtain a configuration with F1-strings and move the D5-branes to reach the canonical brane configuration of the mirror-dual theory. The crucial difference is that  we do not need to move the F1-strings, instead we are allowed to move the D5-branes across the stack of F1-strings if this is necessary to reach the canonical configuration. When we have to move a D5 across the F1-strings the situation is always the same, namely the D5-brane has an excess of D3-branes on the side toward which it is moving. This means that the F1-strings are on the side of the D3-spikes, along which they can be moved  without obstruction. When the D5-brane crosses the F1-strings the number of D3-branes at the bottom of the strings decreases, as in figure \ref{F1figure}. The initial $k$ strings split into $k=k_1+k_2$ strings, where $k_1$ strings end on the remaining D3-branes, while $k_2$ strings have been moved along the D3-spikes and end now on the D5, away from the main brane configuration. There are many possible final brane configurations, associated to the various choices of splittings (constrained by the s-rule) as D5-branes are moved across the F1-strings, and each of these final configuration realizes a Wilson loop in a certain node of the mirror theory, combined with flavor Wilson loops inserted by the strings ending on the D5-branes. Physically $q$ F1-strings ending on a D5 insert a Wilson loop with charge $q$ under the $U(1)$ flavor symmetry acting on the associated hypermultiplet.  
The mirror symmetry prediction is then that the initial Vortex loop is mapped to the sum of the Wilson loops realized by the possible final brane configurations. We will provide explicit mirror map predictions using this algorithm in the next section.

The simplest situations arise when no D5-branes must be moved across   F1-strings, in which case the Vortex loop is mapped to a single Wilson loop in the mirror theory and both loops are labeled by the same representation $\cR$ of a certain $U(N)$ gauge group factor. This is the case for instance in circular quivers with nodes of equal rank, for which mirror symmetry is simply  implemented by $S$-duality on the brane configuration and no D5-brane moves are required  to reach the mirror dual configuration (see section \ref{ssec:CircLoops}).

 \bigskip

In this section we have found a systematic algorithm to construct the mirror map between Wilson loop and Vortex loop operators in mirror dual theories. This can be applied to construct the mirror map for any pair of 3d $\cN=4$  mirror quiver  gauge theories of linear or circular type.  We will apply this algorithm on several explicit examples below.

    \section{Mirror Symmetry and Loop Operators: Examples}
    \label{sec:examples}
    
    In this section we give the explicit mirror map between loop operators using the ideas and tools introduced in the previous section. We provide examples both for linear and circular quivers.  
        
\begin{figure}[th]
\centering
\includegraphics[scale=0.7]{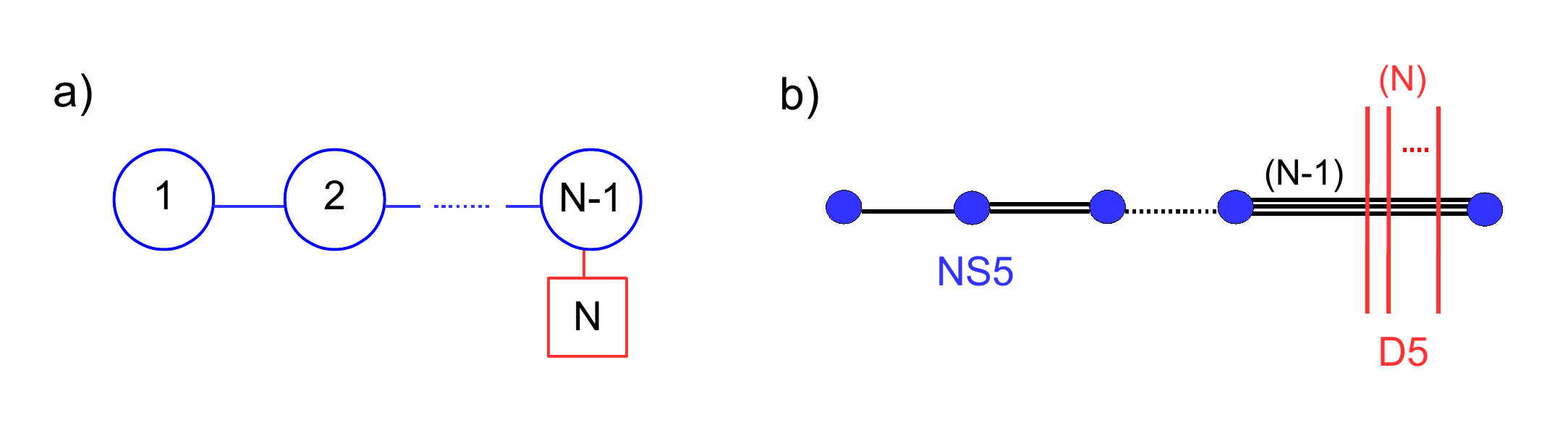}
\vspace{-0.5cm}
\caption{\footnotesize{a) $T[SU(N)]$ quiver. b) Brane realization.}}
\label{TSUN}
\end{figure}

        \smallskip
 We first discuss the action of mirror symmetry of loop operators for a class of 3d $\cN=4$ linear quivers that are self-mirror: that is $T[SU(N)]$ \cite{Gaiotto:2008ak}.
$T[SU(N)]$ is encoded by the quiver diagram of figure \ref{TSUN}-a. The gauge group is $G = \prod_{j=1}^{N-1} U(j)$ and there are $N$ fundamental hypermultiplets in the $U(N-1)$ node. The brane realization of  $T[SU(N)]$ is shown in figure \ref{TSUN}-b.

  Constructing the mirror map for arbitrary loops in $T[SU(N)]$ already incorporates all the subtleties and physical phenomena that emerge in the most general case already discussed in section \ref{ssec:LoopSduality}.  We then provide   duality maps for other examples of linear and circular quivers.

    \subsection{$T[SU(2)]$}
  \label{ssec:TSU2}

\begin{figure}[th]
\centering
\includegraphics[scale=0.7]{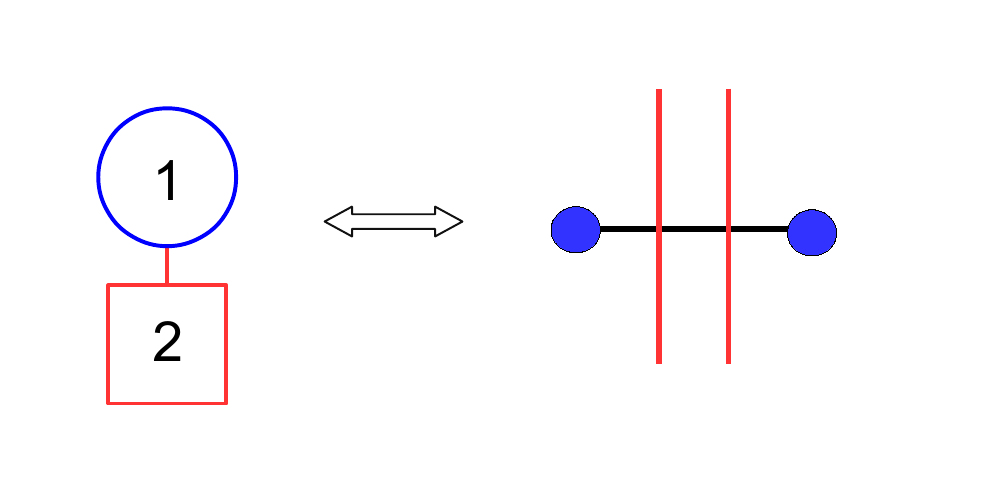}
\vspace{-0.5cm}
\caption{\footnotesize{$T[SU(2)]$ quiver and its brane realization.}}
\label{TSU2}
\end{figure}

\begin{figure}[th]
\centering
\includegraphics[scale=0.6]{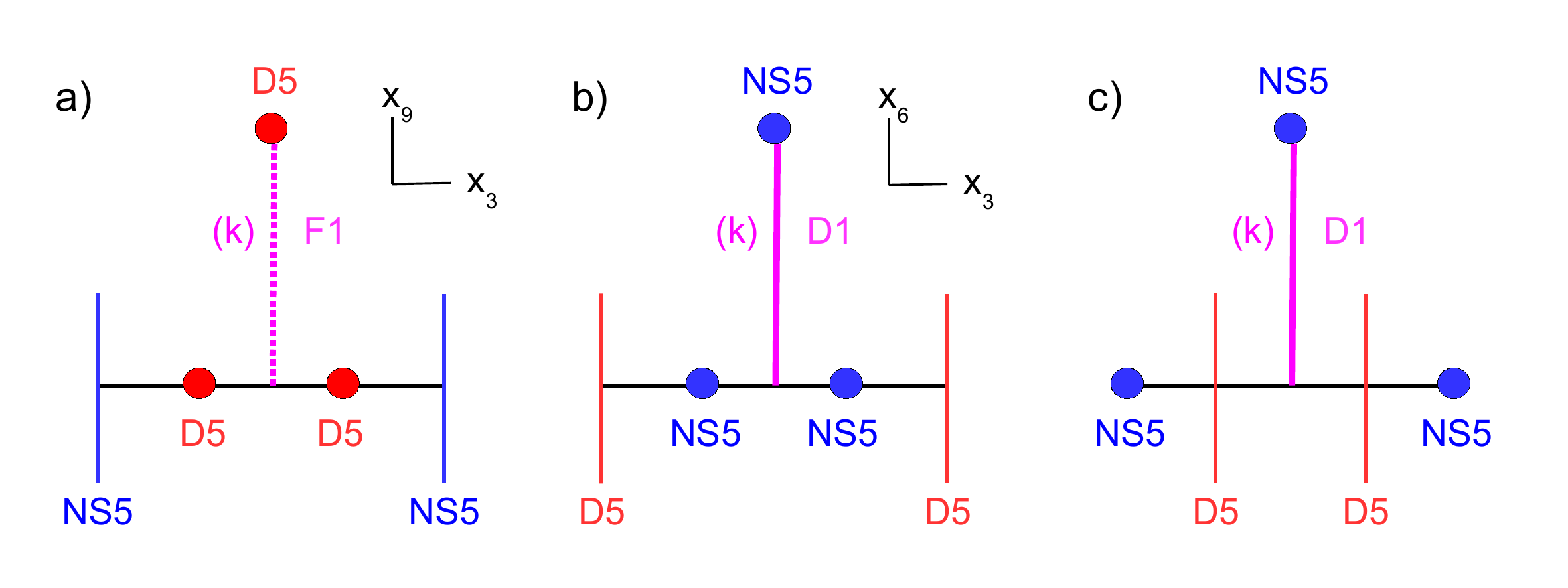}
\caption{\footnotesize{a) Brane realization of a charge $k$ Wilson loop insertion in $T[SU(2)]$. b) Configuration after S-duality (and rotation). c) After Hanany-Witten D5-brane moves: $k$ D1-branes stretched between a NS5-brane and the D3-brane in the mirror $T[SU(2)]$ brane configuration. The D1-branes end between the two D5-branes.  }}
\label{TSU2brane0}
\end{figure}

    Consider a charge $k$ Wilson loop $W_k$ in $T[SU(2)]$, a $U(1)$ gauge theory with two fundamental hypermultiplets (see figure \ref{TSU2}). The brane construction of   $T[SU(2)]$  has a single D3-brane stretched between two NS5-branes which is crossed by two D5-branes.  According to the discussion of section \ref{sec:Branes}, the insertion of the charge $k$ Wilson loop (which is the same as the $\scS_k$ representation in an abelian theory) is realized by adding $k$ F1-strings stretched between the D3-brane and an extra D5-brane far away from the stack (figure \ref{TSU2brane0}-a). The F1-strings (and extra D5-brane) can be moved along the $x^3$ direction along the D3-brane without changing the infrared 3d theory, so we can without loss of generality  place the F1-strings between the two D5-branes.  
    
Acting with S-duality on this enriched brane configuration and moving the D5-branes according to the Hanany-Witten rules we recover the same $T[SU(2)]$ brane configuration but with $k$ D1-branes stretched between the D3-brane and an extra NS5-brane far away from the stack as in figure \ref{TSU2brane0}-c. The D1-branes end on the D3-brane between the two D5-branes. We note that this is the final S-dual brane configuration irrespectively of where we decide  to place the F1-strings relative to the original D5-branes. The presence of the D1-branes is responsible for the insertion of a  supersymmetric  one-dimensional defect in the  $T[SU(2)]$ theory living on the D3-branes. We denote this defect by $V_k$.

\noindent  

      \smallskip
The operator $V_k$ can be described by a 1d $\cN=4$ gauge theory coupled to $T[SU(2)]$. As explained above there are two alternative  3d/1d defect descriptions of $V_k$, that we can read by moving the D1-branes on top of the nearest NS5-brane on the left or on the right.  It is a non-trivial dynamical statement that these two descriptions of $V_k$ do indeed describe the same operator (see section \ref{sec:LoopsOnS3}).
       \smallskip
       
\begin{figure}[th]
\centering
\includegraphics[scale=0.7]{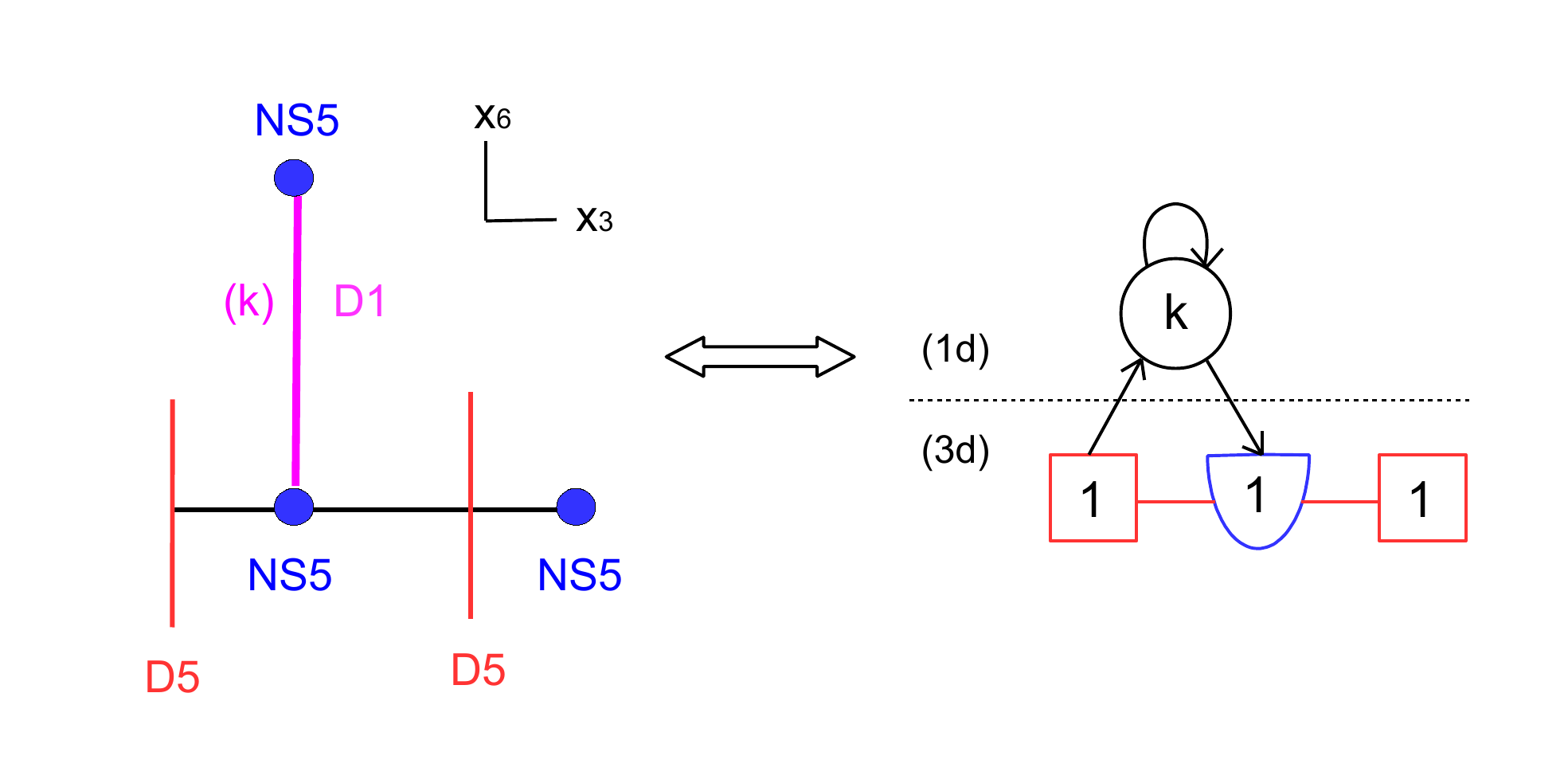}
\vspace{-1cm}
\caption{\footnotesize{Brane configuration after moving the D1-branes on top of the left NS5-brane, as described in the text, and the quiver description of the  3d/1d gauge theory. An arrow arriving to (resp. leaving)   a node denotes a fundamental (resp. anti-fundamental) chiral multiplet for that node.   This 3d/1d defect theory describes the loop operator dual to the abelian Wilson loop in the mirror $T[SU(2)]$ theory.}}
\label{TSU2brane2}
\end{figure}

We start with the description of $V_k$ as a deformation of the theory where the D1-branes end on the left NS5-brane.
 This configuration is reached by first moving the left D5-brane to the left of the NS5-brane, creating a D3-brane by the Hanany-Witten   effect, and then moving the D1-branes (together with the extra NS5-brane far away) until they end on the NS5-brane, thus reaching the configuration shown in figure \ref{TSU2brane2}.  In the IR, the theory supported  on the  brane configuration  is a 3d/1d $\scN=4$   gauge theory coupled to $T[SU(2)]$, which can be read from the rules described in the previous section. 
It is described by the   1d $\cN=4$ quiver gauge theory  of figure \ref{TSU2brane2},   a $U(k)$ gauge theory with an adjoint chiral multiplet, one fundamental and one anti-fundamental chiral multiplet coupled to $T[SU(2)]$ as shown in figure \ref{TSU2brane2}.

      \smallskip
The way the 1d $\cN=4$ quiver gauge theory couples to $T[SU(2)]$ can be read from the brane configuration. The $U(1)^-$ flavor symmetry acting on the anti-fundamental chiral multiplet
is gauged with the 3d $\cN=4$ $U(1)$ vector multiplet, and the $U(1)^+$ flavor symmetry acting on the fundamental chiral multiplet is identified with the $U(1)$ flavour symmetry  acting on the 3d hypermultiplet in $T[SU(2)]$ associated to the left D5-brane. The operator $V_k$ is described by the coupled 3d/1d theory summarized by the mixed 3d/1d quiver diagram in
 \ref{TSU2brane2}. As explained in section \ref{sec:loops}, the identification of 1d with 3d flavour symmetries is implemented with a superpotential coupling \rf{supercoupl} between the relevant defect chiral multiplets  and bulk hypermultiplet. In the initial configuration (figure \ref{TSU2brane0}-c) the D1-branes   end on the D3-branes  instead of the left NS5-brane. The deformation corresponding to moving the D1-branes along the D3-brane to go back to the initial configuration corresponds to turning on an FI term with coupling $\eta \propto \Delta x^3 > 0$ in the SQM, where $\Delta x^3$ is the difference of the positions along $x^3$ between the left NS5-brane and the D1-branes.

      \smallskip

\begin{figure}[th]
\centering
\includegraphics[scale=0.7]{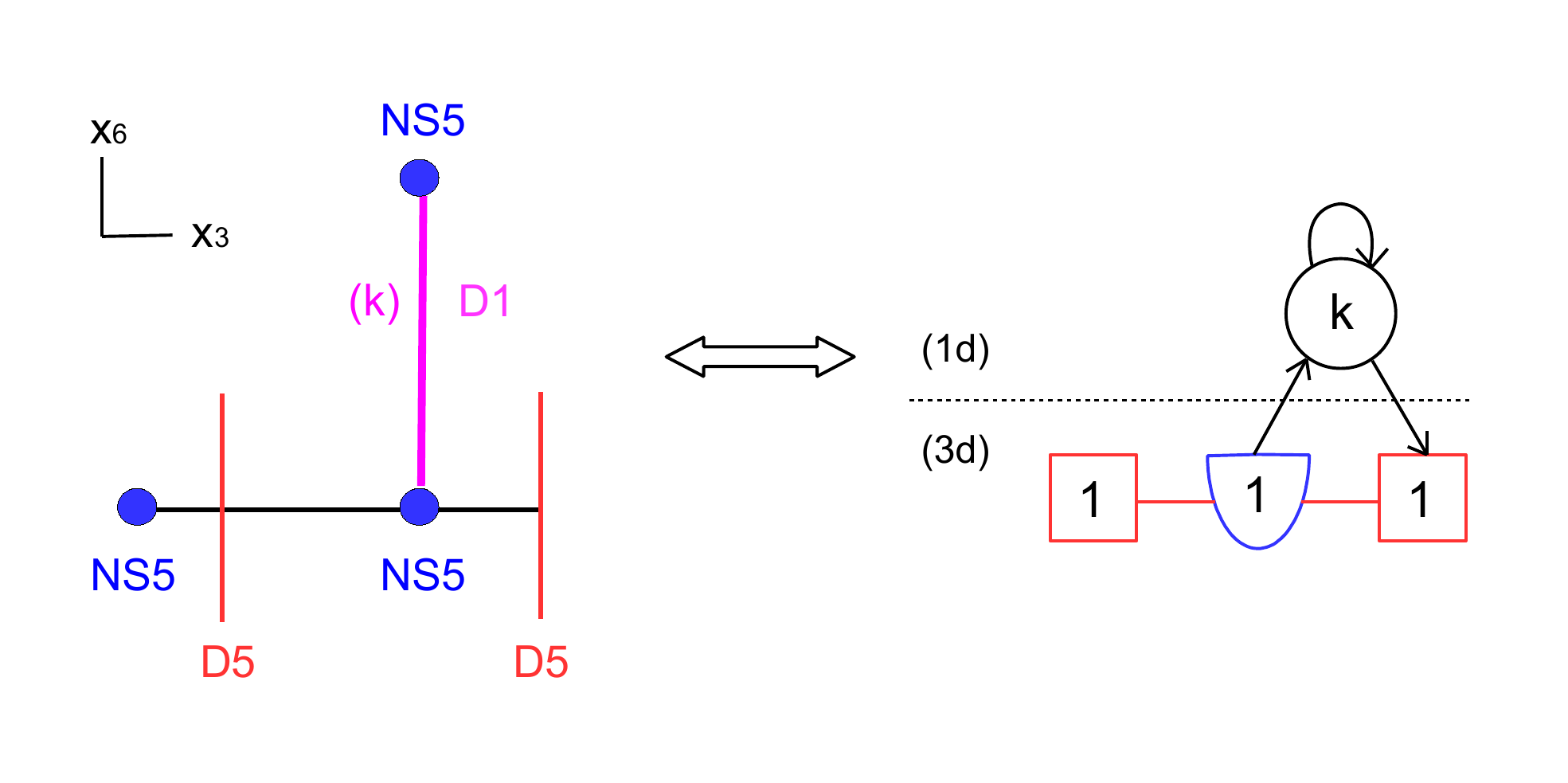}
\vspace{-1cm}
\caption{\footnotesize{Alternative 3d/1d defect theory, read by moving the D1-branes to the right NS5-brane.}}
\label{TSU2brane3}
\end{figure}

Alternatively, we can describe the  operator $V_k$ as  a deformation of a configuration where the D1-branes end on the NS5-brane  on the right. Following the same steps as before we reach the brane configuration of figure \ref{TSU2brane3}, and we find a description of the infrared loop operator $V_k$ as a different coupling of the same 1d $\cN=4$ quiver gauge theory   to $T[SU(2)]$.  Now the $U(1)^+$ flavor symmetry acting on the fundamental chiral multiplet
is gauged with the 3d $\cN=4$ $U(1)$ vector multiplet of $T[SU(2)]$, and the $U(1)^-$ flavor symmetry acting on the anti-fundamental chiral multiplet is identified with the $U(1)$ flavour symmetry  acting on the 3d hypermultiplet in $T[SU(2)]$ associated to the right D5-brane. The operator $V_k$ is described by the coupled 3d/1d theory summarized by the mixed 3d/1d quiver diagram in \ref{TSU2brane3}. 
In the initial configuration (figure \ref{TSU2brane0}-c) the D1-branes   end on the D3-branes  instead of the right NS5-brane. 
The deformation corresponding to moving the D1-branes along the D3-brane to go back to the initial configuration corresponds to turning on a negative FI parameter     $\eta \propto \Delta x^3 < 0$ in the SQM, where $\Delta x^3$ is the difference of the positions along $x^3$ between the right NS5-brane and the D1-branes. 
 
       \smallskip
We conclude that there are two different UV descriptions of the operator $V_k$ mirror dual to the charge $k$ Wilson loop  $W_k$ in $T[SU(2)]$.   In section \ref{ssec:Hoppingduality}  we  explicitly show that the exact expectation value of $V_k$ on $S^3$ computed using the two different UV definitions define the same operator and in section \ref{sssec:TSU2} that it matches the exact expectation value of $W_k$ under the exchange of mass and FI parameters
\beq
\vev{W_k} \ \ \xleftrightarrow{\text{\ mirror \ }} \ \ \vev{V_k}\,.
\label{TSU2Map}
\eeq

\smallskip 

\begin{figure}[th]
\centering
\includegraphics[scale=0.7]{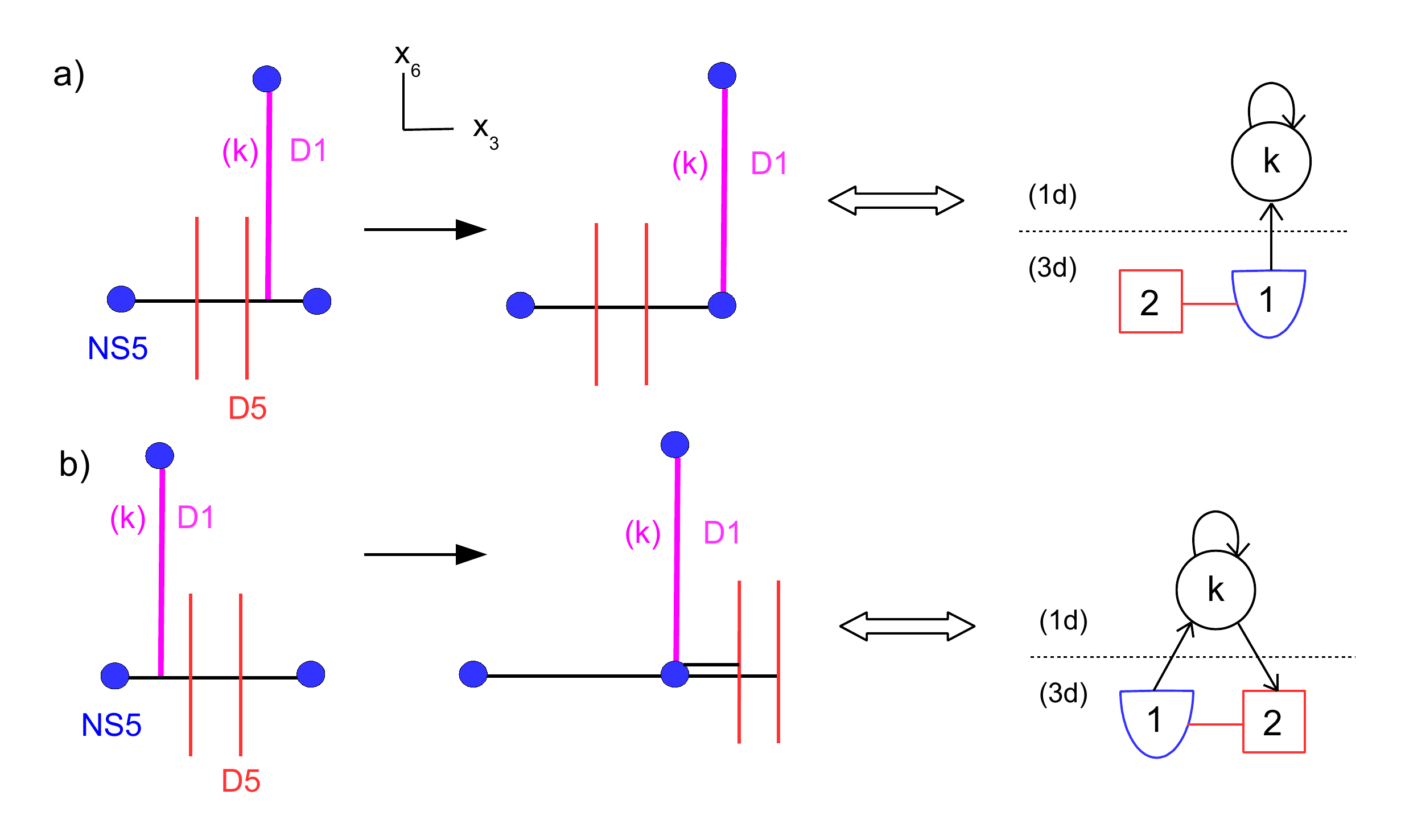}
\vspace{-0.5 cm}
\caption{\footnotesize{ Brane realizations of other $T[SU(2)]$ Vortex loops and associated  3d/1d quiver theory read by moving the D1-branes to the right NS5-brane. a) Vortex loop realized with D1-branes placed to the right of the D5-branes. b) Vortex loop realized with D1-branes placed to the left of the D5-branes.}}
\label{TSU2brane4}
\end{figure}

We  can also consider the Vortex loop operators realized by D1-branes ending on the D3-brane to the left or to the right of the two D5-branes. The corresponding 3d/1d quivers realizing these Vortex loops can be read by moving the D1-branes on top of the NS5-brane on the right, as shown in figure \ref{TSU2brane4}, or on the left.  We analyze the mirror of such Vortex loop operators  for the general $T[SU(N)]$ theory in section \ref{ssec:MirrorTSUNsimple}. They are mirror to flavour Wilson loops (see discussion around equation \rf{MirrorMapVp}).

\subsection{Wilson Loops In The $U(N-1)$ Node Of $T[SU(N)]$}
\label{ssec:MirrorTSUNsimple}

We now turn to   Wilson loops in a non-abelian gauge theory. We  start with   Wilson loop operators  in the $U(N-1)$  node of the self-mirror $T[SU(N)]$ theory. As we shall see, a Wilson loop  for this (last) node maps directly to a single Vortex loop operator, described by a specific 3d/1d gauge theory that  we construct.  Later we show that the expectation value of a Wilson loop operator in another node in $T[SU(N)]$ maps under mirror symmetry to a specific combination of Vortex loop operators.
 
\begin{figure}[th]
\centering
\includegraphics[scale=0.65]{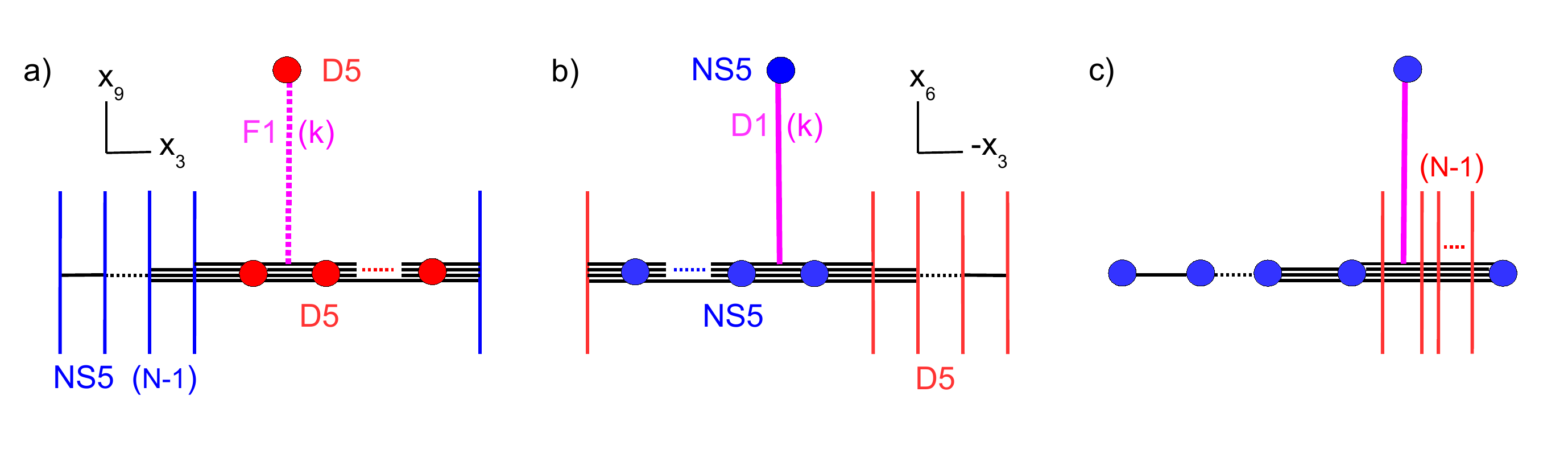}
\vspace{-1cm}
\caption{\footnotesize{a) Brane configuration realizing a $\scS_k$ Wilson loop in $T[SU(N)]$. b) S-dual brane configuration, shown in a different orientation (in particular the $x^3$ orientation is reversed). c) Brane configuration after D5-brane rearranging, realizing a 3d/1d defect theory.}}
\label{SdualBraneTSUN}
\end{figure}

       \smallskip
 We start by considering a Wilson loop  in the $k$-symmetric representation of $U(N-1)$, which we denote by $W^{(N-1)}_{\scS_k}$. The brane realization of this Wilson loop insertion comprises the D3-NS5-D5 system realizing the $T[SU(N)]$ theory in figure \ref{TSUN} with   $k$ additional F1-strings stretched between the $N-1$ D3-branes supporting the $U(N-1)$ node and an extra D5-brane away from the stack, as depicted in figure \ref{SdualBraneTSUN}-a. We are free to place the F1-strings anywhere relative to the $N$ D5-branes that intersect the D3-branes, as this results in   the same Wilson loop operator in the IR.
 
       \smallskip
 
The mirror dual loop operator to $W^{(N-1)}_{\scS_k}$ is obtained by analyzing  the S-dual configuration (figure \ref{SdualBraneTSUN}-b), where we have reversed the $x^3$ direction for convenience.\footnote{This extra reflection allows us to bring the branes (in the absence of F1 or D1) back to their initial configuration, showing that the theory $T[SU(N)]$ is self-mirror.} First we move the D5-branes in the direction of D3-brane excess to reach the 
$T[SU(N)]$ brane configuration, now enriched  with $k$ D1-branes stretched between  the $N-1$ D3-branes associated to the $U(N-1)$ node and an extra NS5-brane far away from the stack (see figure \ref{SdualBraneTSUN}-c). The D1-branes  are positioned with one D5 to their left   and $N-1$ D5-branes to  their right. This final configuration is the same irrespective of where the original F1-strings were placed relative to the D5-branes. In order to reach this configuration we are careful not to let D5-branes cross the D1-branes in the rearrangement, since this would affect the infrared theory living on the D1-branes. This is   S-dual to the statement that we do not let F1-strings move across NS5-branes, as this   clearly changes the Wilson loop inserted. The  position of the D1-branes relative to the D5-branes is   crucial  in determining the Vortex loop operator just as   the position of the F1-strings relative to the NS5-branes is crucial in determining the Wilson loop operator. 
On the other hand, we let the D1-branes move freely   across an NS5-brane  when the number of D3-branes is the same on both sides of the NS5-brane. This  is S-dual  to the fact that F1-strings can be freely moved  across D5-branes  when the number of D3-branes is the same on both sides of the D5-branes. We denote the loop operator defined by the S-dual brane configuration by $V_{N-1, \cS_k}$, where the subscript $N-1$ encodes   the number of D5-branes to the right of the D1-branes.

\begin{figure}[th]
\centering
\includegraphics[scale=0.7]{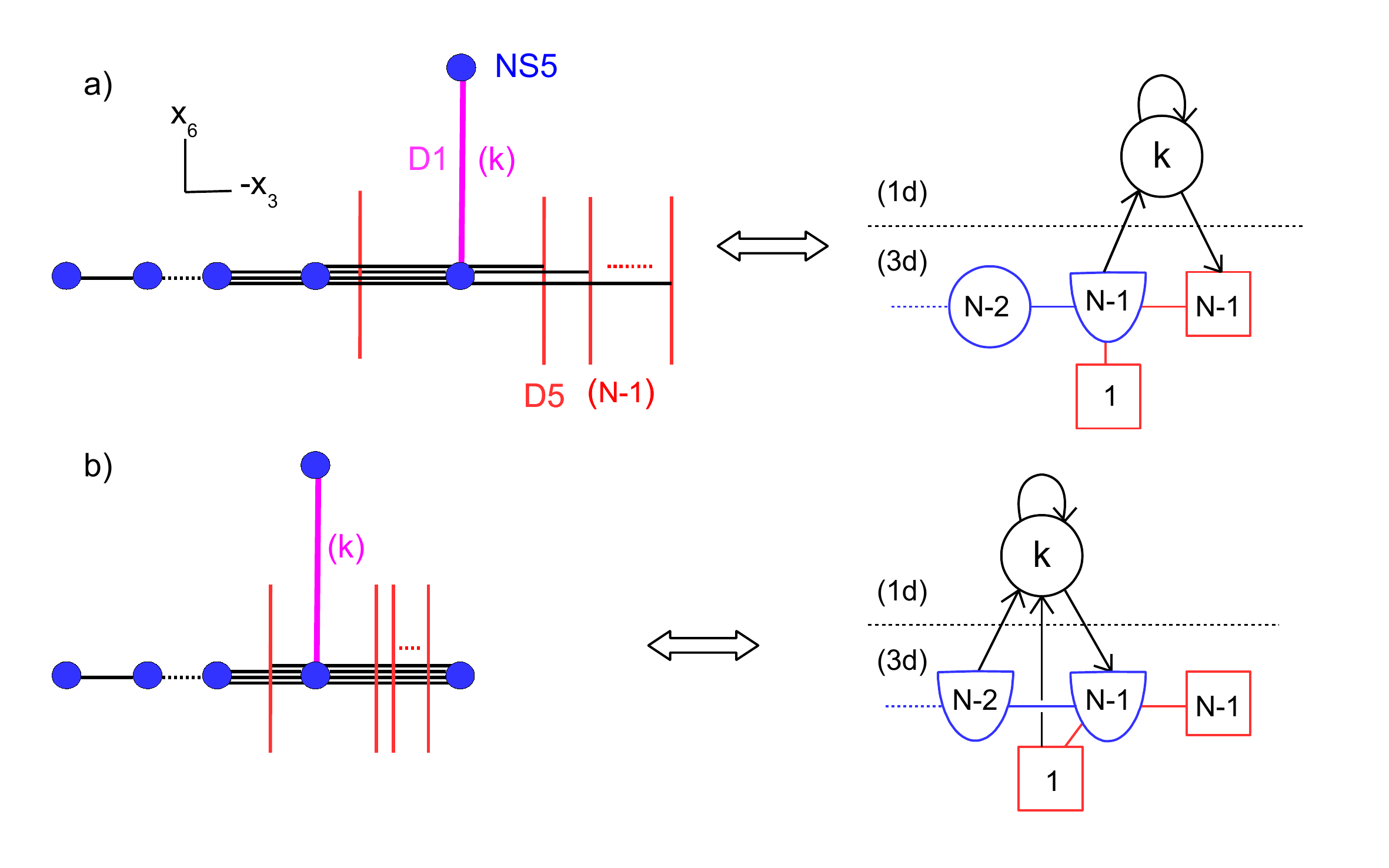}
\vspace{-1cm}
\caption{\footnotesize{a) Brane configuration with D1-branes ending on the right NS5 and associated 3d/1d defect gauge theory. b)  Brane configuration with D1-branes ending on the left NS5-brane and associated dual defect theory.}}
\label{QMTSUN}
\end{figure}

The coupled 3d/1d theory living on   the final brane configuration can be read again in two ways,
depending on which of the nearby NS5-branes we  let the D1-branes end, as explained in section \ref{ssec:LoopSduality}.
One possibility is to move the $N-1$ D5-branes  to the right  and let the D1-branes end on the NS5-brane on their right   as shown in figure \ref{QMTSUN}-a. The associated 1d $\cN=4$ gauge theory is a 
 $U(k)$ vector multiplet, with an adjoint chiral multiplet,   $N-1$ fundamental and $N-1$ anti-fundamental chiral multiplets. This 1d $\cN=4$ theory is captured by the 1d part of quiver diagram in \ref{QMTSUN}-a.
  
         \smallskip
 
The way the 1d gauge theory couples to the 3d $T[SU(N)]$ theory is also captured by the brane configuration in  figure 
 \ref{QMTSUN}-a. The $U(N-1)$ flavour symmetry rotating the fundamental chirals is gauged with the 3d $U(N-1)$ vector multiplet symmetry and the $U(N-1)$ symmetry rotating the anti-fundamental chirals is identified with the $U(N-1)$ flavor symmetry acting on the 3d hypermultiplets associated to the $N-1$ D5-branes. In order to describe the actual  operator $V_{N-1,\cS_k}$ the 1d $\cN=4$ gauge theory must be deformed  with a negative  FI parameter. The coupled 3d/1d description of the Vortex loop operator $V_{N-1,\cS_k}$ mirror to $W^{(N-1)}_{\scS_k}$  is captured by the quiver diagram in figure  \ref{QMTSUN}-a.

       \smallskip

Another description of  $V_{N-1,\cS_k}$  can also be read from the configuration where the D1-branes end on the nearest NS5-brane to their left, with the (single) leftmost  D5-brane pushed to the left  of that NS5-brane. The 1d $\cN=4$ gauge theory living on the D1-branes is the same as before, but the way it couples to $T[SU(N)]$ is rather different.
 The $U(N-1)$ flavour symmetry rotating the anti-fundamental chirals  is gauged with the 3d $U(N-1)$ vector multiplet. The $U(N-1)$ flavour symmetry rotating the fundamental chirals  is broken to $U(N-2) \times U(1)$. The   $U(N-2)$ flavour symmetry is gauged with  the 3d $U(N-2)$ vector multiplet and the remaining $U(1)$ flavour symmetry is identified with the $U(1)$ flavor symmetry acting on the 3d hypermultiplet associated to the leftmost  D5-brane. Now, the 1d $\cN=4$ gauge  theory has to be deformed with a positive  FI parameter as the D1-branes end to the right of the reference NS5-brane.
  The coupled 3d/1d description of the Vortex loop operator $V_{N-1,\cS_k}$ mirror to $W^{(N-1)}_{\scS_k}$  is captured by the quiver diagram in figure 
 \ref{QMTSUN}-b.

In summary, we obtain the following mirror map  of loop operators in $T[SU(N)]$ theory
\beq
\vev{W^{(N-1)}_{\scS_k}} \ \xleftrightarrow{\text{\ mirror \ }} \ \vev{V_{N-1,\cS_k}}\,.
\label{symmee}
\eeq

 \begin{figure}[th]
\centering
\includegraphics[scale=0.7]{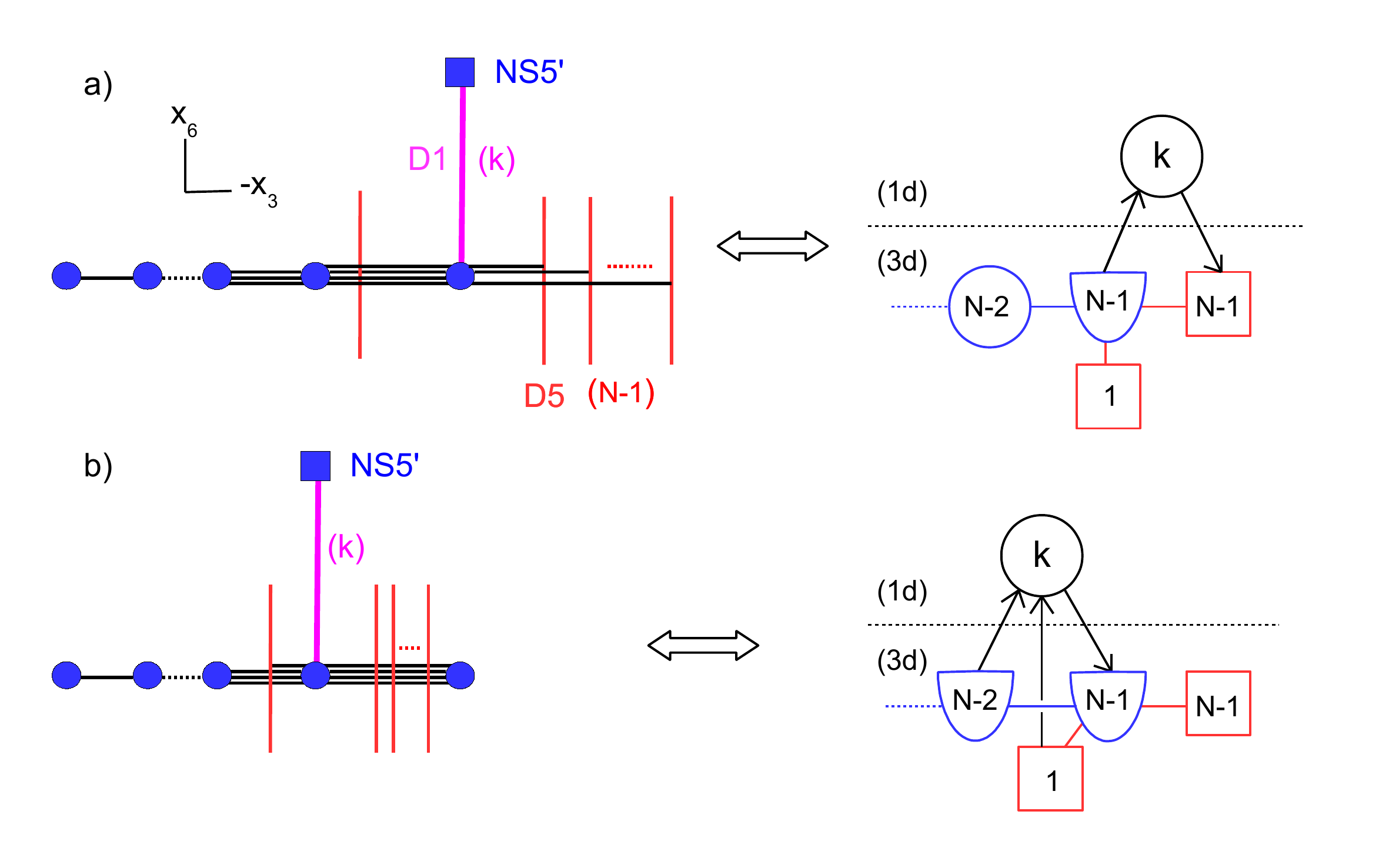}
\vspace{-0.5cm}
\caption{\footnotesize{Brane configuration with $k$ D1-strings ending on a single D5'-brane and associated left and right 3d/1d quivers. }}
\label{QMTSUN2}
\end{figure}

\smallskip
The situation is almost the same if we start with a Wilson loop in the $k$-antisymmetric representation of $U(N-1)$,
which we denote by $W^{(N-1)}_{\scA_k}$. In this case the Wilson loop in realized with $k$ F1-strings stretched between an extra D5'-brane and the $N-1$ D3-branes. After S-duality and brane rearranging we obtain the same brane configuration as in figure \ref{SdualBraneTSUN}-c, except that D1-branes end on an NS5'-brane instead of a NS5-brane. The two possible 3d/1d defect theories describing the mirror loop operator, which we denote by $V_{N-1,\cA_k}$, are given in figure \ref{QMTSUN2}. They are the same as in figure \ref{QMTSUN}, except that there is no adjoint chiral multiplet in the SQM. In this case we get the following duality map
\beq
\vev{W^{(N-1)}_{\cA_k}} \ \xleftrightarrow{\text{\ mirror \ }} \ \vev{V_{N-1,\cA_k}}\,.
\label{antisymmee}
\eeq
In section \ref{ssec:TSUN} we explicitly show by computing the exact expectation value of these loop operators on $S^3$ that  \rf{symmee} and \rf{antisymmee} hold  under the exchange of mass and FI parameters.

\smallskip

\begin{figure}[th]
\centering
\includegraphics[scale=0.6]{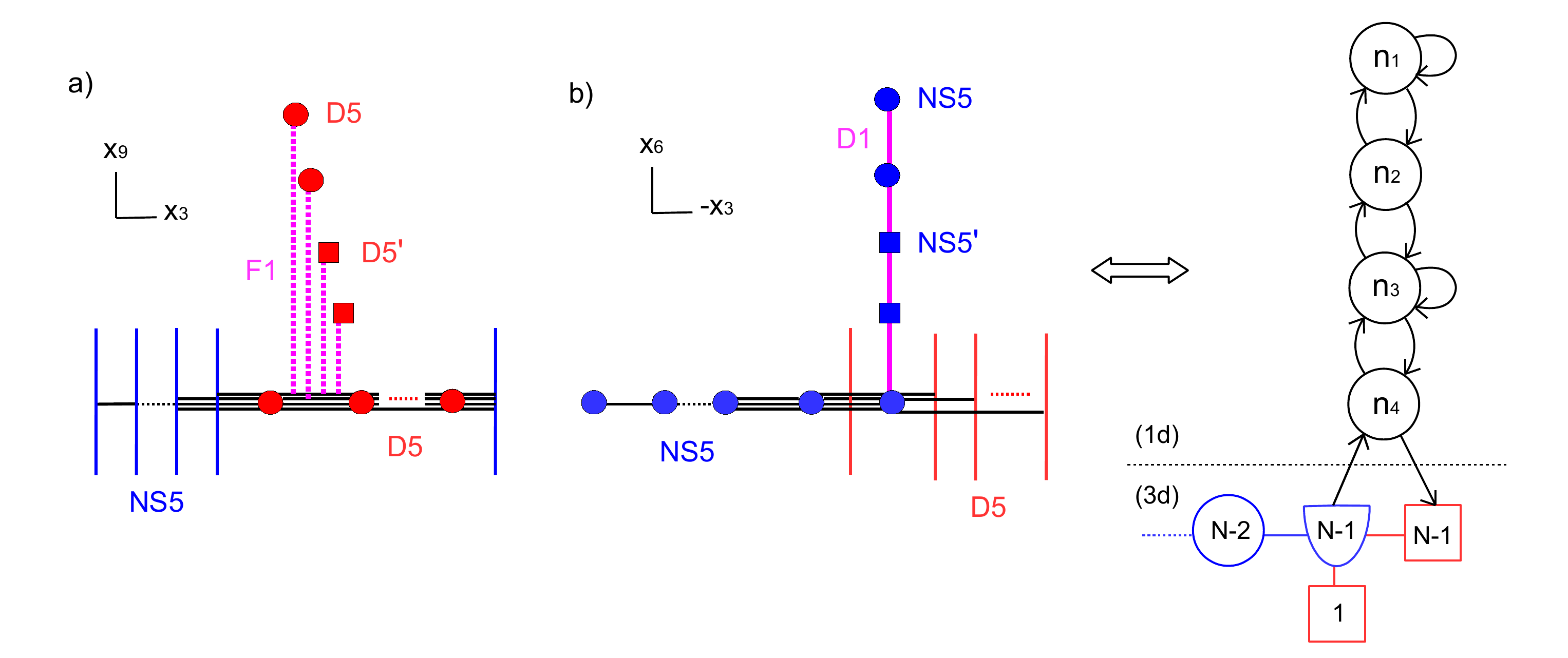}
\vspace{-1cm}
\caption{\footnotesize{a) Brane configuration inserting a Wilson loop in the $\scS_{k^{(1)}} \otimes \scS_{k^{(2)}} \otimes \scA_{l^{(1)}} \otimes \scA_{l^{(2)}}$ representation of $U(N-1)$.  b)  S-dual brane configuration with D1-branes ending on the right NS5-brane and associated dual 3d/1d defect theory with $U(n_1) \times U(n_2) \times U(n_3) \times U(n_4)$ SQM. The ranks are $n_1 = k^{(1)}$ , $n_2 = k^{(1)} + k^{(2)}$, $n_3 = k^{(1)} + k^{(2)}  + l^{(1)}$, $n_4 = k^{(1)} + k^{(2)} + l^{(1)} + l^{(2)}$. }}
\label{ProdRepTSUN}
\end{figure}

The case of the Wilson loop in a tensor product of symmetric and anti-symmetric representations $\scR=\otimes_{a=1}^{d} \scS_{k^{(a)}}   \otimes_{b=1}^{d'} \scA_{l^{(b)}}$ of $U(N-1)$ in $T[SU(N)]$ is analogous. We denote such a Wilson loop by $W^{(N-1)}_\scR$.  As explained in section \ref{ssec:LoopSduality}, the Wilson loop insertion is realized with a collection of $d$ D5-branes and $d'$ D5'-branes, with $k^{(a)}$ F1-strings suspended between the $N-1$  D3-branes and the $a$-th D5-brane, and $l^{(b)}$ F1-strings suspended between the $N-1$ D3-branes and the $b$-th D5'-brane. The case  $d=d'=2$ is displayed in figure \ref{ProdRepTSUN}-a.
After S-duality and D5-brane moves, we obtain a configuration with $d$ NS5-branes and $d'$ NS5'-branes, with $k^{(a)}$ D1-branes suspended between the $N-1$  D3-branes and the $a$-th NS5-brane, and with $l^{(b)}$ D1-branes suspended between the $N-1$ D3-branes and the $b$-th NS5'-brane, realizing the mirror dual loop operator, which we denote by $V_{N-1,\scR}$. In the initial configuration, the relative positions of the D5-branes  and D5'-branes in the $x^9$ direction were irrelevant  for the insertion of the Wilson loop operator. In the S-dual picture, the   positions of the NS5-branes  and NS5'-branes along the $x^6$ direction are important, since they affect  the 1d $\cN=4$ gauge theory  living on the D1-branes. The brane picture, however,  implies that all the possible 3d/1d defect field theories obtained from picking different orderings of the NS5  and NS5' brane positions along $x^6$ direction are equivalent/dual descriptions of the loop operator  
$V_{N-1,\scR}$. In section \ref{sssec:QMindex}  we   provide some explicit evidence that these are indeed dual
descriptions of the same operator by computing the exact partition function on $S^3$ of the 3d/1d defect theories capturing $\vev{V_{N-1,\scR}}$ and showing   that they are the same.  

 \smallskip
The 3d/1d defect theories describing the resulting Vortex loop $V_{N-1,\scR}$ can be read from the rules of section \ref{ssec:LoopSduality} by moving the D1-branes to the closest NS5 on the left or on the right. The example of the Vortex loop associated to the representation $\scR = \scS_{k^{(1)}} \otimes \scS_{k^{(2)}} \otimes \scA_{l^{(1)}} \otimes \scA_{l^{(2)}}$ with a specific ordering of the NS5/NS5'-branes along $x^6$ is worked out in figure \ref{ProdRepTSUN}-b.

\smallskip

\noindent
In summary, we obtain the following mirror map  of those loop operators in $T[SU(N)]$ theory: 
\beq
\vev{W^{(N-1)}_{\cR}} \ \xleftrightarrow{\text{\ mirror \ }} \ \vev{V_{N-1,\cR}}\,.
\label{MapTSUNsimple}
\eeq

\medskip

We have found that the mirror of  a Wilson loop  labeled by a representation of the $U(N-1)$ node is  a Vortex loop operator labeled by the same representation.  We now turn to the more involved case of a Wilson loop  in a representation of another gauge group factor  in $T[SU(N)]$.

\subsection{The Other $T[SU(N)]$ Loops}
\label{ssec:OtherLoops}    
    
    We   proceed now with constructing the mirror map for the remaining loop operators in $T[SU(N)]$. This includes the Wilson loop operators in the other nodes of $T[SU(N)]$.  As explained in section \ref{ssec:LoopSduality} it is easier to describe the mirror map starting from the brane configuration realizing   Vortex loops and going to the S-dual picture with F1-strings realizing the mirror Wilson loops. In the brane realization, defining a Wilson loop requires specifying the location of the F1-strings relative to the NS5-branes and a representation $\cR$. Likewise, a Vortex loop is specified by the   location of the D1-branes relative to the D5-branes and by a choice of representation $\cR$ of the gauge group supported on the D3-branes where the D1-branes end.
    
    \smallskip
    
\begin{figure}[th]
\centering
\includegraphics[scale=0.7]{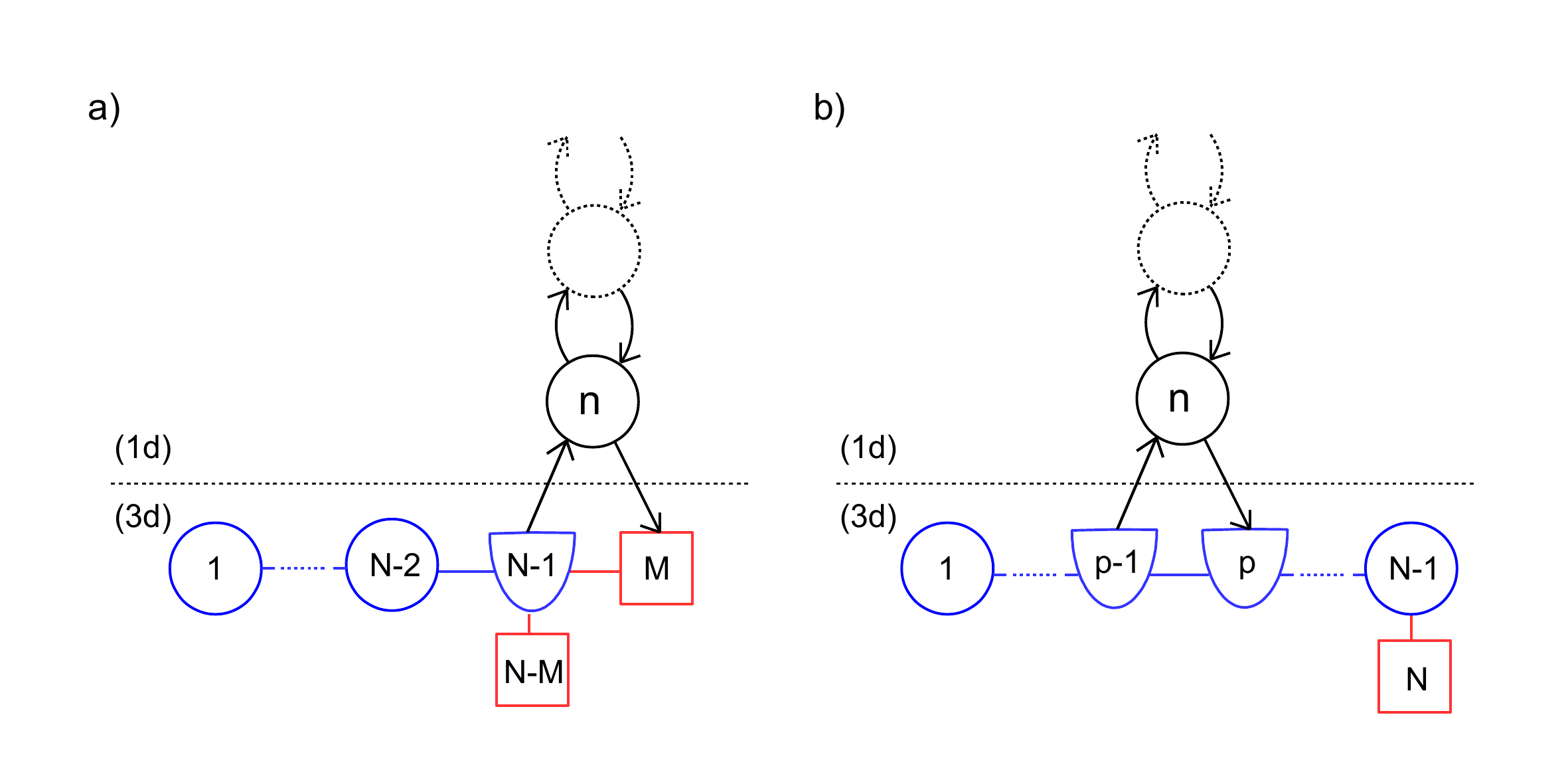}
\vspace{-0.5cm}
\caption{\footnotesize{a) 3d/1d  quiver description of  Vortex loop $V_{M,\scR}$. Here it is derived by moving the D1-strings to the right NS5-brane. b) 3d/1d  quiver description of  Vortex loop $V^{(p)}_\scR$, derived by moving the D1-strings to the left NS5-brane.}}
\label{TSUNQMquivers}
\end{figure}

     In the $T[SU(N)]$ theory  we have Vortex loops labeled by a representation $\scR$ of the $U(N-1)$ gauge node, which are realized with D1-strings ending on the $N-1$ D3-branes supporting that gauge group factor. As just discussed, the relative position of the D1-branes along the $x^3$ direction relative  to the D5-branes  is important. In the brane realization of $T[SU(N)]$, there are $N$ D5-branes crossing the $N-1$ D3-branes, giving rise to the $N$ fundamental hypermultiplets of the $U(N-1)$ node. Each $U(N-1)$ Vortex loop is characterized by a splitting of the $N$ hypermultiplets into
     $N = M + (N-M)$  and corresponds to the arrangement where the D1-branes have $M$ D5-branes to their right and $N-M$ D5-branes to their left. Thus, the data characterizing a $U(N-1)$ Vortex loop is   a representation $\scR$ of $U(N-1)$ and  an integer $0 \le M \le N$. We denote such a Vortex loop by $V_{M,\scR}$. These operators admit a description in terms of a 1d $\cN=4$ quiver gauge theory coupled to $T[SU(N)]$, which 
    is   shown in figure (\ref{TSUNQMquivers}-a) for the case when the  the D1-branes are moved to be  on top of the closest NS5-brane to their right. As already explained in section \ref{ssec:MirrorTSUNsimple} the Vortex loop operator $V_{N-1,\cR}$
gets mapped under mirror symmetry to a Wilson loop $W^{(N-1)}_{\scR}$  in a representation $\cR$ of $U(N-1)$.
    
    \smallskip
This does not exhaust the Vortex loops that can be defined in $T[SU(N)]$. There are 
 Vortex loop operators in $T[SU(N)]$ realized by D1-strings ending on the $p$ D3-branes supporting the $U(p)$ gauge group factor, with $1 \le p \le N-2$. Since there are  no fundamental hypermultiplets for this gauge group factor, and correspondingly no  
 D5-branes crossing the D3-branes supporting the $U(p)$ gauge group factor,   Vortex loop operators in a $U(p)$ node are simply labelled by a representation $\scR$ of $U(p)$. We denote them by $V^{(p)}_\scR$, with $1 \le p \le N-2$. These operators admit a description in terms of a 1d $\cN=4$ quiver gauge theory coupled to $T[SU(N)]$, which 
    is   shown in figure (\ref{TSUNQMquivers}-b) for the case when    the D1-branes are moved to be  on top of the closest NS5-brane to their left.
 
  \smallskip

Let us start by considering a Vortex loop operators $V_{M,\scR}$   with $0 \le M \le N-1$. The case $V_{N,\scR}$ will be analyzed when discussing the Vortex loop operators $V^{(p)}_\scR$, which whom they share a  similar physical mirror interpretation. An operator $V_{M,\scR}$   is realized by $k$ D1-branes   with $N-M$ D5-branes on their left and $M$ D5-branes on their right and ending on the $N-1$ D3-branes supporting the $U(N-1)$ gauge node on one end and on a number of NS5 and NS5'-branes on the other end. $k$ denotes  the total number of boxes in the Young tableau associated to the representation $\scR$, which we also denote by $|\cR|$.
\begin{figure}[th]
\centering
\includegraphics[scale=0.7]{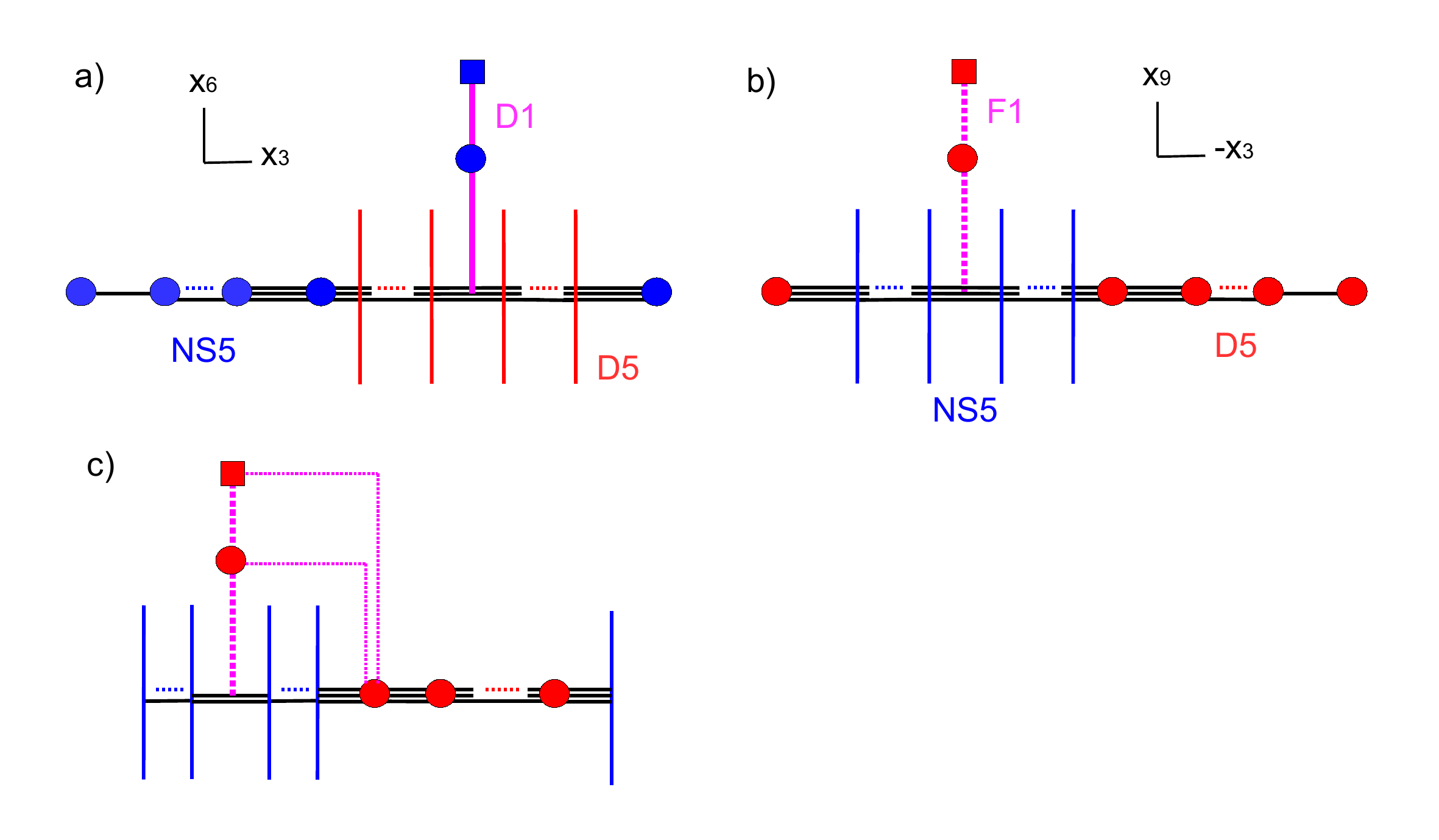}
\vspace{-0.5cm}
\caption{\footnotesize{a) Brane configuration inserting a Vortex loop $V_{M,\scR}$ .  b)  S-dual brane configuration with F1-strings. c) Configuration after moving the D5-branes. Straight strings, ending on D3-branes realize Wilson loops. The other strings, which are also straight despite their pictorial representation,} ending on the left D5-brane, are away from the main configuration and realize flavor Wilson loops.}
\label{TSUNbranegame1}
\end{figure}
Finding the mirror loop operator requires performing S-duality on the brane configuration (followed by a left-right reflection). The resulting configuration with F1-strings is shown in figure (\ref{TSUNbranegame1}-b). In order to  reach the canonical brane configuration of the mirror $T[SU(N)]$ theory, one has to move the D5-branes  until they all lie between the two rightmost NS5-branes. However, except for the configuration realizing $V_{N-1,\scR}$ studied in the previous section, a D5-brane (the leftmost one) has to cross the F1-strings in this process. After this exchange,  the number of D3-branes at the bottom of the F1-stack decreases. 
As explained earlier, when  a D5-brane crosses the stack of $k$ F1-strings, $q \le k$ of them can be moved smoothly from ending on the D3-branes  to ending on the D5-brane. Intuitively this is  possible since the D5-brane has a D3-brane spike
  and the F1-strings can be moved smoothly far from the main brane configuration along this spike. In the final brane picture shown in figure (\ref{TSUNbranegame1}-c) for $0 \le M \le N-1$,  the $k$ original F1-strings are split into $\hat k$ strings ending on the $M$ D3-branes supporting the $U(M)$ gauge group factor  and $q$ strings ending on the leftmost D5-brane, such that  $k=\hat k + q$ \footnote{The orientation of the F1-strings stretched between the D5/D5' away from the main stack and the leftmost D5 seem to break the supersymmetry of the configuration, however this is not the case in the limit when they are sent far away along the transverse $x^{4,5,6}$ directions.}. Such a brane configuration realizes a Wilson loop in a representation $\wat\scR$ of $U(M)$, which we denote by $W^{(M)}_{\wat\scR}$, where  $|\wat\scR |=\hat k$, together with a Wilson loop of charge $q$ under the $U(1)$ flavor symmetry rotating the $N$-th hypermultiplet (associated to the leftmost D5-brane), that we denote by $W^{\rm fl}_{N,q}$.  

\smallskip

There are multiple ways of partitioning the $k$ F1-strings into $\hat k$ strings ending on the D3-branes and $\hat q$ F1-branes attached to the leftmost D5-brane. 
We propose that the  mirror of $V_{M,\scR}$   with $0 \le M \le N-1$ is obtained by summing over the Wilson loops of the mirror dual theory realized by all the possible final brane configurations consistent with the s-rule.

\smallskip

 Our remaining task is to determine the precise sum over Wilson loops $W^{(M)}_{\wat\scR}$ and the representations  $\wat\scR$  of $U(M)$    that appear in the mirror of  $V_{M,\scR}$ .
In section \ref{sec:Branes} we considered brane configurations realizing Wilson loops with  F1-strings  ending on D3-branes on one end and on a set of D5 and D5'-branes on the other end. We argued that the  set of  patterns of  F1-strings  ending on   $N$ D3-branes are in one-to one correspondence with the  weights of  a representation of $U(N)$ associated to the Wilson loop realized by the brane configuration.  The same identification holds for D1-branes ending on D3-branes realizing a Vortex loop. More explicitly, a given weight $w=(w_1, w_2, \cdots, w_N)$ corresponds to a pattern of  strings (both for F1 and D1's) where $w_i$ strings end  on the $i$-th D3-brane in the stack. 
The expectation value of a loop operator, irrespective of whether  it is  a Wilson loop or a Vortex loop, is obtained by summing over all   weights, including degeneracies, which in the case of the Vortex loop are in one-to-one correspondence with the vacua of the corresponding 1d $\cN=4$ SQM that realizes the Vortex loop. In the brane language, this means summing over the string patterns allowed by the brane realization. 
This physical  picture will be made rather explicit  in section \ref{ssec:MM} when we evaluate the exact 1d $\cN=4$ SQM partition function describing a Vortex loop. 

\smallskip

After these preliminaries we can now keep track of what happens to a single-weight pattern for the  Vortex loop $V_{M,\scR}$   associated to a weight $w=(w_1, w_2, \cdots, w_{N-1})$, where $w_j$ D1-strings end on the $j$-th D3-brane, as we go to the mirror dual brane configuration. Implementing $S$-duality and moving the D5-branes as explained, we find the situation where the leftmost  D5-brane crosses the stack of F1-strings. 
A total of $N-1-M$ D3-branes at the end of  the F1-strings disappear in the process of exchanging the branes, and the F1-strings that are  ending on them get attached to the D5-brane. Labelling the D3-branes that disappear by  $j = M+1, \cdots, N-1$, we find  that a  total of $q=w_{M+1} + w_{M+2} + \cdots + w_{N-1}$ F1-strings get attached to the D5-brane. Therefore, the final brane configuration realizes a charge $q$ $U(1)$ flavor Wilson loop  together with a $\hat w = (w_1, w_2, \cdots, w_{M})$ single-weight contribution to a Wilson loop associated  to the remaining F1-strings ending on the $M$ D3-branes supporting the $U(M)$ gauge node. 
Enumerating all the   final patterns associated to all the weights $w$ of $\scR$, we find that they realize the weights of all the representations $({\bf q}, \wat\scR)$, counted with degeneracies, appearing in the decomposition of the representation $\scR$ of $U(N)$ under the subgroup $U(1) \times U(M) $, where $U(M)$ is embedded as $ U(M) \times U(N-M-1) \subset U(N-1)$ and  $U(1)$ is embedded diagonally in $U(N-M-1)$. 

\smallskip

Our mirror symmetry prediction is therefore that the vortex loop $V_{M,\scR}$   gets mapped to the sum of    combined gauge and flavor Wilson loops $W^{\rm fl}_{N,{\bf q}}\, W^{(M)}_{\wat\scR}$, with the sum running over the representations $({\bf q}, \wat\scR)$ appearing in the decomposition of the representation $\scR$ under the subgroup $U(1) \times U(M)$ of $U(N-1)$, where $U(M)$ is embedded as $U(M) \times U(N-M-1) \subset U(N-1)$ and  $U(1)$ is embedded diagonally in $U(N-M-1)$. In formulas 
\begin{align}
U(N-1) \ \rightarrow  \ U(1) \times U(M) \nonumber\\[+2ex]
 \scR  \ \rightarrow \ \bigoplus_{s \in \Delta_M} ({\bf q}_s,\wat\scR_s)\,,
\end{align}
where $\Delta_M$ denotes the set of representations $({\bf q},\wat\scR)$ in the decomposition of $\scR$ counted with degeneracies.
Our mirror map is therefore
\begin{align}
\vev{V_{M, \scR}}  \ \xleftrightarrow{\text{\ mirror \ }}   \  \vev{\sum_{s \in \Delta_M}    W^{\rm fl}_{N,{\bf q}_s}  \  W^{U(M)}_{\wat \scR_s}}   \, \quad 0 \le M \le N-1  \, ,
\label{MirrorMapAllLoops}
\end{align}
Pleasingly, this map  is  reproduced in examples in the exact computations of section \ref{ssec:TSUN}.\footnote{More precisely, this map between loop operators is found after reabsorbing some signs into imaginary shifts of parameters associated with the flavor Wilson loop (see section \ref{sssec:MirrorMapTSUN}).}
The  Vortex loop $V_{0, \scR}$ is   special:  it is mapped to a pure flavor Wilson loop.

\smallskip

\begin{figure}[h]
\centering
\includegraphics[scale=0.7]{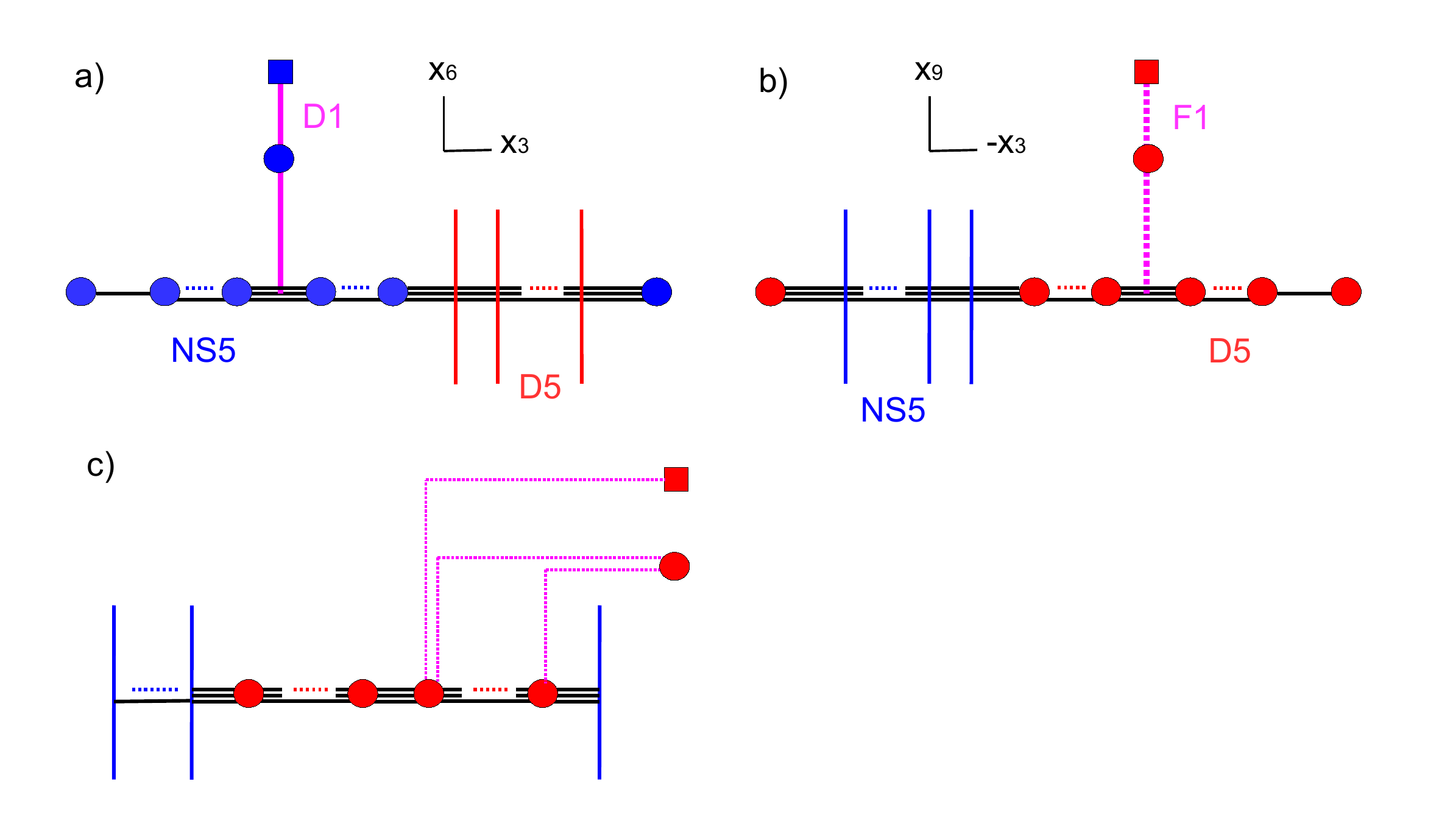}
\vspace{-0.5cm}
\caption{\footnotesize{a) Brane configuration inserting a Vortex loop $V^{(p)}_\scR$ .  b)  S-dual brane configuration. c) Configuration after moving the D5-branes. The F1-strings all end on D5-branes realizing pure flavor Wilson loops.}}
\label{TSUNbranegame2}
\end{figure}

Let us now conclude with the mirror map for the Vortex loops  $V^{(p)}_\scR$ and $V_{N, \scR}$ , which is qualitatively different to the one for $V_{M, \scR}$ with  $0 \le M \le N-1$. 
Let us consider the Vortex loop $V^{(p)}_\scR$, which is realized with $k$ D1-branes ending on the $p$ D3-branes supporting the $U(p)$ gauge node on one end and on a number of NS5 and NS5'-branes on the other, as shown in figure (\ref{TSUNbranegame2}-a). Here $k = |\cR|$ is again the total number of boxes in the Young tableau associated to the representation $\scR$ of $U(p)$.  
In order to  find the mirror loop  we perform once again S-duality on this brane configuration (together with an $x^3$ reflection). The resulting configuration with F1-strings is shown in figure (\ref{TSUNbranegame2}-b). In order to  reach the canonical brane configuration of the mirror $T[SU(N)]$ theory, one has to move the D5-branes. In the process $p$ out the $N$ D5-branes have to cross the stack of F1-strings from the right, after which there are no more D3-branes at the bottom of the initial F1-stack. Therefore,  all F1-strings got attached to these $p$ D5-branes and have been moved away from the main brane configuration along them. This means that the final brane configurations are characterized by having the $k$ F1-strings ending on the $p$ D5-branes on one side and on the D5 and/or D5' branes on the other side, as in figure (\ref{TSUNbranegame2}-c). We interpret this configuration as realizing a non-abelian Wilson loop in the representation $\scR$ of the $U(p)$ flavor symmetry acting on the $p$ hypermultiplets associated to the $p$ D5-branes, which we denote $W^{\rm fl}_{U(p), \scR}$. Therefore, the mirror map is 
\begin{align}
\vev{V^{(p)}_\scR} \ \xleftrightarrow{\text{\ mirror \ }} \  \vev{W^{\rm fl}_{U(p), \scR}}\quad 1 \le p \le N-2  \,.
\label{MirrorMapVp}
\end{align}
Likewise,  the Vortex loop $V_{N,\cR}$ is mirror to a $U(N-1)$ flavour Wilson loop
\begin{align}
\vev{V_{N,\cR}} \ \xleftrightarrow{\text{\ mirror \ }} \  \vev{W^{\rm fl}_{U(N-1), \scR}} \,.
\label{prediiii}
\end{align}

\medskip

\noindent
This concludes our explicit analysis of the mirror map between the loop operators of the self mirror $T[SU(N)]$ theory. 

 \smallskip
The inverted map, that is the map from Wilson loops to Vortex loops,  can be worked out from the above relations. It relates each Wilson loop to a combination of Vortex loops. 
 Let us provide here some illustrative examples. 
The mirror maps \eqref{MirrorMapAllLoops} relating the Vortex loops $V_{N-2,\scS_k}$ and $V_{N-2, \scA_k}$, labeled by the $k$-(anti)symmetric representation of $U(N-1)$, to the Wilson loops of the $U(N-2)$ node of the mirror $T[SU(N)]$ theory, predicted from the brane picture, are 
\begin{align}
& \vev{V_{N-2, \scS_k}}  \  \xleftrightarrow{\text{\ mirror \ }}   \  \vev{\sum_{\hat k =0}^k   W^{\rm fl}_{N, k - \hat k}  \  W^{U(N-2)}_{\scS_{\hat k}}} \, , \\[+3ex]
& \vev{V_{N-2, \scA_k}}  \  \xleftrightarrow{\text{\ mirror \ }}   \    \vev{W^{U(N-2)}_{\scA_{k}} +  W^{\rm fl}_{N,1}  \  W^{U(N-2)}_{\scA_{k -1}} } \, .
\end{align}
Inverting these relations one obtain 
\begin{align}
& \vev{W^{U(N-2)}_{\scS_{k}} }   \ \xleftrightarrow{\text{\ mirror \ }}   \  \vev{ V_{N-2, \scS_k} -  W^{\rm fl}_{N,1}  \  V_{N-2, \scS_{k-1}} } \, , \\[+3ex]
& \vev{W^{U(N-2)}_{\scA_{k}} }   \ \xleftrightarrow{\text{\ mirror \ }}   \    \vev{\sum_{\hat k =0}^k   (-1)^{k - \hat k} \ W^{\rm fl}_{N, k - \hat k}  \  V_{N-2, \scA_{\hat k}}  } \, .
\end{align}
It is tempting to try to derive these relation directly from the brane picture. This would necessitate new rules constraining the brane moves which are rather ad hoc.  It could be interesting to pursue this.

\smallskip

We have found an explicit map relating   Vortex loop operators  to  gauge and flavor Wilson loop operators.
We will check these predictions by exact computations of the expectation value  of the loop operators on $S^3$  in section \ref{ssec:TSUN}.

\subsection{Loops And Mirror Map In Circular Quivers}
\label{ssec:CircLoops}    

Our prescription to derive the mirror map of loop operators described in section \ref{sec:loopsbranes} is used here to find the 
  mirror map for loop operators in 3d $\cN=4$ gauge theories described by circular quivers with equal ranks. The brane realization of this class of quivers is special: after performing S-duality, the branes are already in the canonical brane configuration and no Hanany-Witten moves are required to derive the mirror quiver gauge theory. This immediately implies that a Wilson loop labeled by a representation $\scR$ of the gauge group directly maps to a Vortex loop labeled by the same representation $\scR$, and not to a sum of Vortex loop operators. 
\smallskip

\begin{figure}[th]
\centering
\includegraphics[scale=0.7]{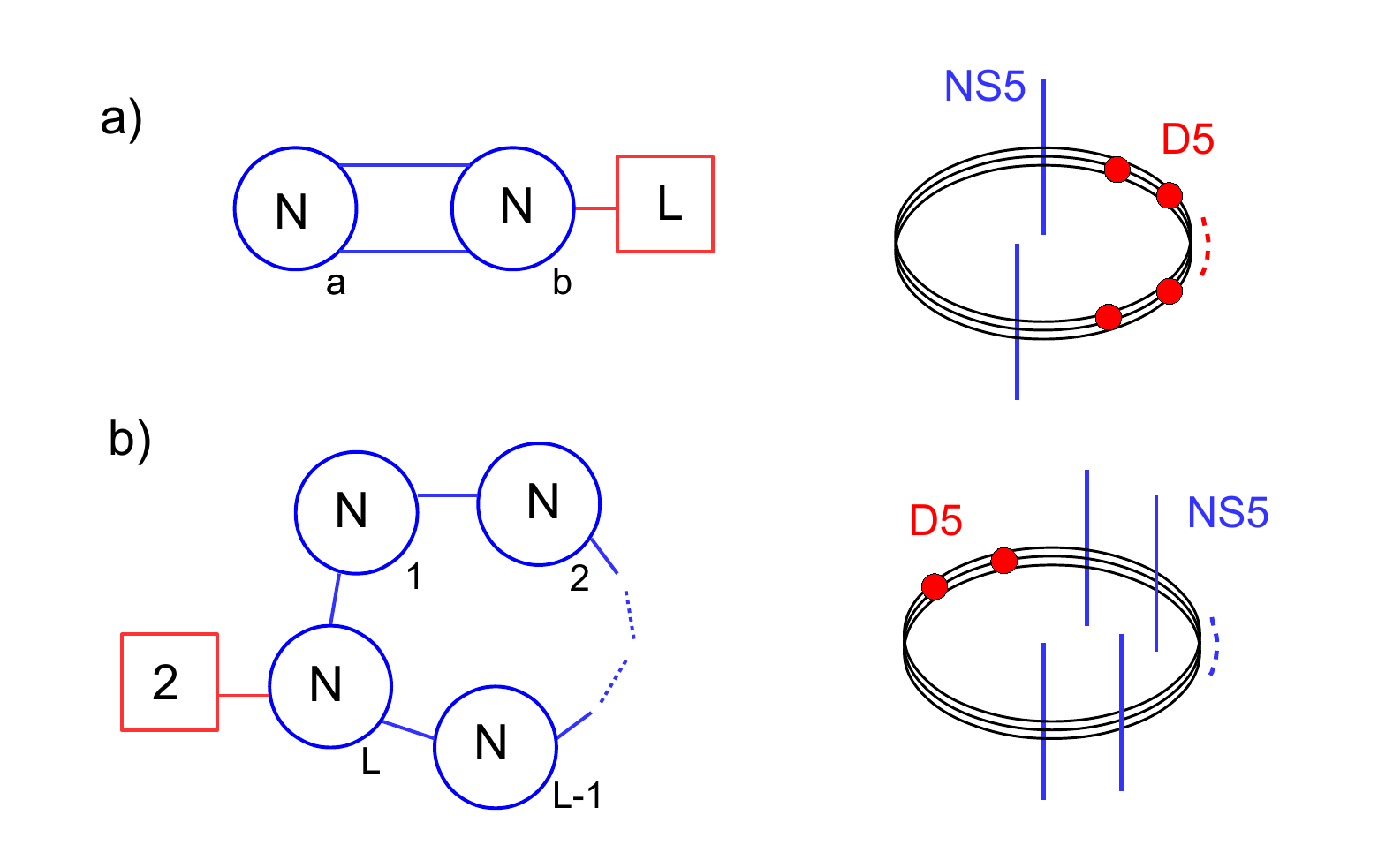}
\vspace{-0.5cm}
\caption{\footnotesize{a) $U(N)_a \times U(N)_b$ circular quiver described in the text and its canonical brane configuration.  b) Mirror dual circular quiver and brane configuration.}}
\label{circquiv2}
\end{figure}

For concreteness, let us consider the circular quiver in figure \ref{circquiv2}-a, which has two $U(N)$ gauge group nodes  $U(N)_a \times U(N)_b$ and $L\geq 2$ fundamental hypermultiplets of $U(N)_b$.\footnote{$L\geq 2$ is required for the UV gauge theory to flow to an irreducible SCFT in the IR.}
The mirror dual quiver is easily found by acting with S-duality and is shown in figure \ref{circquiv2}-b. It has gauge group $\prod_{i=1}^L U(N)_i$ and two fundamental hypermultiplets of  $U(N)_L$.

\begin{figure}[th]
\centering
\includegraphics[scale=0.7]{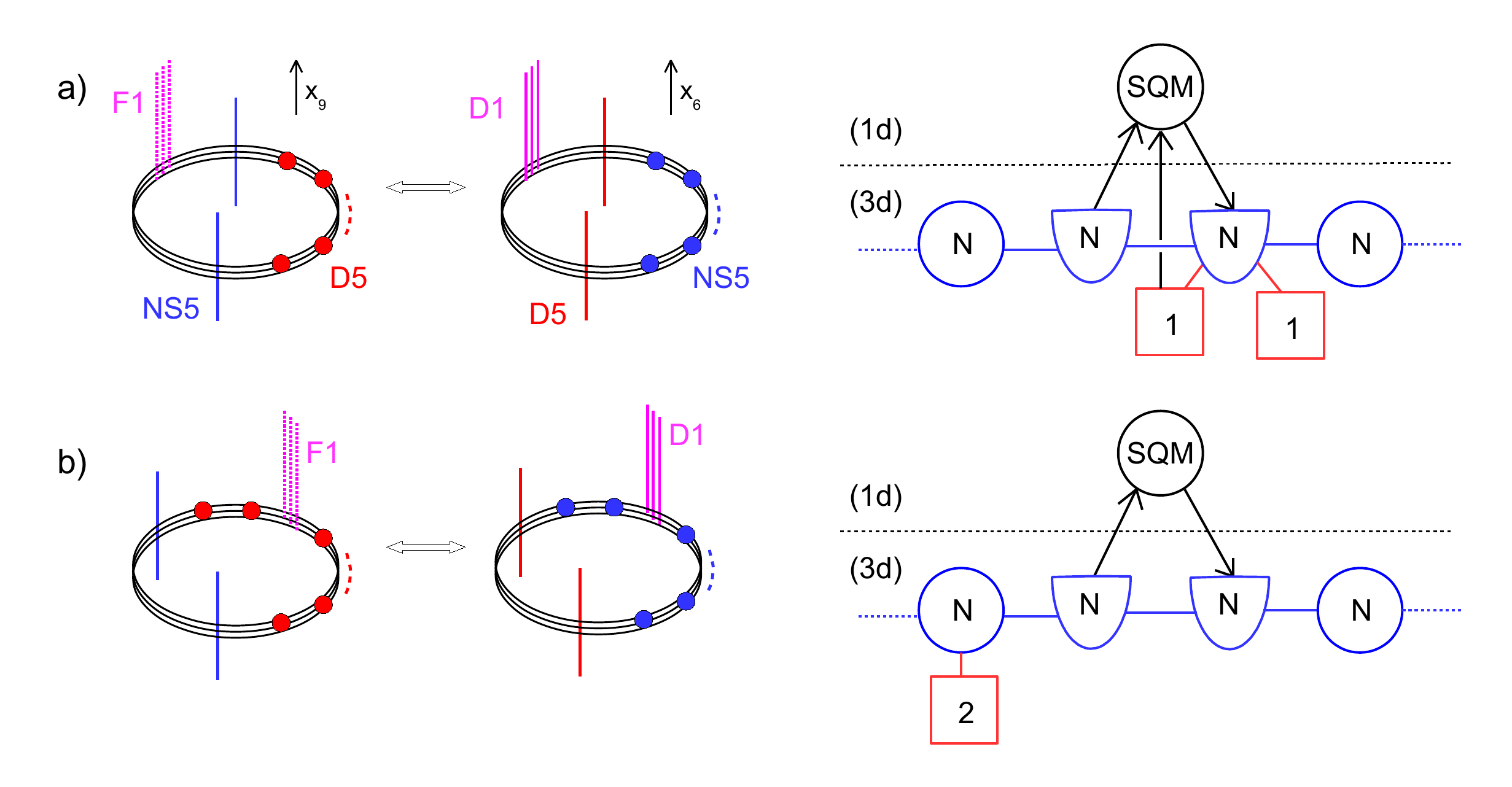}
\vspace{-0.5cm}
\caption{\footnotesize{a) Brane realization of $W^{(a)}_{\scR}$, S-dual brane configuration and 3d/1d defect theory realizing the dual loop $\ti V^{(L)}_{1,\scR}$. The details of the D5/D5' (NS5/NS5') array far from the main stack and the precise SQM$_V$ associated to $\scR$ are left unspecified.  b) Idem for the Wilson loop $W^{(b)}_{\scR}$ and the dual Vortex loop $\ti V^{(m)}_{\scR}$, with $m=2$.}}
\label{loopcircquiv}
\end{figure}

\smallskip
Let us first consider   a Wilson loop  labeled by a representation $\scR$ of $U(N)_a$, which we denote by $W^{(a)}_{\scR}$. This is  realized by the brane configuration in figure \ref{loopcircquiv}-a, where  F1-strings emanate from extra D5 and/or D5'-branes  far away from the main stack and end on the $N$ D3-branes supporting the $U(N)_a$ gauge node. After S-duality the brane configuration is already in its canonical form to read the mirror quiver theory. The F1-strings  become    D1-branes lying between the two D5-branes and ending on the $N$ D3-branes supporting the the $U(N)_L$ node. This configuration realizes a Vortex loop $\ti V^{(L)}_{1,\scR}$, where the index $1$  corresponds to the splitting of the two fundamental hypermultiplets of 
$U(N)_L$ into $2=1+1$ and $\scR$ is now seen as a representation of $U(N)_L$.  The operator $\ti V^{(L)}_{1,\scR}$ can be described by a coupled 3d/1d theory  which when      the D1-branes are  moved on top of the nearest NS5-brane to their left  is  summarized by the mixed 3d/1d quiver diagram in \ref{loopcircquiv}-a. The choice of representation $\cR$ is encoded in the 1d $\cN=4$ quiver gauge theory that is coupled to the 3d theory, as explained in section \ref{ssec:LoopSduality}.

\smallskip

We now consider a Wilson loop $W^{(b)}_{\scR}$ labeled by a representation $\scR$ of $U(N)_b$, realized by the brane configuration of figure \ref{loopcircquiv}-b. Here we can choose to position the F1-strings so that $m$ D5-branes are to their   left and $L-m$ D5-branes are to their   right.  We note that the choice of $m \in \{0, 1, \cdots, L\}$ does not affect the Wilson loop realized by the F1-strings.  $S$-duality maps this setup  to a brane configuration with D1-branes ending on the $N$ D3-branes realizing a Vortex loop $\ti V^{(m)}_{\scR}$ labeled by the same representation $\scR$ for the $U(N)_{m}$ node, for $m = 1, \cdots , L-1$. For $m=0$ and $m=L$ S-duality produces the Vortex loop operators $\ti V^{(L)}_{0,\scR}$  and  $\ti V^{(L)}_{2,\scR}$  labeled by a representation $\cR$ of $U(N)_L$, and where the indices $0$ and $2$ refer to the remaining hypermultiplet splittings $2 = 0 + 2 = 2 +0$. 
The operator $\ti V^{(m)}_{\scR}$ can be described by a coupled 3d/1d theory, which when obtained  by moving the D1-branes on top of the nearest NS5-brane to their left, is summarized by the mixed 3d/1d quiver diagram in figure \ref{loopcircquiv}-b (for $m=2$). We leave as an exercise for the interested reader to identify the the 3d/1d quiver theories describing  $\ti V^{(L)}_{0,\scR}$ and  $\ti V^{(L)}_{2,\scR}$. 

\smallskip
When giving a brane description of the Wilson loop $W^{(b)}_{\scR}$ we had a choice of  where to insert the F1-strings relative to the D5-branes, a choice that does not affect the choice of Wilson loop. This arbitrariness has an interesting counterpart in the    mirror dual theory. Different choices of $m$ result in different 3d/1d quiver gauge theories that describe the same mirror Vortex loop operator in the IR. These are dual descriptions of the same IR operator and are akin to the hopping duality introduced in \cite{Gadde:2013dda} (see also \cite{Gomis:2014eya}) in the context of surface operators in 4d $\cN=2$ gauge theories.

\smallskip
In summary, we find the following mirror maps
\begin{align}
\vev{W^{(a)}_{\scR}}& \ \xleftrightarrow{\text{\ mirror \ }} \ \vev{\ti V^{(L)}_{1,\scR}}\nonumber\\[+2ex]
\vev{W^{(b)}_{\scR}}& \ \xleftrightarrow{\text{\ mirror \ }} \ \vev{\ti V^{(m)}_{\scR}}=\vev{\ti V^{(L)}_{0,\scR} }=\vev{\ti 
V^{(L)}_{2,\scR}} \ , \ m = 1 , \cdots, L-1\,.
\label{CircLoopMap1}
\end{align}

\smallskip

\begin{figure}[th]
\centering
\includegraphics[scale=0.7]{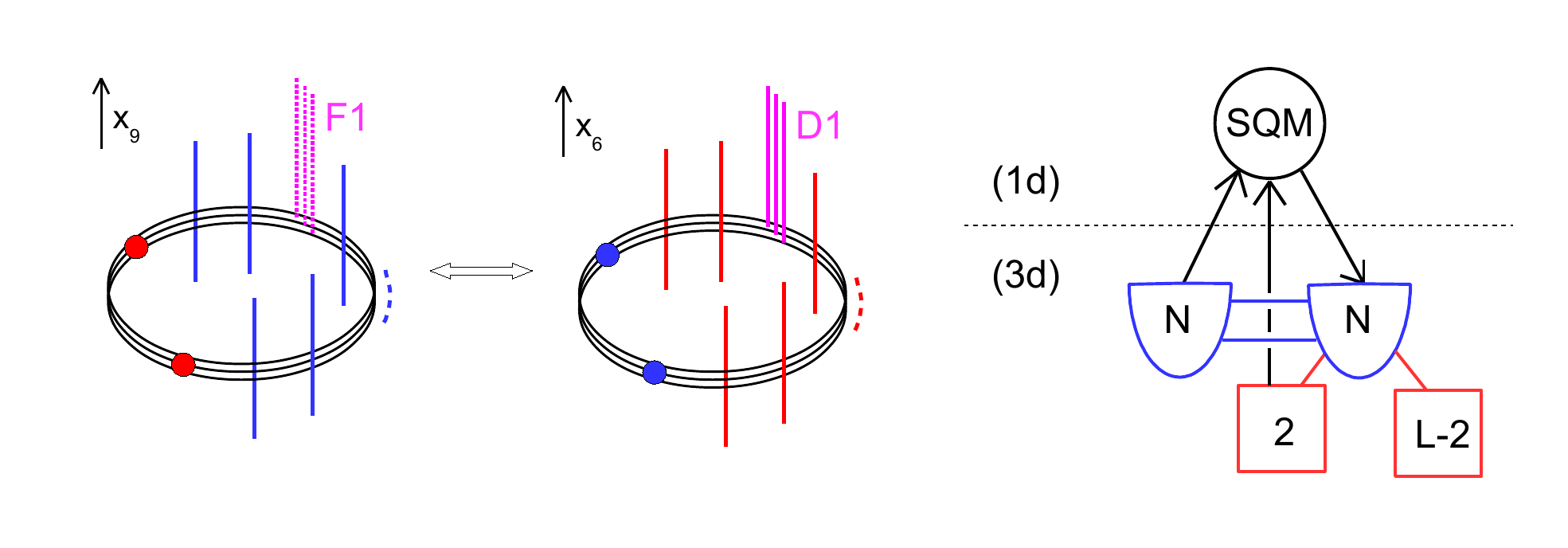}
\vspace{-0.5cm}
\caption{\footnotesize{a) Brane realization of $\ti W^{(m)}_{\scR}$, S-dual brane configuration and 3d/1d defect theory realizing the dual loop $V^{(b)}_{m, \scR}$, for $m=2$.}}
\label{loopcircquiv2}
\end{figure}

The mirror map between   Vortex loops of the initial quiver in figure \ref{loopcircquiv}-a and   Wilson loops of the mirror quiver in \ref{loopcircquiv}-b can be  found in the same way.
The two-node quiver has Vortex loops $V^{(a)}_{\scR}$ in the $U(N)_a$ node and $V^{(b)}_{m, \scR}$ in the $U(N)_b$ node, with $m$ denoting the splitting of hypermultiplets $L = m + (L-m)$ for $m =0 , \cdots, L$. 
The $L$-node quiver has Wilson loops $\ti W^{(m)}_{\scR}$ in the $U(N)_m$ node, with $m = 1 , \cdots , L$. 
The brane realizations of these loop operators are easily mapped under S-duality, as in the example of figure \ref{loopcircquiv2}. The precise mirror map is 
 \begin{align}
\vev{V^{(a)}_{\scR}}=\vev{V^{(b)}_{0, \scR}}=\vev{V^{(b)}_{L, \scR}}  
& \ \xleftrightarrow{\text{\ mirror \ }} \   \vev{\ti W^{(L)}_{\scR}} \nonumber\\[+2ex]
\vev{V^{(b)}_{m, \scR} }   & \ \xleftrightarrow{\text{\ mirror \ }} \ \vev{\ti W^{(m)}_{\scR}} \ , \ m = 1 , \cdots, L-1 \, .
\label{CircLoopMap2}
 \end{align}
We find once again that some Vortex loops have inequivalent descriptions, predicting hopping dualities for the associated 3d/1d theories.

\medskip

Other circular quivers with nodes of equal rank and arbitrary number of fundamental hypermultiplets in each node can be treated similarly, leading to explicit maps between loop operators labeled by the same representation $\cR$.
For circular quivers with varying ranks, one has to rely on the more elaborate analysis presented in section \ref{ssec:LoopSduality}, which still results in explicit mirror maps  between loop operators very similar to those in section \ref{ssec:OtherLoops}.

\subsection{Loops In SQCD with $2N$ Quarks And Its Mirror}
\label{ssec:SQCD}

\begin{figure}[th]
\centering
\includegraphics[scale=0.65]{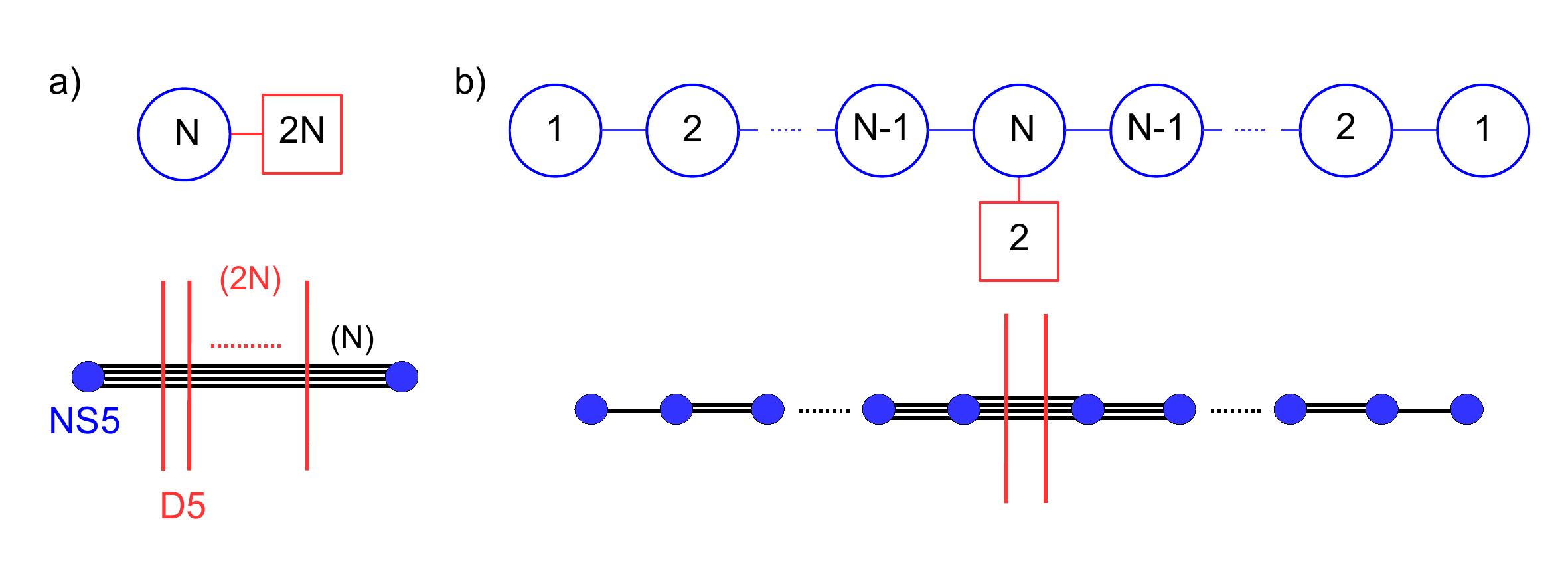}
\vspace{-0.5cm}
\caption{\footnotesize{a) Quiver and brane realization of the $U(N)$ theory with $2N$ fundamental hypermultiplets. b) Mirror theory and brane configuration.}}
\label{SQCDandMirror}
\end{figure}

For our final example we consider 3d $\cN=4$ SQCD, with $U(N)$ gauge group and $2N $ fundamental hypermultiplets, which is the smallest number of fundamental hypermultiplets ($N_f \ge 2 N_c$) required for  SQCD to flow in the IR to an irreducible SCFT.  The quiver  diagram and brane realization of this theory  is given in figure \ref{SQCDandMirror}-a.

\begin{figure}[th]
\centering
\includegraphics[scale=0.65]{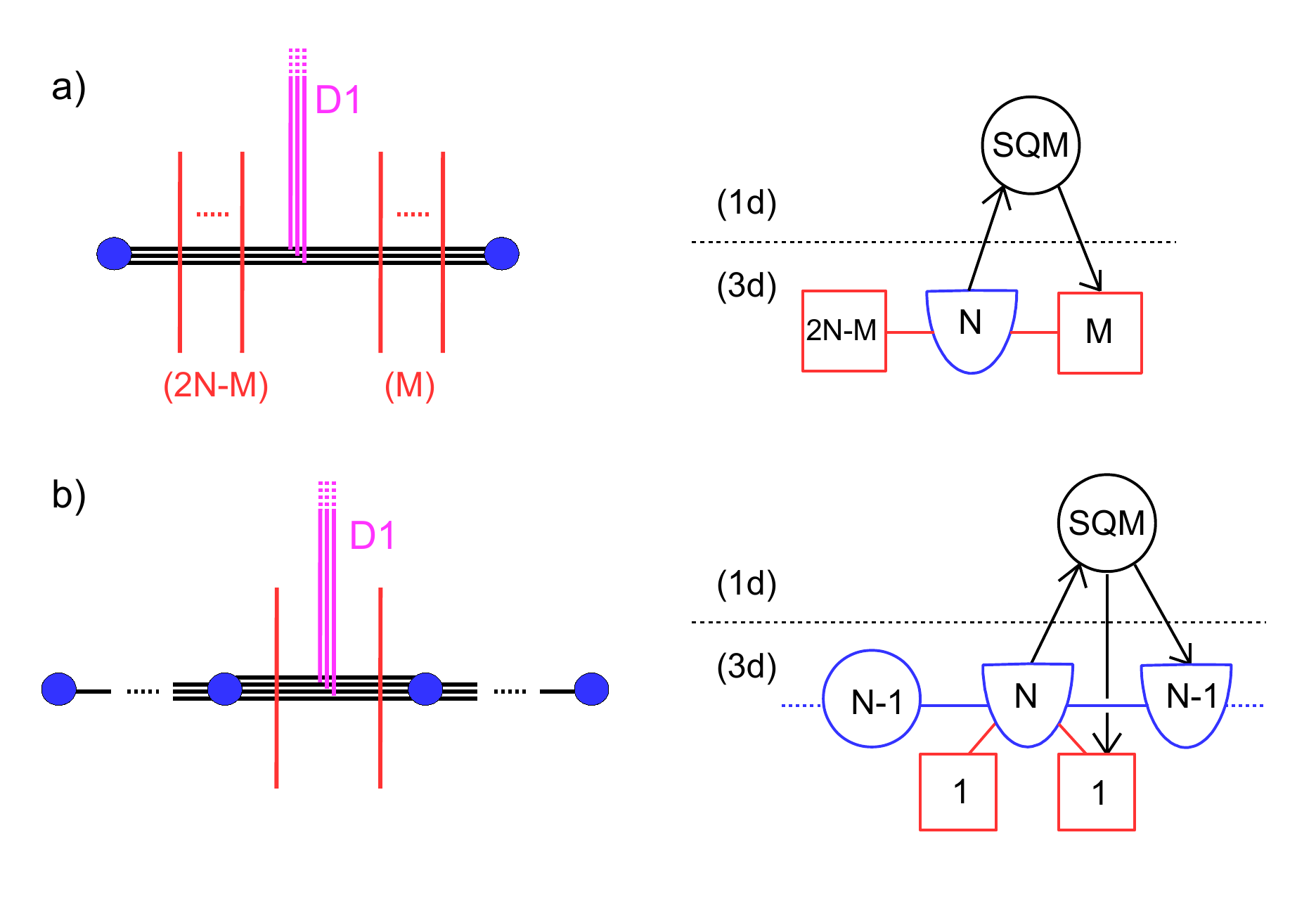}
\vspace{-0.5cm}
\caption{\footnotesize{a) Brane realization of a Vortex loop $V_{M,\scR}$ in the $U(N)$ theory and associated `right' 3d/1d quiver. b) Idem for a Vortex loop $\ti V^{(N)}_{1,\scR}$ of the mirror theory.}}
\label{SQCDVortex}
\end{figure}

This theory has Wilson loops $W_\scR$, labeled by a representation $\scR$ of $U(N)$ and Vortex loops $V_{M,\scR}$, labeled by a representation $\scR$ of $U(N)$ and a splitting $2N = M + (2N-M)$ of the fundamental hypermultiplets. The brane realization and corresponding 3d/1d quiver representation of $V_{M,\scR}$ are shown in figure \ref{SQCDVortex}-a.  As before, the choice of representation $\cR$ is encoded in the 1d $\cN=4$ quiver gauge theory that is coupled to the 3d theory, as explained in section \ref{ssec:LoopSduality}.

\smallskip

The mirror theory is shown is figure \ref{SQCDandMirror}-b. It is a quiver with gauge group
$\prod_{n=1}^{N-1} U(n)_a \times U(N)$ $ \times \prod_{n=1}^{N-1} U(n)_b$ with two fundamental hypermultiplets in the $U(N)$ node. The indices $a$ and $b$ distinguish the nodes on the left and on the right of the central node of the quiver.
This theory has Wilson loops $\ti W^{(a, n)}_\scR$ in the $U(n)_a$ nodes, $\ti W^{(N)}_\scR$ in the $U(N)$ node and $\ti W^{(b, n)}_\scR$ in the $U(n)_b$ nodes. 
We denote the Vortex loops of this mirror theory by $\ti V^{(a,n)}_\scR$ and $\ti V^{(b,n)}_\scR$ for the loops coupled to the $U(N)_{a}$ and $U(N)_{b}$ nodes and $\ti V^{(N)}_{0,\scR}, \ti V^{(N)}_{1,\scR}, \ti V^{(N)}_{2,\scR}$ the Vortex loops of the central $U(N)$ node, labeled by a splitting of the two fundamental hypermultiplets ($2 = 0 +2 = 1+1 = 0 +2$ respectively). We show the brane realization and associated 3d/1d quiver for the Vortex loop $\ti V^{(N)}_{1,\scR}$ in figure \ref{SQCDVortex}-b.

\medskip

\begin{figure}[th]
\centering
\includegraphics[scale=0.65]{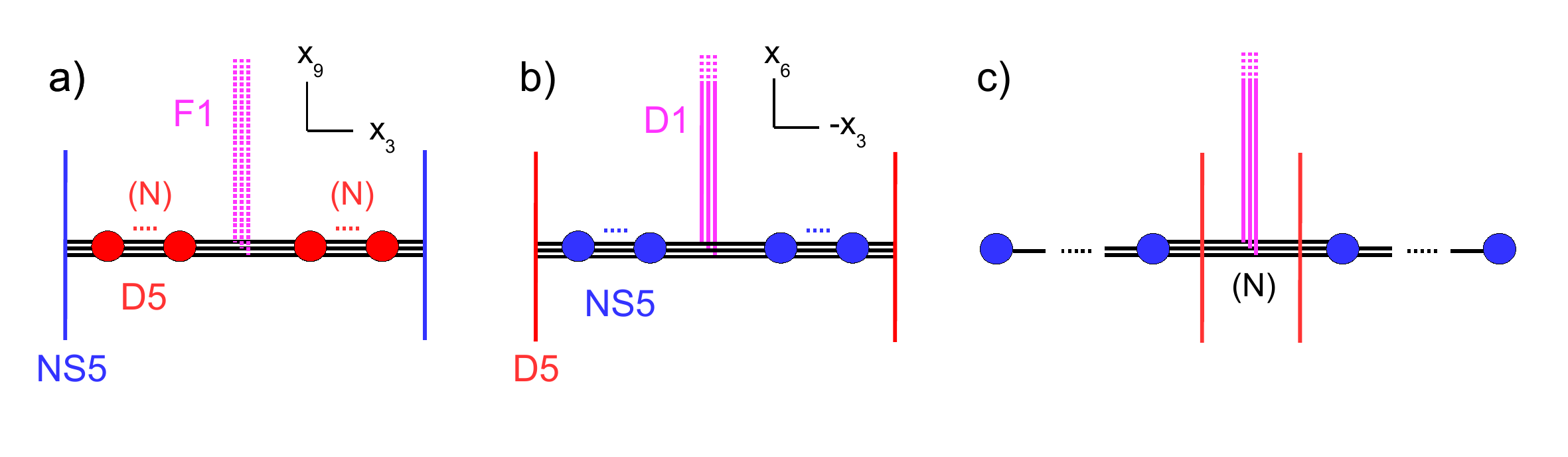}
\vspace{-0.5cm}
\caption{\footnotesize{a) Brane realization of a $U(N)$ Wilson loop $W_{\scR}$. b) S-dual brane configuration with D1-branes (and $x^3$ reflection). c) After D5-brane moves the D1-branes end on the $N$ D3-brane supporting the central $U(N)$ node of the mirror theory and are placed between the two D5s, realizing the Vortex loop $\ti V^{(N)}_{1,\scR}$. }}
\label{SQCDbranemap0}
\end{figure}

We now derive the mirror maps using the brane picture.
Consider first the Wilson loop $W_\scR$ of the SQCD theory. It is realized with F1-strings ending on the $N$ D3-branes supporting the $U(N)$ gauge group, placed among the $2N$ D5-branes at our convenience. We choose to place the strings with $N$ D5-branes on each side. After S-duality and D5-brane rearrangements, we end up with D1-branes ending on the $N$ D3-branes supporting the central $U(N)$ gauge node of the mirror theory, placed between the two D5-branes. This is summarized in figure \ref{SQCDbranemap0}. This configuration realizes the vortex loop $\ti V_{1,\scR}$, which is labeled by the same representation $\scR$ of $U(N)$ as the initial Wilson loop. The mirror map is then simply
\beq
\vev{W_{\cR}} \ \xleftrightarrow{\text{\ mirror \ }} \ \vev{\ti V^{(N)}_{1,\scR}}\,.
\label{MapUNsimple}
\eeq
\smallskip

\begin{figure}[th]
\centering
\includegraphics[scale=0.65]{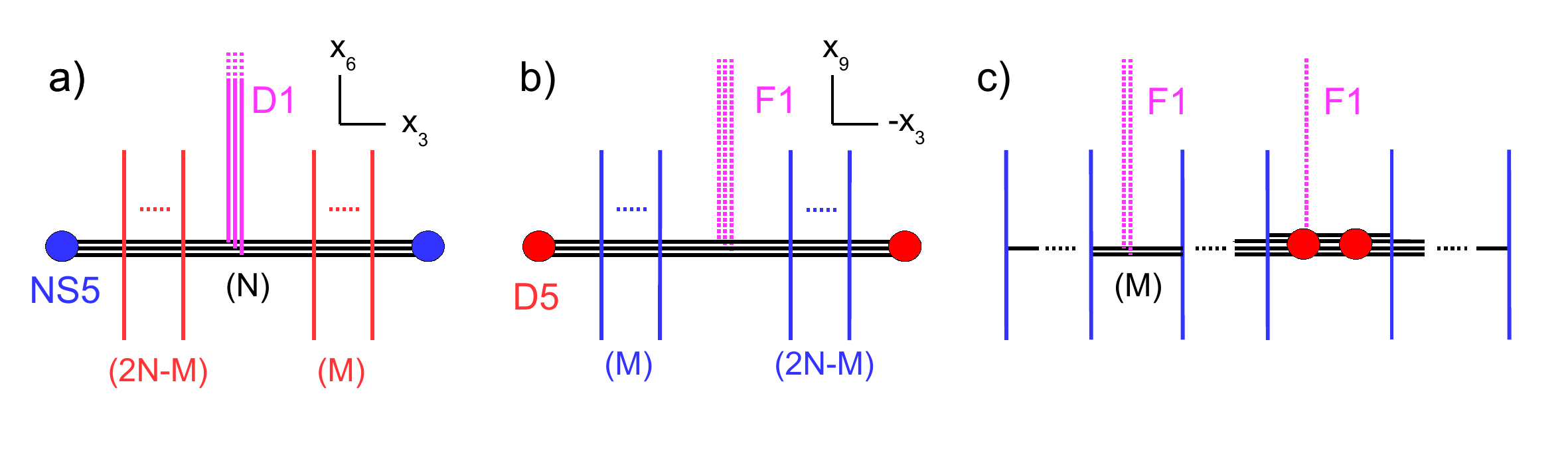}
\vspace{-0.5cm}
\caption{\footnotesize{a) Brane realization of the Vortex loop $V_{M,\scR}$ with $M \le N$. b) S-dual brane configuration with F1-strings (and $x^3 \rightarrow -x^3$). c) After D5-brane moves, part of the strings end on the left D5 (flavor Wilson loop) and the other strings end on the $M$ D3-branes supporting the $U(M)_a$ node (gauge Wilson loop). }}
\label{SQCDbranemap}
\end{figure}

Consider now a Vortex loop $V_{M,\scR}$ of the SQCD theory and assume $M \le N$. The loop is realized with D1-branes ending on the $N$ D3-branes, with $M$ D5-branes on its right and $2N-M$ D5-branes on its left. After S-duality (and $x^3$ reflection), we end up with the configuration in figure \ref{SQCDbranemap}-b. The D5-brane on the left of the picture has to be moved across the F1-strings. This is the same situation as encountered when studying $T[SU(N)]$ Vortex loops. When the D5 crosses the F1-strings the number of D3-branes at the bottom of the strings diminishes and part of the strings get attached to the D5-brane. The final brane configurations have F1-strings ending on the $M$ D3-branes supporting the $U(M)_a$ gauge node and F1-strings ending on the left D5-brane, as in figure \ref{SQCDbranemap}-c. They realize Wilson loops $\ti W^{(a, M)}_{\wat \scR}$ in the $U(M)_a$ node, together with flavor Wilson loops $\ti W^{\rm fl}_{1,{\bf q}}$ of charge ${\bf q}$ under the $U(1)$ flavour symmetry rotating the hypermultiplet associated to the left D5-brane, where the representation $({\bf q}, \wat\scR)$ appears in the decomposition of the representation $\scR$ under the subgroup $U(1) \times U(M)$ of $U(N)$, where $U(M)$ is embedded as $U(M) \times U(N-M) \subset U(N)$ and  $U(1)$ is embedded diagonally in $U(N-M)$.

Following the arguments that we presented for the $T[SU(N)]$ loops in \ref{ssec:OtherLoops}, we arrive at the explicit mirror map 
\begin{align}
\vev{V_{M,\scR}}  \ \xleftrightarrow{\text{\ mirror \ }}  \  \vev{\sum_{s \in \Delta_M}   \ti W^{\rm fl}_{1,{\bf q}_s}  \  \ti W^{(a, M)}_{\wat \scR_s}}   \, , \quad 0 \le M \le N  \, ,
\label{MapUNComplicated}
\end{align}
where $\Delta_M$ denotes the set of representations $({\bf q},\wat\scR)$ in the decomposition of $\scR$ under $U(N) \rightarrow U(1) \times U(M)$, counted with degeneracies.
This is completely analogous to the mirror maps found for the $T[SU(N)]$ theory.

The map relating the Vortex loops $V_{M,\scR}$ with $M \ge N$ to combinations of Wilson loops $\ti W^{\rm fl}_{2,{\bf q}}  \ti W^{b, M}_{\wat \scR}$ is found similarly, with $\ti W^{\rm fl}_{2,{\bf q}} $ denoting a flavor Wilson loop of charge ${\bf q}$ under the $U(1)$ flavour symmetry rotating the hypermultiplet associated to the right D5-brane this time.
Note that the `central' Vortex loop $V_{N,\scR}$ is directly mapped to the Wilson loop $\ti W^{(N)}_{\scR}$ of the central node of the mirror theory:
\begin{align}
\vev{V_{N,\scR}}  \ \xleftrightarrow{\text{\ mirror \ }}  \ \vev{\ti W^{(N)}_{\scR}} \, .
\end{align}

The only loops we did not discuss thus far  are the Vortex loops $\ti V^{(a,n)}_\scR$, $\ti V^{(b,n)}_\scR$ of the $U(n)_a$ and $U(n)_b$ nodes of the mirror theory, and the $\ti V^{(N)}_{0,\scR}, \ti V^{(N)}_{2,\scR}$ Vortex loops of the central $U(N)$ node, which can be naturally renamed as $\ti V^{(a,N)}_\scR$ and $\ti V^{(b,N)}_\scR$ respectively. The analysis of the brane realization of these loops is the same as for the Vortex loops $V^{(p)}_\scR$ of the $T[SU(N)]$ theory (see section \ref{ssec:OtherLoops}) and lead to the conclusion that these loops are mapped to flavor Wilson loops. In particular
\begin{align}
\vev{\ti V^{(a,n)}_\scR} &= \vev{W^{\rm fl}_{U(n)_F, \scR}} \, , \quad n = 1, \cdots , N \, , \\[+2ex]
\vev{\ti V^{(b,n)}_\scR} &= \vev{W^{\rm fl}_{U(n)_{F'}, \scR}} \, , \quad n = 1, \cdots , N \, ,
\end{align}
where $U(n)_F$ is the flavor group acting on the $n$ hypermultiplets associated to the $n$ leftmost D5-branes, $U(n)_{F'}$ is the flavor group acting on the $n$ hypermultiplets associated to the $n$ rightmost D5-branes and $W^{\rm fl}_{U(n), \scR}$ denotes a flavor Wilson loop in the representation $\scR$ of $U(n)$.

\smallskip
This completes the derivation of the mirror map between the loops of the $U(N)$ theory with $2N$ fundamental hypermultiplets and its mirror dual. The analysis generalizes easily to the $U(N)$ theory with $N_f \geq 2N$ fundamental hypermultiplets.

\section{Loop Operators In 3d $\cN=4$ Theories On $S^3$}
\label{sec:LoopsOnS3}

In this section we provide explicit quantitative evidence for our proposed  action of mirror symmetry on loop operators by computing the exact 
expectation value of supersymmetric loop operators on $S^3$ in 3d $\cN=4$ gauge theories. We confirm 
our proposal for  the action of mirror symmetry on loop operators by showing that the expectation value of mirror loop operators match perfectly and indeed give alternative UV descriptions of superconformal line defects in the IR SCFT.
\smallskip

Our first task is to define these observables in $\cN=4$ theories on $S^3$, where exact computations can be performed by supersymmetric localization. Recall that in  $R^3$  the UV Lagrangian definition of a 3d $\cN=4$ SCFT is invariant under those supercharges in $OSp(4|4)$ that close into the isometries of flat space, which generate the 3d $\cN=4$ super-Poincar\'e algebra.\footnote{Super-Euclidean in $R^3$.} While an  $\cN=4$ SCFT  can be placed canonically on the round  $S^3$ by stereographic projection,  its UV  Lagrangian description likewise breaks   $OSp(4|4)$ to a subalgebra, which we now identify.
On $S^3$, the supersymmetries   of a generic UV (massive) 3d $\cN=4$ theory  are generated 
by the supercharges in $OSp(4|4)$ that close into the $SU(2)_l\times SU(2)_r$ isometries of $S^3$  and project  out the remaining $SO(4,1)$  conformal generators on $S^3$. Since we allow arbitrary massive deformations, 
and  mass and FI parameters transform in the  $(\bf{3},\bf{1})$ and  $(\bf{1},\bf{3} )$ representations of   $SU(2)_C\times SU(2)_H$, the R-symmetry  in $OSp(4|4)$ is broken down to its Cartan subalgebra on $S^3$. Under the $SU(2)_l\times SU(2)_r\times U(1)_C\times U(1)_H$ embedding in $SO(4,1)\times SU(2)_C\times SU(2)_H$, the supercharges generating $OSp(4|4)$, which transform in the $(\bf{4},\bf{2},\bf{2})$,  decompose as $(\bf{2},\bf{1})_{++}\oplus (\bf{2},\bf{1})_{+-}\oplus (\bf{2},\bf{1})_{-+}\oplus (\bf{2},\bf{1})_{--}\oplus (\bf{1},\bf{2})_{++}\oplus (\bf{1},\bf{2})_{--}\oplus (\bf{1},\bf{2})_{-+}\oplus  (\bf{1},\bf{2})_{--}$.  The supersymmetries preserved by a UV 3d $\N=4$  theory on $S^3$  are $(\bf{2},\bf{1})_{++}\oplus (\bf{2},\bf{1})_{--}\oplus (\bf{1},\bf{2})_{+-}\oplus (\bf{1},\bf{2})_{-+}$. These supercharges generate the following supergroup\footnote{This supersymmetry  is the fixed locus of an involution in $OSp(4|4)$.} 
 \beq
   SU(2|1)_l\times SU(2|1)_r\,.
   \eeq
   The $U(1)_l$ generator in $SU(2|1)_l$ is $R_C+R_H$  while the $U(1)_r$ generator in $SU(2|1)_l$ is $R_H-R_C$, where $R_C$ and $R_H$ are the Cartan generators of $SU(2)_C$ and $SU(2)_H$ respectively. The isometries of $S^3$ are $SU(2)_l\times SU(2)_r$. Explicitly, the anticommutators of the $SU(2|1)_l\times SU(2|1)_r$ supercharges on an $S^3$ of radius $L$ are 
   \begin{align}
   \{Q_{l\, \alpha}, \overline Q_{l\, \beta}\}&= {2\over L}\left[\left(\gamma^aC^{-1}\right)_{\alpha\beta}J^l_a+{1\over 2} \epsilon_{\alpha\beta} \left(R_C+R_H\right)\right]\\[5pt]
   \{Q_{r\, \alpha}, \overline Q_{r\, \beta}\}&= {2\over L}\left[\left(\gamma^aC^{-1}\right)_{\alpha\beta}J^r_a+{1\over 2} \epsilon_{\alpha\beta} \left(R_C-R_H\right)\right] \,,  \label{susysphere}
   \end{align}
   where $J_a^l$ and $J_a^r$ generate $SU(2)_l$ and $SU(2)_r$ respectively.

   \smallskip
In the flat space $L\rightarrow \infty$ limit, the $SU(2|1)_l\times SU(2|1)_r$ symmetry on the $S^3$ contracts to the 3d $\cN=4$ super-Poincar\'e symmetry. Indeed, writing $J^l_a={1\over 2}(J_a+LP_a)$ and $J^r_a={1\over 2}(J_a-LP_a)$, and  taking the flat space limit we get
 \begin{align}
   \{Q^\infty_{l\, \alpha}, \overline Q^\infty_{l\, \beta}\}&=  \left(\gamma^aC^{-1}\right)_{\alpha\beta}P_a \\[5pt]
   \{Q^\infty_{r\, \alpha}, \overline Q^\infty_{r\, \beta}\}&= - \left(\gamma^aC^{-1}\right)_{\alpha\beta}P_a\,,
   \end{align}
where $P_a$ and $J_a$ are the translation and rotation generators of $R^3$. Comparing with the 3d $\cN=4$ super-Poincar\'e supercharges $Q_{\alpha AA'}$ in \rf{threePoincare} we find the identification: $Q^\infty_{l\, \alpha}=Q_{\alpha 22}, \overline Q^\infty_{l\, \alpha}=Q_{\alpha 11},\newline {Q^\infty_{r\, \alpha}=Q_{\alpha 21}}$ and 
$\overline Q^\infty_{r\, \alpha}=Q_{\alpha 12}$.

\smallskip

      The $SU(2|1)_l$ supersymmetry transformations are generated by  Killing spinors $\epsilon_l$ and $\bar \epsilon_l$
   on $S^3$  obeying \cite{Kapustin:2009kz}\footnote{All these Killing spinors are  the subset of conformal Killing spinors on $S^3$, which obey $\nabla_\mu \epsilon_{AA'}={1\over 3}\gamma_\mu\slashed \nabla \epsilon_{AA'}$, that project out the conformal generators on $S^3$.}
     \beq
\nabla_\mu \epsilon_l={i\over 2L} \gamma_\mu \epsilon_l\qquad \nabla_\mu \bar \epsilon_l={i\over 2L} \gamma_\mu \bar \epsilon_l\,,
\label{KSELeft}
\eeq
while  $SU(2|1)_r$ transformations are  generated by  Killing spinors  $\epsilon_r$ and $\bar \epsilon_r$ on $S^3$ which satisfy \cite{Samsonov:2014pya}
\beq
\nabla_\mu \epsilon_r=-{i\over 2L} \gamma_\mu \epsilon_r\qquad \nabla_\mu \bar \epsilon_r=-{i\over 2L} \gamma_\mu \bar \epsilon_r\,.
\label{KSERight}
\eeq
From the viewpoint of off-shell supersymmetric supergravity backgrounds \cite{Festuccia:2011ws}, these Killing spinor equations arise from 3d $\cN=4$ supergravity in the presence of  a background auxiliary field.\footnote{The background fields in supergravity and conformal compensators are very similar to those required to put 4d $\cN=2$ theories on $S^4$, see \cite{Gomis:2014woa}.} The superisometry of this supergravity background is $SU(2|1)_l\times SU(2|1)_r$.

\smallskip
 We take the left-invariant frame $e^a = L \mu^a$, $a=1,2,3$, where the left-invariant one-forms of $SU(2)$ are defined by $g^{-1} dg = i \mu^a \tau_a$ with $ g = \left(
\begin{array}{cc}
z_1 & z_2 \\
- \bar z_2 & \bar z_1 
\end{array} 
\right)
$ so that $\det g =1$. In this frame $\epsilon_l$ and $\bar \epsilon_l$ are constant while $\epsilon_r$ and $\bar\epsilon_r$ are given by $\epsilon_r = g^{-1} \epsilon_0$, $\bar\epsilon_r = g^{-1} \ti\epsilon_0$, with $\epsilon_0, \ti\epsilon_0$ constant.\footnote{We note that all Killing spinors generating $SU(2|1)_l\times SU(2|1)_r$ are periodic along the Hopf circle.} 
 
 \smallskip
 
 The   explicit 3d $\cN=4$ supersymmetric Lagrangians on $S^3$  can be written down   using the formulae  
  for 3d $\cN=2$ gauge theories on $S^3$ in \cite{Kapustin:2009kz,Hama:2010av,Jafferis:2010un,Hama:2011ea,Samsonov:2014pya}, which are invariant under  $SU(2|1)_l\times SU(2)_r$. We  first decompose the  $\cN=4$ vector multiplet  and hypermultiplet   into $\cN=2$
 vector and chiral multiplets. By  assigning $U(1)_l$ charge 1 and 1/2 to  the chiral multiplets  inside an  $\cN=4$ vector multiplet  and hypermultiplet  respectively, the supersymmetry of such an $\cN=2$ theory on $S^3$  is   enhanced  with extra four     supercharges \cite{Samsonov:2014pya}, which generate the remaining  $SU(2|1)_r$.  Just as in flat space, 3d $\cN=4$ theories on $S^3$ admit canonical relevant  deformations associated to the $G_C\times G_H$ symmetries of the IR SCFT. These are introduced by turning on $SU(2|1)_l\times SU(2|1)_r$ invariant background vector and twisted vector multiplets on $S^3$ for  $G_H$ and $G_C$ respectively.  Supersymmetric backgrounds for  3d $\cN=4$ vector and twisted vector multiples on $S^3$ allow a single scalar in the multiplet to be turned on, instead of the triplet of parameters in flat space. Therefore on $S^3$ there is a single mass parameter and a single  FI for each Cartan generator in $G_H$ and $G_C$ respectively.
 
 \smallskip
 
Following   a similar analysis to  the one in flat space   in section \ref{sec:loops}, there are two inequivalent classes of half-supersymmetric line operators that can be defined  in a UV description of a 3d $\cN=4$ SCFT on $S^3$. They are characterised by two different  embeddings of $SU(1|1)_l\times SU(1|1)_r$ inside $SU(2|1)_l\times SU(2|1)_r$. One class corresponds to supersymmetric Wilson loop operators and the other to Vortex loop operators on $S^3$.

 \subsection{Wilson Loops On $S^3$}
 \label{ssec:Wilsonsphere}

 Consider a Wilson loop operator wrapping a curve $\gamma\in S^3$ 
 \beq
 \hbox{Tr}_R P\exp{\oint_\gamma \left(i A_\mu \dot x^\mu-|\dot x|\sigma\right)d\tau}\, 
 \label{wilsonsphere}
      \eeq       
 in a 3d $\N=4$ gauge theory,   where $\sigma$ is the scalar field in the $\cN=2$ vector multiplet inside the $\cN=4$ vector multiplet. This operator 
    preserves the $SU(2|1)_l\times SU(2|1)_r$ Killing spinors that   are solutions to the following equations on $S^3$ (see appendix \ref{app:Embeddings})\footnote{These equations hold  at the position of the loop.}
   \begin{align}
   \left(\gamma_\mu \dot x^\mu-|\dot x|\right)\epsilon_l&=0\qquad  \left(\gamma_\mu \dot x^\mu+|\dot x|\right)\bar\epsilon_l=0\\[5pt]
  \left(\gamma_\mu \dot x^\mu-|\dot x|\right)\epsilon_r&=0 \qquad \left(\gamma_\mu \dot x^\mu+|\dot x|\right)\bar\epsilon_r=0 \,.
   \end{align}
   By a choice of parametrization of the loop $\gamma\in S^3$ we can always make $|\dot x|=1$. 
 Since in the $SU(2)_l$ left-invariant  frame the Killing spinors $\epsilon_l$ and $\bar\epsilon_l$ are constant,  imposing that the Wilson loop preserves half of the supercharges in $SU(2|1)_l$ requires that 
    \beq
    \dot x^\mu= n^a e_a^\mu\,,
    \label{circlmax}
    \eeq
where $n^a$ is a constant unit  three-vector $n^an^a=1$. 
    A Wilson loop wrapping  the curve $\dot x^\mu= n^a e_a^\mu$ on $S^3$ preserves the following half of the $SU(2|1)_l$ supersymmetries
    \beq
    \gamma_a n^a \epsilon_l=\epsilon_l\qquad  \gamma_a n^a \bar\epsilon_l=-\bar\epsilon_l\,.
    \label{susyhopf}
    \eeq
The integral curves of the vector field $\dot x^m= n^a e_a^{m}$ are great circles on $S^3$.  The choice of unit vector $n^a$ determines a Hopf fibration. Therefore, there is an $S^2=SU(2)/U(1)$  worth of choices of Hopf fibrations. Given a choice of Hopf fibration, there is another $S^2=SU(2)/U(1)$ worth of Hopf circles, labeled by the point on $S^2$ where the Hopf fiber sits. This realizes   the space of maximal circles on $S^3$ as $SO(4)/U(1)\times U(1)=S^2\times S^2$. Note also that all  circles in a choice of Hopf fibration preserve the same supersymmetry in $SU(2|1)_l$.

\smallskip
Without loss of generality we   take $n^a=(0,0,1)$.
The Hopf circle  preserves $J_3^l$ generating $U(1)_l\subset SU(2)_l$ and the $SU(2|1)_l$ supersymetries
\beq
\gamma_3 \epsilon_l=\epsilon_l\qquad \gamma_3 \bar\epsilon_l=-\bar\epsilon_l\,.
\label{Susywsa}
\eeq 
The Hopf circle, which is labeled by a point on $S^2$, also preserves   $U(1)_r \subset SU(2)_r$. 
A Wilson loop supported on this Hopf circle preserves the following supersymmetries in $SU(2|1)_r$
\beq
\gamma_3 \epsilon_r= \epsilon_r\qquad \gamma_3 \bar\epsilon_r= - \bar\epsilon_r\,,
\label{Susywsb}
\eeq
where $\epsilon_r$ and $\bar\epsilon_r$ are non-constant spinors  in the  left-invariant frame. We can without loss of generality take the loop at the North or South pole of $S^2$, in which case $U(1)_r$ is generated by $J_3^r$.  For another Hopf circle at a different point in $S^2$, which can be obtained by the action of an isometry, the preserved $U(1)_r$ and supersymmetry generators are obtained from those at the poles by conjugating  by the action of the isometry. 
   
    \smallskip
  In summary,  the Wilson loop \rf{wilsonsphere} wrapping the Hopf circle at the  North pole of $S^2$ is half-supersymmetric in  a 3d $\cN=4$ theory on $S^3$. It preserves an $SU(1|1)_l\times SU(1|1)_r$ inside the $SU(2|1)_l\times SU(2|1)_r$ supersymmetry of a 3d $\cN=4$ theory on $S^3$. The explicit supersymmetry algebra of the Wilson loop is
  \begin{align}
\{q_l,\bar q_l\} =& \frac 2L \Big[ J_3^l-{1\over 2}\left(R_C+R_H\right) \Big]  \\[6pt]
\{q_r,\bar q_r\} =& \frac 2L \Big[ J_3^r - {1\over 2}\left(R_C-R_H\right) \Big] \,.
\label{wilsonesfera}
\end{align}
 The insertion of this  Wilson loop on $S^3$ can be also realized by coupling the 1d $\cN=4$ fermi multiplet localized on the Hopf circle to the 3d $\cN=4$ theory on $S^3$. Putting the 1d gauged fermi multiplet theory on $S^3$ while preserving $SU(1|1)_l\times SU(1|1)_r$ is  trivial as there are no $1/L$ modifications to the 1d gauged fermi theory in flat space. Integrating out the fermi multiplet inserts  the Wilson loop operator \rf{wilsonsphere}.

 \subsection{Vortex Loops On $S^3$}
 \label{ssec:Vortexsphere}
 
Another family of half-supersymmetric line operators can be defined in  3d $\cN=4$ gauge theories on $S^3$. These loop operators  preserve a different $SU(1|1)_l\times SU(1|1)_r$ inside the $SU(2|1)_l\times SU(2|1)_r$ supersymmetry   on $S^3$. They correspond  to Vortex loop operators. A Vortex loop operator supported on  a Hopf circle at the North or South pole of $S^3$   preserves the following supersymmetries\footnote{A Vortex loop wrapping a different Hopf circle can be obtained by conjugating by a symmetry.} 
\begin{align}
\gamma_3 \epsilon_l&=\epsilon_l\qquad \qquad\gamma_3 \bar\epsilon_l=-\bar\epsilon_l\\
\gamma_3 \epsilon_r&=-\epsilon_r\qquad \qquad\gamma_3 \bar\epsilon_r=\bar\epsilon_r\, ,
\label{SQMVspinors}
\end{align}
Indeed the supersymmetry algebra preserved by a Vortex loop is different to the one preserved by a Wilson loop (cf. \rf{Susywsa}\rf{Susywsb}).
We record in table \ref{table:VortexDefectCharges} the quantum numbers of the preserved supercharges.
 \begin{table}[h]
 \begin{center}
\begin{tabular}{c|c|c|c|c}
  & $J_3^l$ & $J_3^r$ & $R_C$ & $R_H$\\
  \hline
  $q_l$ & -1/2 & 0& -1/2 & -1/2
	        \\[5pt]
	        $\bar q_l$ & 1/2 & 0& 1/2 & 1/2
	        \\[5pt]
  \hline
$q_r$ & 0 & 1/2 & -1/2 & 1/2
	        \\[5pt]
	        $\bar q_r$ & 0 & -1/2 &  1/2 & -1/2
	        \\[5pt]
\end{tabular}
\caption{\footnotesize Supercharges preserving Vortex loop.}
 \label{table:VortexDefectCharges}
\end{center}
\end{table}

 \noindent
We note  that   $J_3^l-{1\over 2}(R_C+R_H)$ and $J_3^r +{1\over 2}(R_C-R_H)$ commute with all four supercharges, a fact   will make good use of  shortly.
The explicit $SU(1|1)_l\times SU(1|1)_r$ anticommutators of the preserved supercharges by a Vortex loop operator are 
\begin{align}
\{q_l,\bar q_l\} =& \frac 2L \Big[ J_3^l-{1\over 2}\left(R_C+R_H\right) \Big]\label{Lefts}   \\[6pt]
\{q_r,\bar q_r\} =& \frac 2L \Big[ J_3^r +{1\over 2}\left(R_C-R_H\right) \Big]\label{Rights}\,.
\end{align}
Note the different relative sign in the last anticommutator in comparison with the supersymmetry algebra \rf{wilsonesfera} for a Wilson loop.

 \smallskip

In order to explicitly define Vortex loop operators on $S^3$ we must learn how to couple in an  $SU(1|1)_l\times SU(1|1)_r$ invariant way the 1d $\cN=4$ SQM$_V$ invariant gauge theories discussed in section \ref{sec:loops} to 3d $\cN=4$ gauge theories on $S^3$. The 1d $\cN=4$ theory is now supported on an $S^1$ and not on a line, modification which plays an important role in what follows. It is instructive to first note that in the flat space limit  where  $L\rightarrow \infty$   the supersymmetry preserved by the loop on $S^3$ reduces to the 1d $\cN=4$ SQM$_V$ supersymmetry  in \rf{SQMV} that was preserved by a Vortex loop in flat space
\begin{align}
\{q_l,\bar q_l\} =& \  H\\[5pt]
\{q_r,\bar q_r\} =& \  -  H\,,
\end{align}
where  the sphere generators $J^l_3={1\over 2}(J_{12}+LH)$ and $J^r_3={1\over 2}(J_{12}-LH)$ become the translation generator $H$ and transverse rotation generator $J_{12}$   in flat space. 
 
 \smallskip

 Coupling a 1d $\cN=4$ SQM$_V$ gauge theory on the loop to a  3d $\cN=4$ theory on $S^3$   requires turning on background fields  in the 1d $\cN=4$ theory in order to make the 3d/1d theory invariant under the $SU(1|1)_l\times SU(1|1)_r$ supersymmetry algebra of the Vortex loop.  This is not surprising. Placing a 3d $\cN=4$ gauge theory on $S^3$  deforms   the   flat space supersymmetry transformations and action. Both modifications can be interpreted as arising  from background 3d $\cN=4$  supergravity  fields  \cite{Festuccia:2011ws}.
 We now show that the 1d $\cN=4$ supersymmetry transformations and action can be deformed in such a way as 
 to yield a supersymmetric Vortex loop operator on $S^3$.  
 
 \smallskip
 
 The starting point is the SQM$_V$ supersymmetry algebra  discussed in section \ref{sec:loops}
 \beq
 \{Q_+,\overline Q_+\}=H\qquad  \{Q_-,\overline Q_-\}=H\,.
 \label{susynow}
 \eeq
 In order to couple the 1d $\cN=4$ theory on $S^3$ supersymmetrically we must deform the 1d $\cN=4$ theory in such a way that the supersymmetry algebra becomes the $SU(1|1)_l\times SU(1|1)_r$ algebra   of the Vortex loop on $S^3$
 \begin{align}
\{q_l,\bar q_l\} =& \frac 2L \Big[ J_3^l-{1\over 2}\left(R_C+R_H\right) \Big]\\[6pt]
\{q_r,\bar q_r\} =& \frac 2L \Big[ J_3^r +{1\over 2}\left(R_C-R_H\right) \Big]\,.
\end{align}
 This can be accomplished in two steps:   
 \begin{itemize}
\item Turning on a fixed background gauge field for the R-symmetry $J_- \equiv J_3^l + J_3^r - R_C$ 
\item Turning on a fixed imaginary mass parameter for the flavour symmetry $G_F\equiv    J_3^l + J_3^r - R_H$
\end{itemize}
 $G_F$ does indeed commute with all four supercharges, as it is the sum   of the bosonic generators in 
\rf{Lefts} and \rf{Rights}, both of   which are central elements. The charge $J_-$ commutes with $q_l$ and $\bar q_l$ but   not with $q_r$ and $\bar q_r$.
We also use in an important way that the 1d $\cN=4$ theory lives on an $S^1$ and non-trivial Wilson lines can be turned on around the circle. Deforming the 1d $\cN=4$ theory by these backgrounds allows to couple a 1d $\cN=4$ theory on a Hopf circle to a 3d $\cN=4$ theory on $S^3$ while preserving $SU(1|1)_l\times SU(1|1)_r$.
 
 \smallskip

The SQM$_V$ supersymmetry algebra \rf{susynow} gets deformed in the presence of    a background  1d $\cN=4$ vector multiplet for a flavour symmetry $G$. We consider a background gauge field $a$ on $S^1$ and  a background scalar field in the vector multiplet corresponding to a real mass $m$.\footnote{Turning on a complex mass associated to the other two scalars in the vector multiplet deforms other commutators.} The deformed SQM$_V$ algebra becomes\footnote{This can be easily understood by dimensional reduction of the 4d $\cN=1$ supersymmetry algebra.}
\beq
 \{Q_+,\overline Q_+\}=H-(a+i m)G \qquad  \{Q_-,\overline Q_-\}=H-(a-i m)G \,.
 \label{susynowa}
 \eeq 
 $H$ is the generator of translations on the circle, which we take to have length $\beta=2\pi L $.
 
 \smallskip 
 Turning on a background gauge field for the R-symmetry $J_-$ generically breaks the supersymmetries associated to the supercharged  $Q_-$ and $\bar Q_-$ charged under $J_-$. However for quantized values of the gauge field $a_- = \frac{n}{L}$, $n \in \bZ$,  supersymmetry remains unbroken with $Q_-$ and $\bar Q_-$ generated by the non-constant Killing spinors $\epsilon_- = e^{i \frac{ n \tau}{L}}$ and $\bar\epsilon_- = e^{-i \frac{ n \tau}{L}}$, with $\tau \in [0, 2\pi L]$ parametrizing $S^1$.\footnote{This can be understood as follows. The constant gauge field  $a_-$ can be absorbed into a redefinition of the fields $\phi' = e^{-i r_- a_- \tau} \phi$, with $r_-$ the $J_-$-charge of $\phi$. The non-standard periodicity of the fields around $S^1$ are then preserved by the supersymmetry transformations generated with  $\epsilon_- = e^{i a_- \tau}$ and $\bar\epsilon_- = e^{- i a_- \tau} $. These spinors are globally defined only for $a_- = \frac{n}{L}$, with $n \in \bZ$.}   For these quantized holonomies, the supersymmetry  algebra is   deformed in the same way as if it were   a flavor symmetry.  
In conclusion, turning on the following background fields for $J_-$ and $G_F$ 
\begin{align}
J_- :  a=-{1\over L}\qquad m=0 \nonumber \\[5pt] 
G_F: a=0\qquad \ m={i\over L}\,,
\label{Background}
\end{align}
the SQM$_V$ algebra \rf{susynow} becomes the $SU(1|1)_l\times SU(1|1)_r$ supersymmetry preserved by a line defect in a 3d $\cN=4$ theory on $S^3$. We can identify the generators as follows:\footnote{In the flat space limit, the R-symmetry generators $J_{\pm}$ are identified with $\mp J_{12} - R_C$. They are related to the generators $J_{\pm}$ presented in section \ref{sec:loops} by a shift by the flavor symmetry generator $J_{12} - R_H$, which is the flat space limit of $G_F$. }
\begin{align}
Q_{+} & =  \bar q_l \,, \quad  \bar Q_{+} \ = \  q_l  \,, \quad  Q_{-} \ = \   \bar q_r \,, \quad  \bar Q_{-} \ = \ - q_r \\[+2ex]
H &= -\frac{2}{L} J^3_r \,,  \quad  J_{\pm}  \ = \  \mp (J^3_l + J^3_r) - R_C  \,, \quad G_F \ =  J^3_l + J^3_r - R_H
\end{align}
 Consistently with the deformation by the R-symmetry gauge field $a_- = -\frac 1L$, we find that the Killing spinors generating $q_r$ and $\bar q_r$  behave at the position of the loop as $\bar\epsilon_- = e^{\frac{i \tau}{L}}$ and $\epsilon_- = e^{-\frac{i \tau}{L}}$, respectively. 
Note that these position dependent Killing spinors  are nevertheless periodic along the Hopf circle.
 \smallskip
 
The background fields \eqref{Background} ensure that the 1d $\cN=4$ theory living on a maximal circle of $S^3$  can be coupled supersymmetrically to the bulk 3d $\cN=4$ theory. This can also be understood by looking at the 3d $\N=4$ supersymmetry transformations of the vector multiplet and hypermultiplet  on $S^3$ generated by the four Killing spinors preserving the Vortex loop. These transformations decompose into    SQM$_V$   supersymmetry transformations deformed by the advertised background. To couple the bulk theory to SQM$_V$, preserving the four supercharges, it is then required to turn on this background in the SQM$_V$ supersymmetry transformations and Lagrangian. The derivation of how the supersymmetry transformations on $S^3$ decompose is presented in appendix \ref{app:Embeddings}.

\smallskip
The presence of these background fields can be recast, upon field redefinition, into twisted boundary conditions for the fields in the 1d $\cN=4$ theory around $S^1$, thus affecting the partition function (or supersymmetric index) of the 1d $\cN=4$ theory. This will play a crucial role in the evaluation of the expectation value of Vortex loop operators in section \ref{ssec:MM}.

\medskip

Having defined the half-supersymmetric Wilson and Vortex loop operators  on $S^3$, we now propose to evaluate   their expectation values exactly, using the results of supersymmetric localization and to test the mirror symmetry predictions of section \ref{sec:examples}. 

\bigskip

\subsection{Exact Partition Function Of 3d/1d Theories On $S^3$}
\label{ssec:MM}

In this section we identify the matrix integral representation of the exact expectation value of a half-supersymmetric Vortex (and Wilson) loop operator in an $\cN=4$ theory on $S^3$, which takes into account the coupling of 1d $\cN=4$ SQM on $S^1$ to the $\cN=4$ theory on $S^3$   described in the previous section. This matrix model model is obtained by combining in an interesting way the matrix integral computing the $S^3$ partition function of $\cN=4$ theory on $S^3$ \cite{Kapustin:2009kz} and the matrix model representation of the supersymmetric index of 1d $\cN=4$ SQM found in \cite{Hori:2014tda,Cordova:2014oxa}.

\subsubsection{$S^3$ Partition Function And Wilson loops}

A powerful probe of the dynamics of a strongly coupled  3d $\cN=4$ IR SCFT   emerging at the endpoint of an RG flow from a  UV 3d $\cN=4$ supersymmetric gauge theory is the partition function of the gauge theory on   $S^3$.
The  $S^3$ partition function is a renormalization group invariant observable, and the computation performed in the UV exactly captures   the  partition function of  the IR SCFT. 

\smallskip

The $S^3$ partition function of  3d $\cN=4$ gauge theories can be localized to a finite dimensional matrix integral \cite{Kapustin:2009kz}.  The matrix integral is defined by
integrating over the Cartan subalgebra   of the gauge group   the product of the classical and one-loop contributions in the (exact) saddle point analysis \cite{Kapustin:2009kz}
\begin{align}
Z_{MM}&= \frac{1}{|\scW|} \int_{C} d\sigma\, Z_{\text{class}}\cdot Z_{\text{vector}}\cdot Z_{\text{hyper}}\,,
\label{3dZ}
\end{align}
where $\scW$ is the Weyl group and $C$ is the contour of integration. The classical contribution depends on the FI parameter $\eta$  for each abelian gauge group factor   
\begin{align}
Z_{\text{class}}=e^{2  \pi i\eta  \text{Tr} \sigma}\,.
\label{classFI}
\end{align}
 The 3d $\cN=4$ vector multiplet   contribution is 
\begin{align}
Z_{\text{vector}}= \prod_{\alpha>0} \sh(\alpha\cdot\sigma)^2\,,
\end{align}
while a hypermultiplet in a representation ${\scR}$ of the gauge group with mass $m$ yields
\begin{align}
Z_{\text{hyper}}=\prod_{w\in {\scR}} {1\over \ch(w\cdot \sigma-m)}\,,
\end{align}
where $\alpha$ are the roots of the Lie algebra and $w$ the weights for the representation. 
Throughout   we   use  the following short-hand notation (borrowed from \cite{Benvenuti:2011ga})
\begin{align}
\sh(x) \equiv 2\sinh(\pi x)\, \quad   \quad \ch(x)  \equiv 2\cosh(\pi x)\,, \quad \thh(x)  \equiv \tanh(\pi x)\,.
\label{notation}
\end{align}
In this paper we focus on  gauge theories with unitary gauge groups. For a $U(N)$ gauge group the set of eigenvalues is $\{\sigma_i \}_{1 \le i \le N}$ and $|\scW| = N!$. The vector multiplet and massive fundamental hypermulptiplet factors are expressed in this case by
\begin{align}
Z_{\text{vector}} &=\prod_{i<j} \sh(\sigma_i - \sigma_j)^2 \, , \qquad  Z_{\text{hyper}} \ =   \prod_{j=1}^N \ch(\sigma_j - m)^{-1}
\label{vectorvm}
\end{align}
 By combining these building blocks the exact $S^3$ partition function of an arbitrary 3d $\cN=4$   gauge theory acquires an elegant matrix model representation.   
 
 \smallskip
 
 The contour of integration $C$ of the matrix model that arises from the localization computation is   the real axis. The condition that the matrix integral over  the real axis is convergent \cite{Kapustin:2010xq} is precisely the same as the criterion for the quiver theory to be ``good" or ``ugly" \cite{Gaiotto:2008ak}, that is that for each $U(N)$ gauge group factor  the number of fundamental hypermultiplets $N_f$ obeys $N_f\geq 2N-1$. In this paper we have considered gauge theories that flow in the IR to an irreducible SCFT, so the matrix models  (in the absence of Wilson loops) are convergent.

 \smallskip
 
The exact expectation value of the  supersymmetric  Wilson loop \rf{wilsonsphere}   is computed by inserting  into the matrix model of the theory \rf{3dZ}  the factor
\begin{align}
\label{Wilsonloc}
 \tr_\scR \Big( e^{2\pi \sigma} \Big) \ =
\ \sum_{w \in \scR} \ e^{2 \pi w\cdot \sigma  } \ ,
\end{align}
where $\scR$ is the representation associated to the Wilson loop and $w$ runs over the weights of $\scR$. Wilson loops expectation values are in addition normalized by the partition function. 

\smallskip
We note that  the matrix integrals  defined over the real contour capturing the expectation value of Wilson loops can become divergent  when the charge of the Wilson loop is sufficiently large. For these superficially divergent matrix models  we regulate the divergent  integrals by deforming the contour of integration $C$ of the matrix model away from the real axis. The results that we find using the deformed contours are completely consistent with our mirror symmetry predictions. It may be possible to justify these deformed contours of integration for the divergent matrix integrals by carefully performing the localization computation. 

\begin{figure}[h]
\centering
\includegraphics[scale=0.3]{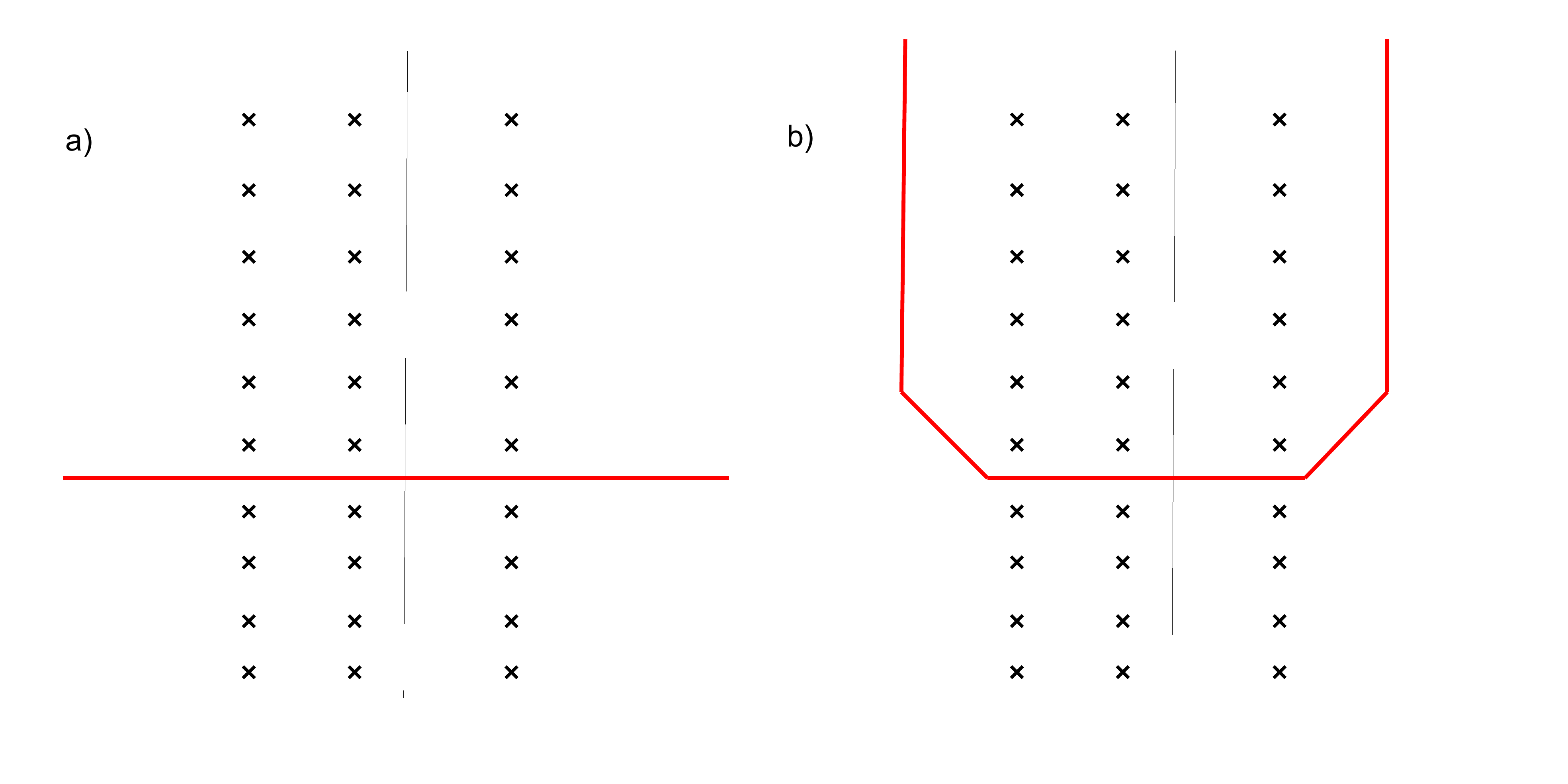}
\vspace{-0.5cm}
\caption{\footnotesize{a) The naive contour (red line) of integration is on the real line. Crosses denote poles of the matrix integrand. b) At positive FI parameter, we deform the contour as depicted, so that it encloses the poles in the upper-half plane.}}
\label{contour}
\end{figure}

More precisely, for positive FI parameter $\eta > 0$, we deform the contour of integration of each eigenvalue so that it encloses the poles of the matrix model integrand in the upper half-plane, as in figure \ref{contour}. This is then the same as closing the real line with a semi-circle going through $+i\infty$. Conversely if $\eta < 0$ we close the contour so that it encloses the poles in the lower half-plane. These choices of contour lead to finite results for arbitrary large values of the Wilson loop charge, as long as $\eta \neq 0$.  For each matrix eigenvalue the Wilson loop charge $q$ combines with the FI parameter $\eta$ in a complex parameter $\eta + i q$ and the evaluation of the integral with our prescription coincides with the analytic continuation to the complex value of the FI parameter.

\subsubsection{1d $\scN=4$ Supersymmetric Quantum Mechanics Partition Function}
 \label{sssec:QMindex}

In this section we introduce and evaluate the  partition function of the 1d $\N=4$ SQM quiver gauge theories on a circle relevant for the  description of Vortex  loop operators in 3d $\cN=4$ gauge theories.
The  1d $\N=4$ SQM partition function admits a Hilbert space interpretation   as an index with fugacities (or chemical potentials). It is defined as a trace over the Hilbert space of the theory \cite{Witten:1982df}
\begin{align}
\scI(z,\mu) &= \tr_{\scH} \ (-1)^F \, e^{2\pi i z J_-} \,  e^{ 2\pi i \mu \Pi}  \ ,
\end{align}
where $F$ is the fermion number, $z$ is a chemical potential for the $U(1)_-$ R-symmetry with generator $J_-$ and $\mu = \{ \mu_j \}$ is a collection of chemical potentials for flavor symmetries with Cartan generators $\Pi_j$. In the path integral representation, the fugacities are associated to twisted boundary conditions for the fields, which can be undone by turning on background vector multiplets for the relevant symmetries. $z$ is associated to the presence of a $U(1)_-$ R-symmetry gauge field, with the relation $z = - L  a_-$, where $2\pi L$ is the length of $S^1$ . A chemical potential $\mu_j$ is associated to the presence of a $U(1)_j$ flavor symmetry gauge field $a_j$ and a real mass parameter $m_j$, through the relation $\mu_j = - L (a_j + i m_j)$.

\smallskip

As explained in section \ref{ssec:Vortexsphere}, coupling the 1d $\cN=4$  theory on a maximal     $S^1\subset S^3$ to a 3d $\cN=4$ theory requires   turning on a constant R-symmetry gauge field corresponding to $z=1$ and a mass deformation corresponding to a chemical potential $\mu = 1$ for the flavor symmetry generated by $G_F=J_3^l-J_3^r-R_H$. Of course, the fugacities for all other flavour symmetries are arbitrary.

\smallskip

The 1d $\cN=4$ SQM theories are deformed by FI terms with parameters $\vec \zeta \equiv \zeta$, taking real values in the center of the gauge algebra. The partition function or index is well-defined when all FI parameters $\zeta_a$ are non-zero.  When the FI parameters vanish
important subtleties arise as the theory may develop a continuous spectrum.

\smallskip

Using supersymmetric localization, the exact index of a  1d $\scN=4$ SQM theory is given by the explicit formula \cite{Hori:2014tda,Cordova:2014oxa}
\begin{align}
\scI(z,\mu)  &= \frac{1}{|W|} \, {\rm JK-Res}_{\zeta} \  g_{\rm vec}(z,u)  \, g_{\rm chi}(z,\mu ,u) \, d^ku \ ,
\label{IndexJK}
\end{align}
where $k$ is the rank of the gauge group $G$, $|W|$ is the order of the Weyl group $W$ of $G$ and $\text{JK-Res}$ denotes the sum over Jeffrey-Kirwan residues \cite{1993alg.geom..7001J} of a meromorphic $k$-form with complex variables $u= \{u_I \}$. The factors 
$g_{\rm vec},g_{\rm chi}$ are the localization one-loop determinants  associated to 1d $\scN=4$ vector and chiral multiplets,
\begin{align}
 g_{\rm vec}(z,u) &=  \lp \frac{\pi}{\sin(\pi z)} \rp^k \,  \prod_{\alpha \in G}  \frac{\sin(-\pi \alpha \cdot u )}{ \sin[\pi (\alpha \cdot u  - z)]}  \\[+2ex]
 g_{\rm chi}(z, \mu ,u) &=   \prod_{w \in \scR} \frac{\sin[-\pi(w \cdot u  + q \cdot \mu  + (\frac{R}{2}-1) z ) ]}{\sin[\pi( w \cdot u  + q \cdot \mu  + \frac{R}{2} z ) ]} \ ,
 \label{gfactors}
\end{align}
where $\alpha$ runs over the roots of the gauge algebra, $w$ runs over the weights of the chiral multiplet representation, $R$ is the chiral multiplet R-charge under $2 J_-$ and $q = \{ q_j \}$ are the chiral multiplet charges under the flavor symmetry generators $\Pi = \{\Pi_j \}$.

\medskip

We now explain briefly how to compute the Jeffrey-Kirwan residues of the meromorphic $k$-form with simple poles   and with FI parameters $\zeta$.
A given set  pole  $u^{\ast}= \{u^{\ast}_I \}$ arises at the intersection of $k$ hyperplanes in $\bC^k$, defined by the equations 
$ w^{(I)} \cdot u^{\ast}  + q^{(I)} \cdot \mu  + \frac{R^{(I)}}{2} z =0$,\footnote{Strictly speaking the hyperplane equations are $ w^{(I)} \cdot u^{\ast}  + q^{(I)} \cdot \mu  + \frac{R^{(I)}}{2} z = n$, with $n \in \bZ$, but in the index computation only   the hyperplane with $n=0$ should be picked.} with $I = 1 ,\cdots , k$, appearing in the denominator  of the chiral multiplet one-loop determinant. In principle we should also consider poles in the vector multiplet one-loop determinant but it happens that one cannot find a collection of $k$ hyperplanes intersecting at a point $u^{\ast}$ if one of the hyperplanes is described by $\alpha \cdot u  - z =0$.\footnote{The pole from the vector multiplet is canceled by a corresponding zero from the chiral multiplet.} Thus   each  pole  $u^{\ast}$ is associated with a set of $k$ weights $\{w^{(I)} \}$, each weight appearing in some chiral multiplet factor.
A set of weights $\{w^{(I)} \}$ defines a cone $C(w^{(I)}) = \{ \sum_I^k c_I w^{(I)} \, | \, c_I > 0 \} \subset \bR^k$.
The Jeffrey-Kirwan residue at $u^{\ast}$ is then given by
\begin{align}
{\rm JK-Res}_{\zeta}[u^{\ast}] \, g(z,\mu ,u) \, d^ku &= 
\left\lbrace
\begin{array}{ll}
{\rm Res}[u^{\ast}]  \,  g(z,\mu ,u)  &  {\rm if } \, -\zeta \in  C(w^{(I)}) \\
0 & {\rm otherwise.}
\end{array}
\right. \,,
\end{align}
where ${\rm Res}[u^{\ast}] g(z,\mu ,u)$ denotes the usual residue at the pole  $u = u^\ast$ and   $\zeta$ is understood as a $k$-component vector.

\smallskip

The total index \eqref{IndexJK} is obtained by summing over all JK-residues at poles $u^\ast$
\begin{align}
\scI(z,\mu)  &= \frac{1}{|W|} \sum_{u^\ast}{\rm JK-Res}_{\zeta}[u^{\ast}] \, g_{\rm vec}(z,u)  \, g_{\rm chi}(z,\mu ,u) \, d^ku \, .
\end{align}

\medskip

We will   give here the result of the evaluation of the index for the class of 1d $\cN=4$ SQM quivers entering our construction of Vortex loop operators mirror to Wilson loops, relegating part of the details of the residue computations to appendix \ref{app:QMindex}.

The class of 1d $\cN=4$ SQM quiver gauge theories we consider is described in figure \ref{VortexQM2}.
They are linear quivers connected by pairs of bifundamental chiral multiplets, with $N_L$ fundamental and $N_R$ anti-fundamental chiral multiplets in the terminating node, which we mean to be the node on the right of the quiver in \ref{VortexQM2} (this is the one that is closest to the 3d quiver in the combined 3d/1d quiver). The gauge group is then $G= \prod_{p=1}^{P} U(n_p)$, where $U(n_P)$ denotes the terminating node. Moreover each node has either zero or one adjoint chiral multiplet and we denote $c^{\rm adj}_p \in \{0,1 \}$ the number of adjoint chiral multiplets in the $U(n_p)$ node. 
As explained in section \ref{ssec:LoopSduality}, the 1d FI parameters of all nodes in the quiver are taken {\bf negative} if the 1d $\cN=4$ theory  is read from the brane configuration with D1-strings moved to the closest NS5-brane on the right. Conversely they are all taken {\bf positive} if the 1d $\cN=4$ theory is read from the brane configuration with D1-strings moved to the closest NS5-brane on the left.\footnote{Less restrictive conditions could be imposed on the signs of the FI parameters in non-terminating nodes. For instance in a two-node quiver with FI parameter $\zeta_2 > 0$ in the terminating node, we could allow for a FI parameter $\zeta_1 > - \zeta_2$, with $\zeta_1 \neq 0$, in the other node. This constraint follows from a careful analysis of the positions of the various NS5-branes in the brane realization of the loop operator. We checked with a few explicit computations that the index does not depend on $\zeta_1$ taking values in this range. At $\zeta_1=0$, corresponding to aligned NS5-branes, we do not know how to evaluate the 1d partition function, but we expect a dramatic change in the result (see discussion around \ref{sec:Branes}). To simplify computations we require  the FI parameters of all the nodes in a quiver to have the same sign ($\zeta_1, \zeta_2 > 0$ in the example). }
 \medskip

\noindent{\bf Superpotentials: } The 1d $\cN=4$   SQM theories have superpotentials constraining the R-charges of the chiral multiplets: {\it cubic superpotential} couplings between  adjoint and  bifundamental chiral multiplets and {\it quartic superpotential} couplings between bifundamental chiral multiplets for the nodes without adjoint chiral multiplet. 
The $N$ fundamental and $M$ anti-fundamental chiral multiplets in the terminating $U(N_P)$ node do not enter into such cubic and quartic superpotentials. Instead they are coupled to a 3d hypermultiplet through a {\it cubic superpotential} as described in section \ref{ssec:LoopSduality}.
This 3d/1d coupling is responsible for the identification of bulk and defect flavor symmetries.
\smallskip

A cubic superpotential imposes the constraint $R_{\rm adj} + R_{\rm bif 1} + R_{\rm bif 2} = 2$ on the R-charges of the fields with respect to the generator $2 J_-$. A quartic superpotential imposes the constraint $R_{\rm bif 1} + R_{\rm bif 2} +R_{\rm bif 3} + R_{\rm bif 4} =2$. The cubic superpotential which couples the fundamental and anti-fundamental chirals to the 3d hypermultiplet imposes the constraints $R_{\rm fund} + R_{\rm a-fund} = 2$, since the bulk hypermultiplet is not charged under $J_- = J^3_l + J^3_r -  R_C$. 

\smallskip
In order to perform the computation of the Jeffrey-Kirwan residues, we only impose that the R-charges obey these superpotential constraints, but otherwise leave them arbitrary, to avoid having to deal with higher order poles. 
It turns out that the final results depend only on the R-charges of the adjoint chiral multiplets, which we need to specify at the end of the computation.  

\smallskip
The R-charges for the adjoint chiral multiplets can be read from the flat space brane realization of the quiver (see section \ref{ssec:LoopSduality}). The complex scalar degrees of freedom of an adjoint chiral living on D1-branes stretched between two NS5-branes correspond to displacements of the D1-branes along the plane $x^1-x^2$.  The scalar transforms as a vector under $SO(2)_{12}$ and is uncharged under $J_{78}$. The adjoint chiral then has charge 2 under $2(J_{12} - J_{78}) $ which is identified with the 1d $\cN=4$ SQM generator $2 J_-$ in the flat space limit. 
On the other hand the complex scalar of adjoint chiral living on D1-branes stretched between two NS5'-branes are associated with displacements of the D1-branes along the plane $x^4-x^5$. It transforms as a vector under $SO(2)_{45}$ and is uncharged under $J_{12}$ and $J_{78}$. The adjoint chiral is then uncharged under $2J_-$. 

The R-charges (under $2J_-$) of the adjoint chiral multiplets are then fixed to be either 2 or 0, depending on whether they arise from a brane construction with NS5 or NS5' branes. In the 1d $\cN=4$ quiver gauge theory description, this prescription is re-expressed as follows. 
Define the integers ${\bf c}_p = P - p - \sum_{q=p}^P c^{\rm adj}_q$.
An adjoint chiral multiplet in the $U(n_p)$ node has R-charge $R_{{\rm adj} , \, p}$ given by
\begin{align}
R_{{\rm adj} , \, p} & = 
\left\lbrace
\begin{array}{cl}
2  & {\rm if } \quad  {\bf c}_p  \quad {\rm odd} \\
0  & {\rm if } \quad  {\bf c}_p  \quad {\rm even}  \, .
\end{array}
\right.
\label{RadjQM}
\end{align}
Similarly one can derive the charge of the adjoint chiral multiplets under the flavor symmetry $G_F$, which is identified in the flat space limit with the 10d rotation generator $J_{12} - J_{45}$.
The charges of the multiplets under the symmetries relevant for the computation of the index and the constraints they obey are summarized in the following table.

 \begin{center}
\begin{tabular}{c|c|c|c}
  & $2J_-$ & $G_F$ & W-constraints \\
  \hline
  adjoint chiral & 2 or 0  & + 1 or -1 & cubic
 \\[5pt]
   bifund. chiral & $r_{\rm bif}$ & $q_{\rm bif}$  & cubic/quartic sp.
 \\[5pt]
fund. chiral  &  $r_+$ &   $q_+$ & 
\\[5pt]
antifund. chiral  &  $r_-$ &   $q_-$ &  $r_+ + r_- = 2$ \ , \  $q_+ + q_-  -1/2= 0$
\\[5pt]
\end{tabular}
 \captionof{table}{Charges and superpotential constraints}
\label{tab:QMCharges}
\end{center}

The last entry in the table, $q_+ + q_- = 1/2$, follows from the superpotential constraint coupling bulk and defect fields and the fact that the 3d hypermultiplet entering in this superpotential has charge $-1/2$ under $G_F = J^3_l + J^3_r - R_H$.

\medskip

As explained in section \ref{ssec:LoopSduality} , each 1d $\cN=4$ quiver gauge theory  obtained by moving the D1-branes to the closest NS5-brane on the right is associated to a tensor product representation $\scR$ of $U(N)$, with $N$ the number of fundamental chiral multiplets, which depends on the ranks of the nodes $n_p$ and on the distribution of adjoint chiral multiplets among the nodes. The representation of $U(N)$ associated to the quiver is $\scR= \otimes_{p=1}^{P} \scR_{k_p}$ with $k_p = n_p - n_{p-1}$ (with $n_0 = 0$) and $\scR_{k_p}$ is given by 
\begin{align}
\scR_{k_p} & = 
\left\lbrace
\begin{array}{cl}
\scS_{k_p} & {\rm if } \quad  {\bf c}_p  \quad {\rm odd} \\
\scA_{k_p} & {\rm if } \quad  {\bf c}_p  \quad {\rm even}  \, .
\end{array}
\right.
\end{align}
$\scS_{k_p}$ and $\scA_{k_p}$ denote the $k_p$-symmetric representation and $k_p$-antisymmetric representation respectively. 

\smallskip
Once the representation $\scR$ is identified we can perform the evaluation of the 1d $\cN=4$ SQM index. As we will explain shortly, we first compute the index at arbitrary values of $z$ and then we take the analytic  continuation to $z \rightarrow 1$ in the combined  3d/1d partition function.\footnote{Poles of the 3d integrand cross the integration contour as $z\rightarrow 1$, which is why we analytically continue to $z=1$.}   This  turns out to be the correct prescription to compute the final exact partition function of the 3d/1d theory describing a Vortex loop operator. 
The $G_F$ chemical potential $\mu$ can be readily set to $\mu=1$.

\smallskip

The details of the computations are given in appendix \ref{app:QMindex}, for single node and two nodes quivers. 
We use the explicit results of our computations to conjecture the index formula for the general class of quivers relevant in this paper. 
 Our computations in  explicit examples  gives us confidence in our formulas \rf{IndexQMfinal_r}\rf{IndexQMfinal_l} below and ultimately result in the
  matching between mirror and Vortex loops extended to an  arbitrary number of nodes.

\smallskip

Let us denote $\scI^r $ and $\scI^l$ the partition functions, or supersymmetric indices, of the 1d $\cN=4$ gauge theory  realized by moving the D1-strings on the closest NS5-brane on the right and on the left respectively.
For the `right' SQM$_V$ theory, which has  $N$ fundamentals with mass  $-\sigma_j$ and $M$ anti-fundamentals of mass  $m_a$ in the terminating node,
we find that  the index $\scI^r $ at arbitrary $z$ is given by a sum of contributions labeled by the weights of the representation $\scR$ of $U(N)$:
\begin{align}
\scI^r (z) = \sum_{ w \in \scR} \scI^r_{w}(z) \, , \qquad 
\scI^r_{w} (z) = \scF(\sigma_i,z)  \, \prod_{a=1}^{M} \prod_{j=1}^{N} \frac{\ch( m_a - \sigma_j)}{\ch( m_a -  \sigma_j - i w_j z)}  \, ,
\label{IndexQMfinal_r}
\end{align}
where $\scF(\sigma_i, z) $ is a factor that will not play any role in our final computation of the expectation value of Vortex loop operators. Let us just mention here that it is a product of the form
\begin{align}
\scF(\sigma_i, z) &= \prod_{\alpha , \kappa } \prod_{i \neq j}^N  \frac{\sin[-\pi (i\sigma_i - i\sigma_j  +  \alpha z \pm z + \kappa/2)]}{\sin[\pi  (i\sigma_i - i \sigma_j + \alpha  z + \kappa/2)]} \, ,
\label{curlyF}
\end{align}
where $\alpha, \kappa$ take real values. In the analytic  continuation to $z=1$, this factor will not affect the evaluation of the coupled 3d/1d matrix model.

\smallskip

For the 1d $\cN=4$  quiver theory obtained by moving the D1-branes to the closest NS5-brane on the left the computation is similar, except for the fact that the FI parameters of the nodes are all {\it positive}. This affects very significantly the index. The answer  is now written as a sum over weights  of a representation $\scR$ of $U(N)$, where now $N$ is the number of {\it anti-fundamental} chiral multiplets in the terminating node and $\scR$ is given by the same tensor product of symmetric and anti-symmetric representation   described above.
Denoting by $-\ti m_a$ the mass  of the $\ti M$ fundamental chiral multiplets and $\sigma_j$ the mass  of the $N$ anti-fundamental chiral multiplets, the index is then given by
\begin{align}
\scI^l (z) = \sum_{ w \in \scR} \scI^l_{w}(z) \, , \qquad 
\scI^l_{w} (z) = \ti\scF(\sigma_i,z)  \, \prod_{a=1}^{\ti M} \prod_{j=1}^{N} \frac{\ch( \ti m_a - \sigma_j)}{\ch( \ti m_a -  \sigma_j + i w_j z)}  \, ,
\label{IndexQMfinal_l}
\end{align}
where $\ti \scF(\sigma_i, z) $ is again a factor that   trivializes when we consider the full 3d/1d matrix model in the analytic continuation to $z=1$.

\medskip
The formulae  \eqref{IndexQMfinal_r} and  \eqref{IndexQMfinal_l} are compatible with the expectations from 1d Seiberg duality. This can be understood as follows. The computation shows that the partition function depends only on the representation $\scR$ and on the masses of the fundamental and  anti-fundamental chiral multiplets in the terminating node. Multiple 1d quivers (except for single node quivers) lead to the same tensor product representation $\scR$ and hence have equal partition functions, namely the quivers read from all possible orderings of the NS5 and NS5'-branes in the $x^6$ direction. These reorderings affect the rank of the gauge nodes and the distribution of adjoint chiral multiplets among them but not the associated representation $\cR$.
These Seiberg-dual quivers are expected to realize the same Vortex loop in the infrared limit and we find indeed that their partition functions match (at least when $z \rightarrow 1$).

 \subsubsection{Matrix Model Computing Vortex Loops}
 \label{subsubsec:3d1dMM}

 Our aim is to compute the exact expectation value of Vortex loops in 3d $\cN=4$ gauge theories on $S^3$. We now explain how this quantity is captured by a matrix model that combines in an interesting way the exact   $S^3$ partition of  3d $\cN=4$ theories   and the exact 
 index of the 1d $\cN=4$ theories defining Vortex loops, both of which we have already introduced.

\smallskip
 A Vortex loop on $S^3$ is defined by a supersymmetric coupling of a 3d $\cN=4$ gauge theory on $S^3$ to a 1d $\cN=4$ gauge theory on an $S^1\subset S^3$. Recall that in order for the 3d/1d defect theory to be supersymmetric, invariant under $SU(1|1)_l\times SU(1|1)_r$, we had to turn on very specific background fields in the SQM. We can compute the exact partition function of the combined 3d/1d theory by choosing the supercharge $q_l$ in $SU(1|1)_l$ to localize the full functional integral. In order to localize the combined 3d/1d partition function, we must add suitable deformation terms for both the 3d $\cN=4$ theory and the 1d $\N=4$ theory. Importantly, the saddle points  for the 3d and 1d fields are not modified. The way the combined 3d/1d partition function encodes the specific couplings between the 3d and 1d theories is by  taking into account that the flavour symmetries of the 1d $\cN=4$ theory are   gauged with 3d $\cN=4$ vector multiplets. At the level of the combined matrix model this means that the 
 1d mass parameters that  are gauged   are replaced by the scalar in the corresponding 3d $\cN=4$ vector multiplet. This scalar can be either dynamical, in which case it corresponds to an eigenvalue in the 3d matrix integral that needs to be integrated over, or it is identified with a mass parameter in the 3d theory. Therefore, the 
 combined 3d/1d partition function is found by convolving the ``gauged" 1d partition function with the 3d partition function. 
 
 \smallskip

The expectation value of a Vortex loop $V$ wrapping a maximal circle of $S^3$
is the partition function of the 3d $\cN=4$  theory coupled to the 1d $\cN=4$ SQM quiver, normalized by the partition function of the 3d theory. It takes the form
\begin{align}
\langle V \rangle \ = \ W^{\rm fl} \times \frac{Z_{3d/1d}}{Z_{3d}} \ = \ W^{\rm fl} \times \frac{1}{Z_{3d}}  \ \lim_{z \rightarrow 1} \  \frac{1}{|\scW|} \int_{C} d\sigma \  F_{\rm 3d}[\sigma, m, \xi] \  \scI[\sigma, m, z] \,.
\label{3d1dMM}
\end{align}  
  $F_{\rm 3d}[\sigma, \xi, m]$ is the matrix model integrand of a 3d $\cN=4$ theory with eigenvalues $\sigma$, hypermultiplet masses $m$ and FI parameters $\xi$, while $\scI[\sigma, m, z]$ is the index of the 1d $\cN=4$ SQM$_V$ defect theory that   realizes the Vortex loop. 
  We note that  the flavour fugacities of the 1d  theory are identified   either with 3d matrix eigenvalues $\sigma$ or with 3d mass parameters $m$, depending on whether the corresponding 1d flavour symmetry is gauged by dynamical or background 3d vector multiplets. 

\smallskip
 
The factor $W^{\rm fl}$ takes the form of an abelian background Wilson loop that we add to the matrix model. It follows from the analysis of the brane realization of the Vortex loop and appears as a necessary ingredient to check successfully mirror maps with Wilson loops. In the brane picture each five-brane is associated to a deformation parameter of the theory (see section \ref{sec:loopsbranes}), masses $m_a$ for D5-branes and  ``FI" parameters $\xi_a$ for NS5-branes. When $q$ F1-strings end on a D5 with parameter $m$, they realize a Wilson loop of charge $q$ for the flavor symmetry associated to the D5-brane, and the corresponding matrix model factor is $e^{2\pi q m}$. Similarly when $q$ D1-branes end on an NS5-brane with parameter $\xi$, they realize a Wilson loop of charge $q$ for the global symmetry associated to the NS5-brane, and the corresponding matrix model factor is $e^{2\pi q \xi}$. Thus when we read the 3d/1d quiver for a Vortex loop by moving the D1-branes on top of the closest  NS5-brane to the left, resp. to the right,  we propose to add to the matrix model the background Wilson loop factor
\begin{align}
\underline{\textrm{left:}} \quad  W^{\rm fl} \ &= \ e^{2\pi |\scR| \, \xi_L}  \, , \quad 
\underline{\textrm{right:}} \quad  W^{\rm fl} \ = \ e^{2\pi |\scR| \, \xi_R}  \, ,
\label{WilsonLine}
\end{align}
where $\xi_L$, resp. $\xi_R$, is the parameter associated to the NS5-brane on the left, resp. on the right, and $|\scR|$ is the total number of boxes in the Young tableau of the representation $\scR$ labeling the Vortex loop. $|\scR|$ coincides with the total number of D1-branes ending on the NS5-brane.
It should be noticed that this Wilson loop factor cannot be associated to a global symmetry acting on the fields of the theory, since the parameter $\xi_{L/R}$ is not a combination of the true FI parameters $\xi_a - \xi_{a+1}$. With this addition, the matrix model computing the Vortex loop depends on all $\xi_a$ parameters. Similarly the matrix model computing the Wilson loop depends on all the masses $m_a$ of fundamental hypermultiplets, although only the parameters $m_a - m_{a+1}$ are associated to actual flavors symmetries.\footnote{This is because the sum of all fundamental hypermultiplet masses is associated to the diagonal $U(1)$ of the flavor symmetry group, which is actually gauged with a dynamical gauge field.} In the presence of a loop operator, the theory seems to admit one extra deformation parameter, mass or FI parameter,  associated to a ``hidden" symmetry, which  act on the fields trivially.

\smallskip

A Vortex loop $V^{(N)}_{M,\scR}$ coupled to a $U(N)$ node of the 3d $\cN=4$ quiver theory is labeled by a representation $\scR$ of $U(N)$ and by a splitting $K= M + (K-M)$ of the $K$ fundamental hypermultiplets of this $U(N)$ gauge node.
Let us consider the Vortex loop realization furnished by  coupling the 3d $\cN=4$ theory to the 1d $\cN=4$ SQM quiver obtained by   moving the D1-branes to the NS5-brane on the right,  illustrated in figure \ref{VortexQM}. In this case the SQM is labeled by a certain representation $\scR$ of $U(N_1)$ and a splitting $M =M_2$, $K-M= M_1$. In the terminating node, the SQM has $N_1$ fundamental chiral multiplets with masses $-\sigma_j$, $j =1, \cdots , N_1$ and $M_2 + N_2$ anti-fundamental chiral multiplets with masses $m_a$, $a = 1 , \cdots , M_2$ and $\ti \sigma_k$, $k = 1, \cdots , N_2$. 
The couplings to the bulk theory identify $\sigma_j$ with the eigenvalues of the 3d $U(N_1)$ gauge node, $\ti\sigma_k$ with the eigenvalues of the 3d $U(N_2)$ gauge node  and $m_a$ with the real masses of the $M_2$ fundamental bulk hypermultiplets.
The identification of bulk and defect parameters works similarly for the SQM   obtained by moving the D1-branes to the left NS5-brane.
\smallskip

We must also explain what we mean by $\lim_{z \rightarrow 1}$. The evaluation of \eqref{3d1dMM} at {\it imaginary} values of $z$ defines a function that admits an analytic continuation to the complex plane. We can thus evaluate the integral for  $z \in i\bR$ and consider the analytic  continuation to $z=1$.\footnote{We note that as we move $z \in i\bR$ no poles cross the integration contour, while if $z \in \bR$
we must deform the contour to enclose the poles that cross the original one.} This is how we extract our final result. 
This recipe is motivated by the observation that plugging $z=1$ directly in the matrix model integrand would trivialize the index (reduce it to be just one) and the coupling to the SQM quiver would not affect at all the total partition function. Instead, the analytic  continuation that we propose takes into account the matrix model   contributions from residues of poles crossing the $\sigma_i$-integration contours, as we vary $z$ continuously from zero to one. The fact that these residues should  be included should  follow from a more detailed analysis of the localization computation. 

\medskip

We can now explain our claim that the factor $\scF(\sigma_i, z)$ appearing in the evaluation of the SQM index \eqref{IndexQMfinal_r} does not contribute to $\langle V^{(N)}_{M, \scR} \rangle$. We noticed that $\scF(\sigma_i, z)$ is a product of terms with the generic form \eqref{curlyF}. In taking the analytic  continuation $z \rightarrow 1$ these factors introduce residues from poles crossing the integration contours proportional to $\sin(\pm \pi z)$. In the limit $z=1$ these extra residues all vanish, implying that the limit $z=1$ can be taken directly in the integrand of \eqref{3d1dMM} for this factor $\scF(\sigma_i, z)$. Noting that $\scF(\sigma_i, z=1) = \pm 1$, we can simply drop this factor from our computations. Our results will be valid only up to an overall sign.

\smallskip
The same argument would not work for the other factors in \eqref{IndexQMfinal_r} because the numerators $\ch( m_a - \sigma_j)$ are exactly canceled by inverse factors from the 3d hypermultiplets contributions! This means that the residues of the poles crossing the integration contour as $z\rightarrow 1$ due to these factors are non-vanishing. 
The discussion is the same for the left index \eqref{IndexQMfinal_l}.

\subsection{ Loops in $T[SU(N)]$}
\label{ssec:TSUN}

In this section we perform the explicit computation of the exact expectation value of Wilson and Vortex loop operators in the $T[SU(N)]$ theory. We show precise agreement with all of our brane-based predictions in section \ref{ssec:OtherLoops}.

\smallskip

We will use the notations
\begin{align}
\prod_j^a \equiv \prod_{j=1}^a \, , \qquad \prod_{j<k}^{a} \equiv \prod_{1 \le j<k \le a}  \, , \qquad \prod_{j,k}^{a,b} \equiv \prod_{j=1}^a \prod_{k=1}^b \, .
\end{align}

\medskip

The $T[SU(N)]$ theory has an $U(N)_J \times U(N)_F$ global symmetry group.\footnote{Strictkly speaking the symmetry is $SU(N)_J \times SU(N)_F$, but it is convenient to make the mapping of parameters under mirror symmetry symmetric to add two extra $U(1)$ symmetries, which act trivially.}
$SU(N)_F$ is the flavor symmetry rotating the fundamental hypermultiplets and $SU(N)_J$ is a global symmetry arising at the infrared fixed point, enhancing the manifest $U(1)^{N-1}$ topological symmetry acting on the dual photons. These global symmetries can be weakly gauged to give masses $ m =(m_1,\ldots,m_N)$ to the fundamental hypermultiplets and FI parameters $\eta_i = \xi_i - \xi_{i+1}$, with $\xi =(\xi_1,\ldots,\xi_N)$, to the $N-1$ nodes.  The $S^3$ partition function of the $T[SU(N)]$ theory deformed by these massive parameters was computed in
 \cite{Benvenuti:2011ga, Nishioka:2011dq} and is given by
\begin{align}
\label{ZT[SU(N)]}
 Z^{T[SU(N)]} &= (-i)^{\frac{N(N-1)}{2}} e^{-2i\pi \xi_N \sum_j m_j}  \frac{ \sum_{\tau\in S_N} (-1)^\tau \, e^{2\pi i\sum_j^N \xi_j m_{\tau(j)}}}{\prod_{j<k}^N \sh(\xi_j - \xi_k)\sh(m_j - m_k)}\,,
\end{align}
where $S_N$ is the group of permutations of $N$ elements  and $(-1)^\tau$ is the signature of the permutation $\tau$. 
$T[SU(N)]$ has the property of being self-mirror. This is realized by virtue of the partition function \rf{ZT[SU(N)]} being invariant under the exchange $m_j\leftrightarrow \xi_j$, except for the phase $e^{-2i\pi \xi_N \sum_j m_j}$, which is unphysical, in the sense that it can be removed by a local counterterm  constructed from the background fields (namely mixed background Chern-Simons terms)    defining 
 the UV   partition function \cite{Closset:2012vg, Closset:2012vp}.  
 \smallskip
 
For the explicit mirror symmetry maps, it will prove useful to define the mirror symmetric quantity
\begin{align}
\scZ^{T[SU(N)]} &\equiv  e^{2i\pi \xi_N \sum_j m_j}  \, Z^{T[SU(N)]}  \ = \  (-i)^{\frac{N(N-1)}{2}}    \frac{ \sum_{\tau\in S_N} (-1)^\tau \, e^{2\pi i\sum_j^N \xi_j m_{\tau(j)}}}{\prod_{j<k}^N \sh(\xi_j - \xi_k)\sh(m_j - m_k)} \,,
\label{curlyZ}
\end{align}
which  can be understood as the partition function expressed in another renormalization scheme and which is manifestly invariant under the exchange of mass and FI parameters $m_j\leftrightarrow \xi_j$.

\smallskip
All our results are nicely expressed   in terms of two shift operators $\mathfrak{S}^{\bf q}$ and $\wat{\mathfrak{S}}^{\bf q}$ acting on the partition function $Z \equiv   Z^{T[SU(N)]}$ of the theory. These operators act by shifting respectively the FI parameters and the masses by {\it imaginary} terms,
\begin{align}
& \mathfrak{S}^{\bf q}Z \ \equiv \ Z_{\xi_j \rightarrow \xi_j - i q_j} \, , \ j=1, \cdots, N \, , \\[+2ex]
& \wat{\mathfrak{S}}^{\bf q} Z \ \equiv \ Z_{m_j \rightarrow m_j - i q_j} \, , \ j=1, \cdots, N \, ,
\end{align}
where ${\bf q} =(q_1, \cdots, q_N)$ is a $N$-component vector.

\subsubsection{T[SU(2)] loops}
\label{sssec:TSU2}

Let us start our analysis with the abelian theory $T[SU(2)]$. The partition function is given by the matrix model
\begin{align}
Z &= \int d\sigma \frac{e^{2i\pi (\xi_1 - \xi_2)\sigma}}{\ch(\sigma - m_2)\ch(\sigma - m_1)} \ .
\end{align}
One can evaluate this integral by closing the contour of integration by a semi-circle going through $i\infty$ (or $-i\infty$) and summing over the residues inside the contour. This yields
\begin{align}
 Z  &= (-i) e^{-2i\pi \xi_2 (m_1+m_2)} \ \frac{e^{2i\pi (\xi_1 m_1 + \xi_2 m_2)} - e^{2i\pi (\xi_1 m_2 + \xi_2 m_1)}}{\sh(\xi_1 - \xi_2)\sh(m_1 - m_2)} \ ,
\end{align}
which agrees with (\ref{ZT[SU(N)]}) for $N=2$.

\smallskip

The matrix model computing the vev of a Wilson loop of charge $q \in \bZ$ is given by
\begin{align}
\label{WTSU2}
\langle W_q \rangle &=  \frac{1}{Z}\int d\sigma  \frac{e^{2i\pi (\xi_1 - \xi_2)\sigma}}{\ch(\sigma - m_2)\ch(\sigma - m_1)}  \ e^{2\pi q \sigma}   \ = \  \frac{\mathfrak{S}^{(q,0)} Z }{Z}  \no\\[+2ex]
&= (-1)^q \ \frac{e^{2\pi q m_1} e^{2i\pi (\xi_1 m_1 + \xi_2 m_2)} - e^{2\pi q m_2} e^{2i\pi (\xi_1 m_2 + \xi_2 m_1)}}{ e^{2i\pi (\xi_1 m_1 + \xi_2 m_2)} -  e^{2i\pi (\xi_1 m_2 + \xi_2 m_1)}}  \, ,
\end{align}
where we evaluated the integral at non-zero $q$ by the analytical continuation $\xi_1 \rightarrow \xi_1 - i q$, following from our choice of deformed contour integral (see discussion after \eqref{Wilsonloc}).\footnote{As explained before, divergent Wilson loops are computed by deforming the contour of integration.}

\smallskip

Let us turn now to the computation of Vortex loops. 
The $T[SU(2)]$ theory has two fundamental hypermultiplets, which give rise to three possible splitting $2= i + (2-i)$, $i=0,1,2$,  defining three possible Vortex loops  $V_{i,q}$.  Using the result for the Vortex loop factor \eqref{IndexQMfinal_r}, computed from its right SQM quiver realization (and ignoring the factor $\scF$ for the reasons mentioned before), inserted in \eqref{3d1dMM} and with extra Wilson line \eqref{WilsonLine}, the matrix models computing   the Vortex loop vevs are found to be\footnote{In section \ref{ssec:TSU2} we denoted for simplicity $V_q\equiv V_{1,q}$.}
\begin{align}
\langle V_{0,q} \rangle &=   \frac{1}{Z} \ \lim_{z \rightarrow 1}  \ \int d\sigma \frac{e^{2i\pi (\xi_1-\xi_2) \sigma}}{\ch(\sigma - m_2) \ch(\sigma - m_1)} \,   e^{2\pi q \xi_2}   \ = \  e^{2\pi q \xi_2} \frac{Z}{Z} \ = \ e^{2\pi q \xi_2}   \\
\langle V_{1,q} \rangle &=   \frac{1}{Z} \ \lim_{z \rightarrow 1}  \ \int d\sigma \frac{e^{2i\pi (\xi_1-\xi_2) \sigma} }{\ch(\sigma - m_2) \ch(\sigma - m_1 + i q z)}  \, e^{2\pi q \xi_2}  \ = \  e^{2\pi q \xi_2} \, \frac{\wat{\mathfrak{S}}^{(q,0)} Z }{Z}   \no\\
&= (-1)^q \ \frac{e^{2\pi q \xi_1} e^{2i\pi (\xi_1 m_1 + \xi_2 m_2)} - e^{2\pi q \xi_2} e^{2i\pi (\xi_1 m_2 + \xi_2 m_1)}}{ e^{2i\pi (\xi_1 m_1 + \xi_2 m_2)} -  e^{2i\pi (\xi_1 m_2 + \xi_2 m_1)}}   \\
\langle V_{2,q} \rangle &=   \frac{1}{Z} \ \lim_{z \rightarrow 1}  \ \int d\sigma \frac{e^{2i\pi (\xi_1-\xi_2) \sigma}}{\ch(\sigma - m_2 + i q z) \ch(\sigma - m_1 + iqz)} \, e^{2\pi q \xi_2}   \ = \  e^{2\pi q \xi_2} \, \frac{\wat{\mathfrak{S}}^{(q,q)} Z }{Z} \ = \ e^{2\pi q \xi_1} \, .
\end{align}
The mirror symmetry prediction \eqref{TSU2Map} is recovered by our exact computations, namely
\begin{align}
\langle W_q \rangle & \ \xleftrightarrow{\text{\ mirror \ }} \ \langle V_{1,q} \rangle \, ,
\end{align}
where $\xleftrightarrow{\text{\ mirror \ }} $ means that the operator vevs are equal under the exchange of masses $m_j$ and FI parameters $\xi_j$, as required by mirror symmetry.

\smallskip
The explicit results also show that the other Vortex loops $V_{0,q}, V_{2,q}$ are mapped to the flavor Wilson loops of the mirror theory, as predicted in section \ref{ssec:OtherLoops}.

\subsubsection{Wilson loops in T[SU(N)]}
\label{sssec:WilsonTSUN}

We turn now to the non-abelian theories by considering Wilson loops in the $T[SU(N)]$ theories. 
We pick a Wilson loop in an arbirary representation $\scR$ of a $U(p)$ node, with $1 \le p \le  N-1$. The matrix model computing it decomposes into a sum of contributions labeled by the weights $w$ of $\scR$
\begin{align}
\label{WpR}
 \langle W^{U(p)}_\scR \rangle &=   \sum_{ w \in \scR} W^{U(p)}(w) \  .
\end{align}
We compute now the matrix model $ W^{U(p)}(w)$ associated to a single weight $w=(w_1,w_2,...,w_p)$. To simplify expressions we   omit the factors depending on (and the integrals over) the eigenvalues $\sigma^{(a)}_j$ with $a>p$ in the matrix model because they do not affect the computation. With this simplification the matrix model is given by

\begin{align}
 W^{U(p)}(w) &= \frac{1}{Z} \ \int \prod_a^{p} \Big[ \ \frac{d^a\sigma^{(a)}}{a!} \ e^{2\pi \sum_j^p w_j \sigma^{(p)}_j} \ \frac{ e^{2i\pi (\xi_a - \xi_{a+1})\sum_j^a \sigma_j^{(a)}} \prod_{j<k}^a \sh(\sigma_j^{(a)}-\sigma_k^{(a)})^2 }{\prod_{j,k}^{a,a+1} \ch(\sigma_j^{(a)} - \sigma_k^{(a+1)}) } \ \Big] \, .
\end{align}
The integration over the the eigenvalues $\sigma^{(a)}_j$ with $a=1,...,p-1$ just reproduces the partition function of $Z^{T[SU(p)]}$ with FI parameters $\xi^{(p)}=(\xi_1, \xi_2, \cdots, \xi_p)$ and mass parameters $\sigma^{(p)} = (\sigma^{(p)}_1, \sigma^{(p)}_2, \cdots , \sigma^{(p)}_p)$. We can use the formula (\ref{ZT[SU(N)]}) to integrate them out (and  relabeling $\sigma^{(p)}_j = \sigma_j$)
\begin{align}
 & W^{U(p)}(w) = \frac{1}{Z} \int \frac{d^p\sigma^{(p)}}{p!} \, Z^{T[SU(p)]}[\xi^{(p)}, \sigma^{(p)}] \  e^{2\pi \sum_j^p w_j \sigma^{(p)}_j} \ \frac{ e^{2i\pi (\xi_p - \xi_{p+1}) \sum_j^p \sigma_j^{(p)}} \prod_{j<k}^p \sh(\sigma_j^{(p)}-\sigma_k^{(p)})^2 }{\prod_{j,k}^{p,p+1} \ch(\sigma_j^{(p)} - \sigma_k^{(p+1)}) } \no\\[+2ex]
& \ \  = \frac{1}{Z} \sum_{\tau \in S_p}  \int \frac{d^p\sigma}{p!} \ \frac{(-1)^\tau (-i)^{\frac{p(p-1)}{2}} e^{2i\pi \sum_j^p \xi_{\tau(j)} \sigma_j}}{\prod_{j<k}^p \sh(\xi_j - \xi_k)} \ e^{2\pi \sum_j^p w_j \sigma_j} \ \frac{ e^{-2i\pi \xi_{p+1}\sum_j^p \sigma_j} \prod_{j<k}^p \sh(\sigma_j-\sigma_k) }{\prod_{j,k}^{p,p+1} \ch(\sigma_j - \sigma_k^{(p+1)}) }  \, .
\end{align}
Relabeling the eigenvalues as $\sigma_j \rightarrow \sigma_{\tau(j)}$ in each integral brings the matrix model to the form
\begin{align}
\label{calcul1}
  W^{U(p)}(w)  &= \frac{1}{Z} \ \frac{1}{p!} \sum_{\tau \in S_p} \ \int d^p\sigma \  \frac{ (-i)^{\frac{p(p-1)}{2}} e^{2i\pi \sum_j^p \xi_{j} \sigma_j}}{\prod_{j<k}^p \sh(\xi_j - \xi_k)} \ e^{2\pi \sum_j^p w_{\tau(j)} \sigma_j} \ \frac{ e^{-2i\pi \xi_{p+1}\sum_j^p \sigma_j} \prod_{j<k}^p \sh(\sigma_j-\sigma_k) }{\prod_{j,k}^{p,p+1} \ch(\sigma_j - \sigma_k^{(p+1)}) } \,.
\end{align}
For each term in the sum over $S_p$ the factor $e^{2\pi \sum_j^p w_{\tau(j)} \sigma_j}$ can be reabsorbed as a shift of the parameters $\xi_j$ by $-iw_{\tau(j)}$, for $j=1, ..., p$, at the cost of an extra factor $(-1)^{(p-1)(w_1+w_2+...+w_p)}$. The matrix model without these imaginary shifts is exactly $Z$ (this can be seen by noticing that $Z$ is the matrix model obtained when $w=0$).
The evaluation of the integrals (with deformed integration contours as explained after \eqref{Wilsonloc}) then coincides with an analytical continuation of $Z$ to complex FI parameters. We obtain 
\begin{align}
W^{U(p)}(w) = \frac{(-1)^{(p-1) w_{\rm tot}}}{p!} \frac{1}{Z}  \sum_{\tau \in S_p} \mathfrak{S}^{w_\tau}  Z\,,
\label{WUp1}
\end{align}
with $w_{\rm tot} = \sum_{j=1}^p w_j$ and $w_\tau = (w_{\tau(1)}, w_{\tau(2)}, \cdots, w_{\tau(p)}, \underbrace{0, \cdots, 0}_{N-p})$.
We need to sum over these single weight contributions to get the final result \eqref{WpR}. Recognizing $S_p$ as the Weyl group $\scW$ of $U(p)$, we can simplify the result using the property
\begin{align}
\sum_{ w \in \scR}  \Big( \frac{1}{p!}  \sum_{\tau \in \scS^p} F[w_\tau]  \Big) 
\ = \ \sum_{ w \in \scR}  \Big( \frac{1}{|\scW|}  \sum_{\tau \in \scW} F[\tau.w]  \Big)
 = \sum_{w \in \scR} F[w] \, .
 \label{SumWeyl}
\end{align}
This leads to our final result
\begin{align}
\langle W^{U(p)}_{\scR} \rangle = (-1)^{(p-1)|\cR|} \, \frac{1}{\scZ} \sum_{w \in \scR} \  \mathfrak{S}^{(w_1,w_2, \cdots, w_p, 0, \cdots, 0)}\scZ\,,
\label{Wpfinal}
\end{align}
where we also used the property that for any weight $w$, $\sum_j^p w_j = |\cR|$, the number of boxes in the Young tableau of $\scR$, and we replaced trivially $Z$ by $\scZ$ defined in \eqref{curlyZ} .

\subsubsection{Vortex loops in T[SU(N)]}
\label{sssec:VortexTSUN}

We now evaluate the matrix models computing the exact expectation value of Vortex loops in $T[SU(N)]$. The Vortex loops are labeled by a representation $\scR$ of a $U(p)$ node and, for the $U(N-1)$ node, by a splitting $N = M + (N-M) $ of the fundamental hypermultiplets (see section \ref{ssec:OtherLoops}). We denote   by  $V^{(p)}_\scR$,  $1 \le p \le N-2$,  the $U(p)$ Vortex loops and by $V_{M,\scR}$ the $U(N-1)$ Vortex loops.  
As for the Wilson loops, the matrix model computing a Vortex loop $V_\scR$, which is labeled (in particular) by a representation $\scR$, decomposes into a sum of contributions labeled by the weights $w$ of $\scR$
\begin{align}
\label{VpR}
 \langle V_\scR \rangle &=   \sum_{ w \in \scR} V(w) \  .
\end{align}
This decomposition in a sum over weights follows from the evaluation of the SQM index \eqref{IndexQMfinal_r}, \eqref{IndexQMfinal_l}. 
We will be computing the matrix model $V(w)$ associated to a single weight $w=(w_1,w_2,...,w_p)$ for each Vortex loop.

\medskip

Let us first discuss   Vortex loops in the $U(p)$ nodes with $1 \le p\le N-2$. The prediction from the brane picture of section \ref{ssec:OtherLoops} is that such a Vortex loop must evaluate to  a background Wilson loop for a global symmetry, depending on the parameters $\xi_j$, $1 \le j \le p$. 
In the special case of the $T[SU(N)]$ theory, the brane picture predicts that the Vortex loop $V^{(p)}_\scR$ will evaluate to a background Wilson loop transforming in a representation $\scR$ of the topological symmetry subgroup $U(p) \subset U(N)_J=G_C$ acting on the Coulomb branch, associated to the deformation parameters $\xi_j$, $1 \le j \le p$. 

\smallskip
To compute the vev of this Vortex loop,  we choose to consider its left-SQM quiver realization. The matrix model is given by \eqref{3d1dMM} with the defect contribution \eqref{IndexQMfinal_l} and additional background loop \eqref{WilsonLine}. 
We simplify the matrix model by replacing the factors depending on the eigenvalues $\sigma^{(a)}_j$ with $a \ge p$, which do not affect the computation, by $\int [\cdots ] $. This gives for the contribution of a single weight $w=(w_1, w_2, \cdots, w_p)$ to the Vortex loop
\begin{align}
V^{(p)}(w) & = \ \lim_{z \rightarrow 1} \ \frac{1}{Z} \ \int [\cdots ] \int \prod_a^{p-1} \Big[ \ \frac{d^a\sigma^{(a)}}{a!}  \ \frac{ e^{2i\pi (\xi_a - \xi_{a+1})\sum_j^a \sigma_j^{(a)}} \prod_{j<k}^a \sh(\sigma_j^{(a)}-\sigma_k^{(a)})^2 }{\prod_{j,k}^{a,a+1} \ch(\sigma_j^{(a)} - \sigma_k^{(a+1)}) } \ \Big]  \no\\
& \phantom{= \frac{1}{Z} } \times \, e^{2\pi w_{\rm tot}\,  \xi_{p}}  \frac{\prod_{j,k}^{p-1,p} \ch(\sigma_j^{(p-1)} - \sigma_k^{(p)}) }{\prod_{j,k}^{p-1,p} \ch(\sigma_j^{(p-1)} - \sigma_k^{(p)} + i w_k z) }
\no\\
& = \ \lim_{z \rightarrow 1} \ \frac{1}{Z} \ \int [\cdots ] \int \prod_a^{p-1} \frac{d^a\sigma^{(a)}}{a!}  \ \prod_a^{p-2} \Big[ \ \frac{d^a\sigma^{(a)}}{a!}  \ \frac{ e^{2i\pi (\xi_a - \xi_{a+1})\sum_j^a \sigma_j^{(a)}} \prod_{j<k}^a \sh(\sigma_j^{(a)}-\sigma_k^{(a)})^2 }{\prod_{j,k}^{a,a+1} \ch(\sigma_j^{(a)} - \sigma_k^{(a+1)}) } \ \Big]  \no\\
& \phantom{= \frac{1}{Z} } \times \, e^{2\pi w_{\rm tot}\,  \xi_{p}} \  \frac{ e^{2i\pi (\xi_{p-1} - \xi_{p})\sum_j^{p-1} \sigma_j^{(p-1)}} \prod_{j<k}^{p-1} \sh(\sigma_j^{(p-1)}-\sigma_k^{(p-1)})^2 }{\prod_{j,k}^{p-1,p} \ch(\sigma_j^{(p-1)} - \sigma_k^{(p)} + i w_k z) }
\, ,
\end{align}
where $w_{\rm tot} = \sum_{k=1}^p w_k =|\scR|$. We remind that $\lim_{z \rightarrow 1} $ means that we compute the matrix model for $z \in i\bR$ and analytically continue the result to $z=1$. For $z \in i\bR$, we recognize in the integrand the matrix model computing the partition function of $T[SU(p)]$ with FI parameters $\xi^{(p)}=(\xi_1, \xi_2, \cdots, \xi_p)$ and shifted mass parameters $\sigma^{(p)} - i w z = (\sigma^{(p)}_1 - i w_1 z, \sigma^{(p)}_2- i w_2 z, \cdots , \sigma^{(p)}_p - i w_p z)$. We can use \eqref{ZT[SU(N)]} to evaluate it and easily perform the analytic  continuation to $z=1$ \footnote{The analytic continuation $z \rightarrow 1$ can be directly taken in the integrand for the factor $ \prod_{j<k}^p\sh^{-1}(\sigma_j^{(p)}-\sigma_k^{(p)} - i w_j z + i w_k z)$, as no pole crosses the contour of integration when $z \rightarrow 1$. This is because the rest of the matrix integrand has $\prod_{j<k}^p\sh(\sigma_j^{(p)}-\sigma_k^{(p)})$ factors, which kill the poles of the first factor. We will rely on this property on several other occasions. Enforcing $z=1$ simplifies the factor to  $ (-1)^{(p-1) w_{\rm tot}} \prod_{j<k}^p\sh^{-1}(\sigma_j^{(p)}-\sigma_k^{(p)})$.}
\begin{align}
V^{(p)}(w) & = \ \lim_{z \rightarrow 1} \ \frac{1}{Z} \ \int [\cdots ]  e^{2\pi w_{\rm tot}\, \xi_{p}} \ Z^{T[SU(p)]}[\xi^{(p)}, \sigma^{(p)} - i w z] \no \\
&= \frac{1}{Z} \ \int [\cdots ] \,  (-i)^{\frac{p(p-1)}{2}} e^{-2i\pi \xi_p \sum_j \sigma^{(p)}_j}  
(-1)^{(p-1) w_{\rm tot}}
\frac{ \sum_{\tau\in S_p} (-1)^\tau \, e^{2\pi \sum_j^p \xi_j w_{\tau(j)}} \, e^{2\pi i\sum_j^p \xi_j \sigma^{(p)}_{\tau(j)}}}{\prod_{j<k}^p \sh(\xi_j - \xi_k)\sh(\sigma^{(p)}_j - \sigma^{(p)}_k)} \, .
\end{align}
Note that the background factor $e^{2\pi w_{\rm tot}\, \xi_{p}}$ canceled in the limit $z=1$.
One can now pull out the sum over permutations $\tau \in S_p$ and play with eigenvalues relabelings $\sigma^{(p)}_{\tau(j)} \leftrightarrow \sigma^{(p)}_{j}$ (knowing that the hidden factors in $[\cdots]$ are invariant under such relabelings). After several operations we can pull out of the integral the factors depending on the weight $w$,
\begin{align}
V^{(p)}(w)  &=  \  (-1)^{(p-1) w_{\rm tot}} \frac{1}{p!} \sum_{\tau\in S_p} e^{2\pi \sum_j^p \xi_j w_{\tau(j)}}  \no\\
& \ \times \frac{1}{Z} \int [\cdots ] \,  (-i)^{\frac{p(p-1)}{2}} e^{-2i\pi \xi_p \sum_j \sigma^{(p)}_j}  
\frac{ \sum_{\tau\in S_p} (-1)^\tau \, e^{2\pi i\sum_j^p \xi_j \sigma^{(p)}_{\tau(j)}}}{\prod_{j<k}^p \sh(\xi_j - \xi_k)\sh(\sigma^{(p)}_j - \sigma^{(p)}_k)} \no\\[+2ex]
&=  \  (-1)^{(p-1) w_{\rm tot}} \frac{1}{p!} \sum_{\tau\in S_p}  e^{2\pi \sum_j^p \xi_j w_{\tau(j)}}  \, 
\frac{1}{Z} \int [\cdots ] \, Z^{T[SU(p)]}[\xi^{(p)}, \sigma^{(p)}] \, .
\end{align}
In the last step we have re-transformed the matrix model integrand into the $Z^{T[SU(p)]}$ partition function, but with unshifted mass parameters. The resulting matrix model is simply the bare $Z^{T[SU(N)]}$ matrix model, which cancels with the normalization $1/Z$, leaving only the $w$ dependent prefactor
\begin{align}
V^{(p)}(w)  &= \  (-1)^{(p-1) w_{\rm tot}} \frac{1}{p!} \sum_{\tau\in S_p}  e^{2\pi \sum_j^p \xi_j w_{\tau(j)}}  \, .
\end{align}
The full Vortex loop vev is obtained by summing over all the single weight contributions. Using \eqref{SumWeyl} we obtain
\begin{align}
\langle V^{(p)}_\scR  \rangle &= \  (-1)^{(p-1) |\cR|}  \sum_{w \in \scR}  e^{2\pi \sum_j^p \xi_j w_j} \, ,
\label{Vortexloopstrivial}
\end{align}
where $|\cR| (= w_{\rm tot})$ is the total number of boxes in the Young tableau associated to the representation  $\scR$. Up to the overall sign factor,
$V^{(p)}_\scR $ evaluates to a background Wilson loop in the representation $\scR$ of the $U(p) \subset U(N)_J$ subgroup of the topological symmetry acting on the Coulomb branch. This is precisely the prediction \eqref{MirrorMapVp}  derived  from the brane picture.

\smallskip

We now turn to the evaluation of the vevs of the Vortex loops $V_{M,\scR}$ of the $U(N-1)$ node. 

\smallskip

\noindent Note that the Vortex loop vev $\langle V_{N,\scR} \rangle$, associated to the splitting of hypermultiplets $N = N + 0$, can be evaluated in exactly the same fashion as the vevs of the $V^{(p)}_\scR$ loops, leading to the result \eqref{Vortexloopstrivial} with $p=N-1$ and reproduces our prediction \rf{prediiii}.

In order to  evaluate the $V_{M,\scR}$ vevs, we are going to consider their ``right" SQM realization, namely we choose to insert in the 3d matrix model  the right SQM index \eqref{IndexQMfinal_r} with the additional background loop \eqref{WilsonLine}. Again we start by considering a single weight contribution $V_{M}(w)$, with $w=(w_1, \cdots, w_{N-1})$, and we simplify the matrix model by replacing the factors irrelevant to the computation by $\int [\cdots ] $,
\begin{align}
V_M(w) 
& =  \ \lim_{z \rightarrow 1} \ \frac{1}{Z} \ \int [\cdots ] \int \frac{d^{N'}\sigma}{N'!} \ \frac{ \prod_{i<j}^{N'} \sh^2(\sigma_i-\sigma_j)}
{\prod_{k=1}^{N}\prod_{j}^{N'} \ch(\sigma_j- m_k)} \no\\[+2ex]
& \phantom{=  \ \lim_{z \rightarrow 1} \ \frac{1}{Z}} \times \
e^{2\pi \xi_N w_{\rm tot}}  \,  \frac{\prod_{k=1}^{M}\prod_{j}^{N'} \ch(m_k - \sigma_j)}{\prod_{k=1}^{M}\prod_{j}^{N'} \ch(m_k- \sigma_j - i w_j z)}\, ,
\end{align}
where $\sigma_j$ denote the eigenvalues of the $U(N-1)$ node and we defined $N'\equiv N-1$ for convenience. The cancellation between numerator and denominator factors leads to
\begin{align}
V_M(w) 
& =  \ \lim_{z \rightarrow 1} \ \frac{1}{Z} \ \int [\cdots ] \int \frac{d^{N'}\sigma}{N'!} \ \frac{ e^{2\pi \xi_N w_{\rm tot}} \ \prod_{i<j}^{N'} \sh^2(\sigma_i-\sigma_j)}
{\prod_{k=M+1}^{N}\prod_{j}^{N'} \ch(\sigma_j- m_k) \prod_{k=1}^{M}\prod_{j}^{N'} \ch(\sigma_j + i w_j z- m_k)}  \, .
\end{align}
We compute this matrix model  by using a generalized Cauchy determinant formula. Let us remind the Cauchy determinant formula
\begin{align}
\label{Cauchyformula}
 \frac{\prod_{i<j}^N \sh(\sigma_i - \sigma_j)\prod_{i<j}^N \sh(\ti \sigma_i - \ti\sigma_j)} {\prod_{i,j}^N \ch(\sigma_i - \ti\sigma_j) } =
 \sum_{ \tau \in S_N} (-1)^{\tau} \frac{1}{\prod_j^N \ch(\sigma_{\tau(j)} - \ti\sigma_j)} \, .
\end{align}
A generalized version of this formula, for $N \geq \ti N$, $\Delta \equiv N-\ti N$, is 
\begin{align}
& \frac{\prod_{i<j}^N \sh(\sigma_i - \sigma_j)\prod_{i<j}^{\ti N} \sh(\ti \sigma_i - \ti\sigma_j)} {\prod_{i}^N \prod_{j}^{\ti N} \ch(\sigma_i - \ti\sigma_j) }   \no\\
 & \qquad = (-1)^{\Delta \ti N} \sum_{ \tau \in S_N} (-1)^{\tau} \prod_{j=1}^{\ti N} \frac{e^{-\pi \Delta (\sigma_{\tau(j)} - \ti\sigma_j)}} {\ch(\sigma_{\tau(j)} - \ti\sigma_j)} \ \prod_{j=\ti N+1}^{N} e^{2\pi \sigma_{\tau(j)} \lp \frac{N+\ti N+1}{2}-j \rp } \, . \label{Cauchyformula2} 
\end{align}
We use this formula in the matrix model to replace the factor depending on the weight $w$, with $\Delta = N'-M$,
\begin{align}
&  \frac{1}{\prod_{k=1}^{M}\prod_{j=1}^{N'} \ch(\sigma_j + i w_j z- m_k)}  \ = \  \frac{(-1)^{\Delta M} }{\prod_{i<j}^{N'} \sh(\sigma_i -\sigma_j + i(w_i - w_j)z ) \prod_{i<j}^M \sh(m_i-m_j) }  \no\\
  & \hspace{4.5cm}  \times  \sum_{ \tau \in S_{N'}} (-1)^{\tau} \prod_{j=1}^M \frac{e^{- \pi \Delta (\sigma_{\tau(j)} + i w_{\tau(j)} z - m_j) }}{\ch(\sigma_{\tau(j)} +i w_{\tau(j)} z  - m_j)} \prod_{j=M+1}^{N'} e^{2\pi (\sigma_{\tau(j)} + i w_{\tau(j)} z) \lp \frac{N'+M+1}{2}-j \rp } \no \, .
\end{align}
We may now plug this result into the matrix model and pull out  the sum over permutation $\tau \in S_{N'}$  
\begin{align}
V_M(w) 
  =  \ \lim_{z \rightarrow 1} \, \sum_{ \tau \in S_{N'}}   \frac{(-1)^{\tau} }{Z} & \ \int [\cdots ] \int \frac{d^{N'}\sigma}{N'!} \ \frac{ \prod_{i<j}^{N'} \sh^2(\sigma_i-\sigma_j) \, \prod_{j=M+1}^{N'} e^{2\pi  i w_{\tau(j)} z  \lp \frac{N'+M+1}{2}-j \rp }}
{\prod_{k=M+1}^{N}\prod_{j}^{N'} \ch(\sigma_j- m_k) \prod_{i<j}^{N'} \sh(\sigma_i -\sigma_j + i(w_i - w_j)z )}  \no\\
&  \times  \frac{(-1)^{\Delta M} \, e^{2\pi \xi_N w_{\rm tot}} }{\prod_{i<j}^M \sh(m_i-m_j) }  \prod_{j=1}^M \frac{e^{- \pi \Delta (\sigma_{\tau(j)} + i w_{\tau(j)} z - m_j) }}{\ch(\sigma_{\tau(j)} +i w_{\tau(j)} z  - m_j)} \prod_{j=M+1}^{N'} e^{2\pi \sigma_{\tau(j)} \lp \frac{N'+M+1}{2}-j \rp }\,.
\label{VMeqn1}
\end{align}
We can simplify the result by taking directly setting $z=1$ in the factors in the first line of \eqref{VMeqn1}, which just produces a sign. This is justified because these factors do not have poles crossing the integration contours as $z$ goes from 0 to 1. We obtain
\begin{align}
V_M(w) 
  =  \ \lim_{z \rightarrow 1} \, \sum_{ \tau \in S_{N'}}   \frac{(-1)^{\tau} }{Z} & \ \int [\cdots ] \int \frac{d^{N'}\sigma}{N'!} \ \frac{ (-1)^{N w_{\rm tot}} (-1)^{(N+M) \sum_{j=M+1}^{N'} w_{\tau(j)} }  \prod_{i<j}^{N'} \sh(\sigma_i-\sigma_j) }
{\prod_{k=M+1}^{N}\prod_{j}^{N'} \ch(\sigma_j- m_k) }  \no\\
&  \times  \frac{(-1)^{\Delta M} \, e^{2\pi \xi_N w_{\rm tot}} }{\prod_{i<j}^M \sh(m_i-m_j) }  \prod_{j=1}^M \frac{e^{- \pi \Delta (\sigma_{\tau(j)} + i w_{\tau(j)} z - m_j) }}{\ch(\sigma_{\tau(j)} +i w_{\tau(j)} z  - m_j)} \prod_{j=M+1}^{N'} e^{2\pi \sigma_{\tau(j)} \lp \frac{N'+M+1}{2}-j \rp }\,.
\label{VMeqn2}
\end{align}
It is now possible to relabel the eigenvalues  $\sigma_{\tau(j)} \leftrightarrow \sigma_{j}$ in each integral and recognize each term in the sum as the same matrix model with shifted masses $m_j \rightarrow m_j - i w_{\tau(j)} z$, for $1 \le j \le p$. The shifted matrix model is simply the bare partition function $Z$, as can be understood by taking $w_j=0$ for all $j$. We have precisely
\begin{align}
V_M(w) 
  &=  \ \lim_{z \rightarrow 1} \, \frac{1}{N'!} \sum_{ \tau \in S_{N'}}   \frac{1}{Z}  \ \int [\cdots ] \int d^{N'}\sigma \ \frac{ (-1)^{N w_{\rm tot}} (-1)^{(N+M) \sum_{j=M+1}^{N'} w_{\tau(j)} }  \prod_{i<j}^{N'} \sh(\sigma_i-\sigma_j) }
{\prod_{k=M+1}^{N}\prod_{j}^{N'} \ch(\sigma_j- m_k) }  \no\\
& \phantom{\ \lim_{z \rightarrow 1} \, \frac{1}{N'!} \sum_{ \tau \in S_{N'}}   \frac{1}{Z}} 
 \times  \frac{(-1)^{\Delta M} \, e^{2\pi \xi_N w_{\rm tot}} }{\prod_{i<j}^M \sh(m_i-m_j) }  \prod_{j=1}^M \frac{e^{- \pi \Delta (\sigma_j + i w_{\tau(j)} z - m_j) }}{\ch(\sigma_{j} +i w_{\tau(j)} z  - m_j)} \prod_{j=M+1}^{N'} e^{2\pi \sigma_{j} \lp \frac{N'+M+1}{2}-j \rp } \no\\
 & = \ e^{2\pi \xi_N w_{\rm tot}} \, (-1)^{M w_{\rm tot}}  \, \frac{1}{N'!} \sum_{ \tau \in S_{N'}}  (-1)^{N' \sum_{j=1}^{M} w_{\tau(j)} } \frac{1}{Z} \wat{\mathfrak{S}}^{w_{M,\tau}}    Z \, ,
\label{VMeqn3}
\end{align}
where $w_{M,\tau} = (w_{\tau(1)}, w_{\tau(2)}, \cdots, w_{\tau(M)}, \underbrace{0, \cdots, 0}_{N-M})$.
Summing over the weights $w$ of the representation $\scR$ and using \eqref{SumWeyl} yields the final result
\begin{align}
\langle V_{M, \scR} \rangle
 & = \ e^{2\pi \xi_N w_{\rm tot}} \, (-1)^{M |\cR|}  \,  \sum_{ w \in \scR}  (-1)^{(N-1) \sum_{j=1}^{M} w_{j} } \frac{1}{Z} \wat{\mathfrak{S}}^{w_{M}}  Z \, ,
\label{VMfinal0}
\end{align}
where $w_{M} = (w_{1}, w_{2}, \cdots, w_{M}, \underbrace{0, \cdots, 0}_{N-M})$. 
To check mirror symmetry, it is useful to re-express the result in terms of $\scZ$, which is manifestly invariant under the exchange of mass and FI parameters,
\smallskip
\begin{align}
\langle V_{M, \scR} \rangle
 & = \ (-1)^{(N+M-1) |\cR|}  \,  \sum_{ w \in \scR}  e^{2\pi \check \xi_N \sum_{j=M+1}^{N-1} w_j} \, \frac{1}{\scZ} \wat{\mathfrak{S}}^{w_{M}}   \scZ \, ,
\label{VMfinal}
\end{align}
where we have reabsorbed $(-1)$ factors into an imaginary shift of $\xi_N$ with $\check \xi_N = \xi_N + i \frac{N-1}{2}$.
We remind the reader that we did not keep track of the overall sign in the evaluation of the SQM index, so that our evaluation of Vortex loops are only valid up to an overall sign.

\smallskip
This completes our evaluation of the matrix models computing the vevs of the $T[SU(N)]$ Vortex loop operators.

\subsubsection{Mirror Map}
\label{sssec:MirrorMapTSUN}

With all the computations out of the way, it remains to confirm the rest of the mirror map predictions of section \ref{sec:examples} between loop operators in $T[SU(N)]$. Let start with the simplest case of the Vortex loop $V_{N-1, \scR}$ discussed in section \ref{ssec:MirrorTSUNsimple}. Its vev is given by \eqref{VMfinal} with $M=N-1$,
\begin{align}
\langle V_{N-1, \scR} \rangle
 & = \  \frac{1}{\scZ} \sum_{ w \in \scR}   \wat{\mathfrak{S}}^{w}  \scZ \, ,
\label{VNminus1}
\end{align}
where we used $\sum_{j=1}^{N-1} w_{j} = w_{\rm tot} = |\cR|$.
Recalling the vev of the Wilson loop in the $U(N-1)$ node given by \eqref{Wpfinal},
\begin{align}
\langle W^{U(N-1)}_{\scR} \rangle = (-1)^{N \, |\cR|} \, \frac{1}{\scZ} \sum_{w \in \scR} \  \mathfrak{S}^{w}  \scZ \, ,
\end{align}
we observe that the two vevs are mapped, up to an irrelevant sign, under the exchange of the masses and FI parameters $m_j \leftrightarrow \xi_j$, 
 \begin{align}
 \langle W^{U(N-1)}_{\scR} \rangle  \ \xleftrightarrow{\text{\ mirror \ }} \  \langle V_{N-1, \scR} \rangle \, .
 \end{align} 
This is precisely the prediction \eqref{MapTSUNsimple} of mirror symmetry obtained from the brane picture.

\medskip

The more complicated mirror symmetry predictions of section \ref{ssec:OtherLoops} are also easily checked with our explicit exact results. The formula \eqref{VMfinal} for the $V_{M, \scR}$ loops vevs expresses a decomposition into contributions labeled by representations $({\bf q},\wat\scR)$ appearing in the decomposition of $\scR$ under the subgroup $U(1) \times U(M) $, where $U(M)$ is embedded as $ U(M) \times U(N-M-1) \subset U(N-1)$ and  $U(1)$ is embedded diagonally in $U(N-M-1)$,
\begin{align}
U(N-1) \ \rightarrow  \ U(1) \times U(M) \, : \qquad \scR  \ \rightarrow \ \bigoplus_{s \in \Delta_M} ({\bf q}_s,\wat\scR_s)\, ,
\end{align}
where $\Delta_M$ denotes the set of representations $({\bf q},\wat\scR)$ in the decomposition of $\scR$ counted with degeneracies. Equation
\eqref{VMfinal} can be re-expressed as
\begin{align}
\langle V_{M, \scR} \rangle
 & = \ (-1)^{M |\cR|}  \,  \sum_{s \in \Delta_M} \Big[ \,    e^{2\pi {\bf q}_s \, \check \xi_N} \, 
 \frac{1}{\scZ} \sum_{\wat w \in \wat\scR_s}  \wat{\mathfrak{S}}^{\wat w}    \scZ  \, \Big] \, .
\end{align}
The factor $ \frac{1}{\scZ} \sum_{\wat w \in \wat\scR_s}  \wat{\mathfrak{S}}^{\wat w}   \scZ $ is labeled by a representation $\wat\scR_s$ of $U(M)$ and is mapped under the exchange of mass and FI parameters  to the vev given in \eqref{Wpfinal} of the $U(M)$ Wilson loop labeled by a representation $\wat \scR_s$. We obtain the explicit mirror symmetry map
\begin{align}
\langle V_{M, \scR} \rangle \ \xleftrightarrow{\text{\ mirror \ }} \   (-1)^{N |\cR|}  \,  \sum_{s \in \Delta_M}   e^{2\pi {\bf q}_s \, \check{\ti m}_N} \, \langle W^{U(M)}_{\wat \scR_s} \rangle  \, ,
\end{align}
where we have used $N_{\wat \scR_s} + {\bf q}_s= |\cR|$ and $\check{\ti m}_N = \xi_N + i \frac{N-1}{2}$ is a fundamental hypermultiplet mass in the mirror dual $T[SU(N)]$, shifted by an imaginary number to reabsorb $(-1)$ factors.
This expresses the fact that the $V_{M, \scR} $ Vortex loop is mapped under mirror symmetry to a linear combination of $W^{U(M)}_{\wat \scR}$ Wilson loops combined with flavor Wilson loops. This reproduces the mirror symmetry prediction \eqref{MirrorMapAllLoops} found by studying carefully the brane realization of the loop operators and their mapping under S-duality, up to the imaginary shift of $\ti m_N = \xi_N$.  This imaginary shift of the flavor parameter in the background Wilson loop is curious and would deserve more investigation.  

\medskip

This completes successfully the checks of mirror symmetry for the $T[SU(N)]$ theory.
We have found that $T[SU(N)]$   Wilson loops and Vortex loops can be expressed as operators acting on the partition function by imaginary shifts of the FI  or mass parameters (see also \cite{Bullimore:2014nla}). More generally, in all $T^{\rho}_{\hat\rho}[SU(N)]$ linear quiver theories it seems possible to express Wilson loops / Vortex loops  as operators acting on the partition function by imaginary shifts of a ``generalized" set of FI parameters/ mass parameters. (we will not provide a proof of this result in this paper).

\subsection{Loops In SQCD}
\label{ssec:SQCDtests}

As our final example, we consider   loops in the $U(N)$ theory with $2N$ fundamental and its mirror dual theory, discussed in section \ref{ssec:SQCD}. We   focus on the prediction \eqref{MapUNsimple}, relating the Wilson loop $W_\scR$ of the $U(N)$ theory to the Vortex loop $\ti V^{(N)}_{1, \scR}$ of the mirror theory. 
The other mirror maps \eqref{MapUNComplicated} can be easily checked with the explicit matrix models computing the vevs of the operators, by computations essentially identical to those presented above for the $T[SU(N)]$ loops.

\smallskip

We denote $\xi_1 - \xi_2$ the FI parameter and $m_a$, $a = 1, \cdots, 2N$, the masses of fundamental hypermultiplets in the $U(N)$ theory, $\ti\xi_a - \ti\xi_{a+1}$, $ a = 1, \cdots, 2N-1$, the FI parameters and $\ti m_1, \ti m_2$ the masses of fundamental hypermultiplets in the mirror theory.
The matrix models computing the $S^3$ partition function of the $U(N)$ theory and its mirror dual are given by
\begin{align}
Z & =  e^{\phi} \int \frac{d^N\sigma}{N!} \, \frac{\prod_{i<j}^N \sh^2(\sigma_i - \sigma_j)}{\prod_j^N \prod_a^{2N} \ch(\sigma_j - m_a)} \, e^{2\pi i (\xi_1 - \xi_2) \sum_j^N \sigma_j} \, , \\[+2ex]
\ti Z &= e^{\ti\phi} \int \frac{d^N\sigma}{N!} \scZ^{T[SU(N)]}\big[ \ti\xi_{1 .. N} ; \sigma_{1 .. N} \big] 
\frac{\prod_{i<j}^N \sh^2(\sigma_i - \sigma_j)}{\prod_j^N  \ch(\sigma_j - \ti m_1) \ch(\sigma_j - \ti m_2)}
\scZ^{T[SU(N)]}\big[ -\ti\xi_{N+1 .. 2N} ; \sigma_{1 .. N} \big] \, ,
\end{align}
where we used to fact that the mirror theory can be decomposed into three pieces: two $T[SU(N)]$ theories (left and right parts of the quiver in figure \ref{SQCDandMirror}-b) whose $SU(N)$ hypermultiplet flavor symmetries are gauged with the $U(N)$ gauge symmetry of the central node (which has two more hypermultiplets by itself). The matrix model is then a combination of these three pieces and can be expressed using two $T[SU(N)]$ partition functions with mass parameters identified with the $U(N)$ node eigenvalues $\sigma_j$ and FI parameters as indicated.
We have also added background Chern-Simons terms given by
\begin{align}
e^{\phi} & = e^{2\pi i \xi_2 \sum_{a=1}^{2N} m_a} \, , \quad e^{\ti\phi} = e^{2\pi i (\ti m_1 + \ti m_2) \sum_{a = N+1}^{2N} \ti \xi_a} \, ,
\end{align}
which are unphysical, as they belong to the set of finite counterterms parametrizing ambiguities of the partition function, but are useful to obtain partition functions which match exactly under mirror symmetry, namely under the identification 
\beq
(\xi_1,\xi_2,m_a) = (\ti m_1, \ti m_2, \ti\xi_a)\,. 
\eeq
This result will follow from our computations.
\smallskip

The matrix model computing the vev of a Wilson loop $W_\scR$ in the $U(N)$ theory is given by
\begin{align}
\vev{W_\scR} &= \sum_{w \in \scR} W(w) \\
W(w) &= \frac{e^{\phi}}{Z} \int \frac{d^N \sigma}{N!} \, e^{2\pi \sum_j^N w_j \sigma_j} \frac{\prod_{i<j}^N \sh^2(\sigma_i - \sigma_j)}{\prod_j^N \prod_a^{2N} \ch(\sigma_j - m_a)} \, e^{2\pi i (\xi_1 - \xi_2) \sum_j^N \sigma_j} \, .
\end{align}
Using twice the Cauchy determinant formula \eqref{Cauchyformula} we find
\begin{align}
W(w) &= \sum_{\tau, \tau' \in S_N} (-1)^{\tau + \tau'}\, \frac{e^{\phi}}{Z}  \frac{1}{\prod[m_a]} \int \frac{d^N \sigma}{N!} \, \frac{e^{2\pi \sum_j^N w_j \sigma_j} \, e^{2\pi i (\xi_1 - \xi_2) \sum_j^N \sigma_j}}{\prod_j^N  \ch(\sigma_j - m_{\tau(j)}) \ch(\sigma_j - m_{\tau'(j) +N}) } \, ,
\end{align}
where we define $\prod[x_a] \equiv \prod_{1 \le a < b \le N} \sh(x_a - x_b) \sh(x_{a+N} - x_{b +N})$.
To compute the integrals we use the identity
\begin{align}
\int d\sigma \frac{ e^{2\pi i \beta \sigma}}{\ch(\sigma - M_1) \ch(\sigma - M_2)} & = ( -i) \, \frac{e^{2\pi i \beta M_1}- e^{2\pi i \beta M_2} }{\sh \beta \, \sh(M_1 - M_2)} \, ,
\label{integral1}
\end{align}
analytically continued to $\beta \in \bC$.  After simplifications we obtain
\begin{align}
W(w) &= \sum_{\tau, \tau' \in S_N} \frac{(-1)^{\tau + \tau'}}{Z} \,  \frac{(-i)^N \, (-1)^{|\scR|}}{N! \prod[m_a] \, \sh^N(\xi_1 -\xi_2)} \no\\ 
 & \quad \times \prod_{j=1}^N \frac{e^{2\pi i (\xi_1 m_{\tau(j)} + \xi_2 m_{\tau'(j)+N} - i w_j m_{\tau(j)}) }  
 -  e^{2\pi i (\xi_2 m_{\tau(j)} + \xi_1 m_{\tau'(j)+N} - i w_j m_{\tau'(j)+N}) } }{\sh (m_{\tau(j)} - m_{\tau'(j)+N})}\, .
 \label{WwSQCD}
\end{align}
We now turn to the mirror dual operator which should be the Vortex loop $\ti V^{(N)}_{1, \scR}$, according to the discussion in section \ref{ssec:SQCD}. We compute its vev from its 3d/1d realization with the left SQM. The matrix model following from \eqref{3d1dMM}, with SQM factor \eqref{IndexQMfinal_l} and background loop \eqref{WilsonLine} can be expressed after cancellations of  factors of ch as
\begin{align}
\vev{\ti V^{(N)}_{1, \scR}} &= \sum_{w \in \scR} \ti V(w) \\
\ti V(w) &= \frac{1}{\ti Z} \lim_{z\rightarrow 1} \, e^{\ti\phi} \, e^{2\pi w_{\rm tot} \ti\xi_N} \int \frac{d^N\sigma}{N!} \, \scZ^{T[SU(N)]}\big[ \ti\xi_{1 .. N} ; \sigma_j - i w_j z \big]  \no\\
& \phantom{= \frac{1}{\ti Z} \lim_{z\rightarrow 1} \, e^{\ti\phi} \, e^{2\pi w_{\rm tot} \ti\xi_N}} \ \times  \frac{\prod_{i<j}^N \sh^2(\sigma_i - \sigma_j)}{\prod_j^N  \ch(\sigma_j - i w_j z - \ti m_2) \ch(\sigma_j - \ti m_1)}
\scZ^{T[SU(N)]}\big[ -\ti\xi_{N+1 .. 2N} ; \sigma_j \big] \, ,
\end{align}
Plugging the explicit values \eqref{ZT[SU(N)]} leads to
\begin{align}
\ti V(w) &= \frac{1}{\ti Z} \lim_{z\rightarrow 1} \, \sum_{\tau, \tau' \in S_N} (-1)^{\tau + \tau'} \, \frac{  e^{\ti\phi} \, e^{2\pi w_{\rm tot} \ti\xi_N} }{\prod[\ti\xi_a]} \int \frac{d^N\sigma}{N!} \, \frac{e^{2\pi i \sum_j^N \sigma_j (\ti\xi_{\tau(j)} - \ti\xi_{\tau'(j)+N}) }}{\prod_j^N  \ch(\sigma_j - i w_j z - \ti m_2) \ch(\sigma_j - \ti m_1)} \no\\
& \phantom{= \frac{1}{\ti Z} \lim_{z\rightarrow 1} \, \sum_{\tau, \tau' \in S_N} (-1)^{\tau + \tau'} \, \frac{  e^{\ti\phi} \, e^{2\pi w_{\rm tot} \ti\xi_N} }{\prod[\ti\xi_a]} } \quad \times e^{2\pi \sum_j^N \ti\xi_{\tau(j)} w_j z} \prod_{i<j}^N \frac{ \sh(\sigma_i - \sigma_j)}{\sh(\sigma_i - \sigma_j - i (w_i - w_j)z)}
\, .
\end{align}
As we did already several times, we can plug directly $z=1$ in the factors on the second lign, since these factors do not have poles as $z$ goes from 0 to 1. The ratio of sh factors then simplifies to $(-1)^{(N-1)w_{\rm tot}}$. The remaining integrals can be performed using \eqref{integral1} and the final result can be analytically continued to $z=1$. After simplification, we obtain
\begin{align}
\ti V(w) &= \sum_{\tau, \tau' \in S_N} \frac{(-1)^{\tau + \tau'}}{\ti Z} \,  \frac{(-i)^N  \, (-1)^{N |\scR|}}{N! \prod[\ti\xi_a] \, \sh^N(\ti m_1 -\ti m_2)} \no\\ 
 & \quad \times \prod_{j=1}^N \frac{e^{2\pi i (\ti m_1 \ti\xi_{\tau(j)} + \ti m_2 \ti\xi_{\tau'(j)+N} - i w_j \ti\xi_{\tau(j)}) }  
 -  e^{2\pi i (\ti m_2 \ti\xi_{\tau(j)} + \ti m_1 \ti\xi_{\tau'(j)+N} - i w_j \ti\xi_{\tau'(j)+N}) } }{\sh (\ti\xi_{\tau(j)} - \ti\xi_{\tau'(j)+N})}\, .
\, .
\end{align}
This matches precisely \eqref{WwSQCD} upon identifying FI and mass parameters $(\xi_1,\xi_2,m_a) = (\ti m_1, \ti m_2, \ti\xi_a)$, up to a sign, which was not carefully analyzed in the computations leading to the matrix models. This confirms the mirror symmetry prediction \rf{MapUNsimple}
\begin{align}
\vev{\ti V^{(N)}_{1, \scR}}  \ \xleftrightarrow{\text{\ mirror \ }} \ \vev{W_\scR} \, .
\end{align}
We notice that the partition functions $Z$ and $\ti Z$ can be found by setting to zero the weight $w$ and removing the normalization factors $Z^{-1}$ and $\ti Z^{-1}$ in the formula for $W(w)$ and $\ti V(w)$ respectively. We observe then that $Z$ and $\ti Z$ are exactly mapped under the identification $(\xi_1,\xi_2,m_a) = (\ti m_1, \ti m_2, \ti\xi_a)$. This would not have happened if we did not add the unphysical phases $e^{\phi}$, $e^{\ti\phi}$ to the matrix models.

\subsection{Hopping Duality}
\label{ssec:Hoppingduality}

We have been claiming several times that each Vortex loop can be realized (at least) by two different 3d/1d defect theories, which are read from the brane realization by moving the D1-branes to the closest NS5 on the left or on the right. 
The equivalence between the two defect theories is called hopping duality, in analogy with \cite{Gadde:2013dda} (see also \cite{Gomis:2014eya}). 
We can show that the $S^3$ partition function of the two defect theories indeed match.

\smallskip

\begin{figure}[th]
\centering
\includegraphics[scale=0.65]{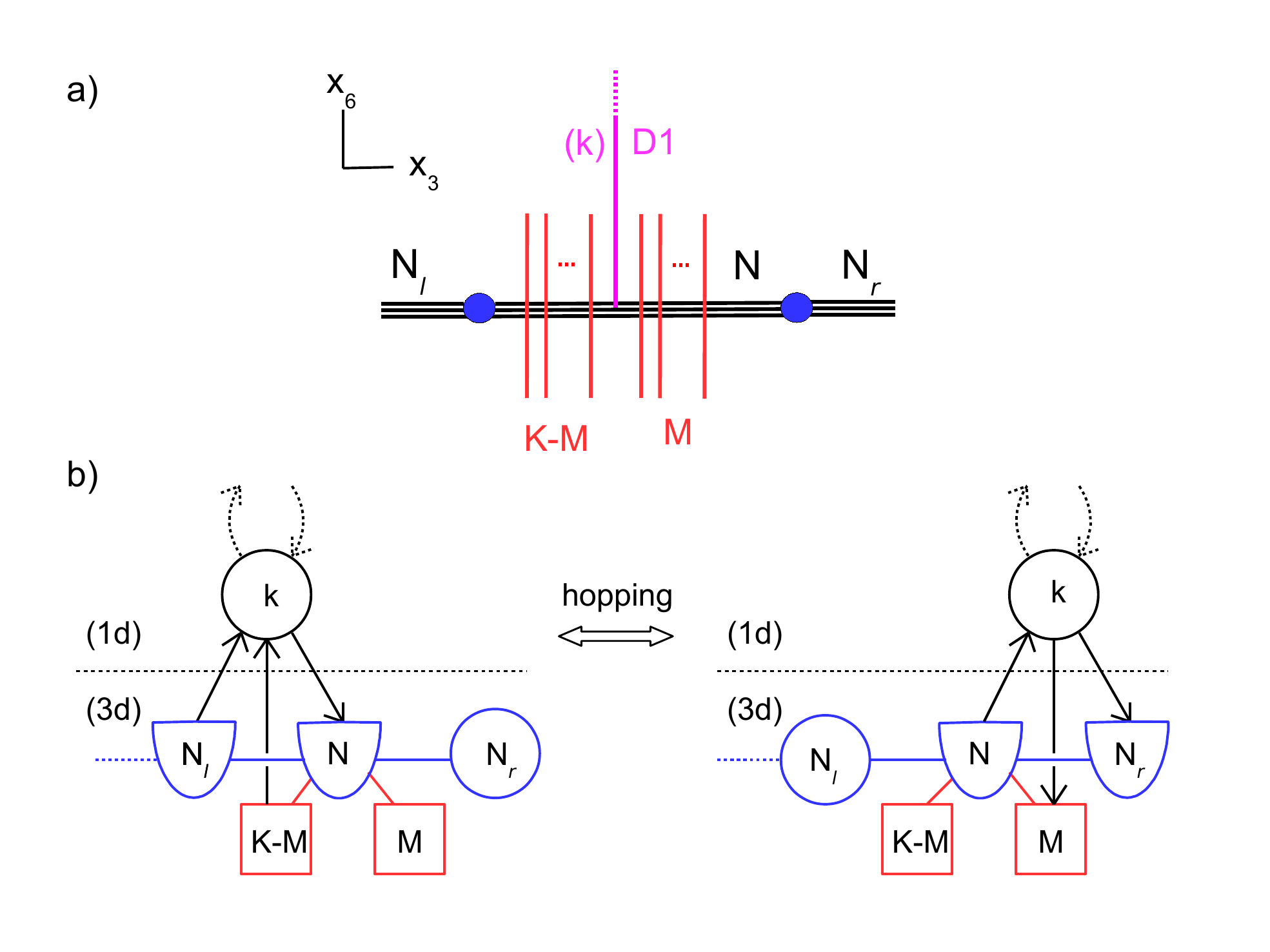}
\vspace{-0.5cm}
\caption{\footnotesize{a) Brane realization of the Vortex loop $V^{(N)}_{M,\scR}$. The $k=|\cR|$ D1-branes end on NS5/NS5'-branes which are not shown in the picture. b) Left and right 3d/1d quivers read from moving the D1-branes to the NS5 on the left or on the right, related by the hopping duality. }}
\label{Hopping}
\end{figure}

Consider a Vortex loop $V^{(N)}_{M,\scR}$ in a certain 3d quiver theory, labeled by a representation $\scR$ of a $U(N)$ node and a splitting $K = M + (K-M)$ of the $K$ fundamental hypermultiplets of that node. It is realized by a brane configuration of figure \ref{Hopping}-a,  with $|\cR|$ D1-branes ending on $N$ D3-branes with $K-M$ D5-branes on the left and $M$ D5-branes on the right. The associated left and right 3d/1d theories are shown in \ref{Hopping}-b.

The matrix model associated to the right 3d/1d theory and computing $\langle V^{(N)}_{M,\scR} \rangle$ is given by \ref{3d1dMM} with the 1d index \ref{IndexQMfinal_r}:
\begin{align}
\langle V^{(N)}_{M,\scR} \rangle_{\rm right} &= \lim_{z \rightarrow 1} \, \sum_{ w \in \scR} \frac{1}{Z}\int \frac{d^N \sigma}{N!} \,  
\frac{e^{2\pi i (\xi_l-\xi_r)\sum_j^N \sigma_j} \, \prod_{i<j}^N \sh^2(\sigma_i - \sigma_j)}{\prod_j^N \big[ \prod_{i}^{N_l} \ch(\sigma^l_i - \sigma_j)  \prod_{a}^K \ch(\sigma_j - m_a) \prod_{k}^{N_r} \ch(\sigma_k - \sigma^r_j) \big] } \no\\
& \phantom{= \lim_{z \rightarrow 1} \, \sum_{ w \in \scR} } . \ e^{2\pi \xi_r |\cR|}   \,  \prod_{j}^{N} \big[\prod_{a=1}^{M} \frac{\ch( m_a - \sigma_j)}{\ch( m_a -  \sigma_j - i w_j z)}\prod_{k=1}^{N_r} \frac{\ch( \sigma^r_k - \sigma_j)}{\ch( \sigma^r_k -  \sigma_j - i w_j z)}  \big] \ \Big[ \cdots \Big] \, ,
\end{align} 
where $[ \, \cdots ]$ indicates the matrix model associated to the other nodes of the 3d quiver, which play no role in the check of the hopping duality, $\xi_l$ and $\xi_r$ are the ``FI parameters" associated to the left and right NS5-branes ($\xi_l - \xi_r$ is the FI parameter of the $U(N)$ node), $m_a$ are the masses of the fundamental hypermultiplets and $\sigma^l_j$ and $\sigma^r_k$ are the eigenvalues of the $U(N_l)$ and $U(N_r)$ nodes standing respectively on the left and on the right of the $U(N)$ node in the quiver diagram. This simplifies to 
\begin{align}
\langle V^{(N)}_{M,\scR} \rangle_{\rm right} &= \lim_{z \rightarrow 1} \, \sum_{ w \in \scR} \frac{1}{Z} \int \frac{d^N \sigma}{N!} \,  
\frac{ e^{2\pi i (\xi_l-\xi_r)\sum_j^N \sigma_j} \, \prod_{i<j}^N \sh^2(\sigma_i - \sigma_j)}{\prod_j^N \big[ \prod_{i}^{N_l} \ch(\sigma^l_i - \sigma_j)  \prod_{a = M+1}^{K} \ch(\sigma_j - m_a) \big] } \no\\
& \phantom{= \lim_{z \rightarrow 1} \, \sum_{ w \in \scR} } . \    \frac{e^{2\pi \xi_r |\cR|}}{\prod_{j}^{N} \big[\prod_{a=1}^{M} \ch( m_a -  \sigma_j - i w_j z) \prod_{k=1}^{N_r} \ch( \sigma^r_k -  \sigma_j - i w_j z)  \big]} \ \Big[ \cdots \Big] \, .
\end{align} 
The meaning of the $z \rightarrow 1$ limit, as explained  after equation \ref{3d1dMM}, is to take the analytical continuation of the matrix model computed with $i z \in \bR$. We can thus perform the change of variable $\sigma_j \rightarrow \sigma_j - i w_j z$, leading to 
\begin{align}
\langle V^{(N)}_{M,\scR} \rangle_{\rm right} &= \lim_{z \rightarrow 1} \, \sum_{ w \in \scR} \frac{1}{Z} \int \frac{d^N \sigma}{N!} \,  
\frac{ e^{2\pi (\xi_l - \xi_r) z  |\cR|}  \, e^{2\pi i (\xi_l-\xi_r)\sum_j^N \sigma_j} \, \prod_{i<j}^N \sh^2(\sigma_i - \sigma_j - i (w_i -w_j) z)}{\prod_j^N \big[ \prod_{i}^{N_l} \ch(\sigma^l_i - \sigma_j + i w_j z)  \prod_{a = M+1}^{K} \ch( m_a - \sigma_j + i w_j z) \big] } \no\\
& \phantom{= \lim_{z \rightarrow 1} \, \sum_{ w \in \scR} } . \    \frac{e^{2\pi \xi_r |\cR|}}{\prod_{j}^{N} \big[\prod_{a=1}^{M} \ch( m_a -  \sigma_j) \prod_{k=1}^{N_r} \ch( \sigma^r_k -  \sigma_j)  \big]} \ \Big[ \cdots \Big] \, .
\end{align} 
The analytical continuation $z \rightarrow 1$ can be taken directly in the integrand for the factors in the numerator, because they do not have poles for $z \in \bC$. The expression then simplifies again and matches the matrix model computing $\langle V^{(N)}_{M,\scR}  \rangle$ from the left 3d/1d theory realization:
\begin{align}
\langle V^{(N)}_{M,\scR} \rangle_{\rm right} &= \lim_{z \rightarrow 1} \, \sum_{ w \in \scR} \frac{1}{Z} \int \frac{d^N \sigma}{N!} \,  
\frac{e^{2\pi \xi_l  |\cR|}  }{\prod_j^N \big[ \prod_{i}^{N_l} \ch(\sigma^l_i - \sigma_j + i w_j z)  \prod_{a = M+1}^{K} \ch( m_a - \sigma_j + i w_j z) \big] } \no\\
& \phantom{= \lim_{z \rightarrow 1} \, \sum_{ w \in \scR} } . \    \frac{e^{2\pi i (\xi_l-\xi_r)\sum_j^N \sigma_j} \, \prod_{i<j}^N \sh^2(\sigma_i - \sigma_j)}{\prod_{j}^{N} \big[\prod_{a=1}^{M} \ch( m_a -  \sigma_j) \prod_{k=1}^{N_r} \ch( \sigma^r_k -  \sigma_j)  \big]} \ \Big[ \cdots \Big] \no\\
& = \ \langle V^{(N)}_{M,\scR} \rangle_{\rm left}   \, ,
\end{align} 
where we have used $\sh(x + i n) = (-1)^n \sh(x)$, for $n \in \bZ$. This is indeed the matrix model associated to the left 3d/1d theory given by  \ref{3d1dMM} with the 1d index \ref{IndexQMfinal_l}, after simplification of some  factors of $\ch$. 
Note that the additional background Wilson loop is important to get a precise match.

\medskip

The hopping duality also explains the equivalence of Vortex loops labeled by representations of different nodes. This occurs for instance for the Vortex loops of circular quivers with nodes of equal ranks described in section \ref{ssec:CircLoops} (see \eqref{CircLoopMap1}, \eqref{CircLoopMap2}). 

\begin{figure}[th]
\centering
\includegraphics[scale=0.65]{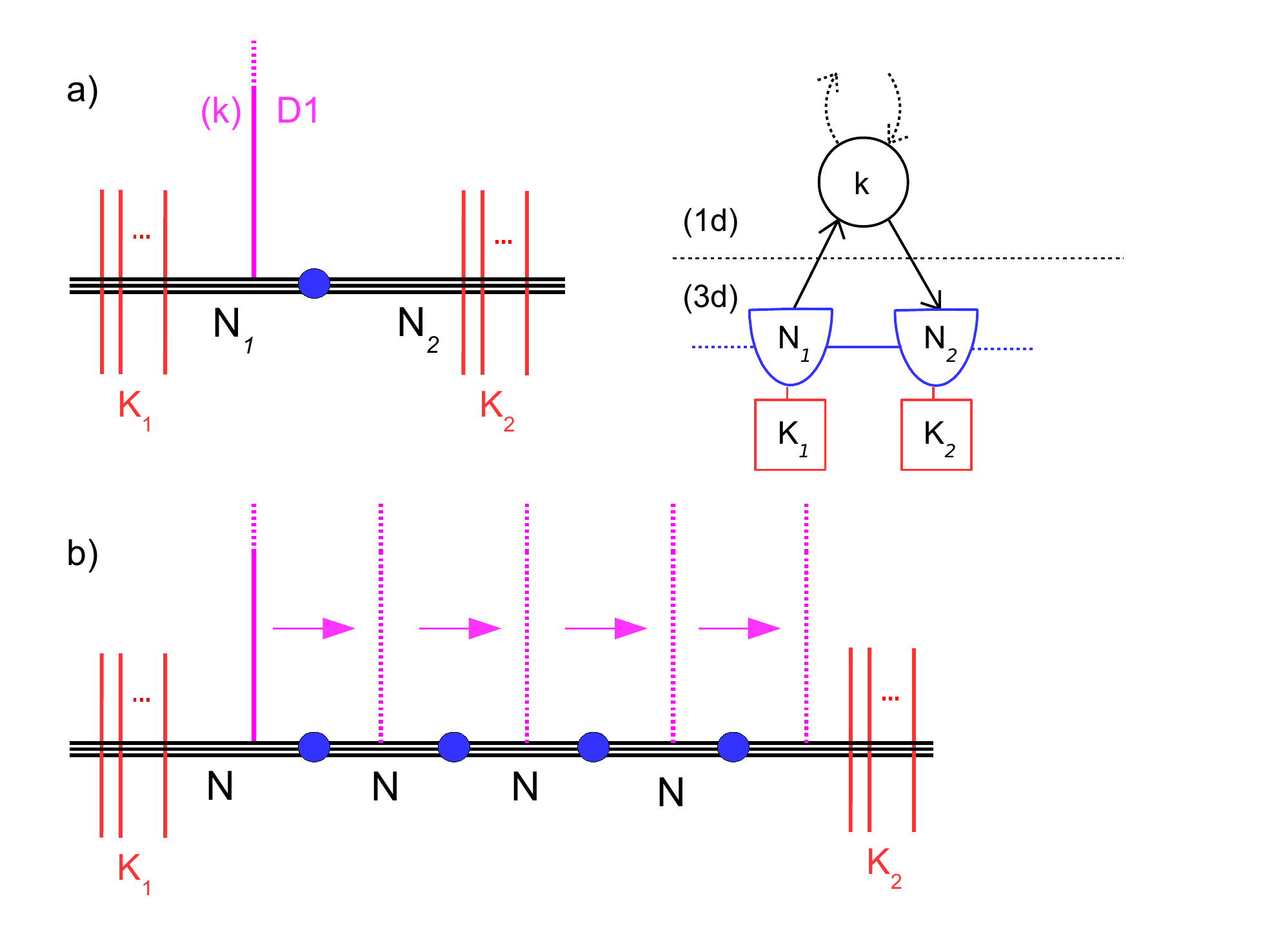}
\vspace{-0.5cm}
\caption{\footnotesize{a) Brane realization of the Vortex loop $V^{(N_1)}_{0,\scR_1}$ (with $k = |\scR_1|$) and associated right 3d/1d theory. b) When the numbers of D3-branes on both sides of N5-branes are equal, namely the ranks of adjacent nodes are equal, the D1-branes can be moved across the NS5-branes. The Vortex loops they realize are all equivalent.}}
\label{Hopping2}
\end{figure}

Consider the quiver in figure \ref{Hopping2}-a with two adjacent nodes $U(N_1)$ and $U(N_2)$, with $K_1$ and $K_2$ fundamental hypermultiplets respectively and assume $N_1 \geq N_2$. The vortex loop $V^{(N_1)}_{0,\scR_1}$, with $\scR_1$ a representation of $U(N_1)$ and the subscript $0$ indicating the splitting $K_1 = 0 + (K_1 -0)$, can be computed from the matrix model associated to the right 3d/1d theory:
\begin{align}
\langle V^{(N_1)}_{0,\scR_1} \rangle_{\rm right} &= \lim_{z \rightarrow 1} \, \sum_{ w \in \scR_1} \frac{1}{Z}\int \frac{d^{N_1} \sigma d^{N_2} \ti\sigma}{N_1! N_2!} \,  \prod_{i<j}^{N_1} \sh^2(\sigma_i - \sigma_j) \prod_{i<j}^{N_2} \sh^2(\ti\sigma_i - \ti\sigma_j) \no\\
& \phantom{= \lim_{z \rightarrow 1} \, \sum_{ w \in \scR_1} \frac{1}{Z} \int \ } \ . \,   e^{2\pi \xi_r | \scR_1|}   \,  \prod_{j=1}^{N_1} \prod_{k=1}^{N_2} \frac{1}{\ch( \ti\sigma_k -  \sigma_j - i w_j z)}  \ \Big[ \cdots \Big] \, ,
\label{VR1}
\end{align}
where we indicated only the matrix factors inserted by the 1d defect and part of sh factors of the two nodes, which will play a role in the check of hopping duality. Moreover the ch factors in the numerator of \eqref{IndexQMfinal_r} have been canceled by the matrix factor of the $U(N_1) \times U(N_2)$ bifundamental hypermultiplet.
This Vortex loop is realized by the brane configuration of figure \ref{Hopping2}-a  with $| \scR_1|$ D1-branes ending on the $N_1$ D3-branes supporting the $U(N-1)$ node and standing to the right of the $K_1$ D5-branes.

\smallskip

In the simplest case when $N_1 = N_2$, we have argued that the D1-branes can be moved to the right of the NS5-brane without changing the loop operator inserted. When D1-branes stand on the right of the NS5-branes, they realize a Vortex loop $V^{(N_2)}_{K_2,\scR_2}$ with $\scR_2$ a representation of $U(N_2)$ and the subscript $K_2$ indicating the splitting of $U(N_2)$ fundamental hypermultiplets $K_2 = K_2 + (K_2 - K_2)$. The left 3d/1d theory  associated to $V^{(N_2)}_{K_2,\scR_2}$ is actually the same as the right 3d/1d theory  associated to $V^{(N_1)}_{0,\scR_1}$, except for one important difference, which is that the FI parameter of the terminating SQM node (the node with $N_1$ fundamental and $N_2$ anti-fundamental chiral multiplets) is positive for $V^{(N_2)}_{K_2,\scR_2}$ and negative for $V^{(N_1)}_{0,\scR_1}$.
The matrix model computing $\langle V^{(N_2)}_{K_2,\scR_2} \rangle$ from the left 3d/1d theory is given by
\begin{align}
\langle V^{(N_2)}_{K_2,\scR_2} \rangle_{\rm left} &= \lim_{z \rightarrow 1} \, \sum_{ w \in \scR_2} \frac{1}{Z}\int \frac{d^{N_1} \sigma d^{N_2} \ti\sigma}{N_1! N_2!} \,   \prod_{i<j}^{N_1} \sh^2(\sigma_i - \sigma_j) \prod_{i<j}^{N_2} \sh^2(\ti\sigma_i - \ti\sigma_j) \no\\
& \phantom{= \lim_{z \rightarrow 1} \, \sum_{ w \in \scR_1} \frac{1}{Z} \int \ } \ . \,   e^{2\pi \xi_r |\scR_2|}   \,  \prod_{j=1}^{N_1} \prod_{k=1}^{N_2} \frac{1}{\ch( \ti\sigma_k - i w_k z -\sigma_j)}  \ \Big[ \cdots \Big] \, .
\label{VR2}
\end{align}
When $N_1 = N_2 \equiv N$ we must find $\langle V^{(N_2)}_{K_2,\scR} \rangle = \langle V^{(N_1)}_{0,\scR} \rangle$, as the D1-branes realizing the loops can be moved freely across the NS5. Let us consider this case first. 
Starting from the matrix model \eqref{VR1} computing $\langle V^{(N_1)}_{0,\scR}  \rangle$, with $N_1 =N$, we can use the same trick as above and make the replacement $\prod_{i<j}^{N} \sh(\sigma_i - \sigma_j) \rightarrow (-1)^{(N-1) |\cR|} \prod_{i<j}^{N} \sh(\sigma_i - \sigma_j + i(w_i - w_j) z)$ in the integrand. This allows us to use the Cauchy determinant formula \eqref{Cauchyformula}
with $\sigma_j \rightarrow \sigma_j + iw_j$, leading to
\begin{align}
\langle  V^{(N_1)}_{0,\scR} \rangle_{\rm right} &= \lim_{z \rightarrow 1} \, \sum_{ w \in \scR}  \sum_{ \tau \in S_N} (-1)^{\tau} \, \frac{1}{Z}\int \frac{d^{N} \sigma d^{N} \ti\sigma}{N!^2} \,  \prod_{i<j}^{N} \sh(\sigma_i - \sigma_j) \prod_{i<j}^{N} \sh(\ti\sigma_i - \ti\sigma_j) \no\\
& \phantom{= \lim_{z \rightarrow 1} \, \sum_{ w \in \scR} \frac{1}{Z} \int \ } \ . \, (-1)^{(N-1) |\cR|} \,   e^{2\pi \xi_r |\cR|}   \,  \frac{ 1}{\prod_j^N \ch(\sigma_{\tau(j)} +i w_{\tau(j)} - \ti\sigma_j)}   \ \Big[ \cdots \Big] \, .
\end{align}
Relabeling $\sigma_j \rightarrow \sigma_{\tau^{-1}(j)}$ and $\ti\sigma_j \rightarrow \ti\sigma_{\tau(j)}$, we obtain
\begin{align}
\langle V^{(N_1)}_{0,\scR} \rangle_{\rm right} &= \lim_{z \rightarrow 1} \, \sum_{ w \in \scR}  \sum_{ \tau \in S_N} (-1)^{\tau} \, \frac{1}{Z}\int \frac{d^{N} \sigma d^{N} \ti\sigma}{N!^2} \,  \prod_{i<j}^{N} \sh(\sigma_i - \sigma_j) \prod_{i<j}^{N} \sh(\ti\sigma_i - \ti\sigma_j) \no\\
& \phantom{= \lim_{z \rightarrow 1} \, \sum_{ w \in \scR} \frac{1}{Z} \int \ } \ . \, (-1)^{(N-1) |\cR|} \,   e^{2\pi \xi_r |\cR|}   \,  \frac{ 1}{\prod_j^N \ch(\sigma_{j} +i w_{\tau(j)} - \ti\sigma_{\tau(j)})}   \ \Big[ \cdots \Big] \no\\
&= \lim_{z \rightarrow 1} \, \sum_{ w \in \scR}  \frac{1}{Z}\int \frac{d^{N} \sigma d^{N} \ti\sigma}{N!^2} \,  \prod_{i<j}^{N} \sh^2(\sigma_i - \sigma_j) \prod_{i<j}^{N} \sh(\ti\sigma_i - \ti\sigma_j) \sh(\ti\sigma_i - \ti\sigma_j - i(w_i - w_j)z ) \no\\
& \phantom{= \lim_{z \rightarrow 1} \, \sum_{ w \in \scR} \frac{1}{Z} \int \ } \ . \, (-1)^{(N-1) |\cR|} \,   e^{2\pi \xi_r |\cR|}   \,  \prod_{j=1}^{N_1} \prod_{k=1}^{N_2} \frac{1}{\ch( \ti\sigma_k - i w_k z -\sigma_j)}  \ \Big[ \cdots \Big] \no\\
&= \langle  V^{(N_2)}_{K_2,\scR} \rangle_{\rm left} \, ,
\end{align}
where we have used the Cauchy identity  to obtain the second equality and we have again enforced $z=1$ in the sh factors of the integrand to reach the matrix model \eqref{VR2} computing $\langle V^{(N_2)}_{K_2,\scR} \rangle$, with $N_2 = N$.
\smallskip

This shows that the Vortex loops realized by D1-branes ending on the left or on the right of an NS5 with equal numbers of D3s on both sides are equivalent. Combining this property with the hopping duality between the left and right 3d/1d quiver realization of a Vortex loop, we prove the equivalence of Vortex loops in circular quivers with nodes of equal rank \eqref{CircLoopMap1}, \eqref{CircLoopMap2}. This is illustrated in figure \ref{Hopping2}-b.

\medskip

When $N_1 > N_2$ the map of Vortex loops is more complicated. It can be found from the brane picture in the same way as we found mirror maps between loop operators. We consider the brane realization of $V^{(N_1)}_{0,\scR_1}$ with $|\scR_1|$ D1-strings ending on the $N_1$ D3-branes as in figure \ref{Hopping2}-a. As $N_1>N_2$, the NS5 has a D3-spike on its left side. We then move the D1-branes to the left, across the NS5-brane.  Some D1-branes can be moved along the D3-spike and get attached to the NS5-brane, realizing flavor Wilson loops for the $U(1)$ global symmetry associated to the NS5-brane. The other D1-branes end on the $N_2$ D3-branes on the right of the NS5 and realize a Vortex loop $V^{(N_2)}_{K_2,\scR_2}$. Recycling the ideas of section \ref{ssec:LoopSduality}, the precise prediction from the brane picture is found to be
\begin{align}
\vev{V^{(N_1)}_{0,\scR_1}} & =  \vev{ \sum_{ s \in \Delta} W^{\rm fl}_{NS5, {\bf q}_s} \  V^{(N_2)}_{K_2,\wat\scR_s} } \, ,
\end{align}
where $\Delta$ is the set of representations $({\bf q}_s, \wat\scR_s)$ appearing in the decomposition of $\scR_1$ under the subgroup $U(1) \times U(N_2) \subset U(N_1 - N_2) \times U(N_2) \subset U(N_1)$, with $U(1)$ embedded diagonally into $U(N_1 - N_2)$.
$W^{\rm fl}_{NS5, {\bf q}}$ denotes the Wilson loop of charge ${\bf q}$ under the $U(1)$ topological symmetry associated to the NS5.

\smallskip

This map can be checked by explicit computations, using the generalized Cauchy formula \eqref{Cauchyformula2}. It implies that the Vortex loops $V^{(N_1)}_{0,\scR_1}$ and $V^{(N_2)}_{K_2,\scR_2}$ are redundant and that in order to describe the mirror map with Wilson loops of the mirror theory, it is sufficient to consider only the loops $V^{(N_1)}_{0,\scR_1}$ or the loops $V^{(N_2)}_{K_2,\scR_2}$.
In general for each pair of consecutive D5-branes in the brane picture, we need only to consider the Vortex loops realized with D1-branes placed between the two D5-branes, with a fixed number of NS5-branes on their left and on their right.   This is the mirror statement to the fact that between two consecutive NS5-branes, we need only consider Wilson loops realized with F1-strings placed between the two NS5s, with a fixed number of D5-branes on their left and on their right.

\medskip 
 
\section*{Acknowledgements}

We would like to thank Kevin Costello, Nadav Drukker, Davide Gaiotto, Heeyeon Kim, Hee-Cheol Kim, Bruno Le Floch, Stefano Cremonesi,  R. Santamaria and E. Witten for discussions. B.A. thanks Perimeter Institute for its generous hospitality during several visits between 2013 and 2015.
JG is    grateful to the KITP for its warm hospitality
during the ``New Methods in Nonperturbative Quantum Field Theory" Program in early 2014, which was supported in part by the National
Science Foundation under Grant No. NSF PHY11-25915. This research was supported in part by Perimeter Institute for Theoretical Physics. Research at Perimeter Institute is supported 
by the Government of Canada through Industry Canada and by the Province of Ontario through the Ministry of Research and Innovation.
J.G. also acknowledges further support from an NSERC Discovery Grant and from an ERA grant by the Province of Ontario.
B.A. acknowledges support by the ERC Starting Grant N. 304806, ``The Gauge/Gravity Duality and Geometry in String Theory.''.

\clearpage



\appendix

\section{Supersymmetry  Transformations  On $S^3$ And SQM Embedding}
\label{app:Embeddings}

We provide here the 3d $\N=4$ supersymmetry transformations of the vector multiplet and hypermultiplet on $S^3$. By restricting to the four supercharges preserved by the SQM$_V$ deformed algebra $SU(1|1)_l\times SU(1|1)_R$ in  \rf{Lefts}\rf{Rights},  we work out the embedding of the 1d $\N=4$ vector and chiral multiplets  inside the 3d $\cN=4$ vector and hypermultiplets.

\smallskip

 The 3d $\N=4$ supersymmetry transformations  are generated by the four two-component Killing spinors $\epsilon^{A A'}$, with $A=1,2$ the index of the ${\bf 2}$ of $SU(2)_C$ and $A'=1,2$ the index of the ${\bf 2}$ of $SU(2)_H$. 
 The transformations of the 3d $\N=4$ abelian vector multiplet $(A_\mu, \Phi_I, \lambda^{A A'}, \mathsf{D}_{I'})$ 
 are\footnote{We derived these transformations by ``covariantizing" the 3d $\N=2$ transformations of \cite{Drukker:2012sr}.} 
\begin{align*}
\delta A_\mu &=  -\frac{i}{2} \lambda_{A A'} \gamma_{\mu} \epsilon^{A A'} \qquad
\delta \Phi_I \ = -\frac 12 \lambda_{A A'} (\tau_{I})^A{}_B \epsilon^{B A'} \\
\delta \lambda^{A A'} &= - \frac{1}{2} F_{\mu\nu} \gamma^{\mu\nu}\epsilon^{A A'} - \mathsf{D}_{I'}(\tau_{I'})^{A'}{}_{B'} \epsilon^{A B'} 
+ i D_\mu \Phi_I \gamma^\mu (\tau_{I})^A{}_B \epsilon^{B A'} + \frac{2i}{3} \Phi_I \gamma^\mu (\tau_{I})^A{}_B D_\mu \epsilon^{B A'} \\
\delta \mathsf{D}_{I'} &= - \frac i2  D_{\mu} \lambda_{A A'} \gamma_{\mu}  (\tau_{I'})^{A'}{}_{B'} \epsilon^{A B'}  - \frac i6  \lambda_{A A'} \gamma_{\mu}   (\tau_{I'})^{A'}{}_{B'} D_{\mu} \epsilon^{A B'}    \, ,
\end{align*}
where $\tau_I = \tau_{I'}$ are the Pauli matrices. $I=1,2,3$ is the index of the ${\bf 3}$ of $SU(2)_C$  while $I'=1,2,3$ is the index of the ${\bf 3}$ of $SU(2)_H$. The Killing spinors $\epsilon^{11} = \bar\epsilon_L$ and $\epsilon^{22} = \epsilon_L$ generate the 3d $\N=2$ supersymmetry transformations in $SU(2|1)_l$, while $\epsilon^{12} = \bar\epsilon_R$ and $\epsilon^{21} = \epsilon_R$ generate  supersymmetry transformations in $SU(2|1)_r$.

\smallskip
The 3d $\N=4$ vector multiplet decomposes into a 3d $\N=2$ vector multiplet $(A_\mu, \Phi_3, \lambda^{11}, \lambda^{22}, \textrm{D}_3) = (A_\mu, \sigma, \lambda, \bar\lambda, D)$ and an adjoint chiral multiplet $(\Phi_1 + i \Phi_2,  \lambda^{12}, \lambda^{21}, \mathsf{D}_1 + i \mathsf{D}_2)= (\phi, \psi, \bar\psi, F)$.

\medskip

\noindent \underline{\bf SQM$_V$ embedding}

\medskip

The  four  Killing spinors generating the deformed SQM$_V$ algebra are (see section \ref{ssec:Vortexsphere})
\begin{align}
\epsilon_l &= \binom{\epsilon_1}{0} \, , \quad  \bar\epsilon_l = \binom{0}{\bar\epsilon_2}
\, , \quad  \epsilon_r = \binom{0}{\tilde\epsilon_2}
\, , \quad  \bar\epsilon_r = \binom{\bar{\tilde \epsilon}_1}{0} \, .
\end{align}
In the left-invariant frame, $\epsilon_l$ and $\bar\epsilon_l $ are constant, while $\epsilon_r$ and $ \bar\epsilon_r $ have a spatial dependent phase $e^{ i \tau/L}$ and $e^{ -i \tau/L}$ respectively, where $\tau \in [0,2\pi L]$ is the coordinate along the loop. 
We recall that the equations solved by the Killing spinors  are
\begin{align}
\nabla_\mu \epsilon_l &= \frac{i}{2L} \gamma_\mu \epsilon_l \, , \qquad \nabla_\mu \bar\epsilon_l = \frac{i}{2L} \gamma_\mu \bar\epsilon_l\,,  \\
\nabla_\mu \epsilon_r &= -\frac{i}{2L} \gamma_\mu \epsilon_r \, , \qquad \nabla_\mu \bar\epsilon_r = -\frac{i}{2L} \gamma_\mu \bar\epsilon_r \, ,
\end{align}
where $L$ is the radius of $S^3$.
The  supersymmetry transformations under the supercharges in deformed  superalgebra SQM$_V$ reduce to\footnote{We do not give the supersymmetry transformations of the auxiliary fields and transverse gauge fields, which combine into another 1d $\scN = 4$ chiral multiplet.}
\begin{align}
\delta A_\tau &= - \frac i2 \left(  \bar\lambda_1 \bar\epsilon_2 + \lambda_2 \epsilon_1 - \bar\psi_2 \bar{\tilde \epsilon}_1 - \psi_1 \tilde\epsilon_2 \right) \no\\
\delta \Phi_3 &= -\frac 12 \left( - \bar\lambda_1 \bar\epsilon_2 - \lambda_2 \epsilon_1 - \bar\psi_2 \bar{\tilde \epsilon}_1 - \psi_1 \tilde\epsilon_2 \right) \no\\
\delta \Phi  &= \bar\lambda_1 \tilde\epsilon_2   + \bar\psi_2 \epsilon_1 
\qquad \delta \bar\Phi = - \lambda_2 \bar{\tilde\epsilon}_1 - \psi_1 \bar\epsilon_2  \no\\
\delta \bar\lambda_1 &= \left(\mathsf{D}_3 - \frac i2 F_{12} + \frac{\Phi_3}{L} \right) \epsilon_1 - i D_{\tau} \Phi_3 \epsilon_1 + i \left( D_\tau \Phi - i \frac{\Phi}{L}  \right) \bar{\tilde\epsilon}_1  \no\\
\delta \lambda_2 &= -\left(\mathsf{D}_3 - \frac i2 F_{12} + \frac{\Phi_3}{L} \right) \bar\epsilon_2 - i D_{\tau} \Phi_3 \bar\epsilon_2 - i\left( D_\tau  \bar\Phi + i \frac{\bar\Phi}{L} \right) \tilde\epsilon_2  \no\\
\delta \psi_1 &= \left(\mathsf{D}_3 - \frac i2 F_{12} + \frac{\Phi_3}{L}  \right) \bar{\tilde\epsilon}_1  + i D_{\tau} \Phi_3 \bar{\tilde\epsilon}_1  + i\left( D_\tau \bar\Phi + i \frac{\bar\Phi}{L}  \right) \epsilon_1   \no\\
\delta \bar\psi_2 &= -\left(\mathsf{D}_3 - \frac i2 F_{12} + \frac{\Phi_3}{L}  \right) \tilde\epsilon_2   + i D_{\tau} \Phi_3 \tilde\epsilon_2   - i\left( D_\tau \Phi - i \frac{\Phi}{L}  \right) \bar\epsilon_2 \, ,
\label{QMVecTranfo}
\end{align}
where $\Phi = \Phi_1 + i \Phi_2$ and we have adopted 3d $\N=2$ notations $\lambda_\alpha, \bar\lambda_\alpha, \psi_\alpha, \bar\psi_\alpha$ for the fermions, with $\alpha =1,2$.

\smallskip
These correspond to the   supersymmetry transformations of a 1d $\N=4$ SQM$_V$ abelian vector multiplet   (see for instance \cite{Hori:2014tda}) but  in the presence of a background gauge field $a_- = -\frac{1}{L}$ for the $U(1)_-$ R-symmetry, generated by $J_-$. The presence of the background gauge field $a_- = -\frac{1}{L}$   affects the covariant derivative of $\Phi, \bar\Phi$ in the transformations above. 
The identification  of the   fields in the 3d $\cN=4$ vector multiplet with the fields of the 1d $\N=2$  vector and adjoint chiral multiplet that furnish a 1d $\cN=4$ SQM$_V$ vector  multiplet  is
\begin{align}
  {\rm  1d  \ vector \ multiplet:} \quad  (v_\tau, x_3, \lambda_-, \bar\lambda_-,D)_{1d} \ \sim& \ (A_\tau, \Phi_3, \lambda_2, \bar\lambda_1, \mathsf{D}_3 - \frac i2 F_{12} + \frac{\Phi_3}{L})  \,, \quad  r_- = 0 \\[+2ex]
  {\rm 1d \ chiral \ multiplet:} \quad  (x_1 + i x_2, \psi, \bar\psi)_{1d} \ \sim& \  (\Phi, \bar\psi_2, \psi_1)  \,, \quad  r_- = -1  \,,
\end{align}
where we have indicated the $U(1)_-$ charges of the 1d $\N=2$ multiplets:\footnote{$U(1)_-$ is a global symmetry from the point of view of the 1d $\cN=2$ subalgebra, so that each $\cN=2$ multiplet comes with a $U(1)_-$ charge.}   the 1d $\N=2$ vector multiplet fields have vanishing $U(1)_-$ charge, while the fields $x_1 + i x_2$ and  $\psi$ in the 1d $\N=2$ SQM adjoint chiral multiplet have $U(1)_-$ charge $-1$ and $\bar\psi$ charge 1.

\medskip

We now turn to the 3d $\cN=4$ hypermultiplet supersymmetry transformations. These do not admit an off-shell formulation, so we provide
the {\it on-shell} supersymmetry transformations: 
\begin{align}
\delta \phi^{A'} &=  -\psi_{A} \epsilon^{A A'}   \\
\delta \psi^{A} &=   i \gamma^\mu \epsilon^{A A'} D_\mu \phi_{A'} + i  \Phi_I (\tau_I)^A{}_B \phi_{A'} \epsilon^{B A'} + \frac i3 \gamma^\mu D_\mu \epsilon^{A A'}  \phi_{A'}  \, .
\end{align}
 $(\phi^1,  -\psi^2)$ and $(\phi^2, \psi^1)$ are two on-shell chiral multiplets of 3d $\N=2$ supersymmetry  of R-charge 1/2   transforming under complex conjugate representations of the gauge group.

Restricting to the same four supercharges generating deformed SQM$_V$ as above, we obtain  
\begin{align}
\delta\phi^1 &= (\psi^1)_1 \tilde\epsilon_2 - (\psi^2)_1 \bar{\epsilon}_2 \no\\
\delta (\psi^1)_1 &=  i \bar{\tilde \epsilon}_1 D_\tau  \phi^1  + i (\Phi_3 -  \frac{i}{2L}  )  \phi^1 \bar{\tilde \epsilon}_1 + i \bar\Phi \phi^1 \epsilon_1 - i (D_1 - i D_2) \phi^2 \bar{\epsilon}_2 \no\\
\delta (\psi^2)_1 &=   i \epsilon_1 D_\tau \phi^1  - i (\Phi_3 -  \frac{i}{2L}) \phi^1 \epsilon_1 + i \Phi \phi^1 \bar{\tilde \epsilon}_1 - i (D_1 - i D_2) \phi^2 \tilde\epsilon_2  \no\\
\delta\phi^2 &= (\psi^1)_2 \epsilon_1 - (\psi^2)_2 \bar{\tilde\epsilon}_1 \no\\
\delta (\psi^1)_2 &= - i \bar \epsilon_2 D_\tau  \phi^2  + i (\Phi_3 +  \frac{i}{2L}  )  \phi^2 \bar\epsilon_2 + i \bar\Phi \phi^2 \tilde\epsilon_2 - i (D_1 + i D_2) \phi^1 \bar{\tilde\epsilon}_1 \no\\
\delta (\psi^2)_2 &= - i \tilde\epsilon_2 D_\tau \phi^2  - i (\Phi_3 +  \frac{i}{2L}) \phi^2 \tilde\epsilon_2 + i \Phi \phi^2 \bar\epsilon_2 - i (D_1 + i D_2) \phi^1 \epsilon_1 \, .
\label{QMHypTranfo}
\end{align}

This matches the on-shell deformed supersymmetry transformations of two 1d $\N=4$ chiral multiplets $(\phi, \psi_+, \psi_-)_{q_F} \sim  (\phi^1, (\psi^1)_1, (\psi^2)_1)_{-\frac 12}$ and $(\phi^2, (\psi^1)_2, (\psi^2)_2)_{\frac 12}$. The shift of $\Phi_3$ in the supersymmetry transformations is identified with a real mass deformation with complex parameter $m_F = \frac{i}{L}  $ for a flavor symmetry $G_F$, such that the 1d $\cN=4$ chiral multiplets have $G_F$ charge $q_F =\mp \frac 12$.
The term $(D_1 + i D_2) \phi^1$ in the on-shell 1d $\N=4$ supersymmetry transformations arises from a superpotential coupling. In the off-shell transformations it gets   replaced with a complex auxiliary field.

\smallskip

We note that the mass deformation obtained by giving a background to the $G_F$ flavour symmetry is not visible anywhere in the transformations \ref{QMVecTranfo}, which means that the adjoint chiral multiplet with bottom component $\Phi =\Phi_1 + i \Phi_2$  is not charged under $G_F$. This allows us to identify this  flavor symmetry with 
\begin{align}
G_F &=   J_3^l + J_3^r  - R_H   \, .
\end{align}
Moreover the background gauge field $a_- = -\frac{1}{L}$ for $J_-$ is not visible in the transformations \rf{QMHypTranfo} of the chiral multiplets with bottom components $\phi^1$ and $\phi^2$, which means that these multiplets are not charged under $J_-$. We then have the identification
\begin{align}
J_- &=   J_3^l + J_3^r  - R_V   \, .
\end{align}
This identification of generators allows us to match the deformed SQM$_W$ supersymmetry algebra with the $SU(1|1)_l\times SU(1|1)_r$ subalgebra preserved by the defect, as explained in section \ref{ssec:Vortexsphere}.

\section{Evaluations of the SQM Index}
\label{app:QMindex}

In this appendix we compute the 1d $\scN=4$ SQM index, or partition function on $S^1$, for some quiver gauge theories with generic $U(1)_-$ R-symmetry background $z$ and $U(1)_F$ flavor  chemical potential $\mu =1$, as defined in section \ref{sssec:QMindex}. The R-symmetry and flavor symmetry charges are summarized in table \ref{tab:QMCharges} and the specific adjoint R-charges are given in equation \eqref{RadjQM}.

\subsection{$U(k)$ Theory With $N$ Fundamental And $M$ Anti-fundamental Chirals}
\label{ssec:UkQM}

We consider a 1d $\cN=4$ SQM gauge theory  with  $U(k)$ gauge group and $N$ fundamental chiral multiplets  with real masses $-\sigma_j$, R-charges $r_+$ and $G_F$ flavor charge $q_+$, and $M$ anti-fundamental chiral multiplets  with masses $m_a$, R-charges $r_-$ and $G_F$ flavor charge $q_-$. Mass parameters are in units of the inverse $S^1$ radius. As explained in the main text the charges obey the superpotential constraints $r_- + r_+ = 2$ and $q_- + q_+ = 1/2$.
We introduce the complex parameters $\hat\sigma_j = \sigma_j + i q_+ + i \frac{r_+}{2} z$ and $\hat m_a = m_a -i q_-  - i\frac{r_-}{2} z$.  Importantly we take a negative FI parameter $\zeta < 0$. This corresponds to the choice of FI parameter when the 1d $\cN=4$ theory is   read by moving the D1-branes to the right NS5-brane.

\smallskip
The partition function is given by 
\begin{align}
\scI &= {\rm JK-Res}_{\zeta} \ \frac{1}{k!} \lp \frac{\pi}{\sin(\pi z)} \rp^k \ 
\prod_{I \neq J}^k \frac{\sin[-\pi(u_I -u_J) ]}{\sin[\pi(u_I - u_J- z ) ]}  \no\\
& \hspace{25mm} . \prod_{I =1}^k \lp \prod_{j=1}^N \frac{\sin[-\pi(u_I -i\hat\sigma_j - z ) ]}{\sin[\pi(u_I - i\hat\sigma_j ) ]}  \ 
\prod_{a=1}^M \frac{\sin[-\pi(-u_I + i \hat m_a -z) ]}{\sin[\pi(-u_I + i \hat m_a ) ]}  \ du_I  \rp  \\
&= \frac{1}{k!} \lp 2i \sin(\pi z) \rp^{-k} \  \oint \prod_{I =1}^k du_I  \ 
\prod_{I \neq J}^k \frac{\sin[-\pi(u_I -u_J) ]}{\sin[\pi(u_I - u_J- z ) ]}  \no\\
& \hspace{25mm} . \prod_{I =1}^k \lp \prod_{j=1}^N \frac{\sin[-\pi(u_I -i\hat\sigma_j - z ) ]}{\sin[\pi(u_I - i\hat\sigma_j ) ]}  \ 
\prod_{a=1}^M \frac{\sin[-\pi(-u_I + i \hat m_a -z) ]}{\sin[\pi(-u_I + i \hat m_a ) ]}  \rp   \ ,
\end{align}
where the integration contour is defined so that it picks the residues at the poles from the fundamental chiral multiplet factors, namely factors in the product $\prod_{j=1}^N$. In particular it does not pick the poles from factors in the product $\prod_{I \neq J}^k$ and $\prod_{a=1}^M$. This follows from the definition of ${\rm JK-Res}_{\zeta}$ for $\zeta<0$, reviewed in the main text. 
This integral has poles at $u^{\ast}_I =  i \hat \sigma_j $, $j=1, \cdots , N$, however, due to the $\sin[-\pi(u_I -u_J) ]$ factors, we get a non-zero residue only when each $u^{\ast}_I$ hits a different $i \hat \sigma_j $, in particular we have non-zero residues only when $k \le N$. 
A non-vanishing residue at $u^{\ast} = \{u^{\ast}_I\}_{1\le I \le k}$ is then associated to a decomposition $k = \sum_{k=1}^N  k_j$, with $k_1, \cdots , k_N \in \{ 0,1\}$, where $k_j = 1$ if  $u^{\ast}_I = i \hat \sigma_j$ for a given $I$ and $k_j = 0$ otherwise. This residue contribution appears with a multiplicity $k!$, corresponding to permutations of the $u^{\ast}_I$. The partition function is then expressed as
\begin{align}
\scI &= \sum_{
\begin{array}{c}
k_1, \cdots , k_N \in \{ 0,1\} \\
\sum_{k=1}^N k_j = k
\end{array}  }
\scI_{(k_j)} \ ,
\end{align}
with
\begin{align}
\scI_{(k_j)} &= (-1)^{(N+M)k}  \prod_{i < j}^N \frac{\sh[\hat\sigma_i - \hat\sigma_j + i(k_i - k_j)z  ]}{\sh[\hat\sigma_i - \hat\sigma_j ]}  
\ \prod_{a=1}^M \prod_{j=1}^N  \frac{\sh (\hat\sigma_j - \hat m_a - i  z  ) }{\sh (\hat\sigma_j - \hat m_a + i (k_j-1) z)}  \ .
\end{align}
The partition function vanishes when $k > N$, consistent with the fact that there are no supersymmetry vacua in that range.

Plugging the constraints $r_+ + r_- = 2$ and $q_+ + q_- = 1/2$ we obtain
\begin{align}
\scI_{(k_j)} &= (-1)^{(N+M)k}  \prod_{i < j}^N \frac{\sh[\sigma_i - \sigma_j + i(k_i - k_j)z  ]}{\sh[\sigma_i - \sigma_j ]}  
\ \prod_{j=1}^N \prod_{a=1}^M  \frac{\ch (\sigma_j - m_a ) }{\ch (\sigma_j -  m_a + i k_j z )}  \ .
\label{I_UkQM}
\end{align}
This result matches the formula \eqref{IndexQMfinal_r}  giving the index as a sum over the weights of the representation $\scA_k$ associated to the 1d $\cN=4$ gauge theory.

\subsection{$U(k)$ Theory With $N$ Fundamental, $M$ Anti-fundamental Chirals And One Adjoint Chiral}
\label{ssec:UkAdjQM}

We consider a 1d $\cN=4$ SQM  gauge theory with  $U(k)$ gauge group and  $N$ fundamental chiral multiplets  with real masses $-\sigma_j$, R-charges $r_+$ and $G_F$ flavor charge $q_+$, $M$ anti-fundamental chiral multiplets  with masses $m_a$, R-charges $r_-$ and $G_F$ flavor charge $q_-$ and an adjoint chiral multiplet with R-charge $R_{\rm adj}=2$ and flavor charge $q_{\rm adj}=1$. Mass parameters are in units of the inverse $S^1$ radius. The charges obey the superpotential constraints $r_- + r_+ = 2$ and $q_- + q_+ = 1/2$.
We introduce the complex parameters $\hat\sigma_j = \sigma_j + i q_+ + i \frac{r_+}{2} z$ and $\hat m_a = m_a -i q_-  - i\frac{r_-}{2} z$.  Importantly we have a negative FI parameter $\zeta < 0$. 

\smallskip
In order to  avoid higher order poles in the computation we keep $R_{\rm adj}$ generic and only set it to $2$ at the end of the computation.
The partition function or index is given by 
\begin{align}
\scI &= {\rm JK-Res}_{\zeta} \  \frac{1}{k!} \lp \frac{\pi}{\sin(\pi z)} \rp^k \ 
\prod_{I \neq J}^k \frac{\sin[-\pi(u_I -u_J) ]}{\sin[\pi(u_I - u_J- z ) ]} \, 
\prod_{I , J}^k \frac{\sin[-\pi(u_I -u_J + \frac{R_{\rm adj}}{2}z - z)  ]}{\sin[\pi(u_I - u_J + \frac{R_{\rm adj}}{2}z ) ]}  \no\\
& \hspace{25mm} . \prod_{I =1}^k \lp \prod_{j=1}^N \frac{\sin[-\pi(u_I -i\hat\sigma_j - z ) ]}{\sin[\pi(u_I - i\hat\sigma_j ) ]}  \ 
\prod_{a=1}^M \frac{\sin[-\pi(-u_I + i \hat m_a -z) ]}{\sin[\pi(-u_I + i \hat m_a ) ]}  \ du_I  \rp  \\
&= \frac{1}{k!} \lp 2i \sin(\pi z) \rp^{-k} \  \oint \prod_{I =1}^k du_I  \ 
\prod_{I \neq J}^k \frac{\sin[-\pi(u_I -u_J) ]}{\sin[\pi(u_I - u_J- z ) ]}  \, 
\prod_{I , J}^k \frac{\sin[-\pi(u_I -u_J + \frac{R_{\rm adj}}{2}z - z)  ]}{\sin[\pi(u_I - u_J + \frac{R_{\rm adj}}{2}z ) ]}  \no\\
& \hspace{25mm} . \prod_{I =1}^k \lp \prod_{j=1}^N \frac{\sin[-\pi(u_I -i\hat\sigma_j - z ) ]}{\sin[\pi(u_I - i\hat\sigma_j ) ]}  \ 
\prod_{a=1}^M \frac{\sin[-\pi(-u_I + i \hat m_a -z) ]}{\sin[\pi(-u_I + i \hat m_a ) ]}  \rp   \ ,
\end{align}
where the integration contour is defined by picking the residues of the poles from the fundamental chiral mutliplets factors, namely the factors in the product $\prod_{j=1}^N$ and from "half" of the adjoint chiral multiplet factors $\prod_{I , J}^k$ according to the ${\rm JK-Res}_{\zeta}$ prescription with $\zeta<0$. Concretely the integral is a sum over residues, each contribution corresponding to a pole $u^{\ast} = \{u^{\ast}_I\}_{1\le I \le k}$ described by  taking a decomposition of $k$ into $N$ non-negative integers, $k=\sum_{i=1}^N k_i$, $k_i \ge 0$, and picking  $u^{\ast}_I \to u^{\ast}_{i, s_i} = i \hat \sigma_i - s_i \frac{R_{\rm adj}}{2}z$ with $s_i = 0, \cdots , k_i-1$. The arrow $\to$ indicates a mapping between the index $I$ into the index $(i,s_i)$. The residue contribution coming from a given pole $u^{\ast} = \{u^{\ast}_{i, s_i}\}$ arises with $k!$ degeneracy, associated to permutations of the $u^{\ast}_I$. The partition function is then given by a sum over the residue contributions $\scI_{(k_i)}$ associated to possible decompositions $k=\sum_{i=1}^N k_i$, counted with $k!$ degeneracy:
\begin{align}
\scI &= \sum_{
\begin{array}{c}
k_1, \cdots , k_N \ge 0\\
\sum_{i=1}^N k_i = k
\end{array}  }
\scI_{(k_i)} \ .
\end{align}
The explicit evaluation of the residues leads to
\begin{align}
\scI_{(k_i)} &= (-1)^{(N+M)k}  \prod_{i \neq j}^N \prod_{s=0}^{k_j-1} \frac{\sh[\hat\sigma_i - \hat\sigma_j + i \frac{R_{\rm adj}}{2}(k_i - s-1) z  ]}{\sh[\hat\sigma_i - \hat\sigma_j + i \frac{R_{\rm adj}}{2}(k_i - s-1) z + i z]}  
\ \prod_{i=1}^N \prod_{s=0}^{k_i-1} \prod_{a=1}^M  \frac{\sh [ \hat\sigma_i - \hat m_a + i \frac{R_{\rm adj}}{2} s z  - iz ] }{\sh [ \hat\sigma_i - \hat m_a + i \frac{R_{\rm adj}}{2} s z ] } \no\\
\prod_{i \neq j}^N  & \prod_{s=0}^{k_i-1} \frac{\sh[\hat\sigma_i - \hat\sigma_j + i \frac{R_{\rm adj}}{2}s z +iz  ]}{\sh[\hat\sigma_i - \hat\sigma_j + i \frac{R_{\rm adj}}{2} s z]}  
\prod_{i \neq j}^N \prod_{s=0}^{k_i-1} \frac{\sh[\hat\sigma_i - \hat\sigma_j + i \frac{R_{\rm adj}}{2}(s+1) z - iz  ]}{\sh[\hat\sigma_i - \hat\sigma_j + i \frac{R_{\rm adj}}{2} (s+1) z]} 
\prod_{i=1}^N \prod_{s=0}^{k_i-1} \frac{\sh[ i \frac{R_{\rm adj}}{2} (s+1) z - i z ]}{\sh[ i \frac{R_{\rm adj}}{2} (s+1) z ]} \, .
\end{align}
If we plug the adjoint R-charge $R_{\rm adj} = 2$ then the result simplifies to 
\begin{align}
\scI_{(k_i)} &= (-1)^{(N+M)k}  \prod_{i \neq j}^N \frac{\sh[\hat\sigma_i - \hat\sigma_j + i(k_i - k_j) z  ]}{\sh[\hat\sigma_i - \hat\sigma_j + i k_i z]}  
\ \prod_{i=1}^N \prod_{a=1}^M  \frac{\sh [ \hat\sigma_i - \hat m_a - i z  ] }{\sh [ \hat\sigma_i - \hat m_a + i (k_i-1) z ] }   \prod_{i=1}^N \frac{\sh[ \epsilon ]}{\sh[ i k_i z + \epsilon ]}  \ ,
\label{IndexEpsilon}
\end{align}
where we have introduced a regulating mass parameter $\epsilon$ for vanishing factors.\footnote{$\epsilon$ can be introduced as a very small real mass for the adjoint chiral multiplet. In this case $\epsilon$ would appear in the other factors as well. We do not introduce it in the other factors since it would complicate significantly the discussion, without changing the final result in the 3d/1d matrix models.}   Let us write $\scI_{\epsilon}(z) = \prod_{i=1}^N \frac{\sh[ \epsilon]}{\sh[ i k_i z + \epsilon]} $. In the limit $z \rightarrow 1$ (at finite $\epsilon$) this factor become trivial.  

\smallskip
Using the contraints $r_+ + r_- = 2$ and $q_+ + q_- = 1/2$, we obtain 
\begin{align}
\scI_{(k_i)} &= (-1)^{(N+M)k}  \prod_{i \neq j}^N \frac{\sh[\sigma_i - \sigma_j + i(k_i - k_j) z  ]}{\sh[\sigma_i - \sigma_j + i k_i z]}  
\ \prod_{i=1}^N \prod_{a=1}^M  \frac{\ch [ \sigma_i -  m_a  ] }{\ch [ \sigma_i -  m_a + i k_i  z ] }  \ \scI_{\epsilon}(z)\ .
\label{I_UkAdjQM}
\end{align}
Here again the result  matches the formula \eqref{IndexQMfinal_r}  giving the index as a sum over the weights of the representation $\scS_k$ associated to the  1d $\cN=4$  theory.

\subsection{Two-Node Quiver}
\label{ssec:2NodeQuiver}

\begin{figure}[h]
\centering
\includegraphics[scale=0.65]{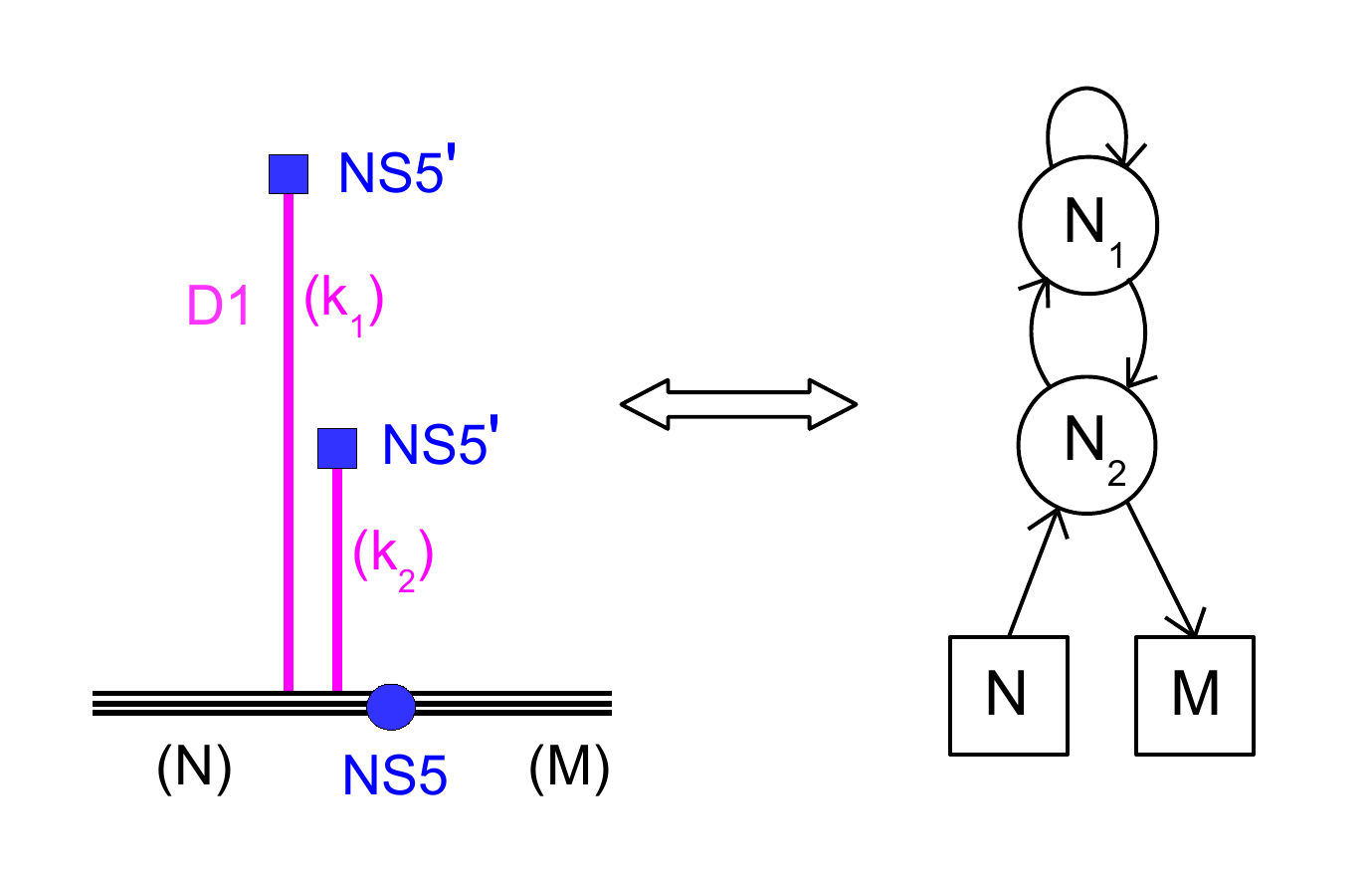}
\vspace{-0.5cm}
\caption{\footnotesize{Brane realization of a Vortex loop associated with the representation $\scA_{k_1} \otimes \scA_{k_2}$ and SQM two-node quiver read by moving the D1-branes to the NS5-brane on the right. Here $N_1 = k_1$ and $N_2 = k_1 + k_2$.}}
\label{QMquiver3}
\end{figure}

We consider a $U(N_1) \times U(N_2)$ quiver gauge theory with bifundamental chiral multiplets of R-charges $\wat r, \ti r$ and $G_F$-charges $\wat q, \ti q$. In addition we have $N$ $U(N_2)$-fundamental chiral multiplets  with real masses $-\sigma_j$, R-charges $r_+$ and $G_F$-charge $q_+$, $M$ $U(N_2)$-anti-fundamental chiral multiplets  with masses $m_a$, R-charges $r_-$ and $G_F$-charge $q_-$ and an adjoint chiral multiplet with R-charge $R_{\rm adj}=0$ and $G_F$-charge $q_{\rm adj}=-1$. The charges obey the superpotential constraints $r_- + r_+ = 2$, $q_- + q_+ = 1/2$, $\wat r + \ti r + R_{\rm adj} =2$ and $\wat q + \ti q = 1$. The FI parameters are both taken negative $\zeta_1, \zeta_2 < 0$. We also assume $N_2 > N_1$. This setup corresponds to  the 1d $\cN=4$ theory obtained by moving the D1-branes to right NS5-brane, with $k_1 = N_1$ D1-branes ending on an extra NS5'-brane and $k_2 = N_2 - N_1$ D1-branes ending on a second extra NS5'-brane (see figure \ref{QMquiver3}). The representation associated to this quiver is $\scR = \scA_{k_1} \otimes \scA_{k_2}$.

\medskip
We introduce the complex parameters $\hat\sigma_j = \sigma_j + i q_+ + i \frac{r_+}{2} z$ and $\hat m_a = m_a -i q_-  - i\frac{r_-}{2} z$. 
in order to avoid higher order poles in the computation we keep $R_{\rm adj}$ generic and only set it to $0$ at the end.
The partition function or index is given by
\begin{align*}
\scI \ =  \    & \oint \prod_{I =1}^{N_1} du_I \prod_{J =1}^{N_2} dv_J  \ \frac{1}{N_1! N_2!}  \lp \frac{\pi}{\sin(\pi z)} \rp^{N_1+N_2}\  \prod_{I \neq J}^{N_1} \frac{\sin[-\pi(u_I -u_J) ]}{\sin[\pi(u_I - u_J- z ) ]}
\prod_{I \neq J}^{N_2} \frac{\sin[-\pi(v_I -v_J) ]}{\sin[\pi(v_I - v_J- z ) ]}  \\
&    \prod_{I , J}^{N_1} \frac{\sin[-\pi(u_I -u_J + \frac{R_{\rm adj}}{2}z - z)  ]}{\sin[\pi(u_I - u_J + \frac{R_{\rm adj}}{2}z ) ]} 
\prod_{I =1}^{N_1} \lp \prod_{j=1}^N \frac{\sin[-\pi(v_I -i\hat\sigma_j - z ) ]}{\sin[\pi(v_I - i\hat\sigma_j ) ]}  \ 
\prod_{a=1}^M \frac{\sin[-\pi(-v_I + i \hat m_a -z) ]}{\sin[\pi(-v_I + i \hat m_a ) ]}  \rp  \\
&   \prod_{I}^{N_1} \prod_{J}^{N_2} \frac{\sin[-\pi(u_I -v_J + \wat q + \frac{\wat r }{2}z - z)  ]}{\sin[\pi(u_I - v_J + \wat q + \frac{\wat r }{2}z ) ]} \frac{\sin[-\pi(v_J -u_I + \ti q + \frac{\ti r }{2}z - z)  ]}{\sin[\pi(v_J -u_I + \ti q + \frac{\ti r }{2}z) ]}    \, .
\end{align*}
The integration contour picks the residues at the poles selected by the JK-recipe as explained in the main text. 
The recipe allows to take poles from the the fundamental chiral mulitplet factors, as well as from "half" of the bifundamental and adjoint chiral multiplet factors. However one can realize, for instance, that picking a pole $u^{\ast}_I = u^{\ast}_J - \frac{R_{\rm adj}}{2}z $ from the adjoint factor and a pole $u^{\ast}_J = v^{\ast}_K - \wat q - \frac{\wat r }{2}z$ from the bifundamental factor, leads to $u^{\ast}_I = v^{\ast}_K + \ti q + \frac{\ti r }{2}z - z -1$, where we have used the superpotential constraints, and in this case $u^{\ast}_I$ has an extra zero from the bifundamental factor, canceling the pole from the adjoint factor. This kind of reasoning leads to the conclusion that we cannot take a pole from the adjoint factor, as it yields a vanishing contribution.

\smallskip

A careful analysis along these lines leads to the following sets of poles $(u^{\ast}, v^{\ast})$ contributing to $\scI$:
\begin{itemize}
\item With $k_1 = N_1$ and $k_2 = N_2 - N_1$,  a pole $\{u^{\ast}_I, v^{\ast}_J\}$ is characterized by a choice of  decomposition
\begin{align*}
k_1 = \sum_{j=1}^N k^{(1)}_j \, , \quad k_2 = \sum_{j=1}^N k^{(2)}_j \, ,  \quad {\rm with} \quad  k^{(\alpha)}_j \in \{0,1\} \, .
\end{align*}
\item The explicit single-integral poles are given by
\begin{align*}
& u^{\ast}_I \ \rightarrow  \ i \hat\sigma_j -  s_j \lp \wat q + \frac{\wat r}{2} \rp \, {\rm for} \quad  0 \le s_j \le k^{(1)}_j  \, , \\
& v^{\ast}_J \ \rightarrow  \  i \hat\sigma_j - s_j \lp \wat q + \ti q + \frac{\wat r + \ti r }{2} \rp \, {\rm for} \quad  0 \le s_j \le k^{(1)}_j  + k^{(2)}_j  \, ,
\end{align*}
with $\rightarrow$ denoting a mapping between the relevant indices. The range of the $s_j$ is such that it correctly gives the $N_1$ $u^{\ast}_I $-poles and the $N_2$ $v^{\ast}_I $-poles.

\item Each residue contribution to the index appears with the degeneracy $k_1 ! (k_1 + k_2) ! = N_1! N_2!$.
\end{itemize}

The evaluation of the index gives
\begin{align}
\scI &= \sum_{
\begin{array}{c}
k^{(1)}_j \in \{ 0,1\} \\
\sum_j^N k^{(1)}_j = k_1
\end{array}  } \ 
\sum_{
\begin{array}{c}
k^{(2)}_j \in \{ 0,1\} \\
\sum_j^N k^{(2)}_j = k_2
\end{array}  }
\scI_{(k^{(1)}, k^{(2)})} \ ,
\label{Index2nodes}
\end{align}
with
\begin{align}
\scI_{(k^{(1)}, k^{(2)})} &=  \scF(\sigma_i, z)
\ \prod_{a=1}^M  \prod_{j=1}^N \prod_{s_j=0}^{k^{(1)}_j + k^{(2)}_j -1}  \frac{\ch \left[ \sigma_j - m_a + i \lp 1- \frac{R_{\rm adj}}{2} \rp  s_j  z \right] }{\ch \left[ \sigma_j -  m_a + i \lp 1 - \frac{R_{\rm adj}}{2} \rp  s_j  z  + iz \right] }  \, ,
\end{align}
where we have used the superpotential constraints. $\scF(\sigma_i, z) $ is a complicated expression which will not be relevant for our analysis of mirror symmetry. It is expressed as a product of the form
\begin{align}
\scF(\sigma_i, z) &= \prod_{\alpha , \kappa } \prod_{i \neq j}^N  \frac{\sin[-\pi (i\sigma_i - i\sigma_j  +  \alpha z \pm z + \kappa/2)]}{\sin[\pi  (i\sigma_i - i \sigma_j + \alpha  z + \kappa/2)]} \, ,
\end{align}
where $\alpha, \kappa$ take real values. As in \eqref{IndexEpsilon}, there are terms which require a regularization by a small mass deformation $\epsilon$ and which evaluates to $\pm 1$ as $z\rightarrow 1$.

\smallskip

Plugging the adjoint R-charge $R_{\rm adj} = 0$, the result simplifies to
\begin{align}
\scI_{(k^{(1)}, k^{(2)})} &= \scF(\sigma_i, z)
\ \prod_{a=1}^M  \prod_{j=1}^N  \frac{\ch [ \sigma_j - m_a] }{\ch [ \sigma_j -  m_a + i \big( k^{(1)}_j + k^{(2)}_j  \big)  z ] }  \, .
\label{Index2nodesWeights}
\end{align}
This result reproduces correctly \eqref{IndexQMfinal_r} as a sum over the weights of the representation  $\scA_{k_1} \otimes \scA_{k_2}$.

\addtocontents{toc}{\protect\setcounter{tocdepth}{1}}

\clearpage

\bibliography{WilsonVortexbib}

\providecommand{\href}[2]{#2}\begingroup\raggedright\begin{thebibliography}{10}

\bibitem{Wilson:1974sk}
K.~G. Wilson, {\it {Confinement of Quarks}},  {\em Phys.Rev.} {\bf D10} (1974)
  2445--2459.

\bibitem{Intriligator:1996ex}
K.~A. Intriligator and N.~Seiberg, {\it {Mirror symmetry in three-dimensional
  gauge theories}},  {\em Phys.Lett.} {\bf B387} (1996) 513--519,
  [\href{http://xxx.lanl.gov/abs/hep-th/9607207}{{\tt hep-th/9607207}}].

\bibitem{Kapustin:2012iw}
A.~Kapustin, B.~Willett, and I.~Yaakov, {\it {Exact results for supersymmetric
  abelian vortex loops in 2+1 dimensions}},
  \href{http://xxx.lanl.gov/abs/1211.2861}{{\tt arXiv:1211.2861}}.

\bibitem{Drukker:2012sr}
N.~Drukker, T.~Okuda, and F.~Passerini, {\it {Exact results for vortex loop
  operators in 3d supersymmetric theories}},
  \href{http://xxx.lanl.gov/abs/1211.3409}{{\tt arXiv:1211.3409}}.

\bibitem{ASG}
B.~Assel, J.~Gomis, and R.~C. Santamaria. unpublished.

\bibitem{Drukker:2008jm}
N.~Drukker, J.~Gomis, and D.~Young, {\it {Vortex Loop Operators, M2-branes and
  Holography}},  {\em JHEP} {\bf 0903} (2009) 004,
  [\href{http://xxx.lanl.gov/abs/0810.4344}{{\tt arXiv:0810.4344}}].

\bibitem{Moore:1989yh}
G.~W. Moore and N.~Seiberg, {\it {Taming the Conformal Zoo}},  {\em Phys.Lett.}
  {\bf B220} (1989) 422.

\bibitem{Gomis:2014eya}
J.~Gomis and B.~Le~Floch, {\it {M2-brane surface operators and gauge theory
  dualities in Toda}},  \href{http://xxx.lanl.gov/abs/1407.1852}{{\tt
  arXiv:1407.1852}}.

\bibitem{Hanany:1996ie}
A.~Hanany and E.~Witten, {\it {Type IIB superstrings, BPS monopoles, and
  three-dimensional gauge dynamics}},  {\em Nucl.Phys.} {\bf B492} (1997)
  152--190, [\href{http://xxx.lanl.gov/abs/hep-th/9611230}{{\tt
  hep-th/9611230}}].

\bibitem{Gomis:2006sb}
J.~Gomis and F.~Passerini, {\it {Holographic Wilson Loops}},  {\em JHEP} {\bf
  0608} (2006) 074, [\href{http://xxx.lanl.gov/abs/hep-th/0604007}{{\tt
  hep-th/0604007}}].

\bibitem{Gomis:2006im}
J.~Gomis and F.~Passerini, {\it {Wilson Loops as D3-Branes}},  {\em JHEP} {\bf
  0701} (2007) 097, [\href{http://xxx.lanl.gov/abs/hep-th/0612022}{{\tt
  hep-th/0612022}}].

\bibitem{Maldacena:1998im}
J.~M. Maldacena, {\it {Wilson loops in large N field theories}},  {\em
  Phys.Rev.Lett.} {\bf 80} (1998) 4859--4862,
  [\href{http://xxx.lanl.gov/abs/hep-th/9803002}{{\tt hep-th/9803002}}].

\bibitem{Rey:1998ik}
S.-J. Rey and J.-T. Yee, {\it {Macroscopic strings as heavy quarks in large N
  gauge theory and anti-de Sitter supergravity}},  {\em Eur.Phys.J.} {\bf C22}
  (2001) 379--394, [\href{http://xxx.lanl.gov/abs/hep-th/9803001}{{\tt
  hep-th/9803001}}].

\bibitem{Kapustin:2010xq}
A.~Kapustin, B.~Willett, and I.~Yaakov, {\it {Nonperturbative Tests of
  Three-Dimensional Dualities}},  {\em JHEP} {\bf 1010} (2010) 013,
  [\href{http://xxx.lanl.gov/abs/1003.5694}{{\tt arXiv:1003.5694}}].

\bibitem{Benvenuti:2011ga}
S.~Benvenuti and S.~Pasquetti, {\it {3D-partition functions on the sphere:
  exact evaluation and mirror symmetry}},
  \href{http://xxx.lanl.gov/abs/1105.2551}{{\tt arXiv:1105.2551}}.

\bibitem{Assel:2014awa}
B.~Assel, {\it {Hanany-Witten effect and SL(2, $\mathbb{Z}$) dualities in
  matrix models}},  {\em JHEP} {\bf 1410} (2014) 117,
  [\href{http://xxx.lanl.gov/abs/1406.5194}{{\tt arXiv:1406.5194}}].

\bibitem{Kapustin:2009kz}
A.~Kapustin, B.~Willett, and I.~Yaakov, {\it {Exact Results for Wilson Loops in
  Superconformal Chern-Simons Theories with Matter}},  {\em JHEP} {\bf 1003}
  (2010) 089, [\href{http://xxx.lanl.gov/abs/0909.4559}{{\tt
  arXiv:0909.4559}}].

\bibitem{Hori:2014tda}
K.~Hori, H.~Kim, and P.~Yi, {\it {Witten Index and Wall Crossing}},  {\em JHEP}
  {\bf 1501} (2015) 124, [\href{http://xxx.lanl.gov/abs/1407.2567}{{\tt
  arXiv:1407.2567}}].

\bibitem{Cordova:2014oxa}
C.~Cordova and S.-H. Shao, {\it {An Index Formula for Supersymmetric Quantum
  Mechanics}},  \href{http://xxx.lanl.gov/abs/1406.7853}{{\tt
  arXiv:1406.7853}}.

\bibitem{Seiberg:1994pq}
N.~Seiberg, {\it {Electric - magnetic duality in supersymmetric nonAbelian
  gauge theories}},  {\em Nucl.Phys.} {\bf B435} (1995) 129--146,
  [\href{http://xxx.lanl.gov/abs/hep-th/9411149}{{\tt hep-th/9411149}}].

\bibitem{Gadde:2013dda}
A.~Gadde and S.~Gukov, {\it {2d Index and Surface operators}},  {\em JHEP} {\bf
  1403} (2014) 080, [\href{http://xxx.lanl.gov/abs/1305.0266}{{\tt
  arXiv:1305.0266}}].

\bibitem{Buchbinder:2007ar}
E.~I. Buchbinder, J.~Gomis, and F.~Passerini, {\it {Holographic gauge theories
  in background fields and surface operators}},  {\em JHEP} {\bf 0712} (2007)
  101, [\href{http://xxx.lanl.gov/abs/0710.5170}{{\tt arXiv:0710.5170}}].

\bibitem{Zarembo:2002an}
K.~Zarembo, {\it {Supersymmetric Wilson loops}},  {\em Nucl.Phys.} {\bf B643}
  (2002) 157--171, [\href{http://xxx.lanl.gov/abs/hep-th/0205160}{{\tt
  hep-th/0205160}}].

\bibitem{Rozansky:1996bq}
L.~Rozansky and E.~Witten, {\it {HyperKahler geometry and invariants of three
  manifolds}},  {\em Selecta Math.} {\bf 3} (1997) 401--458,
  [\href{http://xxx.lanl.gov/abs/hep-th/9612216}{{\tt hep-th/9612216}}].

\bibitem{Kapustin:2006pk}
A.~Kapustin and E.~Witten, {\it {Electric-Magnetic Duality And The Geometric
  Langlands Program}},  {\em Commun.Num.Theor.Phys.} {\bf 1} (2007) 1--236,
  [\href{http://xxx.lanl.gov/abs/hep-th/0604151}{{\tt hep-th/0604151}}].

\bibitem{Gukov:2006jk}
S.~Gukov and E.~Witten, {\it {Gauge Theory, Ramification, And The Geometric
  Langlands Program}},  \href{http://xxx.lanl.gov/abs/hep-th/0612073}{{\tt
  hep-th/0612073}}.

\bibitem{Constable:2002xt}
N.~R. Constable, J.~Erdmenger, Z.~Guralnik, and I.~Kirsch, {\it {Intersecting
  D-3 branes and holography}},  {\em Phys.Rev.} {\bf D68} (2003) 106007,
  [\href{http://xxx.lanl.gov/abs/hep-th/0211222}{{\tt hep-th/0211222}}].

\bibitem{Gaiotto:2009fs}
D.~Gaiotto, {\it {Surface Operators in N = 2 4d Gauge Theories}},  {\em JHEP}
  {\bf 1211} (2012) 090, [\href{http://xxx.lanl.gov/abs/0911.1316}{{\tt
  arXiv:0911.1316}}].

\bibitem{Gaiotto:2008ak}
D.~Gaiotto and E.~Witten, {\it {S-Duality of Boundary Conditions In N=4 Super
  Yang-Mills Theory}},  \href{http://xxx.lanl.gov/abs/0807.3720}{{\tt
  arXiv:0807.3720}}.

\bibitem{Assel:2011xz}
B.~Assel, C.~Bachas, J.~Estes, and J.~Gomis, {\it {Holographic Duals of D=3 N=4
  Superconformal Field Theories}},  {\em JHEP} {\bf 1108} (2011) 087,
  [\href{http://xxx.lanl.gov/abs/1106.4253}{{\tt arXiv:1106.4253}}].

\bibitem{Assel:2012cj}
B.~Assel, C.~Bachas, J.~Estes, and J.~Gomis, {\it {IIB Duals of D=3 N=4
  Circular Quivers}},  {\em JHEP} {\bf 1212} (2012) 044,
  [\href{http://xxx.lanl.gov/abs/1210.2590}{{\tt arXiv:1210.2590}}].

\bibitem{deBoer:1996mp}
J.~de~Boer, K.~Hori, H.~Ooguri, and Y.~Oz, {\it {Mirror symmetry in
  three-dimensional gauge theories, quivers and D-branes}},  {\em Nucl.Phys.}
  {\bf B493} (1997) 101--147,
  [\href{http://xxx.lanl.gov/abs/hep-th/9611063}{{\tt hep-th/9611063}}].

\bibitem{Drukker:2005kx}
N.~Drukker and B.~Fiol, {\it {All-genus calculation of Wilson loops using
  D-branes}},  {\em JHEP} {\bf 0502} (2005) 010,
  [\href{http://xxx.lanl.gov/abs/hep-th/0501109}{{\tt hep-th/0501109}}].

\bibitem{Yamaguchi:2006tq}
S.~Yamaguchi, {\it {Wilson loops of anti-symmetric representation and
  D5-branes}},  {\em JHEP} {\bf 0605} (2006) 037,
  [\href{http://xxx.lanl.gov/abs/hep-th/0603208}{{\tt hep-th/0603208}}].

\bibitem{Danielsson:1997wq}
U.~Danielsson, G.~Ferretti, and I.~R. Klebanov, {\it {Creation of fundamental
  strings by crossing D-branes}},  {\em Phys.Rev.Lett.} {\bf 79} (1997)
  1984--1987, [\href{http://xxx.lanl.gov/abs/hep-th/9705084}{{\tt
  hep-th/9705084}}].

\bibitem{Bachas:1997kn}
C.~P. Bachas, M.~B. Green, and A.~Schwimmer, {\it {(8,0) quantum mechanics and
  symmetry enhancement in type I' superstrings}},  {\em JHEP} {\bf 9801} (1998)
  006, [\href{http://xxx.lanl.gov/abs/hep-th/9712086}{{\tt hep-th/9712086}}].

\bibitem{Callan:1997kz}
C.~G. Callan and J.~M. Maldacena, {\it {Brane death and dynamics from the
  Born-Infeld action}},  {\em Nucl.Phys.} {\bf B513} (1998) 198--212,
  [\href{http://xxx.lanl.gov/abs/hep-th/9708147}{{\tt hep-th/9708147}}].

\bibitem{Kim:2015fba}
H.~Kim, S.-J. Lee, and P.~Yi, {\it {Mutation, Witten Index, and Quiver
  Invariant}},  \href{http://xxx.lanl.gov/abs/1504.0006}{{\tt
  arXiv:1504.0006}}.

\bibitem{Samsonov:2014pya}
I.~Samsonov and D.~Sorokin, {\it {Superfield theories on $S^3$ and their
  localization}},  {\em JHEP} {\bf 1404} (2014) 102,
  [\href{http://xxx.lanl.gov/abs/1401.7952}{{\tt arXiv:1401.7952}}].

\bibitem{Festuccia:2011ws}
G.~Festuccia and N.~Seiberg, {\it {Rigid Supersymmetric Theories in Curved
  Superspace}},  {\em JHEP} {\bf 1106} (2011) 114,
  [\href{http://xxx.lanl.gov/abs/1105.0689}{{\tt arXiv:1105.0689}}].

\bibitem{Gomis:2014woa}
J.~Gomis and N.~Ishtiaque, {\it {K?hler potential and ambiguities in 4d $
  \mathcal{N} $ = 2 SCFTs}},  {\em JHEP} {\bf 1504} (2015) 169,
  [\href{http://xxx.lanl.gov/abs/1409.5325}{{\tt arXiv:1409.5325}}].

\bibitem{Hama:2010av}
N.~Hama, K.~Hosomichi, and S.~Lee, {\it {Notes on SUSY Gauge Theories on
  Three-Sphere}},  {\em JHEP} {\bf 1103} (2011) 127,
  [\href{http://xxx.lanl.gov/abs/1012.3512}{{\tt arXiv:1012.3512}}].

\bibitem{Jafferis:2010un}
D.~L. Jafferis, {\it {The Exact Superconformal R-Symmetry Extremizes Z}},  {\em
  JHEP} {\bf 1205} (2012) 159, [\href{http://xxx.lanl.gov/abs/1012.3210}{{\tt
  arXiv:1012.3210}}].

\bibitem{Hama:2011ea}
N.~Hama, K.~Hosomichi, and S.~Lee, {\it {SUSY Gauge Theories on Squashed
  Three-Spheres}},  {\em JHEP} {\bf 1105} (2011) 014,
  [\href{http://xxx.lanl.gov/abs/1102.4716}{{\tt arXiv:1102.4716}}].

\bibitem{Witten:1982df}
E.~Witten, {\it {Constraints on Supersymmetry Breaking}},  {\em Nucl.Phys.}
  {\bf B202} (1982) 253.

\bibitem{1993alg.geom..7001J}
L.~C. {Jeffrey} and F.~C. {Kirwan}, {\it {Localization for nonabelian group
  actions}},  in {\em Topology 34 no. 2, (1995) 291327}.
\newblock arXiv:alg-geom/9307001.

\bibitem{Nishioka:2011dq}
T.~Nishioka, Y.~Tachikawa, and M.~Yamazaki, {\it {3d Partition Function as
  Overlap of Wavefunctions}},  {\em JHEP} {\bf 1108} (2011) 003,
  [\href{http://xxx.lanl.gov/abs/1105.4390}{{\tt arXiv:1105.4390}}].

\bibitem{Closset:2012vg}
C.~Closset, T.~T. Dumitrescu, G.~Festuccia, Z.~Komargodski, and N.~Seiberg,
  {\it {Contact Terms, Unitarity, and F-Maximization in Three-Dimensional
  Superconformal Theories}},  {\em JHEP} {\bf 1210} (2012) 053,
  [\href{http://xxx.lanl.gov/abs/1205.4142}{{\tt arXiv:1205.4142}}].

\bibitem{Closset:2012vp}
C.~Closset, T.~T. Dumitrescu, G.~Festuccia, Z.~Komargodski, and N.~Seiberg,
  {\it {Comments on Chern-Simons Contact Terms in Three Dimensions}},  {\em
  JHEP} {\bf 1209} (2012) 091, [\href{http://xxx.lanl.gov/abs/1206.5218}{{\tt
  arXiv:1206.5218}}].

\bibitem{Bullimore:2014nla}
M.~Bullimore, M.~Fluder, L.~Hollands, and P.~Richmond, {\it {The superconformal
  index and an elliptic algebra of surface defects}},  {\em JHEP} {\bf 1410}
  (2014) 62, [\href{http://xxx.lanl.gov/abs/1401.3379}{{\tt arXiv:1401.3379}}].

\end{thebibliography}\endgroup
\end{document}